\documentclass[aps,prb,twocolumn,superscriptaddress,nofootinbib,showpacs,reprint]{revtex4-1}
\usepackage{amsfonts}
\usepackage{amsmath}
\usepackage{amsthm}
\usepackage{braket}
\usepackage{graphicx}
\usepackage{hyperref}
\usepackage[dvipsnames]{xcolor}
\usepackage{amssymb}
\usepackage{dsfont}
\usepackage{verbatim}
\usepackage{amsmath,amscd}
\usepackage[all]{xy}
\usepackage{multirow}
\newcommand{\eX}{\dimexpr\fontcharht\font`X\relax}

\definecolor{darkblue}{RGB}{0,0,127} 
\definecolor{darkgreen}{RGB}{0,180,0}
\hypersetup{
	colorlinks,
	linkcolor=darkblue,
	citecolor=darkgreen,
	filecolor=red,
	urlcolor=blue,
	pdftitle={Anomalies and entanglement renormalization},
	pdfauthor={Jacob C. Bridgeman, Dominic J. Williamson}
}

\graphicspath{{./Figures/}}

\usepackage{tikz,pgfplots}\pgfplotsset{compat=newest}
\usetikzlibrary{external,calc}
\newlength\figureheight
\newlength\figurewidth

\newcommand{\includeTikz}[2]{
\includegraphics{#1}
}

\newcommand{\fref}[1]{Fig.~\ref{Fig:#1}}

\newcommand{\mathscr}[1]{\ensuremath{\mathcal{#1}}}

\newcommand{\rg}[1]{\ensuremath{\mathcal{L}_{(#1)}}}

\newcommand{\G}{\ensuremath{\mathcal{G}}}
\newcommand{\Uone}{\ensuremath{\mathsf{U}(1)}}
\newcommand{\coho}[1]{\ensuremath{\mathcal{H}^{#1}(\G,\Uone)}}
\newcommand{\cohom}[2]{\ensuremath{\mathcal{H}^{#1}(#2,\Uone)}}

\newcommand{\disent}[1]{\ensuremath{\mathcal{D}_{#1}}}
\newcommand{\ghz}[1]{\ensuremath{\ket{\text{GHZ}_{#1}}}}

\newcommand{\trans}[1]{\ensuremath{\tau_{#1}}}
\newcommand{\dehn}[1]{\ensuremath{{T}_{#1}}}
\newcommand{\centralizer}[1]{\ensuremath{\mathcal{Z}_{#1}}}
\newcommand{\sso}[2]{\ensuremath{\mathcal{S}_{#1}({#2})}}
\newcommand{\order}[1]{\ensuremath{n_{#1}}}
\newcommand{\crossing}[2]{\ensuremath{{M^{(#1)}_{#2}}}}
\newcommand{\conjclass}[1]{\ensuremath{\mathcal{C}_{#1}}}

\newcommand{\N}{\ensuremath{\mathbb{Z}^+}}
\newcommand{\Z}{\ensuremath{\mathbb{Z}}}
\newcommand{\zt}{\ensuremath{\mathbb{Z}_2}}

\newcommand{\zN}{\ensuremath{\mathbb{Z}_N}}
\newcommand{\proj}[1]{\ensuremath{P}_{#1}}

\newcommand{\rx}[1]{\textcolor{red}{X_{#1}}}

\newcommand{\rz}[1]{\textcolor{red}{Z_{#1}}}
\newcommand{\bx}[1]{\textcolor{blue}{\tilde{X}_{#1}}}

\newcommand{\bz}[1]{\textcolor{blue}{\tilde{Z}_{#1}}}

\newcommand{\rzd}[1]{\textcolor{red}{Z^\dagger_{#1}}}

\newcommand{\bzd}[1]{\textcolor{blue}{\tilde Z^\dagger_{#1}}}

\definecolor{tensorblue}{rgb}{0.8,0.8,1}
\definecolor{tensorred}{rgb}{1,0.5,0.5}
\definecolor{tensorpurp}{rgb}{1,0.5,1}

\tikzset{nonesty/.style={fill=none,draw=none}}
\tikzset{ten/.style={fill=tensorblue}}
\tikzset{tenred/.style={fill=tensorred}}
\tikzset{tengreen/.style={fill=green!50!black!50}}
\tikzset{tenpurp/.style={fill=tensorpurp}}
\tikzset{tengrey/.style={fill=black!20}}
\tikzset{tenorange/.style={fill=orange!30}}
\tikzset{u/.style={fill=blue!20,draw=black}}
\tikzset{w/.style={fill=green!50!black!50,draw=black}}

\tikzset{external/system call={pdflatex \tikzexternalcheckshellescape -halt-on-error -interaction=batchmode -jobname "\image" "\texsource" ; rm "\image".log ; rm "\image".dpth ; rm "\image"Notes.bib ; rm "\image".md5 }}

\begin{document}

  \title{Anomalies and entanglement renormalization}
  \author{Jacob C. \surname{Bridgeman}}
  \affiliation{Centre for Engineered Quantum Systems, School of Physics, The University of Sydney, Sydney, Australia}
  \author{Dominic J. \surname{Williamson}}
  \affiliation{Vienna Center for Quantum Technology, University of Vienna, Boltzmanngasse
  5, 1090 Vienna, Austria}

\begin{abstract}
    We study 't~Hooft anomalies of discrete groups in the framework of (1+1)-dimensional multiscale entanglement renormalization ansatz states on the lattice. Using matrix product operators, general topological restrictions on conformal data are derived. An ansatz class allowing for optimization of MERA with an anomalous symmetry is introduced. 
    We utilize this class to numerically study a family of Hamiltonians with a symmetric critical line. 
    Conformal data is obtained for all irreducible projective representations of each anomalous symmetry twist, corresponding to definite topological sectors. 
    It is numerically demonstrated that this line is a protected gapless phase. Finally, we implement a duality transformation between a pair of critical lines using our subclass of MERA.
\end{abstract}

  \maketitle

Quantum many-body models of strongly interacting spins display surprisingly complex emergent physics. 
Understanding general classes of collective behaviors corresponds to understanding which phases of matter can be realized through local interactions. 
The universal behavior of phases, and their transitions, is determined by the fixed points under renormalization group (RG) flows\cite{Kadanoff:1966wm,RevModPhys.47.773}. 

Symmetries play a fundamental role in the understanding of phases, due to constraints they impose on RG. Indeed, the conventional classification of phases describes how a symmetry can be broken\cite{landau1965course}.  
Distinct quantum phases emerge even without a broken symmetry\cite{wegner1971duality,kosterlitz1973ordering,PhysRevB.40.7387,einarsson,doi:10.1142/S0217979290000139}. In the absence of intrinsic topological order, these phases are known as \emph{symmetry protected topological} (SPT) phases\cite{PhysRevLett.50.1153,gu2009tensor,pollmann2010entanglement,chen2013symmetry}. Despite having no topological order and no local order parameter, 
SPT phases are resources for quantum computation\cite{PhysRevLett.105.040501,1367-2630-15-2-025020,else2012symmetry,1367-2630-14-11-113016,williamson2015symmetry}.

On the lattice, symmetries are usually assumed to act independently on each site. More exotic symmetries, which cannot be made on-site, have recently been studied in chains of anyons\cite{goldenchain,PhysRevB.86.155111,PhysRevB.87.235120, PhysRevB.82.115126,PhysRevB.82.125118} 
and at the boundary of SPT phases\cite{czxmodel,PhysRevB.89.195122,PhysRevB.91.195134,williamson2014matrix,PhysRevD.88.045013,PhysRevLett.112.231602,kapustin2014symmetry,PhysRevB.90.235137,kapustin2014anomalies,wang2013lattice}. In fact, a classification of SPTs can be obtained by considering possible boundary actions of the symmetry. Equivalence classes of such symmetries are labeled by the \emph{'t~Hooft anomalies}\cite{hooft1980naturalness} of a discrete group. Such anomaly labels are preserved by symmetric RG transformations, so restrict the possible fixed points\cite{PhysRevLett.118.021601}.
  
\emph{Tensor network} methods\cite{VerstraeteMurgCirac2008,Orus2014,TNReview} allow anomalous symmetries to be realized directly on the lattice. 
In $(1+1)$ dimensions, \emph{matrix product operators} (MPOs) capture all 't~Hooft anomalies of discrete groups\cite{czxmodel,williamson2014matrix,PhysRevB.89.195122,PhysRevB.91.195134}. Within the framework of tensor networks, phases are classified at the level of states. For example, \emph{matrix product states} (MPS) have proven particularly successful for the study of gapped spin chains\cite{PhysRevLett.69.2863,PhysRevLett.75.3537,0295-5075-43-4-457,PhysRevLett.59.799,Fannes92,0295-5075-24-4-010,MPSrepresentations,1Done,SchuchGarciaCirac11,Cirac2017100}. Despite substantial complications arising for tensor networks in higher dimensions; significant progress has been made, particularly in the study of topological states\cite{doi:10.1143PTP.105.409,peps,Ginjectivity,PhysRevB.92.205307,PhysRevLett.111.236805,Buerschaper14,MPOpaper,Bultinck2017183,ribbons,williamson2016fermionic}.
     
Imposing on-site symmetries on tensor network representations of quantum states is well understood\cite{PhysRevLett.100.167202,Singh2010,1367-2630-12-2-025010}.
Far less effort has been made to study the effect of anomalous group actions on these states. 
Such group actions naturally arise as the effective edge symmetries of $(d+1)$D SPTs\cite{PhysRevLett.112.231602,kapustin2014symmetry,PhysRevB.90.235137}. 
In $(2+1)$D, the edge theory must either spontaneously break this symmetry or be gapless. Since all MPS break the symmetry\cite{czxmodel}, to study gapless, symmetric edge theories we turn to another class of tensor networks known as \emph{multiscale entanglement renormalization ansatz} (MERA)\cite{Vidal2007}. These networks draw on ideas from RG to represent the low energy states of gapless Hamiltonians\cite{Vidal2007,Evenbly2009,Pfeifer2009}.

In this work we define a variational subclass of MERA which can be used to simulate SPT edge physics in a manifestly symmetric way. This subclass allows us to investigate the interplay between RG and anomalies in the framework of tensor networks. We use tensor network methods to derive general consequences of an anomalous symmetry on the conformal field theory (CFT) data of an RG fixed point. 
For a family of Hamiltonians, corresponding to a line of fixed points, we numerically optimize within our variational class to find the lowest energy states and extract conformal data\cite{Ginsparg1988,DiFrancesco1997}. We observe the effects of the anomaly in these results. 
Furthermore, we demonstrate that as a consequence of the anomaly these Hamiltonians admit no relevant, symmetric perturbations. The Hamiltonians therefore support a gapless phase which is protected by an anomalous symmetry.

More generally, RG fixed points may transform non-trivially under an anomalous group action. Our variational class accommodates this possibility, and hence permits the study of gapless models which are not symmetric. 
We utilize this in a numerical simulation of two critical lines that are related by a duality transformation, which we implement at the level of a single tensor.

This paper is organized as follows: In Section~\ref{section:symmetries}, we introduce background material on anomalies, symmetries and tensor networks. In particular, we introduce the 't~Hooft anomaly of a discrete symmetry. We then briefly review the MERA and what it means for it to be symmetric under an on-site group action. The difficulties in enforcing anomalous MPO symmetries locally are then discussed.  
In Section~\ref{section:mpos}, we derive general consequences of an anomalous symmetry on a MERA, which are later utilized in the numerical simulations. We study anomalous symmetry twists and the projective representations under which they transform. From these ingredients, projectors onto definite topological sectors are constructed. Consequences for fields within a sector are discussed. 
In Section~\ref{section:class}, we define a variational subclass of MERA which is later used for manifestly symmetric simulations.  We present a disentangling unitary capable of decoupling a local piece of an anomalous $\zN^3$ group action. This allows the unconstrained variational parameters of any symmetric MERA scheme to be isolated, and therefore optimized over.
In Section~\ref{Numerics}, we bring together tools developed in the preceding sections to simulate a family of Hamiltonians with three critical lines. One of these lines possesses an anomalous symmetry, whilst the other two are dual under the anomalous group action. We present conformal data for these critical lines obtained from a numerically optimized MERA, including two nontrivial topological sectors for the symmetric line. Additionally, we demonstrate that the symmetric line is in fact a protected gapless phase.
In Section~\ref{section:conclusion} we summarize the results and suggest several possible extensions of this work.

We have included several appendices for completeness. In Appendix~\ref{appendix:fullmeradata} we provide conformal data obtained from a symmetric MERA in all topological sectors for the symmetric line of our example model. Additionally, we present fusion rules for these topological sectors computed using a symmetric MERA. In Appendix~\ref{appendix:thirdcoho} we review the notion of third cohomology for an MPO representation of a finite group. In Appendix~\ref{appendix:generalansatz} we provide details of our ansatz for MPO symmetric MERA including example tensors for two MERA schemes. In Appendix~\ref{appendix:gczx} we describe a generalization of the CZX model~\cite{czxmodel} to arbitrary finite groups $\G$, such that the bulk symmetry acts as an MPO duality of $\G$-SPT phases on the boundary.

\section{Symmetries and anomalies in MERA}
\label{section:symmetries}

This section introduces the main tools and concepts utilized in the remainder of this manuscript. We begin by discussing 't~Hooft anomalies of group actions, including some historical context. Lattice realizations of these anomalies, and their influence on tensor network states, are our primary objects of study. Readers unfamiliar with this terminology may skip to Section~\ref{section:lattice} for the definition of anomaly used throughout this work. We then review the MERA, the tensor network designed for critical behavior, and define what it means for it to be symmetric under a unitary group action. We briefly explain how one enforces an on-site symmetry via a local constraint before moving on to discuss the difficulties in enforcing an anomalous symmetry in a similar fashion. 

Recently anomalies have played an important role in the classification and study of topological phases of matter\cite{PhysRevB.85.045104,PhysRevD.88.045013,PhysRevLett.114.031601}. Particularly relevant are 't~Hooft anomalies, which describe obstructions to gauging a global symmetry\cite{hooft1980naturalness}. SPT phases, and their higher symmetry generalizations\cite{kapustin2013higher,Gaiotto2015,thorngren2015higher}, can be classified by the possible 't~Hooft anomalies on their boundaries\cite{PhysRevLett.112.231602,kapustin2014symmetry,PhysRevB.90.235137,kapustin2014anomalies}. Conversely one can think of the possible 't~Hooft anomalies as being classified by what is known as anomaly inflow from one dimension higher\cite{PhysRevD.88.045013,PhysRevLett.112.231602,kapustin2014symmetry,kapustin2014anomalies,wang2013lattice}. 

A global symmetry with an 't~Hooft anomaly has an interesting interplay with the renormalization group (RG). For a connected Lie group symmetry, an 't~Hooft anomaly restricts the possible RG fixed points, even if the symmetry is spontaneously broken\cite{WESS197195,Weinberg:1996kr}. 
In the case of a broken discrete symmetry, this is no longer true. For a symmetry respecting RG flow, however, the 't~Hooft anomaly can not change and hence constrains the possible fixed points\cite{PhysRevLett.112.231602}.

Symmetry actions which can be realized independently on each site have trivial 't~Hooft anomaly because they can be gauged directly on the lattice\cite{levin2012braiding,Gaugingpaper,williamson2014matrix}. Conversely, this gauging procedure cannot be applied directly to symmetries which cannot be made on site. Therefore, we treat the 't~Hooft anomaly as an obstruction to making a symmetry action on-site\cite{PhysRevD.88.045013,wang2013lattice,PhysRevB.91.195134}.

For a discrete symmetry group $\G$ in $(1+1)$D, all 't~Hooft anomalies of bosonic unitary representations occur on the boundaries of $(2+1)$D SPT phases, in other words they arise from anomaly inflow. The anomalies can therefore be classified by $\coho{3}$, the same set of labels as the SPT phases\cite{czxmodel,kapustin2014symmetry,PhysRevB.90.235137}. In the next section, we describe how \emph{matrix product operators} can be utilized to represent these anomalous actions.

\subsection{Symmetries on the lattice}\label{section:lattice}

In this work, we consider unitary representations of finite groups on the lattice. We say a state $\ket{\psi}$ is \emph{symmetric} under a group $\G$ if $U_g\ket{\psi}=\ket{\psi}$ for all $g\in \G$, where $U_g$ is some unitary representation of the group. 

The symmetry is \emph{on-site} if the representation can be decomposed as $U_g=\otimes_{j=1}^N (u_g)_j$, where each $(u_g)_j$ is a (local) unitary representation. 

Although group actions are usually considered to be on-site, this is not the most general way a symmetry can be represented. A more general class of group actions can be represented by matrix product operators (MPOs).
Using the conventional tensor network notation\cite{VerstraeteMurgCirac2008,Orus2014,TNReview}, these are denoted
\begin{align}
U_g&=
\begin{array}{c}
\includeTikz{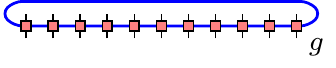}{
	\begin{tikzpicture}[scale=.5]
	\draw[thick,blue,rounded corners=2.5] (2,0)--(3,0)--(3.25,.25)--(3,.5)--(-3,.5)--(-3.25,.25)--(-3,0)--(2,0);
\foreach \x in {-5,-4,...,5}{
	\draw[shift={(\x*.55,0)}](0,-.25)--(0,.25);
	\filldraw[tenred,shift={(\x*.55,0)}] (-0.1,-0.1)--(0.1,-0.1)--(0.1,0.1)--(-0.1,0.1)--cycle;
	}
\node[below right] at(5*.55,0) {\textcolor{black}{\footnotesize$g$}};
	\end{tikzpicture}}
\end{array},
\end{align}
where $g$ next to the MPO indicates which group element it represents. 
We refer to the dimension of the horizontal indices as the \emph{bond dimension} of the MPO. The on-site case corresponds to bond dimension 1, whilst arbitrary bond dimension allows representation of any unitary. We consider the case of a constant bond dimension in the length of the MPO.

To form a representation, the MPOs must obey
\begin{align}
\begin{array}{c}
\includeTikz{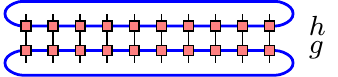}{
	\begin{tikzpicture}[scale=.5]
	\begin{scope}[yscale=-1]
	\draw[thick,blue,rounded corners=2.5] (1,0)--(2.5,0)--(2.75,.25)--(2.5,.5)--(-3,.5)--(-3.25,.25)--(-3,0)--(1,0);
	\end{scope}
	\foreach \x in {-5,-4,...,4}{
		\draw[shift={(\x*.55,0)}](0,-.25)--(0,.25);
		\filldraw[tenred,shift={(\x*.55,0)}] (-0.1,-0.1)--(0.1,-0.1)--(0.1,0.1)--(-0.1,0.1)--cycle;
	}
	\node[right] at(5*.55,0) {\textcolor{black}{\footnotesize$g$}};
	\begin{scope}[shift={(0,.5)}]
	\draw[thick,blue,rounded corners=2.5] (1,0)--(2.5,0)--(2.75,.25)--(2.5,.5)--(-3,.5)--(-3.25,.25)--(-3,0)--(1,0);
		\foreach \x in {-5,-4,...,4}{
			\draw[shift={(\x*.55,0)}](0,-.25)--(0,.25);
			\filldraw[tenred,shift={(\x*.55,0)}] (-0.1,-0.1)--(0.1,-0.1)--(0.1,0.1)--(-0.1,0.1)--cycle;
		}
		\node[right] at(5*.55,0) {\textcolor{black}{\footnotesize$h$}};
	\end{scope}
	\end{tikzpicture}}
\end{array}
&=
\begin{array}{c}
\includeTikz{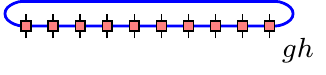}{
	\begin{tikzpicture}[scale=.5]
	\draw[thick,blue,rounded corners=2.5] (1,0)--(2.5,0)--(2.75,.25)--(2.5,.5)--(-3,.5)--(-3.25,.25)--(-3,0)--(1,0);
	\foreach \x in {-5,-4,...,4}{
		\draw[shift={(\x*.55,0)}](0,-.25)--(0,.25);
		\filldraw[tenred,shift={(\x*.55,0)}] (-0.1,-0.1)--(0.1,-0.1)--(0.1,0.1)--(-0.1,0.1)--cycle;
	}
	\node[below right] at(4*.55,0) {\textcolor{black}{\footnotesize$gh$}};
	\end{tikzpicture}}
\end{array},
\end{align}
for all lengths. In contrast to on-site representations, for bond dimensions larger than one this does not hold at the level of the local tensors. Rather there is a tensor $X(g,h)$, referred to as the \emph{reduction tensor}\cite{czxmodel,MPSrepresentations,Fannes92} (Appendix~\ref{appendix:thirdcoho}) such that  
\begin{align}
\begin{array}{c}
\includeTikz{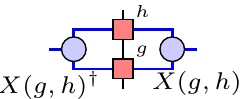}{
	\begin{tikzpicture}[scale=1]
	\clip  (-1.25,-.5)rectangle (1.25,.5);
	\node[above right] at(0,0) {\textcolor{white}{\tiny$gh$}};
	\node[below] at (.75,-.1) {\textcolor{white}{\scriptsize$X(g,h)^\dagger$}};
	\node[below] at (-.75,-.1) {\textcolor{white}{\scriptsize$X(g,h)$}};
	\draw[thick,blue] (0,-.2)--(-.5,-.2)--(-.5,.2)--(.5,.2)--(.5,-.2)--cycle (.5,0)--(.75,0) (-.5,0)--(-.75,0);
	\draw(0,-.4)--(0,.4);
	\filldraw[tenred,shift={(0,.2)}] (-0.1,-0.1)--(0.1,-0.1)--(0.1,0.1)--(-0.1,0.1)--cycle;
	\filldraw[tenred,shift={(0,-.2)}] (-0.1,-0.1)--(0.1,-0.1)--(0.1,0.1)--(-0.1,0.1)--cycle;
		\filldraw[ten](.5,0) circle (.25/2); 
		\filldraw[ten](-.5,0) circle (.25/2);
	\node[above right] at(0,-.2) {\textcolor{black}{\tiny$g$}};
	\node[above right] at(0,.2) {\textcolor{black}{\tiny$h$}};
	\node[below] at (.75,-.1) {\textcolor{black}{\scriptsize$X(g,h)$}};
	\node[below] at (-.75,-.1) {\textcolor{black}{\scriptsize$X(g,h)^\dagger$}};
	\end{tikzpicture}}
\end{array}
&=
\begin{array}{c}
\includeTikz{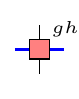}{
	\begin{tikzpicture}[scale=1]
	\clip  (-.4,-.5)rectangle (.4,.5);
	\node[above right] at(0,-.25) {\textcolor{white}{\tiny$g$}};
	\node[above right] at(0,.25) {\textcolor{white}{\tiny$h$}};
	\node[below] at (.75,-.1) {\textcolor{white}{\scriptsize$X(g,h)^\dagger$}};
	\node[below] at (-.75,-.1) {\textcolor{white}{\scriptsize$X(g,h)$}};
	\draw[thick,blue] (-.25,0)--(.25,0);
	\draw(0,-.25)--(0,.25);
	\filldraw[tenred] (-0.1,-0.1)--(0.1,-0.1)--(0.1,0.1)--(-0.1,0.1)--cycle;
	\node[above right] at(0,0) {\textcolor{black}{\tiny$gh$}};
	\end{tikzpicture}}
\end{array}.
\end{align}

The reduction procedure need not be associative. When reducing three tensors, there are two distinct orders of reduction which may differ by a phase $\phi$

\begin{align}
	\begin{array}{c}
		\includeTikz{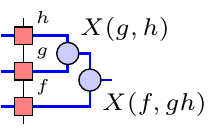}{
			\begin{tikzpicture}[scale=.9]
			\draw[thick,blue]  (-.25,0)--(.5,0)--(.5,.4)--(-.25,.4) (.5,.2)--(.75,.2)--(.75,-.4)--(-.25,-.4) (.75,-.1)--(1,-.1);
			\draw(0,-.6)--(0,.6);
			\draw[thick,blue](-.2,-.4)--(.2,-.4);
			\filldraw[tenred,shift={(0,.4)}] (-0.1,-0.1)--(0.1,-0.1)--(0.1,0.1)--(-0.1,0.1)--cycle;
			\filldraw[tenred,shift={(0,0)}] (-0.1,-0.1)--(0.1,-0.1)--(0.1,0.1)--(-0.1,0.1)--cycle;
			\filldraw[tenred,shift={(0,-.4)}] (-0.1,-0.1)--(0.1,-0.1)--(0.1,0.1)--(-0.1,0.1)--cycle;
			\filldraw[ten](.5,.2) circle (.25/2); 
			\filldraw[ten](.75,-.1) circle (.25/2);
			\node[above right] at(0,-.4) {\textcolor{black}{\tiny$f$}};
			\node[above right] at(0,0) {\textcolor{black}{\tiny$g$}};
			\node[above right] at(0,.4) {\textcolor{black}{\tiny$h$}};
			\node[above right] at (.5,.2) {\textcolor{black}{\scriptsize$X(g,h)$}};
			\node[below right] at (.75,-.1) {\textcolor{black}{\scriptsize$X(f,gh)$}};
			\end{tikzpicture}}
	\end{array}
	&=\phi(f,g,h)
	\begin{array}{c}
		\includeTikz{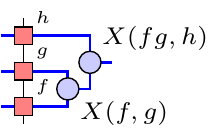}{
			\begin{tikzpicture}[scale=.9]
			\begin{scope}[yscale=-1]
			\draw[thick,blue]  (-.25,0)--(.5,0)--(.5,.4)--(-.25,.4) (.5,.2)--(.75,.2)--(.75,-.4)--(-.25,-.4) (.75,-.1)--(1,-.1);
			\draw(0,-.6)--(0,.6);
			\draw[thick,blue](-.2,-.4)--(.2,-.4);
			\filldraw[tenred,shift={(0,.4)}] (-0.1,-0.1)--(0.1,-0.1)--(0.1,0.1)--(-0.1,0.1)--cycle;
			\filldraw[tenred,shift={(0,0)}] (-0.1,-0.1)--(0.1,-0.1)--(0.1,0.1)--(-0.1,0.1)--cycle;
			\filldraw[tenred,shift={(0,-.4)}] (-0.1,-0.1)--(0.1,-0.1)--(0.1,0.1)--(-0.1,0.1)--cycle;
			\filldraw[ten](.5,.2) circle (.25/2); 
			\filldraw[ten](.75,-.1) circle (.25/2);
			\node[above right] at(0,-.4) {\textcolor{black}{\tiny$h$}};
			\node[above right] at(0,0) {\textcolor{black}{\tiny$g$}};
			\node[above right] at(0,.4) {\textcolor{black}{\tiny$f$}};
			\node[below right] at (.5,.2) {\textcolor{black}{\scriptsize$X(f,g)$}};
			\node[above right] at (.75,-.1) {\textcolor{black}{\scriptsize$X(fg,h)$}};
			\end{scope}
			\end{tikzpicture}}
	\end{array}.
\end{align}

As discussed in Appendix~\ref{appendix:thirdcoho}, $\phi$ is a 3-cocycle with $[\phi]\in\coho{3}$. 
Since on-site representations are locally associative they have a trivial cocycle. 
Hence a nontrivial $[\phi]$ indicates an obstruction to making the symmetry action on-site.  
We can therefore regard a nontrivial $[\phi]$ as a nontrivial 't~Hooft anomaly for $\G$ in $(1+1)$D. 
We remark that each class of 't~Hooft anomaly can be realized using MPOs in this way\cite{Buerschaper14,williamson2014matrix}.

\subsection{MERA and symmetry}
\begin{figure}[t]
	\includeTikz{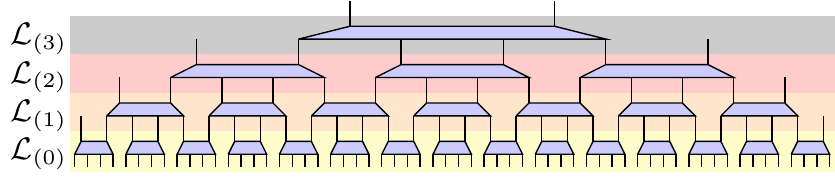}{
		\begin{tikzpicture}[scale=.52]
		\begin{scope}[shift={(0,-.25)}]
		\def\dx{.25}
		\def\dxx{.5}
		\def\dxxx{1}
		\def\dxxxx{2}
		\fill[yellow!20] (-.1,-.35)--(59*\dx+.1,-.35)--(59*\dx+.1,.5)--(-.1,.5)--cycle;
		\fill[orange!20,shift={(0,.8)}] (-.1,-.35)--(59*\dx+.1,-.35)--(59*\dx+.1,.4)--(-.1,.4)--cycle;
		\fill[red!20,shift={(0,1.55)}] (-.1,-.35)--(59*\dx+.1,-.35)--(59*\dx+.1,.4)--(-.1,.4)--cycle;
		\fill[black!20,shift={(0,2.3)}] (-.1,-.35)--(59*\dx+.1,-.35)--(59*\dx+.1,.4)--(-.1,.4)--cycle;
		\foreach \i in {0,...,59}{
			\draw(\i*\dx,0)--(\i*\dx,-.25);
		}
		\foreach \i in {0,4,...,56}{
			\filldraw[ten](\i*\dx,0)--({(\i+3)*\dx},0)--({(\i+2.5)*\dx},\dx)--({(\i+.5)*\dx},\dx)--(\i*\dx,0);
		}
		\foreach \i in {0,...,29}{
			\draw[shift={(.5*\dx,.75)}] (\i*\dxx,0)--(\i*\dxx,-.5);
		}
		\foreach \i in {0,4,...,26}{
			\filldraw[ten,shift={(2.5*\dx,.75)}](\i*\dxx,0)--({(\i+3)*\dxx},0)--({(\i+2.5)*\dxx},\dx)--({(\i+.5)*\dxx},\dx)--(\i*\dxx,0);
		}
		\foreach \i in {0,...,13}{
			\draw[shift={(.5*\dxx+2.5*\dx,1.5)}] (\i*\dxxx,0)--(\i*\dxxx,-.5);
		}
		\foreach \i in {0,4,...,10}{
			\filldraw[ten,shift={(.5*\dxx+6.5*\dx,1.5)}](\i*\dxxx,0)--({(\i+3)*\dxxx},0)--({(\i+2.5)*\dxxx},\dx)--({(\i+.5)*\dxxx},\dx)--(\i*\dxxx,0);
		}
		\foreach \i in {0,...,5}{
			\draw[shift={(.5*\dxxx+3.75*\dxx,2.25)}] (\i*\dxxxx,0)--(\i*\dxxxx,-.5);
		}
		\foreach \i in {0}{
			\filldraw[ten,shift={(.5*\dxxx+7.75*\dxx,2.25)}](\i*\dxxxx,0)--({(\i+3)*\dxxxx},0)--({(\i+2.5)*\dxxxx},\dx)--({(\i+.5)*\dxxxx},\dx)--(\i*\dxxxx,0);
		}
		\foreach \i in {0,...,1}{
			\draw[shift={(.5*\dxxxx+4.25*\dxxx+.5*\dx,3)}] (2*\i*\dxxxx,0)--(2*\i*\dxxxx,-.5);
		}
		\end{scope}
		\node at (-.7,-.25) {$\rg{0}$};
		\node at (-.7,.5) {$\rg{1}$};
		\node at (-.7,1.25) {$\rg{2}$};
		\node at (-.7,2) {$\rg{3}$};
		\end{tikzpicture}}
	\caption{The MERA represents a quantum state using layers of isometric tensors. Together, these tensors define a quantum circuit of logarithmic depth which can be used to prepare an entangled state from a product state. If the tensors are chosen appropriately, the network is thought to be able to accurately represent the ground state of gapless one-dimensional Hamiltonians. 
Throughout the paper we use a convention such that tensor network diagrams read bottom-to-top correspond to matrix multiplication read left-to-right. 
	}\label{Fig:fourtotwoMERA}
\end{figure}
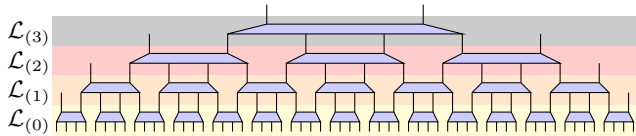
In its most general form\cite{Vidal2007,Evenbly2009}, the MERA can be thought of as a series of locality preserving isometric maps
\begin{align}
\rg{i}:\left(\mathbb{C}^{d_{i+1}}\right)^{\otimes N_{i+1}} \to \left(\mathbb{C}^{d_i}\right)^{\otimes N_i},
\end{align}
where $d_{i+1}^{N_{i+1}}\leq d_i^{N_i}$. Since the size of the lattice decreases at each step, these maps can be thought of as enacting a renormalization group on the real-space lattice. At the base (layer 0), the high energy, short-wavelength, lattice scale Hamiltonian $H^{(0)}$ is defined, with subsequent layers defining increasingly low-energy, long-wavelength effective theories 
\begin{align}
	H^{(i+1)}:=\rg{i}^\dagger H^{(i)}\rg{i}.
\end{align}
To correctly describe the physical RG fixed points, the MERA layers must be chosen to preserve the low-energy physics of $H^{(0)}$.

For concreteness, in this discussion we specialize to the MERA depicted in Fig.~\ref{Fig:fourtotwoMERA}, which we refer to as the 4:2 MERA. This MERA is built from a single kind of tensor, an isometry from 4 sites to 2 sites. In general, these tensors may all contain distinct coefficients, although space-time symmetries such as scale invariance can be imposed by, for example, forcing the tensors on each layer to be identical. We remark that our results are not specific to this choice, rather they work for all MERA schemes. In particular, in Appendix~\ref{appendix:generalansatz}, we describe how the results apply to the commonly used ternary MERA\cite{Evenbly2009,Pfeifer2009}.

In the MERA the fundamental constraint that a symmetry is preserved under renormalization is that each coarse-graining circuit acts as an intertwiner of $\G$ representations. That is, the renormalized symmetry 
\begin{align}
\label{renormalizedu}
U_g^{(i+1)}:=\rg{i}^\dagger U_g^{(i)}\rg{i},
\end{align} is again a representation of $\G$. 
When this condition is satisfied the third cohomology anomaly label of the symmetry does not change along the renormalization group flow\cite{PhysRevLett.112.231602,kapustin2014anomalies}. 
Hence the presence of an anomaly does not introduce any additional constraints on the renormalization process (which is to be expected for a discrete group).

For both practical and physically motivated reasons it is common to require further restrictions on the form of a symmetry throughout renormalization. 
For example, at a scale invariant renormalization group fixed point, the symmetry is also required to be scale invariant\cite{Pfeifer2009}.  
Furthermore, along an RG flow one may require that the bond dimension of an MPO symmetry remain constant, or grow subexponentially with the renormalization step. An extreme case is that of an on-site symmetry where the bond dimension is always required to be one, such that the symmetry remains strictly on-site.

\subsection{On-site symmetry}

In the case of a trivial 't~Hooft anomaly, a physical symmetry can be realized by an on-site representation. For a MERA satisfying  Eqn.~\ref{renormalizedu}, the 't~Hooft anomaly is preserved and hence it should remain possible to realize the symmetry in an on-site fashion at each RG step. This additional constraint is imposed by insisting that $U_g^{(i+1)}$ remains an on-site representation. Therefore the symmetry constraint becomes completely local\cite{Singh2010}. 

The symmetry can then be enforced on a MERA state by ensuring that the local tensors are locality preserving \emph{intertwiners} for the group action
\begin{align}\label{eqn:localsymmetry}
\begin{array}{c}
\includeTikz{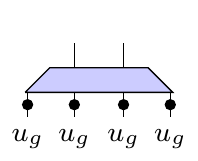}{
\begin{tikzpicture}[scale=.5]
\filldraw[fill=black,draw=black] (.05,-.25) circle (.1);\node[below] at(.05,-.5) {\textcolor{black}{\footnotesize$u_g$}};
\filldraw[fill=black,draw=black] (1,-.25) circle (.1);\node[below] at(1,-.5) {\textcolor{black}{\footnotesize$u_g$}};
\filldraw[fill=black,draw=black] (2,-.25) circle (.1);\node[below] at(2,-.5) {\textcolor{black}{\footnotesize$u_g$}};
\filldraw[fill=black,draw=black] (2.95,-.25) circle (.1);\node[below] at(2.95,-.5) {\textcolor{black}{\footnotesize$u_g$}};
\filldraw[fill=white,draw=white] (1,.75) circle (.1);\node[above] at(1,1) {\textcolor{white}{\footnotesize$v_g$}};
\filldraw[fill=white,draw=white] (2,.75) circle (.1);\node[above] at(2,1) {\textcolor{white}{\footnotesize$v_g$}};
\draw(.05,-.5)--(.05,0);\draw(2.95,-.5)--(2.95,0);
\draw(1,-.5)--(1,1);\draw(2,-.5)--(2,1);
\filldraw[ten](0,0)--(3,0)--(2.5,.5)--(.5,.5)--(0,0);
\end{tikzpicture}}
\end{array}
&=
\begin{array}{c}
\includeTikz{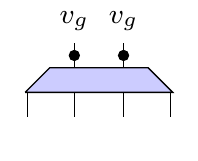}{
\begin{tikzpicture}[scale=.5]
\filldraw[fill=white,draw=white] (.05,-.25) circle (.1);\node[below] at(.05,-.5) {\textcolor{white}{\footnotesize$u_g$}};
\filldraw[fill=white,draw=white] (1,-.25) circle (.1);\node[below] at(1,-.5) {\textcolor{white}{\footnotesize$u_g$}};
\filldraw[fill=white,draw=white] (2,-.25) circle (.1);\node[below] at(2,-.5) {\textcolor{white}{\footnotesize$u_g$}};
\filldraw[fill=white,draw=white] (2.95,-.25) circle (.1);\node[below] at(2.95,-.5) {\textcolor{white}{\footnotesize$u_g$}};
\filldraw[fill=black,draw=black] (1,.75) circle (.1);\node[above] at(1,1) {\textcolor{black}{\footnotesize$v_g$}};
\filldraw[fill=black,draw=black] (2,.75) circle (.1);\node[above] at(2,1) {\textcolor{black}{\footnotesize$v_g$}};
\draw(.05,-.5)--(.05,0);\draw(2.95,-.5)--(2.95,0);
\draw(1,-.5)--(1,1);\draw(2,-.5)--(2,1);
\filldraw[ten](0,0)--(3,0)--(2.5,.5)--(.5,.5)--(0,0);
\end{tikzpicture}}
\end{array},
\end{align}
where the representation on each bond may be distinct. Standard results in representation theory allow one to impose the conditions Eqn.~\ref{eqn:localsymmetry}.

\subsection{Anomalous MPO symmetries}

\begin{figure}[t]
	\includeTikz{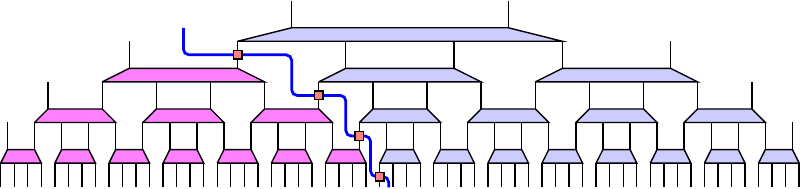}{
		\begin{tikzpicture}[scale=.55]
		\begin{scope}[shift={(0,-.25)}]
		\def\dx{.25}
		\def\dxx{.5}
		\def\dxxx{1}
		\def\dxxxx{2}
		\foreach \i in {0,...,59}{
			\draw(\i*\dx,0)--(\i*\dx,-.45);
		}
		\foreach \i in {0,4,...,27}{
			\filldraw[tenpurp](\i*\dx,0)--({(\i+3)*\dx},0)--({(\i+2.5)*\dx},\dx)--({(\i+.5)*\dx},\dx)--(\i*\dx,0);
		}
		\foreach \i in {28,32,...,56}{
			\filldraw[ten](\i*\dx,0)--({(\i+3)*\dx},0)--({(\i+2.5)*\dx},\dx)--({(\i+.5)*\dx},\dx)--(\i*\dx,0);
		}
		\foreach \i in {0,...,29}{
			\draw[shift={(.5*\dx,.75)}] (\i*\dxx,0)--(\i*\dxx,-.5);
		}
		\foreach \i in {0,4,...,11}{
			\filldraw[tenpurp,shift={(2.5*\dx,.75)}](\i*\dxx,0)--({(\i+3)*\dxx},0)--({(\i+2.5)*\dxx},\dx)--({(\i+.5)*\dxx},\dx)--(\i*\dxx,0);
		}
		\foreach \i in {12,16,...,26}{
			\filldraw[ten,shift={(2.5*\dx,.75)}](\i*\dxx,0)--({(\i+3)*\dxx},0)--({(\i+2.5)*\dxx},\dx)--({(\i+.5)*\dxx},\dx)--(\i*\dxx,0);
		}
		\foreach \i in {0,...,13}{
			\draw[shift={(.5*\dxx+2.5*\dx,1.5)}] (\i*\dxxx,0)--(\i*\dxxx,-.5);
		}
		\foreach \i in {0}{
			\filldraw[tenpurp,shift={(.5*\dxx+6.5*\dx,1.5)}](\i*\dxxx,0)--({(\i+3)*\dxxx},0)--({(\i+2.5)*\dxxx},\dx)--({(\i+.5)*\dxxx},\dx)--(\i*\dxxx,0);
		}
		\foreach \i in {4,8}{
			\filldraw[ten,shift={(.5*\dxx+6.5*\dx,1.5)}](\i*\dxxx,0)--({(\i+3)*\dxxx},0)--({(\i+2.5)*\dxxx},\dx)--({(\i+.5)*\dxxx},\dx)--(\i*\dxxx,0);
		}
		\foreach \i in {0,...,5}{
			\draw[shift={(.5*\dxxx+3.75*\dxx,2.25)}] (\i*\dxxxx,0)--(\i*\dxxxx,-.5);
		}
		\foreach \i in {0}{
			\filldraw[ten,shift={(.5*\dxxx+7.75*\dxx,2.25)}](\i*\dxxxx,0)--({(\i+3)*\dxxxx},0)--({(\i+2.5)*\dxxxx},\dx)--({(\i+.5)*\dxxxx},\dx)--(\i*\dxxxx,0);
		}
		\foreach \i in {0,...,1}{
			\draw[shift={(.5*\dxxxx+4.25*\dxxx+.5*\dx,3)}] (2*\i*\dxxxx,0)--(2*\i*\dxxxx,-.5);
		}
		\def\x{.08}
		\draw[thick,blue,rounded corners=20*\x](28*\dx+\dx/1.5,-.45)--(28*\dx+\dx/1.5,-.25)--(28*\dx-\dx/1.5,-.25)--(28*\dx-\dx/1.5,.5)--(2.5*\dx+11.5*\dxx,.5)--(2.5*\dx+11.5*\dxx,1.25)--
		(.5*\dxx+2.5*\dx+4.5*\dxxx,1.25)--(.5*\dxx+2.5*\dx+4.5*\dxxx,2)--(.5*\dxxx+7.75*\dxx-0.5*\dxxxx,2)--(.5*\dxxx+7.75*\dxx-0.5*\dxxxx,2.5);
		\filldraw[tenred,shift={(28*\dx,-.25)}](-\x,-\x)--(-\x,\x)--(\x,\x)--(\x,-\x)--cycle;
		\filldraw[tenred,shift={(2.5*\dx+12*\dxx,.5)}](-\x,-\x)--(-\x,\x)--(\x,\x)--(\x,-\x)--cycle;
		\filldraw[tenred,shift={(.5*\dxx+2.5*\dx+5*\dxxx,1.25)}](-\x,-\x)--(-\x,\x)--(\x,\x)--(\x,-\x)--cycle;
		\filldraw[tenred,shift={(.5*\dxxx+7.75*\dxx+0*\dxxxx,2)}](-\x,-\x)--(-\x,\x)--(\x,\x)--(\x,-\x)--cycle;
		\end{scope}
		\end{tikzpicture}}
	\caption{By applying the MPO to a half infinite chain, one can insert a domain wall between two dual theories. If the MPO acts as a symmetry, this corresponds to putting the theory on boundary conditions which have been twisted by the group element.}\label{Fig:SymmetryTwistMERA}
\end{figure}

Generally $(2+1)$D SPT states are gapped in the bulk (on a closed manifold), but, on a manifold with a boundary they either spontaneously break the `protecting' symmetry, or possess gapless excitations in the vicinity of the boundary\cite{czxmodel}. Since the low energy physics is confined to the edge, it is interesting to consider the low energy, effective edge theory. When restricting the on-site bulk symmetry to the edge, it becomes anomalous with anomaly label $[\phi]\in\coho{3}$ matching the bulk SPT\cite{czxmodel,PhysRevB.90.235137,PhysRevB.91.195134}. An on-site representation of the bulk symmetry cannot be recovered by any local operations on the edge.

Since anomalous symmetries cannot be made on-site, the condition in Eqn.~\ref{renormalizedu} is no longer strictly local. If the bond dimension of an MPO is allowed to grow at each renormalization step, the only constraint in Eqn.~\ref{renormalizedu} is that the symmetry remains a global representation.

So long as this constraint is satisfied, the nontrivial anomaly label $[\phi]\in\coho{3}$ of an MPO representation, discussed in Appendix~\ref{appendix:thirdcoho}, is invariant under renormalization\cite{PhysRevLett.112.231602}. 

For anomalous symmetries the natural analogue to Eqn.~\ref{eqn:localsymmetry} is 
\begin{align}\label{eqn:anomsymmetry}
\begin{array}{c}
\includeTikz{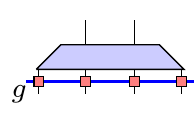}{
	\begin{tikzpicture}[scale=.5]
	\draw[thick,blue] (-.2,-.25)--(3.2,-.25);
	\draw[nonesty,blue] (.8,.75)--(2.2,.75);
	\draw(.05,-.5)--(.05,0);\draw(2.95,-.5)--(2.95,0);
	\draw(1,-.5)--(1,1);\draw(2,-.5)--(2,1);
		\filldraw[tenred,shift={(.05,-.25)}] (-0.1,-0.1)--(0.1,-0.1)--(0.1,0.1)--(-0.1,0.1)--cycle;\node[left] at(.05,-.5) {\textcolor{black}{\footnotesize$g$}};
		\filldraw[tenred,shift={(1,-.25)}] (-0.1,-0.1)--(0.1,-0.1)--(0.1,0.1)--(-0.1,0.1)--cycle;
		\filldraw[tenred,shift={(2,-.25)}] (-0.1,-0.1)--(0.1,-0.1)--(0.1,0.1)--(-0.1,0.1)--cycle;
		\filldraw[tenred,shift={(2.95,-.25)}] (-0.1,-0.1)--(0.1,-0.1)--(0.1,0.1)--(-0.1,0.1)--cycle;
		\filldraw[nonesty,shift={(1,.75)}] (-0.1,-0.1)--(0.1,-0.1)--(0.1,0.1)--(-0.1,0.1)--cycle;\node[left] at(1,1) {\textcolor{white}{\footnotesize$g$}};
		\filldraw[nonesty,shift={(2,.75)}] (-0.1,-0.1)--(0.1,-0.1)--(0.1,0.1)--(-0.1,0.1)--cycle;
	\filldraw[ten](0,0)--(3,0)--(2.5,.5)--(.5,.5)--(0,0);
	\end{tikzpicture}}
\end{array}
&=
\begin{array}{c}
\includeTikz{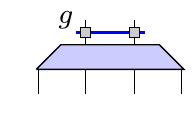}{
	\begin{tikzpicture}[scale=.5]
	\draw[nonesty,blue] (-.2,-.25)--(3.2,-.25);
	\draw[thick,blue] (.8,.75)--(2.2,.75);
	\draw(.05,-.5)--(.05,0);\draw(2.95,-.5)--(2.95,0);
	\draw(1,-.5)--(1,1);\draw(2,-.5)--(2,1);
	\filldraw[nonesty,shift={(.05,-.25)}] (-0.1,-0.1)--(0.1,-0.1)--(0.1,0.1)--(-0.1,0.1)--cycle;\node[left] at(.05,-.5) {\textcolor{white}{\footnotesize$g$}};
	\filldraw[nonesty,shift={(1,-.25)}] (-0.1,-0.1)--(0.1,-0.1)--(0.1,0.1)--(-0.1,0.1)--cycle;
	\filldraw[nonesty,shift={(2,-.25)}] (-0.1,-0.1)--(0.1,-0.1)--(0.1,0.1)--(-0.1,0.1)--cycle;
	\filldraw[nonesty,shift={(2.95,-.25)}] (-0.1,-0.1)--(0.1,-0.1)--(0.1,0.1)--(-0.1,0.1)--cycle;
	\filldraw[tengrey,shift={(1,.75)}] (-0.1,-0.1)--(0.1,-0.1)--(0.1,0.1)--(-0.1,0.1)--cycle;\node[left] at(1,1) {\textcolor{black}{\footnotesize$g$}};
	\filldraw[tengrey,shift={(2,.75)}] (-0.1,-0.1)--(0.1,-0.1)--(0.1,0.1)--(-0.1,0.1)--cycle;
	\filldraw[ten](0,0)--(3,0)--(2.5,.5)--(.5,.5)--(0,0);
	\end{tikzpicture}}
\end{array},
\end{align}
which is a sufficient condition for a symmetric MERA, but is not necessarily implied by Eqn.~\ref{renormalizedu}. 

We remark that this condition does not correspond to a local group action unless further assumptions are made. Consequently conventional techniques from representation theory do not suffice to enforce the constraint. Despite this, in Section~\ref{section:class} we define a class of MERA which allow Eqn.~\ref{eqn:anomsymmetry} to be imposed via a strictly local condition.

Although Eqn.~\ref{eqn:anomsymmetry} generically allows the MPO to change one may wish to insist that the MPO is fixed under the RG. For instance, at an RG fixed point where identical tensors are used at each layer of the MERA.

Unlike an on-site symmetry, an MPO can act as a duality transformation between a pair of critical models. This can be realized in MERA by allowing the MERA tensors themselves to change in Eqn.~\ref{eqn:anomsymmetry}. We demonstrate such an action in Section~\ref{section:duality}.
One can also use the MPO to create a domain wall between the two critical theories by applying the MPO to a half-infinite chain. In the case where the dual theories coincide (i.e. the MPO acts as a symmetry) this corresponds to a symmetry twist (topological defect) or twisted boundary condition. This will be the subject of Section~\ref{section:mpos}.

\subsection{Physical data from MERA}\label{section:physicaldata}

Once a MERA has been obtained, a variety of physical data can be extracted. The most straightforward of these is the energy of the MERA, which simply requires evaluation of $\bra{\psi}H\ket{\psi}$. 

For a MERA representing the ground state of a gapless Hamiltonian, one can also extract a variety of data about the associated conformal field theory (CFT)\cite{Ginsparg1988,DiFrancesco1997}. One can compute the central charge as discussed in Refs.~\onlinecite{Pfeifer2009,Evenbly2011} using the scaling of entanglement entropy in the state.
One can also obtain the scaling dimensions of the associated CFT\cite{Pfeifer2009,Evenbly2011} by seeking eigenoperators of the scaling superoperator
\begin{align}\label{eqn:scaling}
\sso{1}{\begin{array}{c}\includeTikz{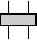}{}\end{array}}&=
\begin{array}{c}
\includeTikz{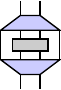}{
	\begin{tikzpicture}[scale=.3]
	\begin{scope}
	\draw(0,-.5)--(0,0);
	\draw(.6666,-.4)--(.6666,1);
	\draw(1.3333,-.4)--(1.3333,1);
	\draw(2,-.5)--(2,0);
	\filldraw[ten](0,0)--(2,0)--(1.3333,.5)--(.66666,.5)--cycle;
	\end{scope}
	\begin{scope}[shift={(0,-1)},yscale=-1]
	\draw(0,-.5)--(0,0);
	\draw(.6666,-.4)--(.6666,1);
	\draw(1.3333,-.4)--(1.3333,1);
	\draw(2,-.5)--(2,0);
	\filldraw[ten](0,0)--(2,0)--(1.3333,.5)--(.66666,.5)--cycle;
	\end{scope}
	\begin{scope}[scale=.4,shift={(1.0,-1.75)}]
	\filldraw[tengrey](0,0)--(3,0)--(3,1)--(0,1)--cycle;
	\end{scope}
	\end{tikzpicture}}
\end{array}
=\lambda
\begin{array}{c}
\includeTikz{ScalingB}{
	\begin{tikzpicture}[scale=.3]
	\begin{scope}
	\draw(.6666,-.4)--(.6666,.15);
	\draw(1.3333,-.4)--(1.3333,.15);
	\end{scope}
	\begin{scope}[shift={(0,-1)},yscale=-1]
	\draw(.6666,-.4)--(.6666,.15);
	\draw(1.3333,-.4)--(1.3333,.15);
	\end{scope}
	\begin{scope}[scale=.4,shift={(1.0,-1.75)}]
	\filldraw[tengrey](0,0)--(3,0)--(3,1)--(0,1)--cycle;
	\end{scope}
	\end{tikzpicture}}
\end{array}.
\end{align}
The scaling dimensions describe the decay of correlations in the theory. We will refer to $\Delta=-\log_2(\lambda)$ as the \emph{scaling dimension} corresponding to a particular \emph{scaling field}.

The scaling fields obtained from the scaling superoperator correspond to local fields in the CFT. Given a symmetric MERA, one can also obtain nonlocal scaling fields by constructing the `symmetry twisted' scaling superoperators
\begin{align}\label{eqn:twistedscaling}
\sso{g}{\begin{array}{c}\includeTikz{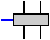}{}\end{array}}&=
\begin{array}{c}
\includeTikz{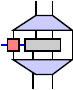}{
	\begin{tikzpicture}[scale=.3]
	\begin{scope}
	\draw(0,-.5)--(0,0);
	\draw(.7666,-.4)--(.7666,1);
	\draw(1.3333,-.4)--(1.3333,1);
	\draw(2,-.5)--(2,0);
	\filldraw[ten](0,0)--(2,0)--(1.3333,.5)--(.76666,.5)--cycle;
	\draw[thick,blue](-.4,-.5)--(.4,-.5);
	\end{scope}
	\begin{scope}[shift={(0,-1)},yscale=-1]
	\draw(0,-.5)--(0,0);
	\draw(.6666,-.3)--(.6666,1);
	\draw(1.3333,-.3)--(1.3333,1);
	\draw(2,-.5)--(2,0);
	\filldraw[ten](0,0)--(2,0)--(1.3333,.5)--(.76666,.5)--cycle;
	\end{scope}
	\begin{scope}[scale=.4,shift={(-.5,-1.75)}]
	\filldraw[tenred](0,0)--(1,0)--(1,1)--(0,1)--cycle;
	\end{scope}
	\begin{scope}[scale=.4,shift={(1.0,-1.75)}]
	\filldraw[tengrey](0,0)--(3,0)--(3,1)--(0,1)--cycle;
	\end{scope}
	\end{tikzpicture}}
\end{array}
=\lambda
\begin{array}{c}
\includeTikz{NLScalingB}{
	\begin{tikzpicture}[scale=.3]
	\begin{scope}
	\draw(.7666,-.4)--(.7666,.15);
	\draw(1.3333,-.4)--(1.3333,.15);
	\draw[thick,blue](-.0,-.5)--(.4,-.5);
	\end{scope}
	\begin{scope}[shift={(0,-1)},yscale=-1]
	\draw(.7666,-.4)--(.7666,.15);
	\draw(1.3333,-.4)--(1.3333,.15);
	\end{scope}
	\begin{scope}[scale=.4,shift={(1.0,-1.75)}]
	\filldraw[tengrey](0,0)--(3,0)--(3,1)--(0,1)--cycle;
	\end{scope}
	\end{tikzpicture}}
\end{array},
\end{align}
where $\includeTikz{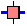}{\begin{tikzpicture}[x=1*\eX,y=1*\eX]
	\def\x{.25}
	\draw[thick,blue](-.5,0)--(.5,0);
	\draw(0,-.5)--(0,.5);
	\filldraw[tenred](-\x,-\x)--(-\x,\x)--(\x,\x)--(\x,-\x)--cycle;
\end{tikzpicture}}$ 
is the symmetry MPO for the group element $g$. These fields correspond to a half infinite symmetry twist, as in Fig.~\ref{Fig:SymmetryTwistMERA}, terminated by a local tensor. Previously, nonlocal scaling operators with a tensor product structure have been obtained in the same way~\cite{Evenbly2010c}, but this more general class involving an anomalous symmetry was not investigated. 
\section{Symmetry twists and topological sectors}
\label{section:mpos}

Once a symmetric MERA is optimized to represent the ground state of a critical model, conformal data can be obtained as discussed in Section~\ref{section:physicaldata}. In this section, we investigate the impact that an anomalous symmetry has on such conformal data. In particular, we use the properties of MPO group representations to obtain possible topological corrections to the conformal spins when a symmetry twist is applied. We observe these corrections in our example model, as shown in Table~\ref{table:CFTid}. Additionally, we construct the projective representations under which the nonlocal scaling fields (as defined in Eqn.~\ref{eqn:twistedscaling}) transform. These allow us to construct projectors onto irreducible topological sectors, extending the usual decomposition into symmetry sectors. We discuss the constraints that this decomposition imposes on the operator product expansion of the CFT. For our example model, we observe these constraints in Table~\ref{table:fusion}.

Throughout this section, for simplicity of presentation, we treat the case of scale invariant MERA with scale invariant MPO symmetry. Furthermore, we assume the technical condition that the MPO representation satisfies the \emph{zipper condition}\cite{williamson2014matrix} 
\begin{align}
\begin{array}{c}
	\includeTikz{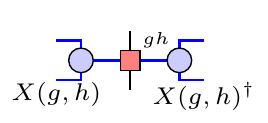}{
		\begin{tikzpicture}[scale=1]
		\node[above right] at(0,-.25) {\textcolor{white}{\tiny$g$}};
		\node[above right] at(0,.25) {\textcolor{white}{\tiny$h$}};
		\draw[thick,blue] (-.5,0)--(.5,0)--(.5,.2)--(.75,.2) (.5,0)--(.5,-.2)--(.75,-.2) (-.5,0)--(-.5,.2)--(-.75,.2) (-.5,0)--(-.5,-.2)--(-.75,-.2);
		\draw(0,-.3)--(0,.3);
		\filldraw[tenred] (-0.1,-0.1)--(0.1,-0.1)--(0.1,0.1)--(-0.1,0.1)--cycle;
		\filldraw[ten](.5,0) circle (.25/2); 
		\filldraw[ten](-.5,0) circle (.25/2);
\node[above right] at(0,0) {\textcolor{black}{\tiny$gh$}};
\node[below] at (.75,-.1) {\textcolor{black}{\scriptsize$X(g,h)^\dagger$}};
\node[below] at (-.75,-.1) {\textcolor{black}{\scriptsize$X(g,h)$}};
		\end{tikzpicture}}
\end{array}
&=
\begin{array}{c}
\includeTikz{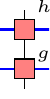}{
	\begin{tikzpicture}[scale=1]
	\clip  (-.25,-.5)rectangle (.25,.5);
	\node[above right] at(0,0) {\textcolor{white}{\tiny$gh$}};
	\node[below] at (.75,-.1) {\textcolor{white}{\scriptsize$X(g,h)^\dagger$}};
	\node[below] at (-.75,-.1) {\textcolor{white}{\scriptsize$X(g,h)$}};
	\draw[thick,blue] (-.3,.2)--(.3,.2) (-.3,-.2)--(.3,-.2);
	\draw(0,-.4)--(0,.4);
	\filldraw[tenred,shift={(0,.2)}] (-0.1,-0.1)--(0.1,-0.1)--(0.1,0.1)--(-0.1,0.1)--cycle;
	\filldraw[tenred,shift={(0,-.2)}] (-0.1,-0.1)--(0.1,-0.1)--(0.1,0.1)--(-0.1,0.1)--cycle;
\node[above right] at(0,-.25) {\textcolor{black}{\tiny$g$}};
\node[above right] at(0,.25) {\textcolor{black}{\tiny$h$}};
	\end{tikzpicture}}
\end{array}.
\end{align} 
These assumptions imply that the MPOs can be deformed freely through a symmetric MERA network. We remark that representative MPOs satisfying the zipper condition have been given for all anomalous discrete symmetries in $(1+1)$D\cite{williamson2014matrix}. 
Additionally, we have suppressed possible orientation dependencies of the MPOs, although this effect is accounted for in our results. For a full treatment of the intricacies that arise due to orientation dependence see {Ref.~\onlinecite{williamson2014matrix}.} 
We note that similar reasoning applies to MPOs not satisfying these simplifying assumptions. 

\subsection{Symmetry twist and topological correction to conformal spin}

For a model described by symmetric Hamiltonian $H$, a symmetry twist can be created by acting with an element of the group on a half-infinite chain. Hamiltonian terms far away from the end of the twist are left invariant and the only remnant is a single twisted Hamiltonian term crossing the end. This is captured by the MERA in Fig.~\ref{Fig:SymmetryTwistMERA} with uniform tensors.

The twisted Hamiltonian term can be used to close a chain into a ring of length $L$. In the case of a trivial (identity) twist this yields periodic boundary conditions. For a nontrivial group element this corresponds to a flux insertion through the ring as there is now a nontrivial monodromy around the ring given by the group element. 

The introduction of an MPO twist by group element $g$ leads to a twisted translation operator 
\begin{align}
\trans{g}&=
\begin{array}{c}
\includeTikz{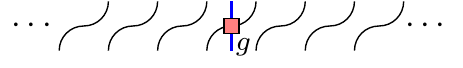}{
	\begin{tikzpicture}
	\def\siz{.075};
	\def\y{.25};
	\foreach \x in {0,...,6}{
		\draw [shift={(.5*\x,0)}] (-.25,-\y) to[out=90,in=180+0](0,0) to[out=0,in=180+90](.25,\y);}
	\draw[thick,blue,shift={(.5*3,0)}](0,-\y)--(0,\y);
	\filldraw[tenred,shift={(.5*3,0)}](-\siz,-\siz)--(-\siz,\siz)--(\siz,\siz)--(\siz,-\siz)--cycle;
	\node at (-.5,0) {$\cdots$};\node at (3.5,0) {$\cdots$};
	\node[below right] at(1.42,0) {\textcolor{black}{\footnotesize$g$}};
	\end{tikzpicture}
},
\end{array} 
\end{align}
which translates the system by one site without moving the end of the twist (previously noted in Refs.~\onlinecite{PhysRevB.94.115125,1751-8121-49-35-354001}). We will see that this leads to corrections to the conformal spin. 

The untwisted translation operator for periodic boundary conditions satisfies ${\trans{1}^L=\openone}$ which implies that local fields have integer conformal spin~\cite{Christe}. 
The twisted translation operator satisfies $ {\trans{g}^L=\dehn{g}}$ where
\begin{align}
\dehn{g}&=
\begin{array}{c}
\includeTikz{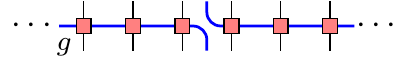}{
	\begin{tikzpicture}
	\def\siz{.075};
	\def\y{.25};
	\draw[thick,blue,rounded corners](-.25,0)--(1.25,0)--(1.25,-\y);
	\draw[thick,blue,rounded corners](2.75,0)--(1.25,0)--(1.25,\y);
	\foreach \x in {0,...,5}{
		\draw[shift={(.5*\x,0)}](0,-\y)--(0,\y);
		\filldraw[tenred,shift={(.5*\x,0)}](-\siz,-\siz)--(-\siz,\siz)--(\siz,\siz)--(\siz,-\siz)--cycle;
	}
	\node[below left] at(0,0) {\textcolor{black}{\footnotesize$g$}};
	\node at (-.5,0) {$\cdots$};\node at (3,0) {$\cdots$};
	\end{tikzpicture}
}
\end{array} 
\end{align}
is the Dehn twist operator. For a faithful on-site representation of $g$ the order of $\dehn{g}$ is simply the order of $g$, denoted $\order{g}$. Hence the conformal spins of $g$-twisted fields may have a topological correction leading them to take values\cite{Christe} in $\frac{1}{\order{g}}\Z$.  
 
We now consider anomalous representations and show that the order of $\dehn{g}$ is $2\order{g}$ in some cases, reflecting a further correction due to the anomaly. We observe this additional correction in our numerical example, as shown in Table~\ref{table:CFTid}.

First we define 
\begin{align}
\crossing{g}{h}&=
	\begin{array}{c}
	\includeTikz{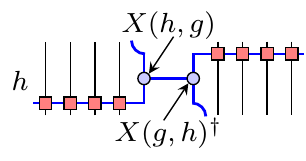}{
		\begin{tikzpicture}[scale=.25]
		\def\x{1}
		\def\y{1}
		\def\siz{.25};
		\draw[thick,blue](-1.5*\x,-\y)--(-1*\x,-\y)--(-1*\x,\y);
		\draw[thick,blue](1*\x,-\y)--(1*\x,\y)--(1.5*\x,\y);
		\draw[thick,blue](-1.5*\x,-\y)--(-5.5*\x,-\y);
		\draw[thick,blue](1.5*\x,\y)--(5.5*\x,\y);
		\draw[thick,blue,rounded corners=3.8](-1*\x,\y)--(-1.5*\x,\y)--(-1.5\x,1.5*\y);
		\draw[thick,blue,rounded corners=3.8](1*\x,-\y)--(1.5*\x,-\y)--(1.5\x,-1.5*\y);
		\draw[thick,blue](-\x,0)--(\x,0);\filldraw[ten](-\x,0) circle (\x/4);\filldraw[ten](\x,0) circle (\x/4);
			\foreach \x in {2,...,5}{
				\draw[shift={(-\x,-\y)}](0,-\y/2)--(0,2.5*\y);
				\filldraw[tenred,shift={(-\x,-\y)}](-\siz,-\siz)--(-\siz,\siz)--(\siz,\siz)--(\siz,-\siz)--cycle;
				\draw[shift={(\x,\y)}](0,\y/2)--(0,-2.5*\y);
				\filldraw[tenred,shift={(\x,\y)}](-\siz,-\siz)--(-\siz,\siz)--(\siz,\siz)--(\siz,-\siz)--cycle;
			}
		\node[above left] at(-5.2*\x,-\y) {\textcolor{black}{\footnotesize$h$}};
		\node[above] at(0,1.1*\y) {\textcolor{black}{\footnotesize$X(h,g)$}};
		\node[below] at(0,-1.1*\y) {\textcolor{black}{\footnotesize$X(g,h)^\dagger$}};
		\draw[-stealth,shift={(.2,.2)}](0,1.5*\y)--(-\x,0);
		\draw[-stealth,shift={(-.2,-.2)}](0,-1.5*\y)--(\x,0);
		\end{tikzpicture}}
	\end{array}.
\end{align} 
which corresponds to the action of $h$ on the $g$ twisted MERA shown in Fig.~\ref{Fig:SymmetryTwistMERA}. 
It was shown in Ref.~\onlinecite{williamson2014matrix} that 
\begin{align}
\dehn{g}\crossing{g}{h}&=\phi(g,h,g)\crossing{g}{gh},
\end{align} 
where $\phi$ is the 3-cocycle of the MPO representation. Applying the Dehn twist $\order{g}$ times results in a phase
\begin{align}
\dehn{g}^{\order{g}}\crossing{g}{1}&=\prod_{i=1}^{\order{g}-1} \phi(g,g^i,g)\crossing{g}{1},
\end{align}
where again $\order{g}$ denotes the order of $g$. 
Since $g$ generates a subgroup  $\Z_{\order{g}}\leqslant\G$ and 
\begin{align}\phi_g(i,j,k):=\phi(g^i,g^j,g^k)\end{align}
 defines a 3-cocycle of $\Z_{\order{g}}$. Denote the relevant cohomology class  by $[\phi_g]\in\cohom{3}{\Z_{\order{g}}}\cong \Z_{\order{g}}$. For simplicity, assume it has been brought into the normal form\cite{propitius} 
\begin{align}
\phi_g(i,j,k)=\omega^{[\phi_g]i(j+k-j\oplus k)/\order{g}},
\end{align}
 where $\omega$ is a primitive $\order{g}^{\text{th}}$ root of unity and $\oplus$ denotes addition modulo $\order{g}$. 
Hence 
\begin{align}
	\prod_{i=1}^{\order{g}-1} \phi(g,g^i,g) &= \omega^{[\phi_g]}
\end{align}
and
\begin{align}
	\dehn{g}^{\order{g}}&= \omega^{[\phi_g]} \openone . 
\end{align}
Consequently an anomaly $[\phi]$ for $g$-twisted fields may induce a further topological correction to their conformal spins. In particular, the correction to the conformal spins take values in 
\begin{align}
	\frac{1}{\order{g}}\Z_{\order{g}}+\frac{[\phi_g]}{\order{g}^2}.
\end{align}

To make this argument we fixed a particular representative of $\phi$, however the topological correction to conformal spin is a gauge invariant quantity and should not depend on this choice.

For the case of $\G=\zt^3$, we observe this anomalous correction in our numerical example, where we see quarter- and three-quarter- integer conformal spins (displayed in Table~\ref{table:CFTid}).

\subsection{Projective representations and topological sectors}

We proceed to construct topological sectors that have a definite topological correction to the conformal spin. These topological sectors are an extension of the usual symmetry sectors used to block diagonalize a Hamiltonian.

Topological sectors are labeled by a conjugacy class $\conjclass{}\subset\G$, indicating twist symmetry twist, and a (projective) irreducible representation (irrep.) $\chi_g^\mu$ of the centralizer of a representative element $g\in\conjclass{}$. The topological sectors are mathematically described by  $\text{D}^\phi(\G)$, the quantum double of the symmetry group $\G$ twisted by the 3-cocycle anomaly $\phi$. This category determines all topological properties of the sectors.

Since the MPO symmetry commutes with the MERA tensors, one can simultaneously diagonalize the twisted scaling superoperator $\sso{g}{\cdot}$ and the action of the symmetry. 
The vector space spanned by $g$-twisted scaling fields  (see Eqn.~\ref{eqn:twistedscaling}) transforms under a projective representation $V^{(g)}_{h}$ of the centralizer $\centralizer{g}$. 
This projective representation has 2-cocycle $\phi^{(g)}$ defined by 
\begin{align}\phi^{(g)}(h,k)&=\frac{\phi(g,h,k)\phi(h,k,g)}{\phi(h,g,k)},\end{align}
which is the slant product of $\phi$. 
The action is explicitly given by\cite{williamson2014matrix} 
\begin{align}
V^{(g)}_{h}&=
\begin{array}{c}
\includeTikz{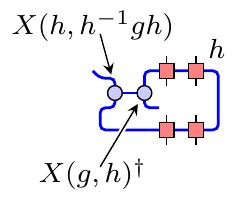}{
	\begin{tikzpicture}[scale=.15]
	\def\siz{.5};
		\draw[thick,blue,rounded corners=1.5](-1.5,-.5)--(-2.5,-.5)--(-2.5,2)--(2.5,2)--(2.5,-2)--(-5.5,-2)--(-5.5,-.5)--(-4.5,-.5)--(-4.5,1.5)--(-5.5,1.5)--(-6,2);
		\draw[thick,blue](-4.5,.5)--(-2.5,.5);
		\filldraw[ten](-4.5,.5) circle (1/2);
		\filldraw[ten](-2.5,.5) circle (1/2);
		\draw[shift={(-1,2)}](0,-1)--(0,1);\draw[shift={(1,2)}](0,-1)--(0,1);
		\draw[shift={(-1,-2)}](0,-1)--(0,1);\draw[shift={(1,-2)}](0,-1)--(0,1);
		\filldraw[tenred,shift={(-1,2)}](-\siz,-\siz)--(-\siz,\siz)--(\siz,\siz)--(\siz,-\siz)--cycle;
		\filldraw[tenred,shift={(1,2)}](-\siz,-\siz)--(-\siz,\siz)--(\siz,\siz)--(\siz,-\siz)--cycle;
		\filldraw[tenred,shift={(-1,-2)}](-\siz,-\siz)--(-\siz,\siz)--(\siz,\siz)--(\siz,-\siz)--cycle;
		\filldraw[tenred,shift={(1,-2)}](-\siz,-\siz)--(-\siz,\siz)--(\siz,\siz)--(\siz,-\siz)--cycle;
		\draw[ultra thick,white](-5.5,-4.5)--(-2.95,-.25);
				\node[above right] at(1,2) {\textcolor{black}{\footnotesize$h$}};
				\node[] at(-6,5) {\textcolor{black}{\footnotesize$X(h,h^{-1}gh)$}};
				\node[] at(-6,-5) {\textcolor{black}{\footnotesize$X(g,h)^\dagger$}};
				\draw[-stealth](-5.5,4.5)--(-4.75,1.75);
				\draw[-stealth](-5.5,-4.5)--(-2.95,-.25);
	\end{tikzpicture}}
\end{array},
\end{align}
where $h^{-1}gh=g$ for $h\in\centralizer{g}$.

The $g$-twisted scaling superoperator commutes with the projective representation
\begin{align} \sso{g}{V^{(g)}_{h}(\cdot)} &= V^{(g)}_{h}(\sso{g}{\cdot}),
\end{align}
and hence can be block diagonalized into projective irreps. 

Topological sectors that contribute a definite correction to the conformal spin can be constructed following the approach of Ref.~\onlinecite{Bultinck2017183}. The first step is to form projectors $\proj{g,\mu}$ onto the projective irreps of $\centralizer{g}$. For a twist $g$ and projective irrep $\mu$ with 2-cocycle $\phi^{(g)}$
\begin{align}\proj{g,\mu}&:= \frac{d_\mu}{|\centralizer{g}|} \sum_{h\in\centralizer{g}} \bar\chi_g^\mu(h) V^{(g)}_{h} ,\end{align}
where $d_\mu$ its dimension,  $\chi^{\mu}_g$ its character and $\bar{\cdot}$ denotes complex conjugation.

The full scaling superoperator, taking into account all sectors, is given by 
\begin{align}\sso{\G}{\cdot}&:=\bigoplus_g \sso{g}{\cdot}.\end{align}
This commutes with the full $|\G|^2$ dimensional algebra spanned by $V^{(g)}_{h}$ (note $V^{(k)}_{l}V^{(g)}_{h}=0$ unless $ k= h^{-1}gh$). This is a ${C}^*$ algebra\cite{Bultinck2017183} and can be diagonalized into blocks. 
The simple central idempotents that project onto each irreducible block are given by
\begin{align}
\label{idempotent}
\proj{\conjclass{g},\mu}:=\sum_{k\in\conjclass{g}}\proj{k,\mu} ,
\end{align}
where $\conjclass{g}$ is the conjugacy class of $g$ in $\G$. 
These projectors block diagonalize $\sso{\G}{\cdot}$ into irreducible topological sectors.  For the numerical example in Appendix~\ref{appendix:fullmeradata}, all conformal data is decomposed into these sectors.

The topological sectors thus constructed have definite topological spin\cite{Bultinck2017183} (correction to conformal spin), which we observe in our example in Table~\ref{table:CFTid}. Additionally, these sectors obey a set of fusion rules, and support a notion of braiding monodromy and exchange statistics. The full set of topological data can be extracted from the idempotents constructed in Eqn.~\ref{idempotent} via the procedure outlined in Ref.~\onlinecite{Bultinck2017183}.
%

In the MERA, with an MPO symmetry, the operator product expansion (OPE)\cite{Ginsparg1988,DiFrancesco1997} for scaling fields $a$ and $b$ in topological sectors labeled $(\mathcal{C}_0,\mu_0)$ and $(\mathcal{C}_1,\mu_1)$ can be computed using\cite{Pfeifer2009,Evenbly2010c}
\begin{align}\label{eqn:OPE}
a\times b&=\sum_{\substack{g\in \mathcal{C}_0\\h\in\mathcal{C}_1}}
\begin{array}{c}
\includeTikz{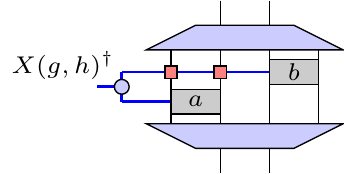}{
	\begin{tikzpicture}[scale=.5]
		\def\a{.25};
	\begin{scope}[shift={(0,.75)}]
		\draw(-1.5,0)--(-1.5,-1);
		\draw(-.5,1)--(-.5,-1);\draw(.5,1)--(.5,-1);
		\draw(1.5,0)--(1.5,-1);
		\filldraw[ten](-2,0)--(2,0)--(1,.5)--(-1,.5)--cycle;
	\end{scope}
	\begin{scope}[shift={(0,-.75)},yscale=-1]
		\draw(-1.5,0)--(-1.5,-1);
		\draw(-.5,1)--(-.5,-1);\draw(.5,1)--(.5,-1);
		\draw(1.5,0)--(1.5,-1);
		\filldraw[ten](-2,0)--(2,0)--(1,.5)--(-1,.5)--cycle;
	\end{scope}
	\begin{scope}[shift={(-1,-.3)}]
	\draw[thick,blue](0,0)--(-1.5,0);
		\filldraw[tengrey](-.5,-.25)--(.5,-.25)--(.5,.25)--(-.5,.25)--cycle;
		\node at (0,0) {\scriptsize$a$};
	\end{scope}
	\begin{scope}[shift={(1,.3)}]
	\draw[thick,blue](0,0)--(-3.5,0);
	\filldraw[tengrey](-.5,-.25)--(.5,-.25)--(.5,.25)--(-.5,.25)--cycle;
		\filldraw[tenred,shift={(-1.5,0)}](-\a/2,-\a/2)--(-\a/2,\a/2)--(\a/2,\a/2)--(\a/2,-\a/2)--cycle;
		\filldraw[tenred,shift={(-2.5,0)}](-\a/2,-\a/2)--(-\a/2,\a/2)--(\a/2,\a/2)--(\a/2,-\a/2)--cycle;
	\node at (0,0) {\scriptsize$b$};
	\end{scope}
	\draw[thick,blue](-2.5,.3)--(-2.5,-.3) (-2.5,0)--(-3,0);
	\draw[ten] (-2.5,0) circle(.15);
	\node[above left] at (-2.4,-.1) {\scriptsize$X(g,h)^\dagger$};
	\end{tikzpicture}}
\end{array}
=\sum_c C_{ab}^c c ,
\end{align}
where the sum is over scaling fields $c$. 
Eqn.~\ref{eqn:OPE} is a tensor network realization of a pair of pants topology with $a$ and $b$ at the feet and $c$ at the waist.
The fusion rules imply topological restrictions on the OPE of scaling fields, generalizing symmetry constraints on the local fields. 
In particular, $C_{ab}^c=0$ unless the sector labeling $c$ appears in the fusion product 
\begin{align}
(\mathcal{C}_0,\mu_0)\times(\mathcal{C}_1,\mu_1)=\sum_{{(\mathcal{C}_2,\mu_2)}} N_{(\mathcal{C}_0,\mu_0)(\mathcal{C}_1,\mu_1)}^{(\mathcal{C}_2,\mu_2)} (\mathcal{C}_2,\mu_2) .
\end{align}

We observe the constraints directly in the numerical MERA in Table~\ref{table:fusion}.

Technically the symmetry twists and their fusion structure are described by the unitary fusion category (UFC) $\text{Vec}_\G^\phi$ while the topological sectors are given by its Drinfeld center $Z(\text{Vec}_\G^\phi)$ --- equivalently the twisted quantum double $\text{D}^\phi(\G)$ --- which is a modular tensor category (MTC)\cite{drinfeld,Moore1989,tubealgebra,DrinfeldCenter,bakalov2001lectures,etingof2005fusion}. The mathematical structure of this MTC determines all topological properties of the fields in each sector, including the topological correction to their conformal spin (equivalently the exchange statistics), topological restriction on the OPE and monodromies (braiding)\cite{FUCHS2002452,FUCHS2002353,PhysRevLett.93.070601,FROHLICH2007354,frohlich2009defect}.  

Interestingly the fusion rules for the topological sectors can be nonabelian, even when the symmetry group is abelian. This requires a nontrivial anomaly $\phi$. This occurs in our numerical example as discussed in Section~\ref{Numerics} and Table~\ref{table:fusion}.

\section{A class of MPO symmetric MERA}
\label{section:class}

To enforce a constraint on a MERA state requires an identification of the remaining variational parameters in such a way that it is possible to optimize over them. In this section we describe an approach that relies on a property of the MPO symmetry: the existence of a local unitary capable of disentangling a contiguous region of each MPO into an inner part that forms a local representation of the symmetry and is decoupled from the original MPO on the outer section. Given such a local representation, conventional techniques can be used to ensure the MERA is symmetric.
 We construct a large class of MPOs with this property and find the resulting constraints on the form of symmetric MERA tensors.

\subsection{Disentangling an MPO}

For scale invariant MERA, where the MPO symmetry is required to be identical at all layers, the goal is to identify a family of MERA circuits which locally coarse grains each MPO to itself. If the MPOs form an on-site symmetry, standard techniques of representation theory allow this to be achieved. For MPOs with bond dimension greater than one it is unclear how to apply these techniques. Our approach involves disentangling a local piece out of each MPO. We can then use representation theory to coarse grain this piece, allowing us to identify the desired family of MERA circuits.

This approach may seem counter-intuitive since no local constant depth circuit is capable of disentangling an MPO representation with a nontrivial third cohomology label into an on-site representation. This does not rule out the possibility of disentangling a contiguous region without decoupling the tensors in its complement. More precisely, there may exist constants $b,k\in\N$ such that for all $n\in k\N$ (where $k$ accounts for possible blocking of sites), and MPOs of arbitrary length $N$, sufficiently larger than $n$, there exists some unitary $\disent{n+2b}$ acting on $n+2b$ sites (where $b$ is a buffer depending on the correlation length of the MPO) such that
\begin{align}\label{eqn:disentangling}
	\begin{array}{c}
		\includeTikz{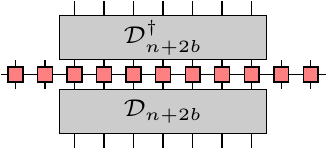}{
			\begin{tikzpicture}[scale=.6]
			\def\a{.25};
			\def\dx{.5};
			\def\maxx{10};
			\draw(-\dx/2,0)--(\maxx*\dx+\dx/2,0);
			\foreach \x in {0,...,\maxx}{
				\draw[shift={(\dx*\x,0)}](0,-\dx/2)--(0,\dx/2);
				\filldraw[tenred,shift={(\dx*\x,0)}](-\a/2,-\a/2)--(-\a/2,\a/2)--(\a/2,\a/2)--(\a/2,-\a/2)--cycle;
			}
			\begin{scope}
			\foreach \x in {2,...,8}{
				\draw[shift={(\dx*\x,4*\a)}](0,-\dx/2)--(0,\dx/2);
			}
			\filldraw[tengrey](2*\dx-\dx/2,\a)--(\maxx*\dx-2*\dx+\dx/2,\a)--(\maxx*\dx-2*\dx+\dx/2,4*\a)--(2*\dx-\dx/2,4*\a)--cycle;
			\node at (5*\dx,2.5*\a){\scriptsize$\mathcal{D}_{n+2b}^\dagger$};
			\end{scope}
			\begin{scope}[yscale=-1]
			\foreach \x in {2,...,8}{
				\draw[shift={(\dx*\x,4*\a)}](0,-\dx/2)--(0,\dx/2);
			}
			\filldraw[tengrey](2*\dx-\dx/2,\a)--(\maxx*\dx-2*\dx+\dx/2,\a)--(\maxx*\dx-2*\dx+\dx/2,4*\a)--(2*\dx-\dx/2,4*\a)--cycle;
			\node at (5*\dx,2.5*\a){\scriptsize$\mathcal{D}_{n+2b}$};
			\end{scope}
			\end{tikzpicture}}
	\end{array}
	&=
	\begin{array}{c}
		\includeTikz{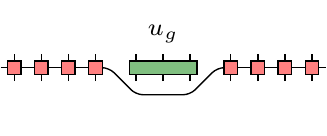}{
			\begin{tikzpicture}[scale=.55]
			\def\a{.25};
			\def\dx{.5};
			\def\maxx{10};
			\begin{scope}
			\foreach \x in {2,...,8}{
				\draw[shift={(\dx*\x,4*\a)},draw=none](0,-\dx/2)--(0,\dx/2);
			}
			\filldraw[fill=none,draw=none](2*\dx-\dx/2,\a)--(\maxx*\dx-2*\dx+\dx/2,\a)--(\maxx*\dx-2*\dx+\dx/2,4*\a)--(2*\dx-\dx/2,4*\a)--cycle;
			\begin{scope}[yscale=-1]
			\foreach \x in {2,...,8}{
				\draw[shift={(\dx*\x,4*\a)},draw=none](0,-\dx/2)--(0,\dx/2);
			}
			\filldraw[fill=none,draw=none](2*\dx-\dx/2,\a)--(\maxx*\dx-2*\dx+\dx/2,\a)--(\maxx*\dx-2*\dx+\dx/2,4*\a)--(2*\dx-\dx/2,4*\a)--cycle;
			\end{scope}
			\end{scope};
			\draw(-\dx/2-\a,0)--(3*\dx-\a,0);
			\draw(4*\dx,0)--(6*\dx,0);
			\draw(7*\dx+\a,0)--(\maxx*\dx+\dx/2+\a,0);
			\draw[rounded corners=2](3*\dx-\a,0)--(3*\dx+\dx/2-\a,0)--(4*\dx,-2*\a)--(6*\dx,-2*\a)--(7*\dx-\dx/2+\a,0)--(7*\dx+\a,0);
			\foreach \x in {0,...,3}{
				\draw[shift={(\dx*\x-\a,0)}](0,-\dx/2)--(0,\dx/2);
				\filldraw[tenred,shift={(\dx*\x-\a,0)}](-\a/2,-\a/2)--(-\a/2,\a/2)--(\a/2,\a/2)--(\a/2,-\a/2)--cycle;
			}
			\foreach \x in {4,...,6}{
				\draw[shift={(\dx*\x,0)}](0,-\dx/2)--(0,\dx/2);
			}
			\filldraw[tengreen](-\a/2+\dx*4,-\a/2)--(-\a/2+\dx*4,\a/2)--(\a/2+\dx*6,\a/2)--(\a/2+\dx*6,-\a/2)--cycle;
			\foreach \x in {7,...,\maxx}{
				\draw[shift={(\dx*\x+\a,0)}](0,-\dx/2)--(0,\dx/2);
				\filldraw[tenred,shift={(\dx*\x+\a,0)}](-\a/2,-\a/2)--(-\a/2,\a/2)--(\a/2,\a/2)--(\a/2,-\a/2)--cycle;
			}
			\node at (5*\dx,2.5*\a){\scriptsize$u_g$};
			\end{tikzpicture}}
	\end{array},
\end{align}
for a local representation $u_g^{(n)}$ acting on $n$ sites. 

This leads to a special form for a MERA tensor that coarse grains $i$ sites into $j$ sites, given by
\begin{align}
	\begin{array}{c}
	\includeTikz{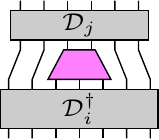}{
		\begin{tikzpicture}[scale=.4]
		\draw(-1.8,-.75)--(-1.8,.75)--(-1.5,1.5)--(-1.5,2.75);
		\draw(-1.2,-.75)--(-1.2,.75)--(-.9,1.5)--(-.9,2.75);
		\draw(1.2,-.75)--(1.2,.75)--(.9,1.5)--(.9,2.75);
		\draw(1.8,-.75)--(1.8,.75)--(1.5,1.5)--(1.5,2.75);
			\draw(-0.6,-.75)--(-0.6,.75);
			\draw(0,-.75)--(0,.75);
			\draw(.6,-.75)--(.6,.75);
			\filldraw[tengrey](-2,-.5)--(2,-.5)--(2,.5)--(-2,.5)--cycle;
			\node at (0,0){\scriptsize$\mathcal{D}_{i}^\dagger$};
		\draw(-0.3,1.5)--(-0.3,2.75);
		\draw(.3,1.5)--(.3,2.75);
		\filldraw[tenpurp](-.8,.75)--(.8,.75)--(.4,1.5)--(-.4,1.5)--cycle;
		\filldraw[tengrey](-1.75,1.75)--(1.75,1.75)--(1.75,2.5)--(-1.75,2.5)--cycle;
		\node at (0,2.125){\scriptsize$\mathcal{D}_{j}$};
		\end{tikzpicture}}
	\end{array}.
\end{align}
In this form the MPO symmetry condition in Eqn.~\ref{eqn:anomsymmetry} becomes 
\begin{align}
\begin{array}{c}
\includeTikz{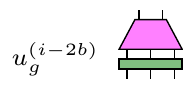}{
	\begin{tikzpicture}[scale=.4]
	\draw[draw=none]  (-.8,-.25)--(.8,-.25)--(.8,2)--(-.8,2)--cycle;
	\draw(-0.3,1.5)--(-0.3,1.75);
	\draw(.3,1.5)--(.3,1.75);
		\draw(-0.6,.5)--(-0.6,.75);
		\draw(0,.5)--(0,.75);
		\draw(0.6,.5)--(0.6,.75);
			\draw(-0.6,0)--(-0.6,.5);
			\draw(0,0)--(0,.5);
			\draw(0.6,0)--(0.6,.5);
	\filldraw[tenpurp](-.8,.75)--(.8,.75)--(.4,1.5)--(-.4,1.5)--cycle;
	\filldraw[tengreen](-.8,.25)--(.8,.25)--(.8,.5)--(-.8,.5)--cycle;
	\node[left] at (-1,.5){\scriptsize$u_g^{(i-2b)}$};
	\end{tikzpicture}}
\end{array}
&=
\begin{array}{c}
\includeTikz{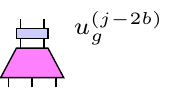}{
	\begin{tikzpicture}[scale=.4]
		\draw[draw=none] (-.8,.25)--(.8,0.25)--(.8,2.5)--(-.8,2.5)--cycle;
	\draw(-0.3,1.5)--(-0.3,1.75);
	\draw(.3,1.5)--(.3,1.75);
	\draw(-0.3,1.5)--(-0.3,2.25);
	\draw(.3,1.5)--(.3,2.25);
	\draw(-0.6,.5)--(-0.6,.75);
	\draw(0.6,.5)--(0.6,.75);
	\draw(0,.5)--(0,.75);
	\filldraw[tenpurp](-.8,.75)--(.8,.75)--(.4,1.5)--(-.4,1.5)--cycle;
	\filldraw[ten](-.4,1.75)--(.4,1.75)--(.4,2)--(-.4,2)--cycle;
	\node[right] at (.75,2){\scriptsize$u_g^{(j-2b)}$};
	\end{tikzpicture}}
\end{array},
\end{align}
which can be handled using standard techniques from representation theory. 
\vspace*{-2mm}
\subsection{A class of anomalous $\zN^3$ MPO symmetries}

We now define a class of anomalous symmetries for the groups $\zN^3$. These symmetries exemplify the role played by an anomalous symmetry both at the boundary of a two dimensional SPT phase and as a duality of distinct one dimensional SPT phases\cite{Chen2013248,Tsui2015330,tsui2015topological,tsui2017conformal}. They occur as the boundary symmetry actions of $\zN^3$ SPTs labeled by a type-III anomaly in two spatial dimensions\cite{propitius}. In addition, they can be seen to act transitively on the set of one dimensional SPT phases with $\zN^2$ symmetry. This particular example is an instance of a more general relation between a two dimensional $\G\times\coho{2}$ SPT and the set of dualities of one dimensional $\G$ SPTs. Further details about the specifics of the $\zN^3$ models, including a fixed point bulk model, bulk to boundary mapping and boundary Hamiltonian, as well as the more general case are contained in Appendix~\ref{appendix:gczx}.

We consider a spin chain with a pair of $N$-dimensional spins at each site. For this discussion, we label the first spin in red and the second in blue. Let $\omega=\exp(2i\pi/N)$ and define the generalized Pauli operators via $ZX=\omega XZ$. Below we work in the basis where $Z$ is the diagonal clock matrix and $X$ is the shift matrix. We define the generalized controlled $X$ and $Z$ operators as
\begin{subequations}
\begin{alignat}{2}
\begin{array}{c}
\includeTikz{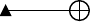}{
	\begin{tikzpicture}
	\filldraw[fill=black,shift={(-.5,-.8)}](-.05,-.05)--(.05,-.05)--(0,.05)--cycle;
	\draw(-.5,-.8)--(.35,-.8);\draw(.25,-.8) circle(.1);
	\draw(.25,-.9)--(.25,-.7);
	\end{tikzpicture}}
\end{array}
&=
\bigl(\begin{array}{c}
\includeTikz{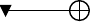}{
	\begin{tikzpicture}
	\filldraw[fill=black,shift={(-.5,-.8)},yscale=-1](-.05,-.05)--(.05,-.05)--(0,.05)--cycle;
	\draw(-.5,-.8)--(.35,-.8);\draw(.25,-.8) circle(.1);
	\draw(.25,-.9)--(.25,-.7);
	\end{tikzpicture}}
\end{array}\bigr)^\dagger=\frac{1}{N}
\sum_{i,j=0}^{N-1}\omega^{ij}Z^i X^j\\
\begin{array}{c}
\includeTikz{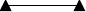}{
	\begin{tikzpicture}
	\filldraw[fill=black,shift={(-.5,-.8)}](-.05,-.05)--(.05,-.05)--(0,.05)--cycle;
	\filldraw[fill=black,shift={(.25,-.8)}](-.05,-.05)--(.05,-.05)--(0,.05)--cycle;
	\draw(-.5,-.8)--(.25,-.8);
	\end{tikzpicture}}
\end{array}
&=
\bigl(\begin{array}{c}
\includeTikz{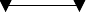}{
	\begin{tikzpicture}
	\filldraw[fill=black,shift={(-.5,-.8)},yscale=-1](-.05,-.05)--(.05,-.05)--(0,.05)--cycle;
	\filldraw[fill=black,shift={(.25,-.8)},yscale=-1](-.05,-.05)--(.05,-.05)--(0,.05)--cycle;
	\draw(-.5,-.8)--(.25,-.8);
	\end{tikzpicture}}
\end{array}\bigr)^\dagger=\frac{1}{N}
\sum_{i,j=0}^{N-1}\omega^{ij}Z^i Z^j
\end{alignat}
\end{subequations}
respectively.

Using the notation $(\alpha_1,\alpha_2,\alpha_3)$ for an element of $\zN^3$, the group action is defined by the generators 
\begin{subequations}\label{eqn:symmetrygroup}
\begin{alignat}{2}
(1,0,0)&\to\bigotimes_{j}\rx{j}\label{eqn:rXsymmetry}\\
(0,1,0)&\to\bigotimes_{j}\bx{j}\label{eqn:bXsymmetry}\\
(0,0,1)&\to\mathcal{C},
\end{alignat}
\end{subequations}
where $\mathcal{C}$ is defined by the (periodic) circuit
\begin{align}
\mathcal{C}&=\begin{array}{c}
\includeTikz{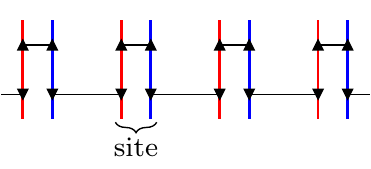}{
\begin{tikzpicture}
\def\dx{.5};
\def\dxa{.3};
\draw[white](0,.7)--(0,-.7);
\foreach \x in {0,2,...,6}{
\draw[red,thick](\x*\dx,-.5)--(\x*\dx,.5);
\filldraw[fill=black,,shift={(\x*\dx,.25)}](-.05,-.05)--(.05,-.05)--(0,.05)--cycle;
\draw[blue,thick](\x*\dx+\dxa,-.5)--(\x*\dx+\dxa,.5);
\filldraw[fill=black,,shift={(\x*\dx+\dxa,.25)}](-.05,-.05)--(.05,-.05)--(0,.05)--cycle;
\draw(\x*\dx,.25)--(\x*\dx+\dxa,.25);
};
\foreach \x in {1,3,...,6}{
\filldraw[fill=black,,shift={(\x*\dx-\dx+\dxa,-.25)},yscale=-1](-.05,-.05)--(.05,-.05)--(0,.05)--cycle;
\filldraw[fill=black,,shift={(\x*\dx+\dx,-.25)},yscale=-1](-.05,-.05)--(.05,-.05)--(0,.05)--cycle;
\draw(\x*\dx-\dx+\dxa,-.25)--(\x*\dx+\dx,-.25);
};
\filldraw[fill=black,,shift={(0,-.25)},yscale=-1](-.05,-.05)--(.05,-.05)--(0,.05)--cycle;
\filldraw[fill=black,,shift={(6*\dx+\dxa,-.25)},yscale=-1](-.05,-.05)--(.05,-.05)--(0,.05)--cycle;
\draw(6*\dx+\dxa,-.25)--(6*\dx+1.75*\dxa,-.25);
\draw(0,-.25)--(-.75*\dxa,-.25);
\draw [decorate,decoration={brace,amplitude=3pt,mirror,raise=1pt},yshift=0pt]
(2*\dx-\dxa/5,-.5) -- (2*\dx+1.2*\dxa,-.5) node [black,midway,below,yshift=-2] {\footnotesize site};
	\end{tikzpicture}}
	\end{array}.
\end{align}

The symmetry operators can be realized using a translationally invariant MPO with on-site tensor defined by
\begin{align}
\begin{array}{c}
\includeTikz{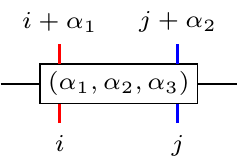}{
\begin{tikzpicture}[scale=.8]
\draw(0,0)--(2,0)--(2,.5)--(0,.5)--cycle;
\node at(1,.25){\scriptsize$(\alpha_1,\alpha_2,\alpha_3)$};
\draw(-.5,.25)--(0,.25);
\draw(2,.25)--(2.5,.25);
\draw[thick,red](.25,-.25)--(.25,0);
\draw[thick,blue](1.75,-.25)--(1.75,0);
\draw[thick,red](.25,.5)--(.25,.75);
\draw[thick,blue](1.75,.5)--(1.75,.75);
\node[below] at (.25,-.25){\scriptsize$i$};\node[below] at (1.75,-.25){\scriptsize$j$};
\node[above] at (.25,.75){\scriptsize$i+\alpha_1$};
\node[above] at (1.75,.75){\scriptsize$j+\alpha_2$};
\end{tikzpicture}}
\end{array}&=
\sum_{k=0}^{N-1}\omega^{j \alpha_3(k-i)}\ket{i}\hspace{-1.1mm}\bra{k},
\end{align}
with all other elements being zero.
The reduction tensor (defined in Appendix~\ref{appendix:thirdcoho}) associated to these MPOs is given by
\begin{align}
\begin{array}{c}
\includeTikz{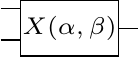}{
\begin{tikzpicture}[scale=.8]
\draw(0,.15)--(1.25,.15)--(1.25,.85)--(0,.85)--cycle;
\draw(-.25,.35)--(0,.35);
\draw(-.25,.75)--(0,.75);
\draw(1.25,.5)--(1.5,.5);
\node at (.625,.5) {\scriptsize$X(\alpha,\beta)$};
\end{tikzpicture}}
\end{array}&=
\sum_{x=0}^{N-1}\omega^{-x\alpha_2\beta_3}\left|\begin{matrix}
x+\alpha_1\\
x
\end{matrix}\right\rangle\hspace{-1.1mm}\bra{x}.
\end{align}
 From this, one can verify that this MPO representation has cocycle $\phi(\alpha,\beta,\gamma)=\omega^{\alpha_1\beta_2\gamma_3}$ which is a representative of the root `type-III' anomaly\cite{propitius}.
 
\subsection{Symmetric MERA tensors}
The disentangling circuit, as defined in Eqn.~\ref{eqn:disentangling}, for this representation is given by 
\begin{align}
\disent{2K}=\prod\limits_{j=1}^{K-1}CX_{1,2j+1}CX_{2K,2j},
\end{align}
and the residual local symmetry is given by 
\begin{align}
u_{(\alpha_1,\alpha_2,\alpha_3)}^{(2K-2)}=\biggl(\prod\limits_{j=1}^{K-1} CZ_{2j,2j+1} \prod\limits_{j=2}^{K-1}CZ^{\dagger}_{2j-1,2j}\biggr)^{\alpha_3}.
\end{align}
For further details see Appendix~\ref{appendix:generalansatz}. This leads to the ansatz for MERA tensors
\begin{align}\label{eqn:42decomp}
\begin{array}{c}
	\includeTikz{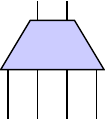}{
		\begin{tikzpicture}[scale=1]
		\def\x{.3}
		\def\dx{.15}
		\def\xi{.25}
		\def\xj{.75}
		\def\ymin{.5}
		\def\ymax{.7}
		\filldraw[ten](\xi-\dx/2,0)--(3*\x+\xi+0.5*\dx,0)--(2*\x+\xi+.5*\dx,.5)--(1*\x+\xi-0.5*\dx,.5)--cycle;
		\foreach \i in {0,...,3}{
			\draw(\i*\x+0*\dx+\xi,0)--(\i*\x+0*\dx+\xi,-\ymin);
		}
		\draw (\x+\xi,.5)--(\x+\xi,\ymax);
		\draw (2*\x+\xi,.5)--(2*\x+\xi,\ymax);
		\end{tikzpicture}
	}
\end{array}
&=
\begin{array}{c}
	\includeTikz{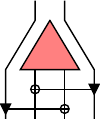}{
		\begin{tikzpicture}[scale=1]
		\def\x{.3}
		\def\dx{.15}
		\def\xi{.25}
		\def\xj{.75}
		\def\ymin{.5}
		\def\ymax{.7}
		\def\dn{.2}
		\filldraw[tenred,yscale=1](0*\x+\xi+\dx,0)--(3*\x+\xi-\dx,0)--(2*\x+\xi-\dx,.5)--(1*\x+\xi+\dx,.5)--cycle;
		\foreach \i in {0,...,3}{
			\draw(\i*\x+0*\dx+\xi,0)--(\i*\x+0*\dx+\xi,-\ymin);
		}
		\draw (\xi,-\ymin)--(\xi,0)--(\x+\xi,.5)--(\x+\xi,\ymax);
		\draw (3*\x+\xi,-\ymin)--(3*\x+\xi,0)--(2*\x+\xi,.5)--(2*\x+\xi,\ymax);
		\begin{scope}[shift={(0,-.2)}]
		\def\n{0};
		\filldraw[black,shift={(3*\x+\xi,-\dn*\n)},yscale=-1](-.05,-.05)--(.05,-.05)--(0,.05)--cycle;
		\draw (1*\x+\xi-.3*\dx,-\dn*\n)--(3*\x+\xi,-\dn*\n);
		\draw (1*\x+\xi,-\dn*\n) circle (.3*\dx);
		\def\n{1};
		\filldraw[black,shift={(\xi+0*\dx,-\dn*\n)},yscale=-1](-.05,-.05)--(.05,-.05)--(0,.05)--cycle;
		\draw (\xi,-\dn*\n)--(2*\x+\xi+.3*\dx,-\dn*\n);
		\draw (2*\x+0*\dx+\xi,-\dn*\n) circle (.3*\dx);
		\end{scope}
		\end{tikzpicture}
	}
\end{array},
\end{align}
 which allows the symmetry to be enforced by a local condition on each tensor.

The symmetry can then be enforced by ensuring the residual tensors obey the local conditions
\begin{align}
\begin{array}{c}
\includeTikz{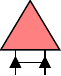}{
	\begin{tikzpicture}[scale=1]
	\def\x{.3}
	\def\dx{.15}
	\def\xi{.25}
	\def\xj{.75}
	\def\ymin{.25}
	\def\ymax{.7}
	\def\dn{.2}
	\filldraw[tenred,yscale=1](0*\x+\xi+\dx,0)--(3*\x+\xi-\dx,0)--(2*\x+\xi-\dx,.5)--(1*\x+\xi+\dx,.5)--cycle;
	\foreach \i in {1,...,2}{
		\draw(\i*\x+0*\dx+\xi,0)--(\i*\x+0*\dx+\xi,-\ymin);
	}
	\filldraw[fill=black,shift={(1*\x+0*\dx+\xi,-.125)}](-.05,-.05)--(.05,-.05)--(0,.05)--cycle;
	\filldraw[fill=black,shift={(2*\x+0*\dx+\xi,-.125)}](-.05,-.05)--(.05,-.05)--(0,.05)--cycle;
	\draw(1*\x+0*\dx+\xi,-.125)--(2*\x+0*\dx+\xi,-.125);
	\end{tikzpicture},
}
\end{array}      
&=
\begin{array}{c}
\includeTikz{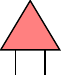}{
	\begin{tikzpicture}[scale=1]
	\def\x{.3}
	\def\dx{.15}
	\def\xi{.25}
	\def\xj{.75}
	\def\ymin{.25}
	\def\ymax{.7}
	\def\dn{.2}
	\filldraw[tenred,yscale=1](0*\x+\xi+\dx,0)--(3*\x+\xi-\dx,0)--(2*\x+\xi-\dx,.5)--(1*\x+\xi+\dx,.5)--cycle;
	\foreach \i in {1,...,2}{
		\draw(\i*\x+0*\dx+\xi,0)--(\i*\x+0*\dx+\xi,-\ymin);
	}
	\filldraw[nonesty,shift={(1*\x+0*\dx+\xi,-.125)}](-.05,-.05)--(.05,-.05)--(0,.05)--cycle;
	\filldraw[nonesty,shift={(2*\x+0*\dx+\xi,-.125)}](-.05,-.05)--(.05,-.05)--(0,.05)--cycle;
	\draw[nonesty](1*\x+0*\dx+\xi,-.125)--(2*\x+0*\dx+\xi,-.125);
	\end{tikzpicture}
}
\end{array},             
\end{align}
which can be achieved using standard techniques of representation theory.  We remark that the on-site $\zN^2$ symmetry is automatically enforced, without any further constraints.

Since the action can be applied locally, this ansatz class can also be used to investigate how the group acts on numerically optimized states which have not been constrained to be invariant. This allows investigation of theories which are dual under anomalous group actions.

The constraint in Eqn.~\ref{eqn:42decomp} was used in an exact renormalization scheme introduced in Ref.~\onlinecite{KY} for the case of a $\zt\times\zt$ symmetry\cite{aroon}. The form of the information transmitted to the next scale of renormalization is extremely restricted in this case. By considering more spins per site we find a less restrictive ansatz, described in Appendix~\ref{appendix:generalansatz}, capable of attaining accurate results as demonstrated in Section~\ref{Numerics}. The scheme described in Ref.~\onlinecite{KY} does not see similar improvement at larger blocking on a model which is unitarily equivalent to the one considered here\cite{aroon}. After blocking at least two spins per site, our ansatz cannot be captured by the approach of Ref.~\onlinecite{KY}.

Analogous circuits exist for all MERA such that the number of ingoing/outgoing $N$-dimensional indices is even. This leads to a family of symmetric MERA with increasing bond dimension and a larger number of variational parameters.
Eqn.~\ref{eqn:42decomp} can also be generalized to other MERA schemes, such as the ternary MERA as discussed in Appendix~\ref{appendix:generalansatz}.

\section{Example: A $\zt^3$ symmetric model}
\label{Numerics}

In this section we focus on the $N=2$ case of the ansatz described in the previous section. We consider a particular Hamiltonian which transforms under the type-III anomalous $\zt^3$ group action. This Hamiltonian has three critical lines, one is symmetric and the other two are dual under the group action. 
We numerically optimize over the ansatz class presented in the previous section along these three lines. 
We present resulting conformal data for the local fields along each line, and for two nontrivial topological sectors along the symmetric line. Furthermore, we numerically implement the duality on the remaining pair of lines.
Finally, we demonstrate that the symmetric line is a gapless phase protected by the anomalous symmetry and translation. 

For a MERA with bond dimension 8 corresponding to three qubits per site, the ansatz for the tensors is 

\begin{align}
\begin{array}{c}
    \includeTikz{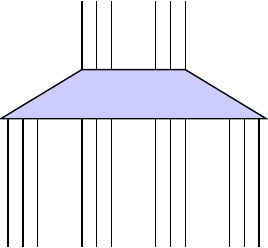}{
    \begin{tikzpicture}[scale=1]
      \def\x{.75}
      \def\dx{.15}
      \def\xi{.25}
      \def\xj{.75}
      \def\ymin{1.3}
      \def\ymax{1.2}
      \filldraw[ten](\xi-\dx/2,0)--(3*\x+\xi+2.5*\dx,0)--(2*\x+\xi+2*\dx,.5)--(1*\x+\xi,.5)--cycle;
      \foreach \i in {0,...,3}{
      \draw(\i*\x+0*\dx+\xi,0)--(\i*\x+0*\dx+\xi,-\ymin);
      \draw(\i*\x+1*\dx+\xi,0)--(\i*\x+1*\dx+\xi,-\ymin);
      \draw(\i*\x+2*\dx+\xi,0)--(\i*\x+2*\dx+\xi,-\ymin);
      }
      \draw (\x+\xi+0*\dx,.5)--(\x+\xi+0*\dx,\ymax);
     \draw (\x+\xi+\dx,.5)--(\x+\xi+\dx,\ymax);
     \draw (\x+\xi+2*\dx,.5)--(\x+\xi+2*\dx,\ymax);
     \draw (2*\x+\xi,.5)--(2*\x+\xi,\ymax);
     \draw (2*\x+\xi+1*\dx,.5)--(2*\x+\xi+1*\dx,\ymax);
     \draw (2*\x+\xi+2*\dx,.5)--(2*\x+\xi+2*\dx,\ymax);
    \end{tikzpicture}
    }
  \end{array}
&=
 \begin{array}{c}
    \includeTikz{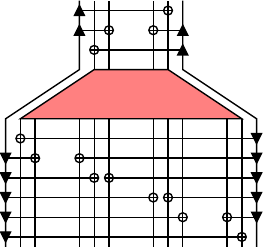}{
    \begin{tikzpicture}[scale=1]
      \def\x{.75}
      \def\dx{.15}
      \def\xi{.25}
      \def\xj{.75}
      \def\ymin{1.3}
      \def\ymax{1.2}
      \def\dn{.2}
      \filldraw[tenred](\xi+\dx,0)--(3*\x+\xi+1*\dx,0)--(2*\x+\xi+1*\dx,.5)--(1*\x+\xi+1*\dx,.5)--cycle;
      \foreach \i in {0,...,3}{
      \draw(\i*\x+0*\dx+\xi,0)--(\i*\x+0*\dx+\xi,-\ymin);
      \draw(\i*\x+1*\dx+\xi,0)--(\i*\x+1*\dx+\xi,-\ymin);
      \draw(\i*\x+2*\dx+\xi,0)--(\i*\x+2*\dx+\xi,-\ymin);
      }
     \draw (\xi,-\ymin)--(\xi,0)--(\x+\xi,.5)--(\x+\xi,\ymax);
     \draw (\x+\xi+\dx,.5)--(\x+\xi+\dx,\ymax);
     \draw (\x+\xi+2*\dx,.5)--(\x+\xi+2*\dx,\ymax);
     \draw (2*\x+\xi,.5)--(2*\x+\xi,\ymax);
     \draw (2*\x+\xi+1*\dx,.5)--(2*\x+\xi+1*\dx,\ymax);
     \draw (3*\x+\xi+2*\dx,-\ymin)--(3*\x+\xi+2*\dx,0)--(2*\x+\xi+2*\dx,.5)--(2*\x+\xi+2*\dx,\ymax);
     \begin{scope}[shift={(0,.7)}]
     \def\n{0};
     \filldraw[black,shift={(2*\x+\xi+2*\dx,\dn*\n)}](-.05,-.05)--(.05,-.05)--(0,.05)--cycle;
     \draw (1*\x+\xi+.7*\dx,\dn*\n)--(2*\x+2*\dx+\xi,\dn*\n);
     \draw (1*\x+\dx+\xi,\dn*\n) circle (.3*\dx);
      \def\n{1};
      \filldraw[black,shift={(\x+\xi+0*\dx,\dn*\n)}](-.05,-.05)--(.05,-.05)--(0,.05)--cycle;
      \draw (1*\x+\xi,\dn*\n)--(1*\x+2.3*\dx+\xi,\dn*\n);
      \draw (1*\x+2*\dx+\xi,\dn*\n) circle (.3*\dx);
      \filldraw[black,shift={(2*\x+\xi+2*\dx,\dn*\n)}](-.05,-.05)--(.05,-.05)--(0,.05)--cycle;
      \draw (2*\x+\xi-.3*\dx,\dn*\n)--(2*\x+2*\dx+\xi,\dn*\n);
      \draw (2*\x+0*\dx+\xi,\dn*\n) circle (.3*\dx);
      \def\n{2};
      \filldraw[black,shift={(\x+\xi+0*\dx,\dn*\n)}](-.05,-.05)--(.05,-.05)--(0,.05)--cycle;
      \draw (1*\x+\xi,\dn*\n)--(2*\x+1.3*\dx+\xi,\dn*\n);
      \draw (2*\x+1*\dx+\xi,\dn*\n) circle (.3*\dx);
     \end{scope}
      \begin{scope}[shift={(0,-.2)}]
	      \def\n{0};
	      \filldraw[black,shift={(3*\x+\xi+2*\dx,-\dn*\n)},yscale=-1](-.05,-.05)--(.05,-.05)--(0,.05)--cycle;
	      \draw (0*\x+\xi+.7*\dx,-\dn*\n)--(3*\x+2*\dx+\xi,-\dn*\n);
	      \draw (0*\x+\dx+\xi,-\dn*\n) circle (.3*\dx);
	      \begin{scope}
		      \def\n{1};
		      \filldraw[black,shift={(\xi+0*\dx,-\dn*\n)},yscale=-1](-.05,-.05)--(.05,-.05)--(0,.05)--cycle;
		      \draw (\xi,-\dn*\n)--(2*\dx+\xi+.3*\dx,-\dn*\n);
		      \draw (+2*\dx+\xi,-\dn*\n) circle (.3*\dx);
		      \filldraw[black,shift={(3*\x+\xi+2*\dx,-\dn*\n)},yscale=-1](-.05,-.05)--(.05,-.05)--(0,.05)--cycle;
		      \draw (\x+\xi-.3*\dx,-\dn*\n)--(3*\x+2*\dx+\xi,-\dn*\n);
		      \draw (\x+\xi,-\dn*\n) circle (.3*\dx);
	      \end{scope}
	      \begin{scope}
	      	      \def\n{2};
	      	      \filldraw[black,shift={(\xi+0*\dx,-\dn*\n)},yscale=-1](-.05,-.05)--(.05,-.05)--(0,.05)--cycle;
	      	      \draw (\xi,-\dn*\n)--(\x+1*\dx+\xi+.3*\dx,-\dn*\n);
	      	      \draw (\x+1*\dx+\xi,-\dn*\n) circle (.3*\dx);
	      	      \filldraw[black,shift={(3*\x+\xi+2*\dx,-\dn*\n)},yscale=-1](-.05,-.05)--(.05,-.05)--(0,.05)--cycle;
	      	      \draw (1*\x+\xi+1.7*\dx,-\dn*\n)--(3*\x+2*\dx+\xi,-\dn*\n);
	      	      \draw (\x+\xi+2*\dx,-\dn*\n) circle (.3*\dx);
	       \end{scope}
	       \begin{scope}
	       	      \def\n{3};
	       	      \filldraw[black,shift={(\xi+0*\dx,-\dn*\n)},yscale=-1](-.05,-.05)--(.05,-.05)--(0,.05)--cycle;
	       	      \draw (\xi,-\dn*\n)--(2*\x+0*\dx+\xi+.3*\dx,-\dn*\n);
	       	      \draw (2*\x+0*\dx+\xi,-\dn*\n) circle (.3*\dx);
	       	      \filldraw[black,shift={(3*\x+\xi+2*\dx,-\dn*\n)},yscale=-1](-.05,-.05)--(.05,-.05)--(0,.05)--cycle;
	       	      \draw (2*\x+\xi+.7*\dx,-\dn*\n)--(3*\x+2*\dx+\xi,-\dn*\n);
	       	      \draw (2*\x+\dx+\xi,-\dn*\n) circle (.3*\dx);
	       \end{scope}
	       \begin{scope}
	       	      \def\n{4};
	       	      \filldraw[black,shift={(\xi+0*\dx,-\dn*\n)},yscale=-1](-.05,-.05)--(.05,-.05)--(0,.05)--cycle;
	       	      \draw (\xi,-\dn*\n)--(2*\x+2*\dx+\xi+.3*\dx,-\dn*\n);
	       	      \draw (2*\x+2*\dx+\xi,-\dn*\n) circle (.3*\dx);
	       	      \filldraw[black,shift={(3*\x+\xi+2*\dx,-\dn*\n)},yscale=-1](-.05,-.05)--(.05,-.05)--(0,.05)--cycle;
	       	      \draw (3*\x+\xi-.3*\dx,-\dn*\n)--(3*\x+2*\dx+\xi,-\dn*\n);
	       	      \draw (3*\x+0*\dx+\xi,-\dn*\n) circle (.3*\dx);
	       \end{scope}
	       \def\n{5};
	       \filldraw[black,shift={(\xi+0*\dx,-\dn*\n)},yscale=-1](-.05,-.05)--(.05,-.05)--(0,.05)--cycle;
	       \draw (\xi,-\dn*\n)--(3*\x+1*\dx+\xi+.3*\dx,-\dn*\n);
	       \draw (3*\x+1*\dx+\xi,-\dn*\n) circle (.3*\dx);
       \end{scope}
    \end{tikzpicture}
    }
  \end{array},\label{eqn:circuitconstraint}
\end{align}
with symmetry constraint
  
\begin{align}
\begin{array}{c}
    \includeTikz{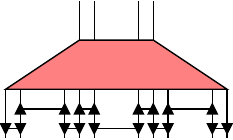}{
    \begin{tikzpicture}[scale=1]
          \def\x{.75}
          \def\dx{.15}
          \def\xi{.25}
          \def\xj{.75}
          \def\ymin{.5}
          \def\ymax{.9}
          \def\dn{.2}
          \filldraw[tenred](\xi+\dx,0)--(3*\x+\xi+1*\dx,0)--(2*\x+\xi+1*\dx,.5)--(1*\x+\xi+1*\dx,.5)--cycle;
          \draw(1*\dx+\xi,0)--(1*\dx+\xi,-\ymin);
          \draw(2*\dx+\xi,0)--(2*\dx+\xi,-\ymin);
          \foreach \i in {1,...,2}{
          \draw(\i*\x+0*\dx+\xi,0)--(\i*\x+0*\dx+\xi,-\ymin);
          \draw(\i*\x+1*\dx+\xi,0)--(\i*\x+1*\dx+\xi,-\ymin);
          \draw(\i*\x+2*\dx+\xi,0)--(\i*\x+2*\dx+\xi,-\ymin);
          }
         \draw(3*\x+0*\dx+\xi,0)--(3*\x+0*\dx+\xi,-\ymin);
         \draw(3*\x+1*\dx+\xi,0)--(3*\x+1*\dx+\xi,-\ymin);
         \draw (\x+\xi+\dx,.5)--(\x+\xi+\dx,\ymax);
         \draw (\x+\xi+2*\dx,.5)--(\x+\xi+2*\dx,\ymax);
         \draw (2*\x+\xi,.5)--(2*\x+\xi,\ymax);
         \draw (2*\x+\xi+1*\dx,.5)--(2*\x+\xi+1*\dx,\ymax);
         \begin{scope}[shift={(0,-2*\dn)}]
		\filldraw[black,shift={(0*\x+\xi+1*\dx,0)},yscale=-1](-.05,-.05)--(.05,-.05)--(0,.05)--cycle;
		\draw(0*\x+\xi+1*\dx,0)--(0*\x+\xi+2*\dx,0);
		\filldraw[black,shift={(0*\x+\xi+2*\dx,0)},yscale=-1](-.05,-.05)--(.05,-.05)--(0,.05)--cycle;
		\filldraw[black,shift={(1*\x+\xi+0*\dx,0)},yscale=-1](-.05,-.05)--(.05,-.05)--(0,.05)--cycle;
		\draw(1*\x+\xi+0*\dx,0)--(1*\x+\xi+1*\dx,0);
		\filldraw[black,shift={(1*\x+\xi+1*\dx,0)},yscale=-1](-.05,-.05)--(.05,-.05)--(0,.05)--cycle;
				\filldraw[black,shift={(1*\x+\xi+2*\dx,0)},yscale=-1](-.05,-.05)--(.05,-.05)--(0,.05)--cycle;
				\draw(1*\x+\xi+2*\dx,0)--(2*\x+\xi+0*\dx,0);
				\filldraw[black,shift={(2*\x+\xi+0*\dx,0)},yscale=-1](-.05,-.05)--(.05,-.05)--(0,.05)--cycle;
				\filldraw[black,shift={(2*\x+\xi+1*\dx,0)},yscale=-1](-.05,-.05)--(.05,-.05)--(0,.05)--cycle;
				\draw(2*\x+\xi+1*\dx,0)--(2*\x+\xi+2*\dx,0);
				\filldraw[black,shift={(2*\x+\xi+2*\dx,0)},yscale=-1](-.05,-.05)--(.05,-.05)--(0,.05)--cycle;
		\filldraw[black,shift={(3*\x+\xi+0*\dx,0)},yscale=-1](-.05,-.05)--(.05,-.05)--(0,.05)--cycle;
				\draw(3*\x+\xi+0*\dx,0)--(3*\x+\xi+1*\dx,0);
				\filldraw[black,shift={(3*\x+\xi+1*\dx,0)},yscale=-1](-.05,-.05)--(.05,-.05)--(0,.05)--cycle;
         \end{scope}
         \begin{scope}[shift={(0,-1*\dn)}]
         		\filldraw[black,shift={(0*\x+\xi+2*\dx,0)}](-.05,-.05)--(.05,-.05)--(0,.05)--cycle;
         		\draw(0*\x+\xi+2*\dx,0)--(1*\x+\xi+0*\dx,0);
         		\filldraw[black,shift={(1*\x+\xi+0*\dx,0)}](-.05,-.05)--(.05,-.05)--(0,.05)--cycle;
         		\filldraw[black,shift={(1*\x+\xi+1*\dx,0)}](-.05,-.05)--(.05,-.05)--(0,.05)--cycle;
         		\draw(1*\x+\xi+1*\dx,0)--(1*\x+\xi+2*\dx,0);
         		\filldraw[black,shift={(1*\x+\xi+2*\dx,0)}](-.05,-.05)--(.05,-.05)--(0,.05)--cycle;
         				\filldraw[black,shift={(2*\x+\xi+0*\dx,0)}](-.05,-.05)--(.05,-.05)--(0,.05)--cycle;
         				\draw(2*\x+\xi+0*\dx,0)--(2*\x+\xi+1*\dx,0);
         				\filldraw[black,shift={(2*\x+\xi+1*\dx,0)}](-.05,-.05)--(.05,-.05)--(0,.05)--cycle;
         				\filldraw[black,shift={(2*\x+\xi+2*\dx,0)}](-.05,-.05)--(.05,-.05)--(0,.05)--cycle;
         				\draw(2*\x+\xi+2*\dx,0)--(3*\x+\xi+0*\dx,0);
         				\filldraw[black,shift={(3*\x+\xi+0*\dx,0)}](-.05,-.05)--(.05,-.05)--(0,.05)--cycle;
                  \end{scope}
        \end{tikzpicture}
    }
  \end{array}
&=
 \begin{array}{c}
    \includeTikz{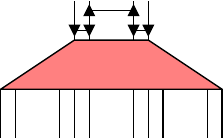}{
    \begin{tikzpicture}[scale=1]
              \def\x{.75}
              \def\dx{.15}
              \def\xi{.25}
              \def\xj{.75}
              \def\ymin{.5}
              \def\ymax{.9}
              \def\dn{.2}
              \filldraw[tenred](\xi+\dx,0)--(3*\x+\xi+1*\dx,0)--(2*\x+\xi+1*\dx,.5)--(1*\x+\xi+1*\dx,.5)--cycle;
              \draw(1*\dx+\xi,0)--(1*\dx+\xi,-\ymin);
              \draw(2*\dx+\xi,0)--(2*\dx+\xi,-\ymin);
              \foreach \i in {1,...,2}{
              \draw(\i*\x+0*\dx+\xi,0)--(\i*\x+0*\dx+\xi,-\ymin);
              \draw(\i*\x+1*\dx+\xi,0)--(\i*\x+1*\dx+\xi,-\ymin);
              \draw(\i*\x+2*\dx+\xi,0)--(\i*\x+2*\dx+\xi,-\ymin);
              }
             \draw(3*\x+0*\dx+\xi,0)--(3*\x+0*\dx+\xi,-\ymin);
             \draw(3*\x+1*\dx+\xi,0)--(3*\x+1*\dx+\xi,-\ymin);
             \draw (\x+\xi+\dx,.5)--(\x+\xi+\dx,\ymax);
             \draw (\x+\xi+2*\dx,.5)--(\x+\xi+2*\dx,\ymax);
             \draw (2*\x+\xi,.5)--(2*\x+\xi,\ymax);
             \draw (2*\x+\xi+1*\dx,.5)--(2*\x+\xi+1*\dx,\ymax);
             \begin{scope}[shift={(0,3*\dn)}]
    		\filldraw[black,shift={(1*\x+\xi+1*\dx,0)},yscale=-1](-.05,-.05)--(.05,-.05)--(0,.05)--cycle;
    		\draw(1*\x+\xi+1*\dx,0)--(1*\x+\xi+2*\dx,0);
    		\filldraw[black,shift={(1*\x+\xi+2*\dx,0)},yscale=-1](-.05,-.05)--(.05,-.05)--(0,.05)--cycle;
    		\filldraw[black,shift={(2*\x+\xi+0*\dx,0)},yscale=-1](-.05,-.05)--(.05,-.05)--(0,.05)--cycle;
    		\draw(2*\x+\xi+0*\dx,0)--(2*\x+\xi+1*\dx,0);
    		\filldraw[black,shift={(2*\x+\xi+1*\dx,0)},yscale=-1](-.05,-.05)--(.05,-.05)--(0,.05)--cycle;
    		\filldraw[black,shift={(1*\x+\xi+2*\dx,\dn)},yscale=1](-.05,-.05)--(.05,-.05)--(0,.05)--cycle;
    		\draw(1*\x+\xi+2*\dx,\dn)--(2*\x+\xi+0*\dx,\dn);
    		\filldraw[black,shift={(2*\x+\xi+0*\dx,\dn)},yscale=1](-.05,-.05)--(.05,-.05)--(0,.05)--cycle;
             \end{scope}
            \end{tikzpicture}
    }
  \end{array}.\label{eqn:CZsymmetry}
\end{align}

This tensor contains all degrees of freedom which are not fixed by the symmetry, so can be optimized over.
\subsection{Family of Hamiltonians}

\begin{figure}
	\includeTikz{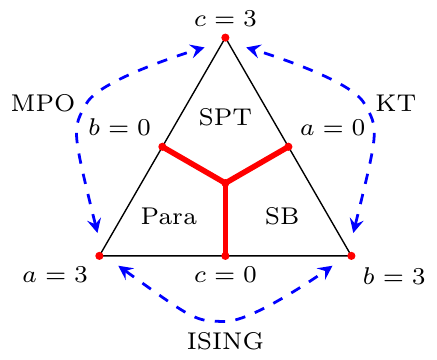}{
		\begin{tikzpicture}[scale=1.28]
		\pgfmathsetmacro{\za}{sqrt(3)};
		\coordinate (a) at (-1,-\za/2);\coordinate(b) at (1,-\za/2);\coordinate (c) at (0,\za/2);
		\coordinate (ce) at (0,-\za/6);
		\coordinate (xab) at ($(a)!.5!(b)$);
		\coordinate (xac) at ($(a)!.5!(c)$);
		\coordinate (xbc) at ($(b)!.5!(c)$);
		\coordinate (eab) at ($(xab)!-.3!(c)$);
		\coordinate (da) at ($(a)!.15!(eab)$);\coordinate (db) at ($(b)!.15!(eab)$);
		\coordinate (eac) at ($(xac)!-.4!(b)$);
		\coordinate (ma) at ($(a)!.15!(eac)$);\coordinate (mc) at ($(c)!.15!(eac)$);
		\coordinate (ebc) at ($(xbc)!-.4!(a)$);
		\coordinate (nb) at ($(b)!.15!(ebc)$);\coordinate (nc) at ($(c)!.15!(ebc)$);
		\draw (a)--(b)--(c)--(a);
		\node[anchor=north east] at (a) {\scriptsize $a=3$};
		\node[anchor=north west] at (b) {\scriptsize $b=3$};
		\node[anchor=south] at (c) {\scriptsize $c=3$};
		\node[anchor=south west] at (xbc) {\scriptsize $a=0$};
		\node[anchor=south east] at (xac) {\scriptsize $b=0$};
		\node[anchor=north] at (xab) {\scriptsize $c=0$};
		\node at ($(a)!.55!(ce)$) {\scriptsize Para};
		\node at ($(b)!.55!(ce)$) {\scriptsize SB};
		\node at ($(c)!.55!(ce)$) {\scriptsize SPT};
		\draw[red,ultra thick] (xab)--(ce);\draw[red,ultra thick] (xac)--(ce);\draw[red,ultra thick] (xbc)--(ce);
		\filldraw[red] (xab) circle (.75pt);\filldraw[red] (xac) circle (.75pt);\filldraw[red] (xbc) circle (.75pt);\filldraw[red] (ce) circle (.75pt);
		\filldraw[red] (a) circle (.75pt);\filldraw[red] (b) circle (.75pt);\filldraw[red] (c) circle (.75pt);
		\draw[stealth-stealth,blue,thick,style=dashed] plot [smooth] coordinates {(da) (eab) (db)};\node[anchor=north] at (eab) {\scriptsize ISING};
		\draw[stealth-stealth,blue,thick,style=dashed] plot [smooth] coordinates {(ma) (eac) (mc)};\node[anchor=east] at (eac) {\scriptsize MPO};
		\draw[stealth-stealth,blue,thick,style=dashed] plot [smooth] coordinates {(nb) (ebc) (nc)};\node[anchor=west] at (ebc) {\scriptsize KT};
		\end{tikzpicture}}
	\caption{Phase diagram of the abc model where $a+b+c=3$. SB=Symmetry breaking, ferromagnetic phase. SPT=$\zt\times\zt$ symmetry protected topological phase. Para=Paramagnetic/disordered phase. RG fixed points are indicated in red, and the dashed blue lines indicate the unitary mappings between the phases. ISING=Ising duality map, KT=(Generalized) Kennedy-Tasaki transformation\cite{PhysRevB.45.304,Else2012b}, MPO=action of (1,1,1) defined in Eqn.~\ref{eqn:symmetrygroup}.}\label{Fig:abcphasediagram}
\end{figure}
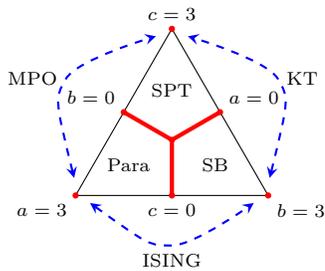  

The Hamiltonian we study is
  \begin{align}
    H=-&a\sum (\rx{j}+\bx{j})- b\sum (\rz{j}\rz{j+1}+\bz{j}\bz{j+1})\nonumber\\
    &-c\sum (\rz{j}\bx{j}\rz{j+1}+\bz{j}\rx{j+1}\bz{j+1}),\label{eqn:hamiltonian}
  \end{align}
for positive values of $(a,b,c)$. Here
 $\rx{j}(\rz{j})$ and $\bx{j}(\bz{j})$ are the qubit Pauli operators action on the first and second qubit on site $j$. This model, which we refer to as the \emph{abc model}, has a rich phase diagram as depicted in \fref{abcphasediagram}, possessing fully symmetric disordered and SPT phases, in addition to a fully symmetry breaking phase. For all values of $(a,b,c)$, this Hamiltonian has an on-site $\zt\times\zt$ symmetry corresponding to Eqn.~\ref{eqn:rXsymmetry} and Eqn.~\ref{eqn:bXsymmetry}, whilst the anomalous action exchanges the terms with strength $a$ and $c$, so is only a symmetry when $a=c$. The SPT phase is protected by the on-site symmetry.

We note that unitarily equivalent models have previously been studied\cite{Yang1987,Baake1987a,Alcaraz1988JPHYS,Alcaraz1988,Bridgemanmasters,Bridgeman2015}. The critical lines in this model can all be exchanged by (nonlocal) unitary transformations, so all are known to be described by a conformal field theory (CFT) with central charge 1. Additionally, the ground state energy along each of these lines is known\cite{Alcaraz1988,Bridgemanmasters,Bridgeman2015}. 

\begin{figure}
	\begin{tabular}{r}
		\includeTikz{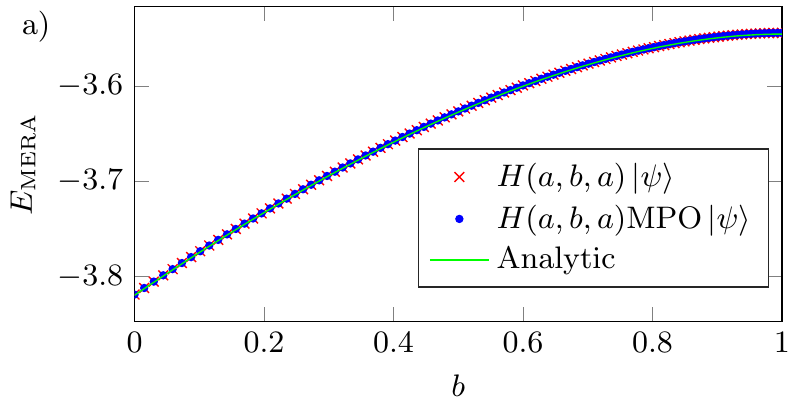}{
			\setlength\figureheight{.37\columnwidth}
			\setlength\figurewidth{.8\columnwidth}
			\input{Graphs/bEnergy}}
		\\
		\includeTikz{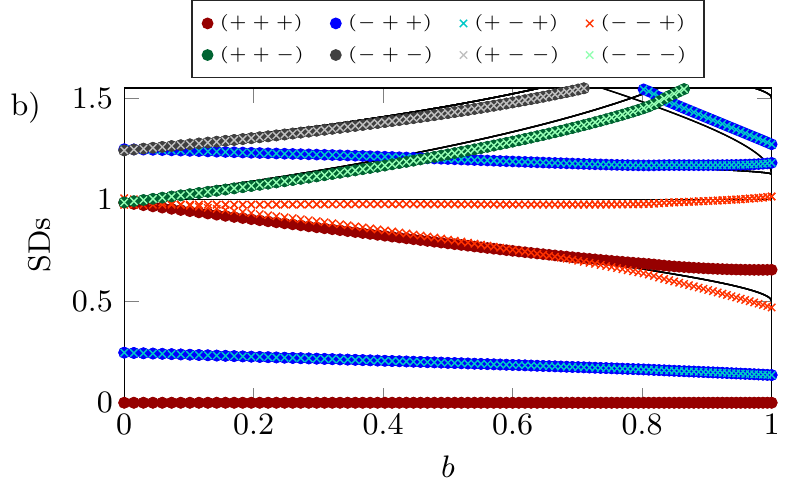}{
			\setlength\figureheight{.37\columnwidth}
			\setlength\figurewidth{.8\columnwidth}
			\input{Graphs/LocalAveraged}}
		\\
		\includeTikz{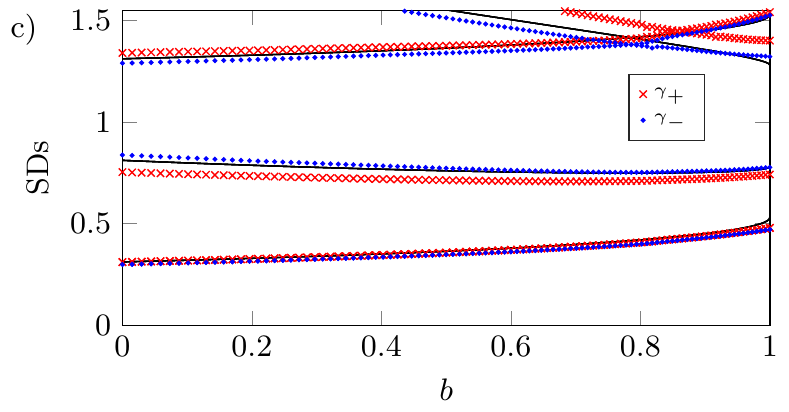}{
			\setlength\figureheight{.37\columnwidth}
			\setlength\figurewidth{.8\columnwidth}
			\input{Graphs/SemionAveraged}}
	\end{tabular}
	\caption{MERA data for the abc model along the `$b$ line'. This line is symmetric under the full $\zt^3$. CFT data, including averaging process, is discussed in more detail in Appendix~\ref{appendix:fullmeradata}.\\
		a) The energy of the optimized MERA state. The state remains a ground state when the anomalous symmetry operator is applied.\\
		b) Scaling dimensions of the associated CFT. These vary continuously with the parameter $b$. Points are averaged MERA data, whilst black lines correspond to Eqn.~\ref{eqn:CFTSD} for integer $e$ and $m$. Distinct colors/markers indicate under which irrep. the fields transform.\\
		c) Scaling dimensions of nonlocal operators corresponding to applying an anomalous symmetry (for group element $(1,1,1)$ defined in Eqn.~\ref{eqn:symmetrygroup}) twist to half of the chain. Points are averaged MERA data, whilst black lines correspond to Eqn.~\ref{eqn:CFTSD} for $e,m\in \mathbb{Z}+1/2$. Distinct colors/markers indicate under which projective irrep. the fields transform.}\label{Fig:bline}
\end{figure}

In \fref{bline}, we study the model with $a=c$ (referred to as the $b$ line) using a MERA with full anomalous symmetry enforced. For convenience, we allow a single transitional layer followed by a scale invariant portion. This leaves a pair of tensors which completely specify the state. After optimizing these residual degrees of freedom ($2\times16376$ real parameters) within this symmetric manifold, we obtain a good approximation to the ground state for all values of $b$, as evidenced by the ground state energy in Fig.~\hyperref[Fig:bline]{\ref*{Fig:bline}a} (relative error $\mathcal{O}(10^{-4})$). When the symmetry operator is applied to the state, we see that the state is unchanged (a property which was explicitly enforced). The central charge remains within 4.2\% of the analytic value for all values of $b$, comparable to that found in Ref.~\onlinecite{Bridgeman2015}.

\subsection{Scaling dimensions and topological sectors}

From our optimized MERA tensors, we have obtained the scaling dimensions of the associated CFT in each symmetry sector using Eqn.~\ref{eqn:scaling}. The data is shown in Fig.~\hyperref[Fig:bline]{\ref*{Fig:bline}b}. As expected, the scaling dimensions vary continuously with the parameter $b$.

The local fields are those of the compactified boson CFT at a radius 
\begin{align}
R^2=\frac{\pi}{2\cos^{-1}(\frac{2b}{b-3})}.
\end{align} The fields can be labeled by a pair of integers, and have scaling dimension $\Delta$ and conformal spin $s$ given by\cite{Ginsparg1988,DiFrancesco1997}
\begin{subequations}\label{eqn:CFTID}
	\begin{alignat}{2}
	\Delta_{e,m}&=\frac{e^2}{R^2}+\frac{m^2 R^2}{4},\label{eqn:CFTSD}\\
	s_{e,m}&=em\label{eqn:CFTspin},\\
	e,m&\in\mathbb{Z}\nonumber.
	\end{alignat}
\end{subequations}

Finally, we investigate the effect of $(1,1,1)$ symmetry twist in Fig.~\hyperref[Fig:bline]{\ref*{Fig:bline}c}. By applying the symmetry to half of the infinite chain we create the twist, and a set of nonlocal (with respect to the original theory) twisted fields can be obtained\cite{Evenbly2010c}. These operators correspond to eigenoperators of the `symmetry twisted' scaling superoperator (Eqn.~\ref{eqn:twistedscaling}). Since the symmetry acts projectively on the twisted fields, they can be decomposed into projective irreps corresponding to definite topological sectors. We can then diagonalize $\sso{g}{\cdot}$ within each sector, allowing us to label the twisted fields by the projective irrep under which they transform.

Again we can compare the numerically calculated twisted scaling dimensions to the analytic results to identify conformal spins of the twisted fields. As displayed in Table~\ref{table:CFTid}, within each topological sector, all conformal spins receive the same correction.

From the MERA data, we can identify the fields with a $(1,1,1)$ twist as carrying scaling dimension and conformal spin given by Eqn.~\ref{eqn:CFTSD} and Eqn.~\ref{eqn:CFTspin} respectively, but with $e,m\in \mathbb{Z}+\frac{1}{2}$, leading to quarter- and three-quarter- integer spins in this sector.

To examine the effect of the anomalous symmetry on the OPE, we computed fusion rules for the topological sectors using Eqn.~\ref{eqn:OPE} for a symmetric MERA tensor. Despite the fact that the symmetry group is abelian, we observe nonabelian fusion for all sectors with nontrivial twist. For example, fusion of sectors with twist $(1,1,1)$ results in only half of the trivial twist sectors.
The full set of fusion rules is given in Table~\ref{table:fusion} (Appendix~\ref{appendix:fullmeradata}).

In this example, the modular tensor category describing the topological sectors is $D^\phi(\zt^3)$. This category is known to be equivalent to $D(\mathsf{D}_4)$, where $\mathsf{D}_4$ is the symmetry group of a square.
The fusion table obtained from MERA matches that of $D^\phi(\zt^3)\cong D(\mathsf{D}_4)$\cite{propitius,DEWILDPROPITIUS1997297,Goff,Juven2014,PhysRevB.95.035131}.

The data for all topological sectors is displayed in full in Appendix~\ref{appendix:fullmeradata}.

\begin{figure}
	\begin{tabular}{r}
		\includeTikz{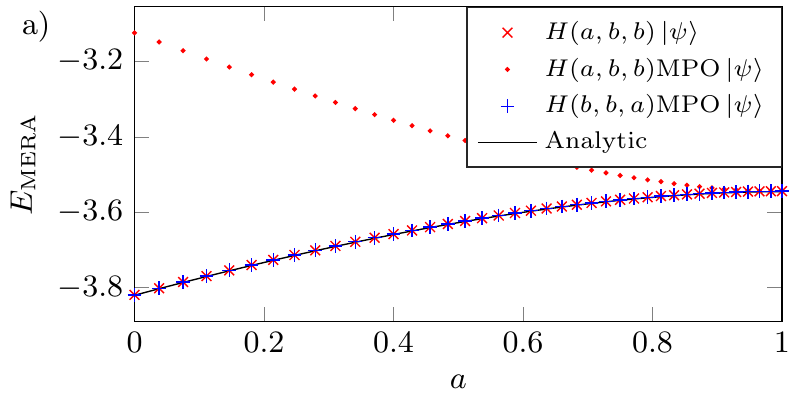}{
			\setlength\figureheight{.37\columnwidth}
			\setlength\figurewidth{.8\columnwidth}
			\input{Graphs/aEnergy}}
		\\
		\includeTikz{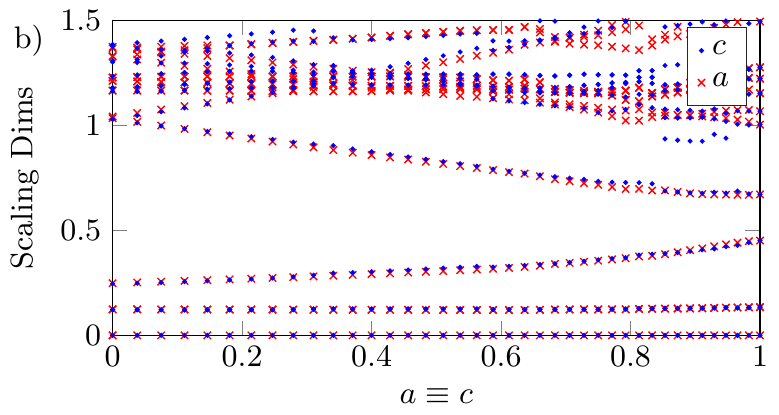}{
			\setlength\figureheight{.37\columnwidth}
			\setlength\figurewidth{.8\columnwidth}
			\input{Graphs/acSD}}
	\end{tabular}
	\caption{MERA data for the abc model along the `$a$' and `$c$' lines. These are exchanged by the symmetry action.\\
		a) Ground state energy of the optimized MERA.  By applying the symmetry operator to a state optimized for the Hamiltonian with $(a,b,b)$, we obtain a state which is the ground state of the Hamiltonian with parameters $(b,b,a)$. This demonstrates that the states are transforming properly.\\
		b) The local fields in the CFTs describing these two lines are identical, but distinct from those on the `$b$' line.}\label{Fig:acline}
\end{figure}

\subsection{Duality and domain walls}\label{section:duality}

We have also studied the `$a$' and `$c$' lines which are not symmetric under the anomalous $\zt$, but are exchanged by its action. We optimize over tensors of the form Eqn.~\ref{eqn:circuitconstraint}, but do not enforce the symmetry constraint on the residual degrees of freedom. 

The ground state energy obtained after optimization along the $b=c$ line is shown in Fig.~\hyperref[Fig:acline]{\ref*{Fig:acline}a}. If the symmetry MPO corresponding to group element $(1,1,1)$ is applied to the optimized state (via local application of Eqn.~\ref{eqn:CZsymmetry}), the result is an excited state. If the energy of this state is measured using the Hamiltonian with parameters $a$ and $c$ switched, we see that it is a ground state. This confirms that the state is transforming as expected under the anomalous action, that is, the MPO is acting as a duality transformation of the `$a$' and `$c$' critical lines.

We also show the scaling dimensions of the CFTs corresponding to the two dual lines (Fig.~\hyperref[Fig:acline]{\ref*{Fig:acline}b}). We observe that the local field content is identical, indicating that the same CFT describes these two lines. This CFT is distinct (in its local content) from that describing the `$b$' line, although it still has central charge 1.

\subsection{An anomaly protected gapless phase}

In Ref.~\onlinecite{czxmodel} it was shown that a phase with anomalous MPO symmetry can either be gapped and spontaneous break the symmetry, or be gapless. 
Furthermore it is known from Refs.~\onlinecite{goldenchain,PhysRevLett.101.050401,gils2009topology,PhysRevB.86.155111,PhysRevB.87.235120} that a topological symmetry, together with translation, can protect a gapless phase. An anomalous MPO symmetry is in fact an example of a topological symmetry. Hence one may suspect that there exist gapless phases protected by such a symmetry. 

Here we demonstrate that under an anomalous $\zt^3$ symmetry, along with translations, the gaplessness of the Hamiltonian along the `$b$' line is protected. That is, there are no translation invariant terms which are both symmetric under the full anomalous symmetry and are relevant in the renormalization group sense, and would therefore gap the Hamiltonian. 

\begin{figure}[t]
  \begin{tabular}{r}
    \includeTikz{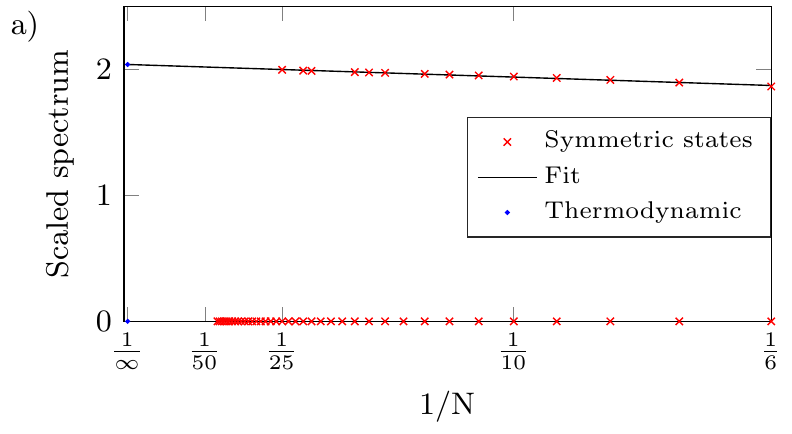}{
    \setlength\figureheight{.37\columnwidth}
    \setlength\figurewidth{.8\columnwidth}
    \input{Graphs/EDFit}}
    \\
	 \includeTikz{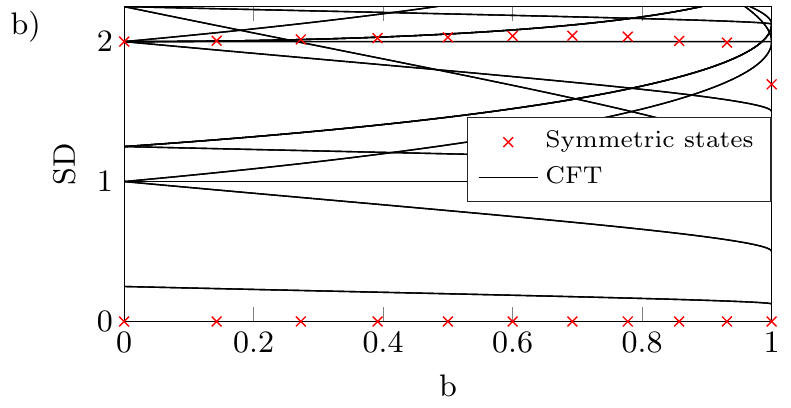}{
	 	\setlength\figureheight{.37\columnwidth}
	 	\setlength\figurewidth{.8\columnwidth}
	 	\input{Graphs/FullySymmetric}}
  \end{tabular}
  \caption{Finite size scaling data for the fully symmetric sector of the model.\\
  	a) After rescaling the spectrum so that the lowest excitation is consistent with the lowest nontrivial primary of the CFT, the fully symmetric states can be extracted. Fitting the data and extrapolating to the thermodynamic limit gives the scaling dimension. \\
  	b) For almost the whole `$b$' line, we observe that there are no fully symmetric states with scaling dimension less than 2 (RG relevant). This implies that no local, symmetric, translationally invariant terms can be added to the Hamiltonian to gap it out, thus the gapless phase is protected.}\label{Fig:ED}
\end{figure}

Since the effect of translations cannot be tested in the MERA framework, we performed a finite size scaling analysis\cite{Christe} to test this. Using the ALPS MPS library\cite{ALPS1,ALPS2}, the lowest 40 eigenstates of the Hamiltonian (Eqn.~\ref{eqn:hamiltonian}) along the `$b$' line were obtained. Bond dimensions were capped at 100 and lengths of between 6 and 55 sites (12-110 qubits) were considered. Scaling dimensions are obtained by first normalizing the Hamiltonian such that the ground state has energy 0 and the first excited state has energy corresponding to the smallest nonzero scaling dimension of the CFT\cite{Gehlen1987}. The energy levels are then fitted as a function of $1/N$ and extrapolated to $N=\infty$. This is shown in Fig.~\hyperref[Fig:ED]{\ref*{Fig:ED}a} for $b=.6$.

The Hamiltonian and symmetry operators were then simultaneously diagonalized within this subspace. In the fully symmetric sector (all symmetries acting as $+1$), the translation operator was diagonalized, allowing the momentum to be extracted.  

Under the combined action of the anomalous symmetry group and translations by a single spin, there are no fully symmetric states with scaling dimension less than 2 (Fig.~\hyperref[Fig:ED]{\ref*{Fig:ED}b}). This implies there are no local symmetric terms which can gap the Hamiltonian, thus the gapless phase is protected. 
We remark that under the operator which translates by a full site; an RG relevant, fully symmetric state with momentum zero does exist and therefore the Hamiltonian can be gapped by a staggered term. A similar effect was observed in Ref.~\onlinecite{goldenchain}.

\section{ Conclusions}\label{section:conclusion}

We have studied anomalous MPO symmetries in the framework of MERA. 
Following Ref.~\onlinecite{PhysRevLett.112.231602}, the third cohomology class of an MPO representation of a finite group was identified with an 't~Hooft anomaly. 

The properties of a fully MPO symmetric MERA were derived, including anomalous symmetry twists and the projective representations they carry. These were used to construct all topological sectors. 
This construction allows the complete set of topological data to be extracted, including a definite topological correction to the conformal spins of the fields in each sector and topological restrictions on the OPE.

A local condition to enforce the symmetry in the MERA was formulated, which allows for optimization of states with an anomalous symmetry. 
This ansatz works by locally disentangling the symmetry action, decoupling degrees of freedom on which the action can be expressed locally.

By way of an example, MERA states were optimized for a Hamiltonian with an anomalous $\zt^3$ symmetry. 
We have obtained accurate energy and conformal data for states optimized over our ansatz class, and demonstrated that the states transform as expected. 
All topological sectors were constructed and the resultant topological data was extracted. 
The conformal data was computed within each topological sector, and the projective action of the symmetry on the scaling fields was found. 
Furthermore, a correction to the conformal spin was identified, and shown to match the topological spin.

We applied the ansatz to study a duality of two critical lines. By extracting conformal data from optimized MERA the local content of the dual CFTs was shown to match. 
It was demonstrated that the action of the MPO mapped MERA ground-states optimized for Hamiltonians along one line to ground-states of the dual Hamiltonians. This required the ability to apply the MPO in a local fashion, which our ansatz permits. 

We performed a finite size scaling analysis of the anomalous $\zt^3$ symmetric line for large system sizes. It was numerically demonstrated that the anomalous MPO symmetry, together with translation, protects a gapless phase. 

There are several extensions of this work which suggest themselves. 
Our restricted MERA ansatz was only constructed for a particular class of anomalous group actions. 
It would be interesting to extend this to other MPOs, such as: nonabelian group representations with different cocycle anomalies, the Ising duality map or the translation operator. 

The most general extension conceivable is to a set of MPOs described by a unitary fusion category\cite{Bultinck2017183,daveprep,buican2017anyonic}. 
While the construction of topological sectors is known in this general case\cite{ tubealgebra,DrinfeldCenter,daveprep,buican2017anyonic,Bultinck2017183,Qalgebra,haah,dyonicspectrum}, an ansatz which allows the symmetry to be enforced locally in the MERA remains to be found. 

It would be interesting to determine which of these general symmetries protects a gapless phase such as the one observed in this work and those in Refs.~\onlinecite{goldenchain,PhysRevLett.101.050401,gils2009topology,PhysRevB.86.155111,PhysRevB.87.235120}. 

One could adapt these results to the recent tensor network renormalization (TNR)\cite{PhysRevLett.115.180405,PhysRevB.95.045117,yang2015loop,bal2017renormalization} scheme, constraining the RG flow to remain MPO symmetric. 
We remark that the Ising duality has previously been studied both numerically, using TNR but without manifestly enforcing the symmetry, in Ref.~\onlinecite{PhysRevB.94.115125}  and theoretically in Ref.~\onlinecite{1751-8121-49-35-354001}. 

It would also be interesting to consider the influence of an MPO symmetry on the entanglement entropy. We remark that by considering MPO symmetries of topologically ordered tensor network states in $(2+1)$D one recovers the topological entanglement entropy\cite{topologicalrenyi,MPOpaper,Bultinck2017183,KitaevPreskill,levinwenentanglement}. 

A particularly interesting future direction is to generalize our MPO symmetric MERA ansatz to a $(2+1)$D MERA describing a topologically ordered state that is symmetric under an anomalous PEPO symmetry. 

\acknowledgments
The authors thank Dave Aasen, Matthias Bal, Stephen Bartlett, Nick Bultinck, Christopher Chubb, Andrew Doherty, Steve Flammia, Micha{\"e}l Mari{\"e}n, Sam Roberts, Thomas Smith, Ryan Thorngren, Frank Verstraete, Guifre Vidal and Juven Wang for useful discussions. 
DW especially thanks Dave Aasen for pointing out the connection between the tube algebra and topological defects in conformal field theories. 
JB acknowledges support from the Australian Research Council via the Center of Excellence in Engineered Quantum Systems (EQuS), project number CE110001013.
The authors acknowledge the University of Sydney HPC service at The University of Sydney for providing HPC resources. 
DW acknowledges The University of Sydney quantum theory group for their hospitality, and the Perimeter Institute Visiting Graduate Fellowship program.

Optimal contraction sequences of the networks used in this work were computed using the netcon package of Ref.~\onlinecite{netcon}.

\bibliographystyle{apsrev_jacob}
\bibliography{Refs}

\begin{thebibliography}{127}%
\makeatletter
\providecommand \@ifxundefined [1]{%
 \@ifx{#1\undefined}
}%
\providecommand \@ifnum [1]{%
 \ifnum #1\expandafter \@firstoftwo
 \else \expandafter \@secondoftwo
 \fi
}%
\providecommand \@ifx [1]{%
 \ifx #1\expandafter \@firstoftwo
 \else \expandafter \@secondoftwo
 \fi
}%
\providecommand \natexlab [1]{#1}%
\providecommand \enquote  [1]{``#1''}%
\providecommand \bibnamefont  [1]{#1}%
\providecommand \bibfnamefont [1]{#1}%
\providecommand \citenamefont [1]{#1}%
\providecommand \href@noop [0]{\@secondoftwo}%
\providecommand \href [0]{\begingroup \@sanitize@url \@href}%
\providecommand \@href[1]{\@@startlink{#1}\@@href}%
\providecommand \@@href[1]{\endgroup#1\@@endlink}%
\providecommand \@sanitize@url [0]{\catcode `\\12\catcode `\$12\catcode
  `\&12\catcode `\#12\catcode `\^12\catcode `\_12\catcode `\%12\relax}%
\providecommand \@@startlink[1]{}%
\providecommand \@@endlink[0]{}%
\providecommand \url  [0]{\begingroup\@sanitize@url \@url }%
\providecommand \@url [1]{\endgroup\@href {#1}{\urlprefix }}%
\providecommand \urlprefix  [0]{URL }%
\providecommand \Eprint [0]{\href }%
\providecommand \doibase [0]{http://dx.doi.org/}%
\providecommand \selectlanguage [0]{\@gobble}%
\providecommand \bibinfo  [0]{\@secondoftwo}%
\providecommand \bibfield  [0]{\@secondoftwo}%
\providecommand \translation [1]{[#1]}%
\providecommand \BibitemOpen [0]{}%
\providecommand \bibitemStop [0]{}%
\providecommand \bibitemNoStop [0]{.\EOS\space}%
\providecommand \EOS [0]{\spacefactor3000\relax}%
\providecommand \BibitemShut  [1]{\csname bibitem#1\endcsname}%
\let\auto@bib@innerbib\@empty
\bibitem [{\citenamefont {Kadanoff}(1966)}]{Kadanoff:1966wm}%
  \BibitemOpen
  \bibfield  {author} {\bibinfo {author} {\bibfnamefont {L.~P.}\ \bibnamefont
  {Kadanoff}},\ }{Scaling laws for Ising models near T$_c$},\ \href {\doibase
  10.1142/9789812798763_0011} {\bibfield  {journal} {\bibinfo  {journal}
  {Physics}\ }\textbf {\bibinfo {volume} {2}},\ \bibinfo {pages} {263}}
  (\bibinfo {year} {1966})\BibitemShut {NoStop}%
\bibitem [{\citenamefont {Wilson}(1975)}]{RevModPhys.47.773}%
  \BibitemOpen
  \bibfield  {author} {\bibinfo {author} {\bibfnamefont {K.~G.}\ \bibnamefont
  {Wilson}},\ }{The renormalization group: Critical phenomena and the Kondo
  problem},\ \href {\doibase 10.1103/RevModPhys.47.773} {\bibfield  {journal}
  {\bibinfo  {journal} {Rev. Mod. Phys.}\ }\textbf {\bibinfo {volume} {47}},\
  \bibinfo {pages} {773}} (\bibinfo {year} {1975})\BibitemShut {NoStop}%
\bibitem [{\citenamefont {Landau}\ and\ \citenamefont
  {Lifshitz}(1965)}]{landau1965course}%
  \BibitemOpen
  \bibfield  {author} {\bibinfo {author} {\bibfnamefont {L.~D.}\ \bibnamefont
  {Landau}}\ and\ \bibinfo {author} {\bibfnamefont {E.~M.}\ \bibnamefont
  {Lifshitz}},\ }\href@noop {} {\emph {\bibinfo {title} {{Course of theoretical
  physics}}}}\ (\bibinfo  {publisher} {Pergamon Press},\ \bibinfo {year}
  {1965})\BibitemShut {NoStop}%
\bibitem [{\citenamefont {Wegner}(1971)}]{wegner1971duality}%
  \BibitemOpen
  \bibfield  {author} {\bibinfo {author} {\bibfnamefont {F.~J.}\ \bibnamefont
  {Wegner}},\ }{Duality in generalized Ising models and phase transitions
  without local order parameters},\ \href {\doibase 10.1063/1.1665530}
  {\bibfield  {journal} {\bibinfo  {journal} {J. Math. Phys.}\ }\textbf
  {\bibinfo {volume} {12}},\ \bibinfo {pages} {2259}} (\bibinfo {year}
  {1971})\BibitemShut {NoStop}%
\bibitem [{\citenamefont {Kosterlitz}\ and\ \citenamefont
  {Thouless}(1973)}]{kosterlitz1973ordering}%
  \BibitemOpen
  \bibfield  {author} {\bibinfo {author} {\bibfnamefont {J.~M.}\ \bibnamefont
  {Kosterlitz}}\ and\ \bibinfo {author} {\bibfnamefont {D.~J.}\ \bibnamefont
  {Thouless}},\ }{Ordering, metastability and phase transitions in
  two-dimensional systems},\ \href {\doibase 10.1088/0022-3719/6/7/010}
  {\bibfield  {journal} {\bibinfo  {journal} {J. Phys. C: Solid State Phys.}\
  }\textbf {\bibinfo {volume} {6}},\ \bibinfo {pages} {1181}} (\bibinfo {year}
  {1973})\BibitemShut {NoStop}%
\bibitem [{\citenamefont {Wen}(1989)}]{PhysRevB.40.7387}%
  \BibitemOpen
  \bibfield  {author} {\bibinfo {author} {\bibfnamefont {X.-G.}\ \bibnamefont
  {Wen}},\ }{Vacuum degeneracy of chiral spin states in compactified space},\
  \href {\doibase 10.1103/PhysRevB.40.7387} {\bibfield  {journal} {\bibinfo
  {journal} {Phys. Rev. B}\ }\textbf {\bibinfo {volume} {40}},\ \bibinfo
  {pages} {7387}} (\bibinfo {year} {1989})\BibitemShut {NoStop}%
\bibitem [{\citenamefont {Einarsson}(1990)}]{einarsson}%
  \BibitemOpen
  \bibfield  {author} {\bibinfo {author} {\bibfnamefont {T.}~\bibnamefont
  {Einarsson}},\ }{Fractional statistics on a torus},\ \href {\doibase
  10.1103/physrevlett.64.1995} {\bibfield  {journal} {\bibinfo  {journal}
  {Phys. Rev. Lett.}\ }\textbf {\bibinfo {volume} {64}},\ \bibinfo {pages}
  {1995}} (\bibinfo {year} {1990})\BibitemShut {NoStop}%
\bibitem [{\citenamefont {Wen}(1990)}]{doi:10.1142/S0217979290000139}%
  \BibitemOpen
  \bibfield  {author} {\bibinfo {author} {\bibfnamefont {X.-G.}\ \bibnamefont
  {Wen}},\ }{Topological Orders in Rigid States},\ \href {\doibase
  10.1142/S0217979290000139} {\bibfield  {journal} {\bibinfo  {journal} {Int.
  J. Mod. Phys. B}\ }\textbf {\bibinfo {volume} {04}},\ \bibinfo {pages} {239}}
  (\bibinfo {year} {1990})\BibitemShut {NoStop}%
\bibitem [{\citenamefont {Haldane}(1983)}]{PhysRevLett.50.1153}%
  \BibitemOpen
  \bibfield  {author} {\bibinfo {author} {\bibfnamefont {F.~D.~M.}\
  \bibnamefont {Haldane}},\ }{Nonlinear Field Theory of Large-Spin Heisenberg
  Antiferromagnets: Semiclassically Quantized Solitons of the One-Dimensional
  Easy-Axis N\'eel State},\ \href {\doibase 10.1103/PhysRevLett.50.1153}
  {\bibfield  {journal} {\bibinfo  {journal} {Phys. Rev. Lett.}\ }\textbf
  {\bibinfo {volume} {50}},\ \bibinfo {pages} {1153}} (\bibinfo {year}
  {1983})\BibitemShut {NoStop}%
\bibitem [{\citenamefont {Gu}\ and\ \citenamefont {Wen}(2009)}]{gu2009tensor}%
  \BibitemOpen
  \bibfield  {author} {\bibinfo {author} {\bibfnamefont {Z.-C.}\ \bibnamefont
  {Gu}}\ and\ \bibinfo {author} {\bibfnamefont {X.-G.}\ \bibnamefont {Wen}},\
  }{Tensor-entanglement-filtering renormalization approach and
  symmetry-protected topological order},\ \href {\doibase
  10.1103/PhysRevB.80.155131} {\bibfield  {journal} {\bibinfo  {journal} {Phys.
  Rev. B}\ }\textbf {\bibinfo {volume} {80}},\ \bibinfo {pages} {155131}},\
  \Eprint {http://arxiv.org/abs/0903.1069} {arXiv:0903.1069}  (\bibinfo {year}
  {2009})\BibitemShut {NoStop}%
\bibitem [{\citenamefont {Pollmann}\ \emph {et~al.}(2010)\citenamefont
  {Pollmann}, \citenamefont {Turner}, \citenamefont {Berg},\ and\ \citenamefont
  {Oshikawa}}]{pollmann2010entanglement}%
  \BibitemOpen
  \bibfield  {author} {\bibinfo {author} {\bibfnamefont {F.}~\bibnamefont
  {Pollmann}}, \bibinfo {author} {\bibfnamefont {A.~M.}\ \bibnamefont
  {Turner}}, \bibinfo {author} {\bibfnamefont {E.}~\bibnamefont {Berg}}, \ and\
  \bibinfo {author} {\bibfnamefont {M.}~\bibnamefont {Oshikawa}},\
  }{Entanglement spectrum of a topological phase in one dimension},\ \href
  {\doibase 10.1103/PhysRevB.81.064439} {\bibfield  {journal} {\bibinfo
  {journal} {Phys. Rev. B}\ }\textbf {\bibinfo {volume} {81}},\ \bibinfo
  {pages} {064439}},\ \Eprint {http://arxiv.org/abs/0910.1811}
  {arXiv:0910.1811}  (\bibinfo {year} {2010})\BibitemShut {NoStop}%
\bibitem [{\citenamefont {Chen}\ \emph
  {et~al.}(2013{\natexlab{a}})\citenamefont {Chen}, \citenamefont {Gu},
  \citenamefont {Liu},\ and\ \citenamefont {Wen}}]{chen2013symmetry}%
  \BibitemOpen
  \bibfield  {author} {\bibinfo {author} {\bibfnamefont {X.}~\bibnamefont
  {Chen}}, \bibinfo {author} {\bibfnamefont {Z.-C.}\ \bibnamefont {Gu}},
  \bibinfo {author} {\bibfnamefont {Z.-X.}\ \bibnamefont {Liu}}, \ and\
  \bibinfo {author} {\bibfnamefont {X.-G.}\ \bibnamefont {Wen}},\ }{Symmetry
  protected topological orders and the group cohomology of their symmetry
  group},\ \href {\doibase 10.1103/PhysRevB.87.155114} {\bibfield  {journal}
  {\bibinfo  {journal} {Phys. Rev. B}\ }\textbf {\bibinfo {volume} {87}},\
  \bibinfo {pages} {155114}},\ \Eprint {http://arxiv.org/abs/1106.4772}
  {arXiv:1106.4772}  (\bibinfo {year} {2013}{\natexlab{a}})\BibitemShut
  {NoStop}%
\bibitem [{\citenamefont {Miyake}(2010)}]{PhysRevLett.105.040501}%
  \BibitemOpen
  \bibfield  {author} {\bibinfo {author} {\bibfnamefont {A.}~\bibnamefont
  {Miyake}},\ }{Quantum Computation on the Edge of a Symmetry-Protected
  Topological Order},\ \href {\doibase 10.1103/PhysRevLett.105.040501}
  {\bibfield  {journal} {\bibinfo  {journal} {Phys. Rev. Lett.}\ }\textbf
  {\bibinfo {volume} {105}},\ \bibinfo {pages} {040501}},\ \Eprint
  {http://arxiv.org/abs/1003.4662} {arXiv:1003.4662}  (\bibinfo {year}
  {2010})\BibitemShut {NoStop}%
\bibitem [{\citenamefont {Renes}\ \emph {et~al.}(2013)\citenamefont {Renes},
  \citenamefont {Miyake}, \citenamefont {Brennen},\ and\ \citenamefont
  {Bartlett}}]{1367-2630-15-2-025020}%
  \BibitemOpen
  \bibfield  {author} {\bibinfo {author} {\bibfnamefont {J.~M.}\ \bibnamefont
  {Renes}}, \bibinfo {author} {\bibfnamefont {A.}~\bibnamefont {Miyake}},
  \bibinfo {author} {\bibfnamefont {G.~K.}\ \bibnamefont {Brennen}}, \ and\
  \bibinfo {author} {\bibfnamefont {S.~D.}\ \bibnamefont {Bartlett}},\
  }{Holonomic quantum computing in symmetry-protected ground states of spin
  chains},\ \href {\doibase 10.1088/1367-2630/15/2/025020} {\bibfield
  {journal} {\bibinfo  {journal} {New J. Phys.}\ }\textbf {\bibinfo {volume}
  {15}},\ \bibinfo {pages} {025020}},\ \Eprint {http://arxiv.org/abs/1103.5076}
  {arXiv:1103.5076}  (\bibinfo {year} {2013})\BibitemShut {NoStop}%
\bibitem [{\citenamefont {Else}\ \emph
  {et~al.}(2012{\natexlab{a}})\citenamefont {Else}, \citenamefont {Schwarz},
  \citenamefont {Bartlett},\ and\ \citenamefont {Doherty}}]{else2012symmetry}%
  \BibitemOpen
  \bibfield  {author} {\bibinfo {author} {\bibfnamefont {D.~V.}\ \bibnamefont
  {Else}}, \bibinfo {author} {\bibfnamefont {I.}~\bibnamefont {Schwarz}},
  \bibinfo {author} {\bibfnamefont {S.~D.}\ \bibnamefont {Bartlett}}, \ and\
  \bibinfo {author} {\bibfnamefont {A.~C.}\ \bibnamefont {Doherty}},\
  }{Symmetry-protected phases for measurement-based quantum computation},\
  \href {\doibase 10.1103/PhysRevLett.108.240505} {\bibfield  {journal}
  {\bibinfo  {journal} {Phys. Rev. Lett.}\ }\textbf {\bibinfo {volume} {108}},\
  \bibinfo {pages} {240505}},\ \Eprint {http://arxiv.org/abs/1201.4877}
  {arXiv:1201.4877}  (\bibinfo {year} {2012}{\natexlab{a}})\BibitemShut
  {NoStop}%
\bibitem [{\citenamefont {Else}\ \emph
  {et~al.}(2012{\natexlab{b}})\citenamefont {Else}, \citenamefont {Bartlett},\
  and\ \citenamefont {Doherty}}]{1367-2630-14-11-113016}%
  \BibitemOpen
  \bibfield  {author} {\bibinfo {author} {\bibfnamefont {D.~V.}\ \bibnamefont
  {Else}}, \bibinfo {author} {\bibfnamefont {S.~D.}\ \bibnamefont {Bartlett}},
  \ and\ \bibinfo {author} {\bibfnamefont {A.~C.}\ \bibnamefont {Doherty}},\
  }{Symmetry protection of measurement-based quantum computation in ground
  states},\ \href {\doibase 10.1088/1367-2630/14/11/113016} {\bibfield
  {journal} {\bibinfo  {journal} {New J. Phys.}\ }\textbf {\bibinfo {volume}
  {14}},\ \bibinfo {pages} {113016}},\ \Eprint {http://arxiv.org/abs/1207.4805}
  {arXiv:1207.4805}  (\bibinfo {year} {2012}{\natexlab{b}})\BibitemShut
  {NoStop}%
\bibitem [{\citenamefont {Williamson}\ and\ \citenamefont
  {Bartlett}(2015)}]{williamson2015symmetry}%
  \BibitemOpen
  \bibfield  {author} {\bibinfo {author} {\bibfnamefont {D.~J.}\ \bibnamefont
  {Williamson}}\ and\ \bibinfo {author} {\bibfnamefont {S.~D.}\ \bibnamefont
  {Bartlett}},\ }{Symmetry-protected adiabatic quantum transistors},\ \href
  {\doibase 10.1088/1367-2630/17/5/053019} {\bibfield  {journal} {\bibinfo
  {journal} {New J. Phys.}\ }\textbf {\bibinfo {volume} {17}},\ \bibinfo
  {pages} {053019}},\ \Eprint {http://arxiv.org/abs/1408.3415}
  {arXiv:1408.3415}  (\bibinfo {year} {2015})\BibitemShut {NoStop}%
\bibitem [{\citenamefont {Feiguin}\ \emph {et~al.}(2007)\citenamefont
  {Feiguin}, \citenamefont {Trebst}, \citenamefont {Ludwig}, \citenamefont
  {Troyer}, \citenamefont {Kitaev}, \citenamefont {Wang},\ and\ \citenamefont
  {Freedman}}]{goldenchain}%
  \BibitemOpen
  \bibfield  {author} {\bibinfo {author} {\bibfnamefont {A.}~\bibnamefont
  {Feiguin}}, \bibinfo {author} {\bibfnamefont {S.}~\bibnamefont {Trebst}},
  \bibinfo {author} {\bibfnamefont {A.~W.~W.}\ \bibnamefont {Ludwig}}, \bibinfo
  {author} {\bibfnamefont {M.}~\bibnamefont {Troyer}}, \bibinfo {author}
  {\bibfnamefont {A.}~\bibnamefont {Kitaev}}, \bibinfo {author} {\bibfnamefont
  {Z.}~\bibnamefont {Wang}}, \ and\ \bibinfo {author} {\bibfnamefont {M.~H.}\
  \bibnamefont {Freedman}},\ }{Interacting anyons in topological quantum
  liquids: The golden chain},\ \href {\doibase 10.1103/PhysRevLett.98.160409}
  {\bibfield  {journal} {\bibinfo  {journal} {Phys. Rev. Lett}\ }\textbf
  {\bibinfo {volume} {95}},\ \bibinfo {pages} {160409}},\ \Eprint
  {http://arxiv.org/abs/cond-mat/0612341} {arXiv:cond-mat/0612341}  (\bibinfo
  {year} {2007})\BibitemShut {NoStop}%
\bibitem [{\citenamefont {Pfeifer}\ \emph {et~al.}(2012)\citenamefont
  {Pfeifer}, \citenamefont {Buerschaper}, \citenamefont {Trebst}, \citenamefont
  {Ludwig}, \citenamefont {Troyer},\ and\ \citenamefont
  {Vidal}}]{PhysRevB.86.155111}%
  \BibitemOpen
  \bibfield  {author} {\bibinfo {author} {\bibfnamefont {R.~N.~C.}\
  \bibnamefont {Pfeifer}}, \bibinfo {author} {\bibfnamefont {O.}~\bibnamefont
  {Buerschaper}}, \bibinfo {author} {\bibfnamefont {S.}~\bibnamefont {Trebst}},
  \bibinfo {author} {\bibfnamefont {A.~W.~W.}\ \bibnamefont {Ludwig}}, \bibinfo
  {author} {\bibfnamefont {M.}~\bibnamefont {Troyer}}, \ and\ \bibinfo {author}
  {\bibfnamefont {G.}~\bibnamefont {Vidal}},\ }{Translation invariance,
  topology, and protection of criticality in chains of interacting anyons},\
  \href {\doibase 10.1103/PhysRevB.86.155111} {\bibfield  {journal} {\bibinfo
  {journal} {Phys. Rev. B}\ }\textbf {\bibinfo {volume} {86}},\ \bibinfo
  {pages} {155111}},\ \Eprint {http://arxiv.org/abs/1005.5486}
  {arXiv:1005.5486}  (\bibinfo {year} {2012})\BibitemShut {NoStop}%
\bibitem [{\citenamefont {Gils}\ \emph {et~al.}(2013)\citenamefont {Gils},
  \citenamefont {Ardonne}, \citenamefont {Trebst}, \citenamefont {Huse},
  \citenamefont {Ludwig}, \citenamefont {Troyer},\ and\ \citenamefont
  {Wang}}]{PhysRevB.87.235120}%
  \BibitemOpen
  \bibfield  {author} {\bibinfo {author} {\bibfnamefont {C.}~\bibnamefont
  {Gils}}, \bibinfo {author} {\bibfnamefont {E.}~\bibnamefont {Ardonne}},
  \bibinfo {author} {\bibfnamefont {S.}~\bibnamefont {Trebst}}, \bibinfo
  {author} {\bibfnamefont {D.~A.}\ \bibnamefont {Huse}}, \bibinfo {author}
  {\bibfnamefont {A.~W.~W.}\ \bibnamefont {Ludwig}}, \bibinfo {author}
  {\bibfnamefont {M.}~\bibnamefont {Troyer}}, \ and\ \bibinfo {author}
  {\bibfnamefont {Z.}~\bibnamefont {Wang}},\ }{Anyonic quantum spin chains:
  Spin-1 generalizations and topological stability},\ \href {\doibase
  10.1103/PhysRevB.87.235120} {\bibfield  {journal} {\bibinfo  {journal} {Phys.
  Rev. B}\ }\textbf {\bibinfo {volume} {87}},\ \bibinfo {pages} {235120}},\
  \Eprint {http://arxiv.org/abs/1303.4290} {arXiv:1303.4290}  (\bibinfo {year}
  {2013})\BibitemShut {NoStop}%
\bibitem [{\citenamefont {Pfeifer}\ \emph {et~al.}(2010)\citenamefont
  {Pfeifer}, \citenamefont {Corboz}, \citenamefont {Buerschaper}, \citenamefont
  {Aguado}, \citenamefont {Troyer},\ and\ \citenamefont
  {Vidal}}]{PhysRevB.82.115126}%
  \BibitemOpen
  \bibfield  {author} {\bibinfo {author} {\bibfnamefont {R.~N.~C.}\
  \bibnamefont {Pfeifer}}, \bibinfo {author} {\bibfnamefont {P.}~\bibnamefont
  {Corboz}}, \bibinfo {author} {\bibfnamefont {O.}~\bibnamefont {Buerschaper}},
  \bibinfo {author} {\bibfnamefont {M.}~\bibnamefont {Aguado}}, \bibinfo
  {author} {\bibfnamefont {M.}~\bibnamefont {Troyer}}, \ and\ \bibinfo {author}
  {\bibfnamefont {G.}~\bibnamefont {Vidal}},\ }{Simulation of anyons with
  tensor network algorithms},\ \href {\doibase 10.1103/PhysRevB.82.115126}
  {\bibfield  {journal} {\bibinfo  {journal} {Phys. Rev. B}\ }\textbf {\bibinfo
  {volume} {82}},\ \bibinfo {pages} {115126}},\ \Eprint
  {http://arxiv.org/abs/1006.3532} {arXiv:1006.3532}  (\bibinfo {year}
  {2010})\BibitemShut {NoStop}%
\bibitem [{\citenamefont {K\"onig}\ and\ \citenamefont
  {Bilgin}(2010)}]{PhysRevB.82.125118}%
  \BibitemOpen
  \bibfield  {author} {\bibinfo {author} {\bibfnamefont {R.}~\bibnamefont
  {K\"onig}}\ and\ \bibinfo {author} {\bibfnamefont {E.}~\bibnamefont
  {Bilgin}},\ }{Anyonic entanglement renormalization},\ \href {\doibase
  10.1103/PhysRevB.82.125118} {\bibfield  {journal} {\bibinfo  {journal} {Phys.
  Rev. B}\ }\textbf {\bibinfo {volume} {82}},\ \bibinfo {pages} {125118}},\
  \Eprint {http://arxiv.org/abs/1006.2478} {arXiv:1006.2478}  (\bibinfo {year}
  {2010})\BibitemShut {NoStop}%
\bibitem [{\citenamefont {Chen}\ \emph
  {et~al.}(2011{\natexlab{a}})\citenamefont {Chen}, \citenamefont {Liu},\ and\
  \citenamefont {Wen}}]{czxmodel}%
  \BibitemOpen
  \bibfield  {author} {\bibinfo {author} {\bibfnamefont {X.}~\bibnamefont
  {Chen}}, \bibinfo {author} {\bibfnamefont {Z.-X.}\ \bibnamefont {Liu}}, \
  and\ \bibinfo {author} {\bibfnamefont {X.-G.}\ \bibnamefont {Wen}},\
  }{Two-dimensional symmetry-protected topological orders and their protected
  gapless edge excitations},\ \href {\doibase 10.1103/PhysRevB.84.235141}
  {\bibfield  {journal} {\bibinfo  {journal} {Phys. Rev. B}\ }\textbf {\bibinfo
  {volume} {84}},\ \bibinfo {pages} {235141}},\ \Eprint
  {http://arxiv.org/abs/1106.4752} {arXiv:1106.4752}  (\bibinfo {year}
  {2011}{\natexlab{a}})\BibitemShut {NoStop}%
\bibitem [{\citenamefont {Santos}\ and\ \citenamefont
  {Wang}(2014)}]{PhysRevB.89.195122}%
  \BibitemOpen
  \bibfield  {author} {\bibinfo {author} {\bibfnamefont {L.~H.}\ \bibnamefont
  {Santos}}\ and\ \bibinfo {author} {\bibfnamefont {J.}~\bibnamefont {Wang}},\
  }{Symmetry-protected many-body Aharonov-Bohm effect},\ \href {\doibase
  10.1103/PhysRevB.89.195122} {\bibfield  {journal} {\bibinfo  {journal} {Phys.
  Rev. B}\ }\textbf {\bibinfo {volume} {89}},\ \bibinfo {pages} {195122}},\
  \Eprint {http://arxiv.org/abs/1310.8291} {arXiv:1310.8291}  (\bibinfo {year}
  {2014})\BibitemShut {NoStop}%
\bibitem [{\citenamefont {Wang}\ \emph
  {et~al.}(2015{\natexlab{a}})\citenamefont {Wang}, \citenamefont {Santos},\
  and\ \citenamefont {Wen}}]{PhysRevB.91.195134}%
  \BibitemOpen
  \bibfield  {author} {\bibinfo {author} {\bibfnamefont {J.~C.}\ \bibnamefont
  {Wang}}, \bibinfo {author} {\bibfnamefont {L.~H.}\ \bibnamefont {Santos}}, \
  and\ \bibinfo {author} {\bibfnamefont {X.-G.}\ \bibnamefont {Wen}},\
  }{Bosonic anomalies, induced fractional quantum numbers, and degenerate zero
  modes: The anomalous edge physics of symmetry-protected topological states},\
  \href {\doibase 10.1103/PhysRevB.91.195134} {\bibfield  {journal} {\bibinfo
  {journal} {Phys. Rev. B}\ }\textbf {\bibinfo {volume} {91}},\ \bibinfo
  {pages} {195134}},\ \Eprint {http://arxiv.org/abs/1403.5256}
  {arXiv:1403.5256}  (\bibinfo {year} {2015}{\natexlab{a}})\BibitemShut
  {NoStop}%
\bibitem [{\citenamefont {Williamson}\ \emph
  {et~al.}(2016{\natexlab{a}})\citenamefont {Williamson}, \citenamefont
  {Bultinck}, \citenamefont {Mari\"en}, \citenamefont {\ifmmode
  \mbox{\c{S}}\else \c{S}\fi{}ahino\ifmmode~\breve{g}\else \u{g}\fi{}lu},
  \citenamefont {Haegeman},\ and\ \citenamefont
  {Verstraete}}]{williamson2014matrix}%
  \BibitemOpen
  \bibfield  {author} {\bibinfo {author} {\bibfnamefont {D.~J.}\ \bibnamefont
  {Williamson}}, \bibinfo {author} {\bibfnamefont {N.}~\bibnamefont
  {Bultinck}}, \bibinfo {author} {\bibfnamefont {M.}~\bibnamefont {Mari\"en}},
  \bibinfo {author} {\bibfnamefont {M.~B.}\ \bibnamefont {\ifmmode
  \mbox{\c{S}}\else \c{S}\fi{}ahino\ifmmode~\breve{g}\else \u{g}\fi{}lu}},
  \bibinfo {author} {\bibfnamefont {J.}~\bibnamefont {Haegeman}}, \ and\
  \bibinfo {author} {\bibfnamefont {F.}~\bibnamefont {Verstraete}},\ }{Matrix
  product operators for symmetry-protected topological phases: Gauging and edge
  theories},\ \href {\doibase 10.1103/PhysRevB.94.205150} {\bibfield  {journal}
  {\bibinfo  {journal} {Phys. Rev. B}\ }\textbf {\bibinfo {volume} {94}},\
  \bibinfo {pages} {205150}},\ \Eprint {http://arxiv.org/abs/1412.5604}
  {arXiv:1412.5604}  (\bibinfo {year} {2016}{\natexlab{a}})\BibitemShut
  {NoStop}%
\bibitem [{\citenamefont {Wen}(2013)}]{PhysRevD.88.045013}%
  \BibitemOpen
  \bibfield  {author} {\bibinfo {author} {\bibfnamefont {X.-G.}\ \bibnamefont
  {Wen}},\ }{Classifying gauge anomalies through symmetry-protected trivial
  orders and classifying gravitational anomalies through topological orders},\
  \href {\doibase 10.1103/PhysRevD.88.045013} {\bibfield  {journal} {\bibinfo
  {journal} {Phys. Rev. D}\ }\textbf {\bibinfo {volume} {88}},\ \bibinfo
  {pages} {045013}},\ \Eprint {http://arxiv.org/abs/1303.1803}
  {arXiv:1303.1803}  (\bibinfo {year} {2013})\BibitemShut {NoStop}%
\bibitem [{\citenamefont {Kapustin}\ and\ \citenamefont
  {Thorngren}(2014{\natexlab{a}})}]{PhysRevLett.112.231602}%
  \BibitemOpen
  \bibfield  {author} {\bibinfo {author} {\bibfnamefont {A.}~\bibnamefont
  {Kapustin}}\ and\ \bibinfo {author} {\bibfnamefont {R.}~\bibnamefont
  {Thorngren}},\ }{Anomalous Discrete Symmetries in Three Dimensions and Group
  Cohomology},\ \href {\doibase 10.1103/PhysRevLett.112.231602} {\bibfield
  {journal} {\bibinfo  {journal} {Phys. Rev. Lett.}\ }\textbf {\bibinfo
  {volume} {112}},\ \bibinfo {pages} {231602}},\ \Eprint
  {http://arxiv.org/abs/1403.0617} {arXiv:1403.0617}  (\bibinfo {year}
  {2014}{\natexlab{a}})\BibitemShut {NoStop}%
\bibitem [{\citenamefont {Kapustin}(2014)}]{kapustin2014symmetry}%
  \BibitemOpen
  \bibfield  {author} {\bibinfo {author} {\bibfnamefont {A.}~\bibnamefont
  {Kapustin}},\ }Symmetry protected topological phases, anomalies, and
  cobordisms: beyond group cohomology,\ \href@noop {} {\ }\Eprint
  {http://arxiv.org/abs/1403.1467} {arXiv:1403.1467}  (\bibinfo {year}
  {2014})\BibitemShut {NoStop}%
\bibitem [{\citenamefont {Else}\ and\ \citenamefont
  {Nayak}(2014)}]{PhysRevB.90.235137}%
  \BibitemOpen
  \bibfield  {author} {\bibinfo {author} {\bibfnamefont {D.~V.}\ \bibnamefont
  {Else}}\ and\ \bibinfo {author} {\bibfnamefont {C.}~\bibnamefont {Nayak}},\
  }{Classifying symmetry-protected topological phases through the anomalous
  action of the symmetry on the edge},\ \href {\doibase
  10.1103/PhysRevB.90.235137} {\bibfield  {journal} {\bibinfo  {journal} {Phys.
  Rev. B}\ }\textbf {\bibinfo {volume} {90}},\ \bibinfo {pages} {235137}},\
  \Eprint {http://arxiv.org/abs/1409.5436} {arXiv:1409.5436}  (\bibinfo {year}
  {2014})\BibitemShut {NoStop}%
\bibitem [{\citenamefont {Kapustin}\ and\ \citenamefont
  {Thorngren}(2014{\natexlab{b}})}]{kapustin2014anomalies}%
  \BibitemOpen
  \bibfield  {author} {\bibinfo {author} {\bibfnamefont {A.}~\bibnamefont
  {Kapustin}}\ and\ \bibinfo {author} {\bibfnamefont {R.}~\bibnamefont
  {Thorngren}},\ }{Anomalies of discrete symmetries in various dimensions and
  group cohomology},\ \href@noop {} {\ }\Eprint
  {http://arxiv.org/abs/1404.3230} {arXiv:1404.3230}  (\bibinfo {year}
  {2014}{\natexlab{b}})\BibitemShut {NoStop}%
\bibitem [{\citenamefont {Wang}\ and\ \citenamefont
  {Wen}(2013)}]{wang2013lattice}%
  \BibitemOpen
  \bibfield  {author} {\bibinfo {author} {\bibfnamefont {J.}~\bibnamefont
  {Wang}}\ and\ \bibinfo {author} {\bibfnamefont {X.-G.}\ \bibnamefont {Wen}},\
  }{A Lattice Non-Perturbative Hamiltonian Construction of 1+ 1D Anomaly-Free
  Chiral Fermions and Bosons-on the equivalence of the anomaly matching
  conditions and the boundary fully gapping rules},\ \href@noop {} {\ }\Eprint
  {http://arxiv.org/abs/1307.7480} {arXiv:1307.7480}  (\bibinfo {year}
  {2013})\BibitemShut {NoStop}%
\bibitem [{\citenamefont {'t~Hooft}(1980)}]{hooft1980naturalness}%
  \BibitemOpen
  \bibfield  {author} {\bibinfo {author} {\bibfnamefont {G.}~\bibnamefont
  {'t~Hooft}},\ }{Naturalness, chiral symmetry, and spontaneous chiral symmetry
  breaking},\ \href {\doibase 10.1007/978-1-4684-7571-5_9} {\bibfield
  {journal} {\bibinfo  {journal} {Recent Developments in Gauge Theories}\ ,\
  \bibinfo {pages} {135}}} (\bibinfo {year} {1980})\BibitemShut {NoStop}%
\bibitem [{\citenamefont {Furuya}\ and\ \citenamefont
  {Oshikawa}(2017)}]{PhysRevLett.118.021601}%
  \BibitemOpen
  \bibfield  {author} {\bibinfo {author} {\bibfnamefont {S.~C.}\ \bibnamefont
  {Furuya}}\ and\ \bibinfo {author} {\bibfnamefont {M.}~\bibnamefont
  {Oshikawa}},\ }{Symmetry Protection of Critical Phases and a Global Anomaly
  in $1+1$ Dimensions},\ \href {\doibase 10.1103/PhysRevLett.118.021601}
  {\bibfield  {journal} {\bibinfo  {journal} {Phys. Rev. Lett.}\ }\textbf
  {\bibinfo {volume} {118}},\ \bibinfo {pages} {021601}},\ \Eprint
  {http://arxiv.org/abs/1503.07292} {arXiv:1503.07292}  (\bibinfo {year}
  {2017})\BibitemShut {NoStop}%
\bibitem [{\citenamefont {Verstraete}\ \emph {et~al.}(2009)\citenamefont
  {Verstraete}, \citenamefont {Cirac},\ and\ \citenamefont
  {Murg}}]{VerstraeteMurgCirac2008}%
  \BibitemOpen
  \bibfield  {author} {\bibinfo {author} {\bibfnamefont {F.}~\bibnamefont
  {Verstraete}}, \bibinfo {author} {\bibfnamefont {J.~I.}\ \bibnamefont
  {Cirac}}, \ and\ \bibinfo {author} {\bibfnamefont {V.}~\bibnamefont {Murg}},\
  }{Matrix Product States, Projected Entangled Pair States, and variational
  renormalization group methods for quantum spin systems},\ \href {\doibase
  10.1080/14789940801912366} {\bibfield  {journal} {\bibinfo  {journal} {Adv.
  Phys.}\ }\textbf {\bibinfo {volume} {57}},\ \bibinfo {pages} {143}},\ \Eprint
  {http://arxiv.org/abs/0907.2796} {arXiv:0907.2796}  (\bibinfo {year}
  {2009})\BibitemShut {NoStop}%
\bibitem [{\citenamefont {Or{\'{u}}s}(2014)}]{Orus2014}%
  \BibitemOpen
  \bibfield  {author} {\bibinfo {author} {\bibfnamefont {R.}~\bibnamefont
  {Or{\'{u}}s}},\ }{A practical introduction to tensor networks: Matrix product
  states and projected entangled pair states},\ \href {\doibase
  10.1016/j.aop.2014.06.013} {\bibfield  {journal} {\bibinfo  {journal} {Ann.
  Phys.}\ }\textbf {\bibinfo {volume} {349}},\ \bibinfo {pages} {117}},\
  \Eprint {http://arxiv.org/abs/1306.2164} {arXiv:1306.2164}  (\bibinfo {year}
  {2014})\BibitemShut {NoStop}%
\bibitem [{\citenamefont {Bridgeman}\ and\ \citenamefont
  {Chubb}(2017)}]{TNReview}%
  \BibitemOpen
  \bibfield  {author} {\bibinfo {author} {\bibfnamefont {J.~C.}\ \bibnamefont
  {Bridgeman}}\ and\ \bibinfo {author} {\bibfnamefont {C.~T.}\ \bibnamefont
  {Chubb}},\ }{Hand-waving and Interpretive Dance: An Introductory Course on
  Tensor Networks},\ \href {\doibase 10.1088/1751-8121/aa6dc3} {\bibfield
  {journal} {\bibinfo  {journal} {J. Phys. A}\ }\textbf {\bibinfo {volume}
  {50}},\ \bibinfo {pages} {223001}},\ \Eprint
  {http://arxiv.org/abs/1603.03039} {arXiv:1603.03039}  (\bibinfo {year}
  {2017})\BibitemShut {NoStop}%
\bibitem [{\citenamefont {White}(1992)}]{PhysRevLett.69.2863}%
  \BibitemOpen
  \bibfield  {author} {\bibinfo {author} {\bibfnamefont {S.~R.}\ \bibnamefont
  {White}},\ }{Density matrix formulation for quantum renormalization groups},\
  \href {\doibase 10.1103/PhysRevLett.69.2863} {\bibfield  {journal} {\bibinfo
  {journal} {Phys. Rev. Lett.}\ }\textbf {\bibinfo {volume} {69}},\ \bibinfo
  {pages} {2863}} (\bibinfo {year} {1992})\BibitemShut {NoStop}%
\bibitem [{\citenamefont {\"Ostlund}\ and\ \citenamefont
  {Rommer}(1995)}]{PhysRevLett.75.3537}%
  \BibitemOpen
  \bibfield  {author} {\bibinfo {author} {\bibfnamefont {S.}~\bibnamefont
  {\"Ostlund}}\ and\ \bibinfo {author} {\bibfnamefont {S.}~\bibnamefont
  {Rommer}},\ }{Thermodynamic Limit of Density Matrix Renormalization},\ \href
  {\doibase 10.1103/PhysRevLett.75.3537} {\bibfield  {journal} {\bibinfo
  {journal} {Phys. Rev. Lett.}\ }\textbf {\bibinfo {volume} {75}},\ \bibinfo
  {pages} {3537}},\ \Eprint {http://arxiv.org/abs/cond-mat/9503107}
  {arXiv:cond-mat/9503107}  (\bibinfo {year} {1995})\BibitemShut {NoStop}%
\bibitem [{\citenamefont {Dukelsky}\ \emph {et~al.}(1998)\citenamefont
  {Dukelsky}, \citenamefont {Mart{\'i}n-Delgado}, \citenamefont {Nishino},\
  and\ \citenamefont {Sierra}}]{0295-5075-43-4-457}%
  \BibitemOpen
  \bibfield  {author} {\bibinfo {author} {\bibfnamefont {J.}~\bibnamefont
  {Dukelsky}}, \bibinfo {author} {\bibfnamefont {M.~A.}\ \bibnamefont
  {Mart{\'i}n-Delgado}}, \bibinfo {author} {\bibfnamefont {T.}~\bibnamefont
  {Nishino}}, \ and\ \bibinfo {author} {\bibfnamefont {G.}~\bibnamefont
  {Sierra}},\ }{Equivalence of the variational matrix product method and the
  density matrix renormalization group applied to spin chains},\ \href
  {\doibase 10.1209/epl/i1998-00381-x} {\bibfield  {journal} {\bibinfo
  {journal} {EPL}\ }\textbf {\bibinfo {volume} {43}},\ \bibinfo {pages}
  {457}},\ \Eprint {http://arxiv.org/abs/cond-mat/9710310}
  {arXiv:cond-mat/9710310}  (\bibinfo {year} {1998})\BibitemShut {NoStop}%
\bibitem [{\citenamefont {Affleck}\ \emph {et~al.}(1987)\citenamefont
  {Affleck}, \citenamefont {Kennedy}, \citenamefont {Lieb},\ and\ \citenamefont
  {Tasaki}}]{PhysRevLett.59.799}%
  \BibitemOpen
  \bibfield  {author} {\bibinfo {author} {\bibfnamefont {I.}~\bibnamefont
  {Affleck}}, \bibinfo {author} {\bibfnamefont {T.}~\bibnamefont {Kennedy}},
  \bibinfo {author} {\bibfnamefont {E.~H.}\ \bibnamefont {Lieb}}, \ and\
  \bibinfo {author} {\bibfnamefont {H.}~\bibnamefont {Tasaki}},\ }{Rigorous
  results on valence-bond ground states in antiferromagnets},\ \href {\doibase
  10.1103/PhysRevLett.59.799} {\bibfield  {journal} {\bibinfo  {journal} {Phys.
  Rev. Lett.}\ }\textbf {\bibinfo {volume} {59}},\ \bibinfo {pages} {799}}
  (\bibinfo {year} {1987})\BibitemShut {NoStop}%
\bibitem [{\citenamefont {Fannes}\ \emph {et~al.}(1992)\citenamefont {Fannes},
  \citenamefont {Nachtergaele},\ and\ \citenamefont {Werner}}]{Fannes92}%
  \BibitemOpen
  \bibfield  {author} {\bibinfo {author} {\bibfnamefont {M.}~\bibnamefont
  {Fannes}}, \bibinfo {author} {\bibfnamefont {B.}~\bibnamefont
  {Nachtergaele}}, \ and\ \bibinfo {author} {\bibfnamefont {R.~F.}\
  \bibnamefont {Werner}},\ }{Finitely correlated states on quantum spin
  chains},\ \href {\doibase 10.1007/BF02099178} {\bibfield  {journal} {\bibinfo
   {journal} {Commun. Math. Phys.}\ }\textbf {\bibinfo {volume} {144}},\
  \bibinfo {pages} {443}} (\bibinfo {year} {1992})\BibitemShut {NoStop}%
\bibitem [{\citenamefont {Klümper}\ \emph {et~al.}(1993)\citenamefont
  {Klümper}, \citenamefont {Schadschneider},\ and\ \citenamefont
  {Zittartz}}]{0295-5075-24-4-010}%
  \BibitemOpen
  \bibfield  {author} {\bibinfo {author} {\bibfnamefont {A.}~\bibnamefont
  {Klümper}}, \bibinfo {author} {\bibfnamefont {A.}~\bibnamefont
  {Schadschneider}}, \ and\ \bibinfo {author} {\bibfnamefont {J.}~\bibnamefont
  {Zittartz}},\ }{Matrix Product Ground States for One-Dimensional Spin-1
  Quantum Antiferromagnets},\ \href
  {http://stacks.iop.org/0295-5075/24/i=4/a=010} {\bibfield  {journal}
  {\bibinfo  {journal} {EPL}\ }\textbf {\bibinfo {volume} {24}},\ \bibinfo
  {pages} {293}},\ \Eprint {http://arxiv.org/abs/cond-mat/9307028}
  {arXiv:cond-mat/9307028}  (\bibinfo {year} {1993})\BibitemShut {NoStop}%
\bibitem [{\citenamefont {P{\'e}rez-Garc{\'i}a}\ \emph
  {et~al.}(2007)\citenamefont {P{\'e}rez-Garc{\'i}a}, \citenamefont
  {Verstraete}, \citenamefont {Wolf},\ and\ \citenamefont
  {Cirac}}]{MPSrepresentations}%
  \BibitemOpen
  \bibfield  {author} {\bibinfo {author} {\bibfnamefont {D.}~\bibnamefont
  {P{\'e}rez-Garc{\'i}a}}, \bibinfo {author} {\bibfnamefont {F.}~\bibnamefont
  {Verstraete}}, \bibinfo {author} {\bibfnamefont {M.~M.}\ \bibnamefont
  {Wolf}}, \ and\ \bibinfo {author} {\bibfnamefont {J.~I.}\ \bibnamefont
  {Cirac}},\ }{Matrix Product State Representations},\ \href
  {http://dl.acm.org/citation.cfm?id=2011832.2011833} {\bibfield  {journal}
  {\bibinfo  {journal} {Quantum Info. Comput.}\ }\textbf {\bibinfo {volume}
  {7}},\ \bibinfo {pages} {401}},\ \Eprint
  {http://arxiv.org/abs/quant-ph/0608197} {arXiv:quant-ph/0608197}  (\bibinfo
  {year} {2007})\BibitemShut {NoStop}%
\bibitem [{\citenamefont {Chen}\ \emph
  {et~al.}(2011{\natexlab{b}})\citenamefont {Chen}, \citenamefont {Gu},\ and\
  \citenamefont {Wen}}]{1Done}%
  \BibitemOpen
  \bibfield  {author} {\bibinfo {author} {\bibfnamefont {X.}~\bibnamefont
  {Chen}}, \bibinfo {author} {\bibfnamefont {Z.-C.}\ \bibnamefont {Gu}}, \ and\
  \bibinfo {author} {\bibfnamefont {X.-G.}\ \bibnamefont {Wen}},\
  }{Classification of gapped symmetric phases in one-dimensional spin
  systems},\ \href {\doibase 10.1103/PhysRevB.83.035107} {\bibfield  {journal}
  {\bibinfo  {journal} {Phys. Rev. B}\ }\textbf {\bibinfo {volume} {83}},\
  \bibinfo {pages} {035107}},\ \Eprint {http://arxiv.org/abs/1008.3745}
  {arXiv:1008.3745}  (\bibinfo {year} {2011}{\natexlab{b}})\BibitemShut
  {NoStop}%
\bibitem [{\citenamefont {Schuch}\ \emph {et~al.}(2011)\citenamefont {Schuch},
  \citenamefont {P\'erez-Garc\'ia},\ and\ \citenamefont
  {Cirac}}]{SchuchGarciaCirac11}%
  \BibitemOpen
  \bibfield  {author} {\bibinfo {author} {\bibfnamefont {N.}~\bibnamefont
  {Schuch}}, \bibinfo {author} {\bibfnamefont {D.}~\bibnamefont
  {P\'erez-Garc\'ia}}, \ and\ \bibinfo {author} {\bibfnamefont
  {I.}~\bibnamefont {Cirac}},\ }{Classifying quantum phases using matrix
  product states and projected entangled pair states},\ \href {\doibase
  10.1103/PhysRevB.84.165139} {\bibfield  {journal} {\bibinfo  {journal} {Phys.
  Rev. B}\ }\textbf {\bibinfo {volume} {84}},\ \bibinfo {pages} {165139}},\
  \Eprint {http://arxiv.org/abs/1010.3732} {arXiv:1010.3732}  (\bibinfo {year}
  {2011})\BibitemShut {NoStop}%
\bibitem [{\citenamefont {Cirac}\ \emph {et~al.}(2017)\citenamefont {Cirac},
  \citenamefont {P{\'e}rez-Garc{\'i}a}, \citenamefont {Schuch},\ and\
  \citenamefont {Verstraete}}]{Cirac2017100}%
  \BibitemOpen
  \bibfield  {author} {\bibinfo {author} {\bibfnamefont {J.}~\bibnamefont
  {Cirac}}, \bibinfo {author} {\bibfnamefont {D.}~\bibnamefont
  {P{\'e}rez-Garc{\'i}a}}, \bibinfo {author} {\bibfnamefont {N.}~\bibnamefont
  {Schuch}}, \ and\ \bibinfo {author} {\bibfnamefont {F.}~\bibnamefont
  {Verstraete}},\ }{Matrix product density operators: Renormalization fixed
  points and boundary theories},\ \href {\doibase 10.1016/j.aop.2016.12.030}
  {\bibfield  {journal} {\bibinfo  {journal} {Ann. Phys.}\ }\textbf {\bibinfo
  {volume} {378}},\ \bibinfo {pages} {100 }},\ \Eprint
  {http://arxiv.org/abs/1606.00608} {arXiv:1606.00608}  (\bibinfo {year}
  {2017})\BibitemShut {NoStop}%
\bibitem [{\citenamefont {Nishino}\ \emph {et~al.}(2001)\citenamefont
  {Nishino}, \citenamefont {Hieida}, \citenamefont {Okunishi}, \citenamefont
  {Maeshima}, \citenamefont {Akutsu},\ and\ \citenamefont
  {Gendiar}}]{doi:10.1143PTP.105.409}%
  \BibitemOpen
  \bibfield  {author} {\bibinfo {author} {\bibfnamefont {T.}~\bibnamefont
  {Nishino}}, \bibinfo {author} {\bibfnamefont {Y.}~\bibnamefont {Hieida}},
  \bibinfo {author} {\bibfnamefont {K.}~\bibnamefont {Okunishi}}, \bibinfo
  {author} {\bibfnamefont {N.}~\bibnamefont {Maeshima}}, \bibinfo {author}
  {\bibfnamefont {Y.}~\bibnamefont {Akutsu}}, \ and\ \bibinfo {author}
  {\bibfnamefont {A.}~\bibnamefont {Gendiar}},\ }{Two-Dimensional Tensor
  Product Variational Formulation},\ \href {\doibase 10.1143/PTP.105.409}
  {\bibfield  {journal} {\bibinfo  {journal} {Progr. Theor. Phys.}\ }\textbf
  {\bibinfo {volume} {105}},\ \bibinfo {pages} {409}},\ \Eprint
  {http://arxiv.org/abs/cond-mat/0011103} {arXiv:cond-mat/0011103}  (\bibinfo
  {year} {2001})\BibitemShut {NoStop}%
\bibitem [{\citenamefont {{Verstraete}}\ and\ \citenamefont
  {{Cirac}}(2004)}]{peps}%
  \BibitemOpen
  \bibfield  {author} {\bibinfo {author} {\bibfnamefont {F.}~\bibnamefont
  {{Verstraete}}}\ and\ \bibinfo {author} {\bibfnamefont {J.~I.}\ \bibnamefont
  {{Cirac}}},\ }{Renormalization algorithms for Quantum-Many Body Systems in
  two and higher dimensions},\ \href@noop {} {\ }\Eprint
  {http://arxiv.org/abs/cond-mat/0407066} {arXiv:cond-mat/0407066}  (\bibinfo
  {year} {2004})\BibitemShut {NoStop}%
\bibitem [{\citenamefont {Schuch}\ \emph {et~al.}(2010)\citenamefont {Schuch},
  \citenamefont {Cirac},\ and\ \citenamefont {Perez-Garcia}}]{Ginjectivity}%
  \BibitemOpen
  \bibfield  {author} {\bibinfo {author} {\bibfnamefont {N.}~\bibnamefont
  {Schuch}}, \bibinfo {author} {\bibfnamefont {I.}~\bibnamefont {Cirac}}, \
  and\ \bibinfo {author} {\bibfnamefont {D.}~\bibnamefont {Perez-Garcia}},\
  }{PEPS as ground states: Degeneracy and topology},\ \href {\doibase
  10.1016/j.aop.2010.05.008} {\bibfield  {journal} {\bibinfo  {journal} {Ann.
  Phys.}\ }\textbf {\bibinfo {volume} {325}},\ \bibinfo {pages} {2153 }},\
  \Eprint {http://arxiv.org/abs/1001.3807} {arXiv:1001.3807}  (\bibinfo {year}
  {2010})\BibitemShut {NoStop}%
\bibitem [{\citenamefont {Dubail}\ and\ \citenamefont
  {Read}(2015)}]{PhysRevB.92.205307}%
  \BibitemOpen
  \bibfield  {author} {\bibinfo {author} {\bibfnamefont {J.}~\bibnamefont
  {Dubail}}\ and\ \bibinfo {author} {\bibfnamefont {N.}~\bibnamefont {Read}},\
  }{Tensor network trial states for chiral topological phases in two dimensions
  and a no-go theorem in any dimension},\ \href {\doibase
  10.1103/PhysRevB.92.205307} {\bibfield  {journal} {\bibinfo  {journal} {Phys.
  Rev. B}\ }\textbf {\bibinfo {volume} {92}},\ \bibinfo {pages} {205307}},\
  \Eprint {http://arxiv.org/abs/1307.7726} {arXiv:1307.7726}  (\bibinfo {year}
  {2015})\BibitemShut {NoStop}%
\bibitem [{\citenamefont {Wahl}\ \emph {et~al.}(2013)\citenamefont {Wahl},
  \citenamefont {Tu}, \citenamefont {Schuch},\ and\ \citenamefont
  {Cirac}}]{PhysRevLett.111.236805}%
  \BibitemOpen
  \bibfield  {author} {\bibinfo {author} {\bibfnamefont {T.~B.}\ \bibnamefont
  {Wahl}}, \bibinfo {author} {\bibfnamefont {H.-H.}\ \bibnamefont {Tu}},
  \bibinfo {author} {\bibfnamefont {N.}~\bibnamefont {Schuch}}, \ and\ \bibinfo
  {author} {\bibfnamefont {J.~I.}\ \bibnamefont {Cirac}},\ }{Projected
  Entangled-Pair States Can Describe Chiral Topological States},\ \href
  {\doibase 10.1103/PhysRevLett.111.236805} {\bibfield  {journal} {\bibinfo
  {journal} {Phys. Rev. Lett.}\ }\textbf {\bibinfo {volume} {111}},\ \bibinfo
  {pages} {236805}},\ \Eprint {http://arxiv.org/abs/1308.0316}
  {arXiv:1308.0316}  (\bibinfo {year} {2013})\BibitemShut {NoStop}%
\bibitem [{\citenamefont {Buerschaper}(2014)}]{Buerschaper14}%
  \BibitemOpen
  \bibfield  {author} {\bibinfo {author} {\bibfnamefont {O.}~\bibnamefont
  {Buerschaper}},\ }{Twisted injectivity in projected entangled pair states and
  the classification of quantum phases},\ \href {\doibase
  10.1016/j.aop.2014.09.007} {\bibfield  {journal} {\bibinfo  {journal} {Ann.
  Phys.}\ }\textbf {\bibinfo {volume} {351}},\ \bibinfo {pages} {447 }},\
  \Eprint {http://arxiv.org/abs/1307.7763} {arXiv:1307.7763}  (\bibinfo {year}
  {2014})\BibitemShut {NoStop}%
\bibitem [{\citenamefont {\ifmmode \mbox{\c{S}}\else
  \c{S}\fi{}ahino\ifmmode~\breve{g}\else \u{g}\fi{}lu}\ \emph
  {et~al.}(2014)\citenamefont {\ifmmode \mbox{\c{S}}\else
  \c{S}\fi{}ahino\ifmmode~\breve{g}\else \u{g}\fi{}lu}, \citenamefont
  {Williamson}, \citenamefont {Bultinck}, \citenamefont {Mari{\"e}n},
  \citenamefont {Haegeman}, \citenamefont {Schuch},\ and\ \citenamefont
  {Verstraete}}]{MPOpaper}%
  \BibitemOpen
  \bibfield  {author} {\bibinfo {author} {\bibfnamefont {M.~B.}\ \bibnamefont
  {\ifmmode \mbox{\c{S}}\else \c{S}\fi{}ahino\ifmmode~\breve{g}\else
  \u{g}\fi{}lu}}, \bibinfo {author} {\bibfnamefont {D.}~\bibnamefont
  {Williamson}}, \bibinfo {author} {\bibfnamefont {N.}~\bibnamefont
  {Bultinck}}, \bibinfo {author} {\bibfnamefont {M.}~\bibnamefont
  {Mari{\"e}n}}, \bibinfo {author} {\bibfnamefont {J.}~\bibnamefont
  {Haegeman}}, \bibinfo {author} {\bibfnamefont {N.}~\bibnamefont {Schuch}}, \
  and\ \bibinfo {author} {\bibfnamefont {F.}~\bibnamefont {Verstraete}},\
  }{Characterizing topological order with matrix product operators},\
  \href@noop {} {\ }\Eprint {http://arxiv.org/abs/1409.2150} {arXiv:1409.2150}
  (\bibinfo {year} {2014})\BibitemShut {NoStop}%
\bibitem [{\citenamefont {Bultinck}\ \emph {et~al.}(2017)\citenamefont
  {Bultinck}, \citenamefont {Mariën}, \citenamefont {Williamson},
  \citenamefont {\ifmmode \mbox{\c{S}}\else
  \c{S}\fi{}ahino\ifmmode~\breve{g}\else \u{g}\fi{}lu}, \citenamefont
  {Haegeman},\ and\ \citenamefont {Verstraete}}]{Bultinck2017183}%
  \BibitemOpen
  \bibfield  {author} {\bibinfo {author} {\bibfnamefont {N.}~\bibnamefont
  {Bultinck}}, \bibinfo {author} {\bibfnamefont {M.}~\bibnamefont {Mariën}},
  \bibinfo {author} {\bibfnamefont {D.}~\bibnamefont {Williamson}}, \bibinfo
  {author} {\bibfnamefont {M.~B.}\ \bibnamefont {\ifmmode \mbox{\c{S}}\else
  \c{S}\fi{}ahino\ifmmode~\breve{g}\else \u{g}\fi{}lu}}, \bibinfo {author}
  {\bibfnamefont {J.}~\bibnamefont {Haegeman}}, \ and\ \bibinfo {author}
  {\bibfnamefont {F.}~\bibnamefont {Verstraete}},\ }{Anyons and matrix product
  operator algebras},\ \href {\doibase 10.1016/j.aop.2017.01.004} {\bibfield
  {journal} {\bibinfo  {journal} {Ann. Phys.}\ }\textbf {\bibinfo {volume}
  {378}},\ \bibinfo {pages} {183 }},\ \Eprint {http://arxiv.org/abs/1511.08090}
  {arXiv:1511.08090}  (\bibinfo {year} {2017})\BibitemShut {NoStop}%
\bibitem [{\citenamefont {Bridgeman}\ \emph {et~al.}(2016)\citenamefont
  {Bridgeman}, \citenamefont {Flammia},\ and\ \citenamefont
  {Poulin}}]{ribbons}%
  \BibitemOpen
  \bibfield  {author} {\bibinfo {author} {\bibfnamefont {J.}~\bibnamefont
  {Bridgeman}}, \bibinfo {author} {\bibfnamefont {S.~T.}\ \bibnamefont
  {Flammia}}, \ and\ \bibinfo {author} {\bibfnamefont {D.}~\bibnamefont
  {Poulin}},\ }{Detecting Topological Order with Ribbon Operators},\ \href
  {\doibase 10.1103/PhysRevB.94.205123} {\bibfield  {journal} {\bibinfo
  {journal} {Phys. Rev. B}\ }\textbf {\bibinfo {volume} {94}},\ \bibinfo
  {pages} {205123}},\ \Eprint {http://arxiv.org/abs/1603.02275}
  {arXiv:1603.02275}  (\bibinfo {year} {2016})\BibitemShut {NoStop}%
\bibitem [{\citenamefont {Williamson}\ \emph
  {et~al.}(2016{\natexlab{b}})\citenamefont {Williamson}, \citenamefont
  {Bultinck}, \citenamefont {Haegeman},\ and\ \citenamefont
  {Verstraete}}]{williamson2016fermionic}%
  \BibitemOpen
  \bibfield  {author} {\bibinfo {author} {\bibfnamefont {D.~J.}\ \bibnamefont
  {Williamson}}, \bibinfo {author} {\bibfnamefont {N.}~\bibnamefont
  {Bultinck}}, \bibinfo {author} {\bibfnamefont {J.}~\bibnamefont {Haegeman}},
  \ and\ \bibinfo {author} {\bibfnamefont {F.}~\bibnamefont {Verstraete}},\
  }{Fermionic Matrix Product Operators and Topological Phases of Matter},\
  \href@noop {} {\ }\Eprint {http://arxiv.org/abs/1609.02897}
  {arXiv:1609.02897}  (\bibinfo {year} {2016}{\natexlab{b}})\BibitemShut
  {NoStop}%
\bibitem [{\citenamefont {P\'erez-Garc\'{\i}a}\ \emph
  {et~al.}(2008)\citenamefont {P\'erez-Garc\'{\i}a}, \citenamefont {Wolf},
  \citenamefont {Sanz}, \citenamefont {Verstraete},\ and\ \citenamefont
  {Cirac}}]{PhysRevLett.100.167202}%
  \BibitemOpen
  \bibfield  {author} {\bibinfo {author} {\bibfnamefont {D.}~\bibnamefont
  {P\'erez-Garc\'{\i}a}}, \bibinfo {author} {\bibfnamefont {M.~M.}\
  \bibnamefont {Wolf}}, \bibinfo {author} {\bibfnamefont {M.}~\bibnamefont
  {Sanz}}, \bibinfo {author} {\bibfnamefont {F.}~\bibnamefont {Verstraete}}, \
  and\ \bibinfo {author} {\bibfnamefont {J.~I.}\ \bibnamefont {Cirac}},\
  }{String Order and Symmetries in Quantum Spin Lattices},\ \href {\doibase
  10.1103/PhysRevLett.100.167202} {\bibfield  {journal} {\bibinfo  {journal}
  {Phys. Rev. Lett.}\ }\textbf {\bibinfo {volume} {100}},\ \bibinfo {pages}
  {167202}},\ \Eprint {http://arxiv.org/abs/0802.0447} {arXiv:0802.0447}
  (\bibinfo {year} {2008})\BibitemShut {NoStop}%
\bibitem [{\citenamefont {Singh}\ \emph {et~al.}(2010)\citenamefont {Singh},
  \citenamefont {Pfeifer},\ and\ \citenamefont {Vidal}}]{Singh2010}%
  \BibitemOpen
  \bibfield  {author} {\bibinfo {author} {\bibfnamefont {S.}~\bibnamefont
  {Singh}}, \bibinfo {author} {\bibfnamefont {R.}~\bibnamefont {Pfeifer}}, \
  and\ \bibinfo {author} {\bibfnamefont {G.}~\bibnamefont {Vidal}},\ }{Tensor
  network decompositions in the presence of a global symmetry},\ \href
  {\doibase 10.1103/PhysRevA.82.050301} {\bibfield  {journal} {\bibinfo
  {journal} {Phys. Rev. A}\ }\textbf {\bibinfo {volume} {82}},\ \bibinfo
  {pages} {050301}},\ \Eprint {http://arxiv.org/abs/0907.2994}
  {arXiv:0907.2994}  (\bibinfo {year} {2010})\BibitemShut {NoStop}%
\bibitem [{\citenamefont {P{\'e}rez-Garc{\'i}a}\ \emph
  {et~al.}(2010)\citenamefont {P{\'e}rez-Garc{\'i}a}, \citenamefont {Sanz},
  \citenamefont {Gonz{\'a}lez-Guill{\'e}n}, \citenamefont {Wolf},\ and\
  \citenamefont {Cirac}}]{1367-2630-12-2-025010}%
  \BibitemOpen
  \bibfield  {author} {\bibinfo {author} {\bibfnamefont {D.}~\bibnamefont
  {P{\'e}rez-Garc{\'i}a}}, \bibinfo {author} {\bibfnamefont {M.}~\bibnamefont
  {Sanz}}, \bibinfo {author} {\bibfnamefont {C.~E.}\ \bibnamefont
  {Gonz{\'a}lez-Guill{\'e}n}}, \bibinfo {author} {\bibfnamefont {M.~M.}\
  \bibnamefont {Wolf}}, \ and\ \bibinfo {author} {\bibfnamefont {J.~I.}\
  \bibnamefont {Cirac}},\ }{Characterizing symmetries in a projected entangled
  pair state},\ \href {http://stacks.iop.org/1367-2630/12/i=2/a=025010}
  {\bibfield  {journal} {\bibinfo  {journal} {New J. Phys.}\ }\textbf {\bibinfo
  {volume} {12}},\ \bibinfo {pages} {025010}},\ \Eprint
  {http://arxiv.org/abs/0908.1674} {arXiv:0908.1674}  (\bibinfo {year}
  {2010})\BibitemShut {NoStop}%
\bibitem [{\citenamefont {Vidal}(2007)}]{Vidal2007}%
  \BibitemOpen
  \bibfield  {author} {\bibinfo {author} {\bibfnamefont {G.}~\bibnamefont
  {Vidal}},\ }{Entanglement Renormalization},\ \href {\doibase
  10.1103/PhysRevLett.99.220405} {\bibfield  {journal} {\bibinfo  {journal}
  {Phys. Rev. Lett.}\ }\textbf {\bibinfo {volume} {99}},\ \bibinfo {pages}
  {220405}},\ \Eprint {http://arxiv.org/abs/cond-mat/0512165}
  {arXiv:cond-mat/0512165}  (\bibinfo {year} {2007})\BibitemShut {NoStop}%
\bibitem [{\citenamefont {Evenbly}\ and\ \citenamefont
  {Vidal}(2009)}]{Evenbly2009}%
  \BibitemOpen
  \bibfield  {author} {\bibinfo {author} {\bibfnamefont {G.}~\bibnamefont
  {Evenbly}}\ and\ \bibinfo {author} {\bibfnamefont {G.}~\bibnamefont
  {Vidal}},\ }{Algorithms for entanglement renormalization},\ \href {\doibase
  10.1103/PhysRevB.79.144108} {\bibfield  {journal} {\bibinfo  {journal} {Phys.
  Rev. B}\ }\textbf {\bibinfo {volume} {79}},\ \bibinfo {pages} {144108}},\
  \Eprint {http://arxiv.org/abs/0707.1454} {arXiv:0707.1454}  (\bibinfo {year}
  {2009})\BibitemShut {NoStop}%
\bibitem [{\citenamefont {Pfeifer}\ \emph {et~al.}(2009)\citenamefont
  {Pfeifer}, \citenamefont {Evenbly},\ and\ \citenamefont
  {Vidal}}]{Pfeifer2009}%
  \BibitemOpen
  \bibfield  {author} {\bibinfo {author} {\bibfnamefont {R.}~\bibnamefont
  {Pfeifer}}, \bibinfo {author} {\bibfnamefont {G.}~\bibnamefont {Evenbly}}, \
  and\ \bibinfo {author} {\bibfnamefont {G.}~\bibnamefont {Vidal}},\
  }{Entanglement renormalization, scale invariance, and quantum criticality},\
  \href {\doibase 10.1103/PhysRevA.79.040301} {\bibfield  {journal} {\bibinfo
  {journal} {Phys. Rev. A}\ }\textbf {\bibinfo {volume} {79}},\ \bibinfo
  {pages} {040301}},\ \Eprint {http://arxiv.org/abs/0810.0580}
  {arXiv:0810.0580}  (\bibinfo {year} {2009})\BibitemShut {NoStop}%
\bibitem [{\citenamefont {Ginsparg}(1990)}]{Ginsparg1988}%
  \BibitemOpen
  \bibfield  {author} {\bibinfo {author} {\bibfnamefont {P.}~\bibnamefont
  {Ginsparg}},\ }in\ \href@noop {} {\emph {\bibinfo {booktitle} {{Fields,
  Strings and Critical Phenomena}}}},\ \bibinfo {series and number} {Les
  Houches 1988, Session XLIX},\ \bibinfo {editor} {edited by\ \bibinfo {editor}
  {\bibfnamefont {E.}~\bibnamefont {Br\'ezin}}\ and\ \bibinfo {editor}
  {\bibfnamefont {J.~Z.}\ \bibnamefont {Justin}}}\ (\bibinfo  {publisher}
  {North-Holland},\ \bibinfo {address} {Amsterdam},\ \bibinfo {year} {1990})\
  \Eprint {http://arxiv.org/abs/hep-th/9108028} {arXiv:hep-th/9108028}
  \BibitemShut {NoStop}%
\bibitem [{\citenamefont {{Di Francesco}}(1997)}]{DiFrancesco1997}%
  \BibitemOpen
  \bibfield  {author} {\bibinfo {author} {\bibfnamefont {P.}~\bibnamefont {{Di
  Francesco}}},\ }\href@noop {} {\emph {\bibinfo {title} {{Conformal field
  theory}}}}\ (\bibinfo  {publisher} {Springer},\ \bibinfo {address} {New
  York},\ \bibinfo {year} {1997})\BibitemShut {NoStop}%
\bibitem [{\citenamefont {Ryu}\ \emph {et~al.}(2012)\citenamefont {Ryu},
  \citenamefont {Moore},\ and\ \citenamefont {Ludwig}}]{PhysRevB.85.045104}%
  \BibitemOpen
  \bibfield  {author} {\bibinfo {author} {\bibfnamefont {S.}~\bibnamefont
  {Ryu}}, \bibinfo {author} {\bibfnamefont {J.~E.}\ \bibnamefont {Moore}}, \
  and\ \bibinfo {author} {\bibfnamefont {A.~W.~W.}\ \bibnamefont {Ludwig}},\
  }{Electromagnetic and gravitational responses and anomalies in topological
  insulators and superconductors},\ \href {\doibase 10.1103/PhysRevB.85.045104}
  {\bibfield  {journal} {\bibinfo  {journal} {Phys. Rev. B}\ }\textbf {\bibinfo
  {volume} {85}},\ \bibinfo {pages} {045104}},\ \Eprint
  {http://arxiv.org/abs/1010.0936} {arXiv:1010.0936}  (\bibinfo {year}
  {2012})\BibitemShut {NoStop}%
\bibitem [{\citenamefont {Wang}\ \emph
  {et~al.}(2015{\natexlab{b}})\citenamefont {Wang}, \citenamefont {Gu},\ and\
  \citenamefont {Wen}}]{PhysRevLett.114.031601}%
  \BibitemOpen
  \bibfield  {author} {\bibinfo {author} {\bibfnamefont {J.~C.}\ \bibnamefont
  {Wang}}, \bibinfo {author} {\bibfnamefont {Z.-C.}\ \bibnamefont {Gu}}, \ and\
  \bibinfo {author} {\bibfnamefont {X.-G.}\ \bibnamefont {Wen}},\
  }{Field-Theory Representation of Gauge-Gravity Symmetry-Protected Topological
  Invariants, Group Cohomology, and Beyond},\ \href {\doibase
  10.1103/PhysRevLett.114.031601} {\bibfield  {journal} {\bibinfo  {journal}
  {Phys. Rev. Lett.}\ }\textbf {\bibinfo {volume} {114}},\ \bibinfo {pages}
  {031601}},\ \Eprint {http://arxiv.org/abs/1405.7689} {arXiv:1405.7689}
  (\bibinfo {year} {2015}{\natexlab{b}})\BibitemShut {NoStop}%
\bibitem [{\citenamefont {Kapustin}\ and\ \citenamefont
  {Thorngren}(2013)}]{kapustin2013higher}%
  \BibitemOpen
  \bibfield  {author} {\bibinfo {author} {\bibfnamefont {A.}~\bibnamefont
  {Kapustin}}\ and\ \bibinfo {author} {\bibfnamefont {R.}~\bibnamefont
  {Thorngren}},\ }{Higher symmetry and gapped phases of gauge theories},\
  \href@noop {} {\ }\Eprint {http://arxiv.org/abs/1309.4721} {arXiv:1309.4721}
  (\bibinfo {year} {2013})\BibitemShut {NoStop}%
\bibitem [{\citenamefont {Gaiotto}\ \emph {et~al.}(2015)\citenamefont
  {Gaiotto}, \citenamefont {Kapustin}, \citenamefont {Seiberg},\ and\
  \citenamefont {Willett}}]{Gaiotto2015}%
  \BibitemOpen
  \bibfield  {author} {\bibinfo {author} {\bibfnamefont {D.}~\bibnamefont
  {Gaiotto}}, \bibinfo {author} {\bibfnamefont {A.}~\bibnamefont {Kapustin}},
  \bibinfo {author} {\bibfnamefont {N.}~\bibnamefont {Seiberg}}, \ and\
  \bibinfo {author} {\bibfnamefont {B.}~\bibnamefont {Willett}},\ }{Generalized
  global symmetries},\ \href {\doibase 10.1007/JHEP02(2015)172} {\bibfield
  {journal} {\bibinfo  {journal} {J. High Energy Phys.}\ }\textbf {\bibinfo
  {volume} {2015}},\ \bibinfo {pages} {172}},\ \Eprint
  {http://arxiv.org/abs/1412.5148} {arXiv:1412.5148}  (\bibinfo {year}
  {2015})\BibitemShut {NoStop}%
\bibitem [{\citenamefont {Thorngren}\ and\ \citenamefont {von
  Keyserlingk}(2015)}]{thorngren2015higher}%
  \BibitemOpen
  \bibfield  {author} {\bibinfo {author} {\bibfnamefont {R.}~\bibnamefont
  {Thorngren}}\ and\ \bibinfo {author} {\bibfnamefont {C.}~\bibnamefont {von
  Keyserlingk}},\ }{Higher {SPT}'s and a generalization of anomaly in-flow},\
  \href@noop {} {\ }\Eprint {http://arxiv.org/abs/1511.02929}
  {arXiv:1511.02929}  (\bibinfo {year} {2015})\BibitemShut {NoStop}%
\bibitem [{\citenamefont {Wess}\ and\ \citenamefont
  {Zumino}(1971)}]{WESS197195}%
  \BibitemOpen
  \bibfield  {author} {\bibinfo {author} {\bibfnamefont {J.}~\bibnamefont
  {Wess}}\ and\ \bibinfo {author} {\bibfnamefont {B.}~\bibnamefont {Zumino}},\
  }{Consequences of anomalous ward identities},\ \href {\doibase
  10.1016/0370-2693(71)90582-X} {\bibfield  {journal} {\bibinfo  {journal}
  {Phys. Lett. B}\ }\textbf {\bibinfo {volume} {37}},\ \bibinfo {pages} {95 }}
  (\bibinfo {year} {1971})\BibitemShut {NoStop}%
\bibitem [{\citenamefont {Weinberg}(2013)}]{Weinberg:1996kr}%
  \BibitemOpen
  \bibfield  {author} {\bibinfo {author} {\bibfnamefont {S.}~\bibnamefont
  {Weinberg}},\ }\href {\doibase 10.1063/1.881660} {\emph {\bibinfo {title}
  {{The quantum theory of fields. Vol. 2: Modern applications}}}}\ (\bibinfo
  {publisher} {Cambridge University Press},\ \bibinfo {year}
  {2013})\BibitemShut {NoStop}%
\bibitem [{\citenamefont {Levin}\ and\ \citenamefont
  {Gu}(2012)}]{levin2012braiding}%
  \BibitemOpen
  \bibfield  {author} {\bibinfo {author} {\bibfnamefont {M.}~\bibnamefont
  {Levin}}\ and\ \bibinfo {author} {\bibfnamefont {Z.-C.}\ \bibnamefont {Gu}},\
  }{Braiding statistics approach to symmetry-protected topological phases},\
  \href {\doibase 10.1103/PhysRevB.86.115109} {\bibfield  {journal} {\bibinfo
  {journal} {Phys. Rev. B}\ }\textbf {\bibinfo {volume} {86}},\ \bibinfo
  {pages} {115109}},\ \Eprint {http://arxiv.org/abs/1202.3120}
  {arXiv:1202.3120}  (\bibinfo {year} {2012})\BibitemShut {NoStop}%
\bibitem [{\citenamefont {Haegeman}\ \emph {et~al.}(2015)\citenamefont
  {Haegeman}, \citenamefont {Van~Acoleyen}, \citenamefont {Schuch},
  \citenamefont {Cirac},\ and\ \citenamefont {Verstraete}}]{Gaugingpaper}%
  \BibitemOpen
  \bibfield  {author} {\bibinfo {author} {\bibfnamefont {J.}~\bibnamefont
  {Haegeman}}, \bibinfo {author} {\bibfnamefont {K.}~\bibnamefont
  {Van~Acoleyen}}, \bibinfo {author} {\bibfnamefont {N.}~\bibnamefont
  {Schuch}}, \bibinfo {author} {\bibfnamefont {J.~I.}\ \bibnamefont {Cirac}}, \
  and\ \bibinfo {author} {\bibfnamefont {F.}~\bibnamefont {Verstraete}},\
  }{Gauging quantum states: from global to local symmetries in many-body
  systems},\ \href {\doibase 10.1103/PhysRevX.5.011024} {\bibfield  {journal}
  {\bibinfo  {journal} {Phys. Rev. X}\ }\textbf {\bibinfo {volume} {5}},\
  \bibinfo {pages} {011024}},\ \Eprint {http://arxiv.org/abs/1407.1025}
  {arXiv:1407.1025}  (\bibinfo {year} {2015})\BibitemShut {NoStop}%
\bibitem [{\citenamefont {Evenbly}\ and\ \citenamefont
  {Vidal}(2013)}]{Evenbly2011}%
  \BibitemOpen
  \bibfield  {author} {\bibinfo {author} {\bibfnamefont {G.}~\bibnamefont
  {Evenbly}}\ and\ \bibinfo {author} {\bibfnamefont {G.}~\bibnamefont
  {Vidal}},\ }in\ \href@noop {} {\emph {\bibinfo {booktitle} {Strongly
  Correlated Systems-Numerical Methods}}},\ \bibinfo {series} {Springer Series
  in Solid-State Sciences}, Vol.\ \bibinfo {volume} {176},\ \bibinfo {editor}
  {edited by\ \bibinfo {editor} {\bibfnamefont {A.}~\bibnamefont {Avella}}\
  and\ \bibinfo {editor} {\bibfnamefont {F.}~\bibnamefont {Mancini}}}\
  (\bibinfo  {publisher} {Springer},\ \bibinfo {address} {Berlin New York},\
  \bibinfo {year} {2013})\ pp.\ \bibinfo {pages} {99--130},\ \Eprint
  {http://arxiv.org/abs/1109.5334} {arXiv:1109.5334} \BibitemShut {NoStop}%
\bibitem [{\citenamefont {Evenbly}\ \emph {et~al.}(2010)\citenamefont
  {Evenbly}, \citenamefont {Corboz},\ and\ \citenamefont
  {Vidal}}]{Evenbly2010c}%
  \BibitemOpen
  \bibfield  {author} {\bibinfo {author} {\bibfnamefont {G.}~\bibnamefont
  {Evenbly}}, \bibinfo {author} {\bibfnamefont {P.}~\bibnamefont {Corboz}}, \
  and\ \bibinfo {author} {\bibfnamefont {G.}~\bibnamefont {Vidal}},\ }{Nonlocal
  scaling operators with entanglement renormalization},\ \href {\doibase
  10.1103/PhysRevB.82.132411} {\bibfield  {journal} {\bibinfo  {journal} {Phys.
  Rev. B}\ }\textbf {\bibinfo {volume} {82}},\ \bibinfo {pages} {132411}},\
  \Eprint {http://arxiv.org/abs/0912.2166} {arXiv:0912.2166}  (\bibinfo {year}
  {2010})\BibitemShut {NoStop}%
\bibitem [{\citenamefont {Hauru}\ \emph {et~al.}(2016)\citenamefont {Hauru},
  \citenamefont {Evenbly}, \citenamefont {Ho}, \citenamefont {Gaiotto},\ and\
  \citenamefont {Vidal}}]{PhysRevB.94.115125}%
  \BibitemOpen
  \bibfield  {author} {\bibinfo {author} {\bibfnamefont {M.}~\bibnamefont
  {Hauru}}, \bibinfo {author} {\bibfnamefont {G.}~\bibnamefont {Evenbly}},
  \bibinfo {author} {\bibfnamefont {W.~W.}\ \bibnamefont {Ho}}, \bibinfo
  {author} {\bibfnamefont {D.}~\bibnamefont {Gaiotto}}, \ and\ \bibinfo
  {author} {\bibfnamefont {G.}~\bibnamefont {Vidal}},\ }{Topological conformal
  defects with tensor networks},\ \href {\doibase 10.1103/PhysRevB.94.115125}
  {\bibfield  {journal} {\bibinfo  {journal} {Phys. Rev. B}\ }\textbf {\bibinfo
  {volume} {94}},\ \bibinfo {pages} {115125}},\ \Eprint
  {http://arxiv.org/abs/1512.03846} {arXiv:1512.03846}  (\bibinfo {year}
  {2016})\BibitemShut {NoStop}%
\bibitem [{\citenamefont {Aasen}\ \emph {et~al.}(2016)\citenamefont {Aasen},
  \citenamefont {Mong},\ and\ \citenamefont
  {Fendley}}]{1751-8121-49-35-354001}%
  \BibitemOpen
  \bibfield  {author} {\bibinfo {author} {\bibfnamefont {D.}~\bibnamefont
  {Aasen}}, \bibinfo {author} {\bibfnamefont {R.~S.~K.}\ \bibnamefont {Mong}},
  \ and\ \bibinfo {author} {\bibfnamefont {P.}~\bibnamefont {Fendley}},\
  }{Topological Defects on the Lattice I: The Ising model},\ \href {\doibase
  10.1088/1751-8113/49/35/354001} {\bibfield  {journal} {\bibinfo  {journal}
  {J. Phys. A}\ }\textbf {\bibinfo {volume} {49}},\ \bibinfo {pages}
  {354001}},\ \Eprint {http://arxiv.org/abs/1601.07185} {arXiv:1601.07185}
  (\bibinfo {year} {2016})\BibitemShut {NoStop}%
\bibitem [{\citenamefont {Christe}\ and\ \citenamefont
  {Henkel}(1993)}]{Christe}%
  \BibitemOpen
  \bibfield  {author} {\bibinfo {author} {\bibfnamefont {P.}~\bibnamefont
  {Christe}}\ and\ \bibinfo {author} {\bibfnamefont {M.}~\bibnamefont
  {Henkel}},\ }\href@noop {} {\emph {\bibinfo {title} {{Introduction to
  Conformal Invariance and Its Applications to Critical Phenomena}}}}\
  (\bibinfo  {publisher} {Springer-Verlag},\ \bibinfo {address} {Berlin},\
  \bibinfo {year} {1993})\ \Eprint {http://arxiv.org/abs/cond-mat/9304035}
  {arXiv:cond-mat/9304035} \BibitemShut {NoStop}%
\bibitem [{\citenamefont {{de Wild Propitius}}(1995)}]{propitius}%
  \BibitemOpen
  \bibfield  {author} {\bibinfo {author} {\bibfnamefont {M.}~\bibnamefont {{de
  Wild Propitius}}},\ }\emph {\bibinfo {title} {{Topological interactions in
  broken gauge theories}}},\ \href@noop {} {Ph.D. thesis},\ \bibinfo  {school}
  {University of Amsterdam},\ \Eprint {http://arxiv.org/abs/hep-th/9511195}
  {arXiv:hep-th/9511195}  (\bibinfo {year} {1995})\BibitemShut {NoStop}%
\bibitem [{\citenamefont {Drinfel’d}(1986)}]{drinfeld}%
  \BibitemOpen
  \bibfield  {author} {\bibinfo {author} {\bibfnamefont {V.}~\bibnamefont
  {Drinfel’d}},\ }Quantum groups,\ \href
  {http://www.mathunion.org/ICM/ICM1986.1/Main/icm1986.1.0798.0820.ocr.pdf}
  {\bibfield  {journal} {\bibinfo  {journal} {Proc. Int. Congr. Math.}\
  }\textbf {\bibinfo {volume} {1}},\ \bibinfo {pages} {798}} (\bibinfo {year}
  {1986})\BibitemShut {NoStop}%
\bibitem [{\citenamefont {Moore}\ and\ \citenamefont
  {Seiberg}(1989)}]{Moore1989}%
  \BibitemOpen
  \bibfield  {author} {\bibinfo {author} {\bibfnamefont {G.}~\bibnamefont
  {Moore}}\ and\ \bibinfo {author} {\bibfnamefont {N.}~\bibnamefont
  {Seiberg}},\ }Classical and quantum conformal field theory,\ \href {\doibase
  10.1007/BF01238857} {\bibfield  {journal} {\bibinfo  {journal} {Commun. Math.
  Phys.}\ }\textbf {\bibinfo {volume} {123}},\ \bibinfo {pages} {177}}
  (\bibinfo {year} {1989})\BibitemShut {NoStop}%
\bibitem [{\citenamefont {Evans}\ and\ \citenamefont
  {Kawahigashi}(1995)}]{tubealgebra}%
  \BibitemOpen
  \bibfield  {author} {\bibinfo {author} {\bibfnamefont {D.}~\bibnamefont
  {Evans}}\ and\ \bibinfo {author} {\bibfnamefont {Y.}~\bibnamefont
  {Kawahigashi}},\ }{On Ocneanu's theory of asymptotic inclusions for
  subfactors, topological quantum field theories and quantum doubles},\ \href
  {\doibase 10.1142/s0129167x95000468} {\bibfield  {journal} {\bibinfo
  {journal} {Int. J. Math.}\ }\textbf {\bibinfo {volume} {6}},\ \bibinfo
  {pages} {205}} (\bibinfo {year} {1995})\BibitemShut {NoStop}%
\bibitem [{\citenamefont {Muger}(2003)}]{DrinfeldCenter}%
  \BibitemOpen
  \bibfield  {author} {\bibinfo {author} {\bibfnamefont {M.}~\bibnamefont
  {Muger}},\ }{From subfactors to categories and topology II: The quantum
  double of tensor categories and subfactors},\ \href {\doibase
  10.1016/S0022-4049(02)00248-7} {\bibfield  {journal} {\bibinfo  {journal} {J.
  Pure Appl. Algebr.}\ }\textbf {\bibinfo {volume} {180}},\ \bibinfo {pages}
  {159 }},\ \Eprint {http://arxiv.org/abs/math/0111205} {arXiv:math/0111205}
  (\bibinfo {year} {2003})\BibitemShut {NoStop}%
\bibitem [{\citenamefont {Bakalov}\ and\ \citenamefont
  {Kirillov}(2001)}]{bakalov2001lectures}%
  \BibitemOpen
  \bibfield  {author} {\bibinfo {author} {\bibfnamefont {B.}~\bibnamefont
  {Bakalov}}\ and\ \bibinfo {author} {\bibfnamefont {A.~A.}\ \bibnamefont
  {Kirillov}},\ }\href {\doibase 10.1090/ulect/021} {\emph {\bibinfo {title}
  {{Lectures on tensor categories and modular functors}}}},\ Vol.~\bibinfo
  {volume} {21}\ (\bibinfo  {publisher} {American Mathematical Soc.},\ \bibinfo
  {year} {2001})\BibitemShut {NoStop}%
\bibitem [{\citenamefont {Etingof}\ \emph {et~al.}(2005)\citenamefont
  {Etingof}, \citenamefont {Nikshych},\ and\ \citenamefont
  {Ostrik}}]{etingof2005fusion}%
  \BibitemOpen
  \bibfield  {author} {\bibinfo {author} {\bibfnamefont {P.}~\bibnamefont
  {Etingof}}, \bibinfo {author} {\bibfnamefont {D.}~\bibnamefont {Nikshych}}, \
  and\ \bibinfo {author} {\bibfnamefont {V.}~\bibnamefont {Ostrik}},\ }{On
  fusion categories},\ \href {\doibase 10.4007/annals.2005.162.581} {\bibfield
  {journal} {\bibinfo  {journal} {Ann. Math.}\ ,\ \bibinfo {pages}
  {581}}}\Eprint {http://arxiv.org/abs/math/0203060} {arXiv:math/0203060}
  (\bibinfo {year} {2005})\BibitemShut {NoStop}%
\bibitem [{\citenamefont {Fuchs}\ \emph
  {et~al.}(2002{\natexlab{a}})\citenamefont {Fuchs}, \citenamefont {Runkel},\
  and\ \citenamefont {Schweigert}}]{FUCHS2002452}%
  \BibitemOpen
  \bibfield  {author} {\bibinfo {author} {\bibfnamefont {J.}~\bibnamefont
  {Fuchs}}, \bibinfo {author} {\bibfnamefont {I.}~\bibnamefont {Runkel}}, \
  and\ \bibinfo {author} {\bibfnamefont {C.}~\bibnamefont {Schweigert}},\
  }Conformal correlation functions, frobenius algebras and triangulations,\
  \href {\doibase 10.1016/S0550-3213(01)00638-1} {\bibfield  {journal}
  {\bibinfo  {journal} {Nucl. Phys. B}\ }\textbf {\bibinfo {volume} {624}},\
  \bibinfo {pages} {452 }},\ \Eprint {http://arxiv.org/abs/hep-th/0110133}
  {arXiv:hep-th/0110133}  (\bibinfo {year} {2002}{\natexlab{a}})\BibitemShut
  {NoStop}%
\bibitem [{\citenamefont {Fuchs}\ \emph
  {et~al.}(2002{\natexlab{b}})\citenamefont {Fuchs}, \citenamefont {Runkel},\
  and\ \citenamefont {Schweigert}}]{FUCHS2002353}%
  \BibitemOpen
  \bibfield  {author} {\bibinfo {author} {\bibfnamefont {J.}~\bibnamefont
  {Fuchs}}, \bibinfo {author} {\bibfnamefont {I.}~\bibnamefont {Runkel}}, \
  and\ \bibinfo {author} {\bibfnamefont {C.}~\bibnamefont {Schweigert}},\ }Tft
  construction of rcft correlators i: partition functions,\ \href {\doibase
  10.1016/S0550-3213(02)00744-7} {\bibfield  {journal} {\bibinfo  {journal}
  {Nucl. Phys. B}\ }\textbf {\bibinfo {volume} {646}},\ \bibinfo {pages} {353
  }},\ \Eprint {http://arxiv.org/abs/hep-th/0204148} {arXiv:hep-th/0204148}
  (\bibinfo {year} {2002}{\natexlab{b}})\BibitemShut {NoStop}%
\bibitem [{\citenamefont {Fr{\"o}hlich}\ \emph {et~al.}(2004)\citenamefont
  {Fr{\"o}hlich}, \citenamefont {Fuchs}, \citenamefont {Runkel},\ and\
  \citenamefont {Schweigert}}]{PhysRevLett.93.070601}%
  \BibitemOpen
  \bibfield  {author} {\bibinfo {author} {\bibfnamefont {J.}~\bibnamefont
  {Fr{\"o}hlich}}, \bibinfo {author} {\bibfnamefont {J.}~\bibnamefont {Fuchs}},
  \bibinfo {author} {\bibfnamefont {I.}~\bibnamefont {Runkel}}, \ and\ \bibinfo
  {author} {\bibfnamefont {C.}~\bibnamefont {Schweigert}},\ }Kramers-wannier
  duality from conformal defects,\ \href {\doibase
  10.1103/PhysRevLett.93.070601} {\bibfield  {journal} {\bibinfo  {journal}
  {Phys. Rev. Lett.}\ }\textbf {\bibinfo {volume} {93}},\ \bibinfo {pages}
  {070601}},\ \Eprint {http://arxiv.org/abs/cond-mat/0404051}
  {arXiv:cond-mat/0404051}  (\bibinfo {year} {2004})\BibitemShut {NoStop}%
\bibitem [{\citenamefont {Fr{\"o}hlich}\ \emph {et~al.}(2007)\citenamefont
  {Fr{\"o}hlich}, \citenamefont {Fuchs}, \citenamefont {Runkel},\ and\
  \citenamefont {Schweigert}}]{FROHLICH2007354}%
  \BibitemOpen
  \bibfield  {author} {\bibinfo {author} {\bibfnamefont {J.}~\bibnamefont
  {Fr{\"o}hlich}}, \bibinfo {author} {\bibfnamefont {J.}~\bibnamefont {Fuchs}},
  \bibinfo {author} {\bibfnamefont {I.}~\bibnamefont {Runkel}}, \ and\ \bibinfo
  {author} {\bibfnamefont {C.}~\bibnamefont {Schweigert}},\ }Duality and
  defects in rational conformal field theory,\ \href {\doibase
  10.1016/j.nuclphysb.2006.11.017} {\bibfield  {journal} {\bibinfo  {journal}
  {Nucl. Phys. B}\ }\textbf {\bibinfo {volume} {763}},\ \bibinfo {pages} {354
  }},\ \Eprint {http://arxiv.org/abs/hep-th/0607247} {arXiv:hep-th/0607247}
  (\bibinfo {year} {2007})\BibitemShut {NoStop}%
\bibitem [{\citenamefont {Fr{\"o}hlich}\ \emph {et~al.}(2009)\citenamefont
  {Fr{\"o}hlich}, \citenamefont {Fuchs}, \citenamefont {Runkel},\ and\
  \citenamefont {Schweigert}}]{frohlich2009defect}%
  \BibitemOpen
  \bibfield  {author} {\bibinfo {author} {\bibfnamefont {J.}~\bibnamefont
  {Fr{\"o}hlich}}, \bibinfo {author} {\bibfnamefont {J.}~\bibnamefont {Fuchs}},
  \bibinfo {author} {\bibfnamefont {I.}~\bibnamefont {Runkel}}, \ and\ \bibinfo
  {author} {\bibfnamefont {C.}~\bibnamefont {Schweigert}},\ }Defect lines,
  dualities, and generalised orbifolds,\ \href@noop {} {\ }\Eprint
  {http://arxiv.org/abs/0909.5013} {arXiv:0909.5013}  (\bibinfo {year}
  {2009})\BibitemShut {NoStop}%
\bibitem [{\citenamefont {Chen}\ \emph
  {et~al.}(2013{\natexlab{b}})\citenamefont {Chen}, \citenamefont {Wang},
  \citenamefont {Lu},\ and\ \citenamefont {Lee}}]{Chen2013248}%
  \BibitemOpen
  \bibfield  {author} {\bibinfo {author} {\bibfnamefont {X.}~\bibnamefont
  {Chen}}, \bibinfo {author} {\bibfnamefont {F.}~\bibnamefont {Wang}}, \bibinfo
  {author} {\bibfnamefont {Y.-M.}\ \bibnamefont {Lu}}, \ and\ \bibinfo {author}
  {\bibfnamefont {D.-H.}\ \bibnamefont {Lee}},\ }{Critical theories of phase
  transition between symmetry protected topological states and their relation
  to the gapless boundary theories},\ \href {\doibase
  10.1016/j.nuclphysb.2013.04.015} {\bibfield  {journal} {\bibinfo  {journal}
  {Nucl. Phys. B}\ }\textbf {\bibinfo {volume} {873}},\ \bibinfo {pages} {248
  }},\ \Eprint {http://arxiv.org/abs/1302.3121} {arXiv:1302.3121}  (\bibinfo
  {year} {2013}{\natexlab{b}})\BibitemShut {NoStop}%
\bibitem [{\citenamefont {Tsui}\ \emph
  {et~al.}(2015{\natexlab{a}})\citenamefont {Tsui}, \citenamefont {Jiang},
  \citenamefont {Lu},\ and\ \citenamefont {Lee}}]{Tsui2015330}%
  \BibitemOpen
  \bibfield  {author} {\bibinfo {author} {\bibfnamefont {L.}~\bibnamefont
  {Tsui}}, \bibinfo {author} {\bibfnamefont {H.-C.}\ \bibnamefont {Jiang}},
  \bibinfo {author} {\bibfnamefont {Y.-M.}\ \bibnamefont {Lu}}, \ and\ \bibinfo
  {author} {\bibfnamefont {D.-H.}\ \bibnamefont {Lee}},\ }{Quantum phase
  transitions between a class of symmetry protected topological states},\ \href
  {\doibase 10.1016/j.nuclphysb.2015.04.020} {\bibfield  {journal} {\bibinfo
  {journal} {Nucl. Phys. B}\ }\textbf {\bibinfo {volume} {896}},\ \bibinfo
  {pages} {330 }},\ \Eprint {http://arxiv.org/abs/1503.06794}
  {arXiv:1503.06794}  (\bibinfo {year} {2015}{\natexlab{a}})\BibitemShut
  {NoStop}%
\bibitem [{\citenamefont {Tsui}\ \emph
  {et~al.}(2015{\natexlab{b}})\citenamefont {Tsui}, \citenamefont {Wang},\ and\
  \citenamefont {Lee}}]{tsui2015topological}%
  \BibitemOpen
  \bibfield  {author} {\bibinfo {author} {\bibfnamefont {L.}~\bibnamefont
  {Tsui}}, \bibinfo {author} {\bibfnamefont {F.}~\bibnamefont {Wang}}, \ and\
  \bibinfo {author} {\bibfnamefont {D.-H.}\ \bibnamefont {Lee}},\ }Topological
  versus landau-like phase transitions,\ \href@noop {} {\ }\Eprint
  {http://arxiv.org/abs/1511.07460} {arXiv:1511.07460}  (\bibinfo {year}
  {2015}{\natexlab{b}})\BibitemShut {NoStop}%
\bibitem [{\citenamefont {Tsui}\ \emph {et~al.}(2017)\citenamefont {Tsui},
  \citenamefont {Huang}, \citenamefont {Jiang},\ and\ \citenamefont
  {Lee}}]{tsui2017conformal}%
  \BibitemOpen
  \bibfield  {author} {\bibinfo {author} {\bibfnamefont {L.}~\bibnamefont
  {Tsui}}, \bibinfo {author} {\bibfnamefont {Y.-T.}\ \bibnamefont {Huang}},
  \bibinfo {author} {\bibfnamefont {H.-C.}\ \bibnamefont {Jiang}}, \ and\
  \bibinfo {author} {\bibfnamefont {D.-H.}\ \bibnamefont {Lee}},\ }{The phase
  transitions between ${Z_n \times× Z_n}$ bosonic topological phases in 1 +
  1D, and a constraint on the central charge for the critical points between
  bosonic symmetry protected topological phases},\ \href {\doibase
  10.1016/j.nuclphysb.2017.03.021} {\bibfield  {journal} {\bibinfo  {journal}
  {Nucl. Phys. B}\ }\textbf {\bibinfo {volume} {919}},\ \bibinfo {pages}
  {470}},\ \Eprint {http://arxiv.org/abs/1701.00834} {arXiv:1701.00834}
  (\bibinfo {year} {2017})\BibitemShut {NoStop}%
\bibitem [{\citenamefont {Kubica}\ and\ \citenamefont {Yoshida}(2014)}]{KY}%
  \BibitemOpen
  \bibfield  {author} {\bibinfo {author} {\bibfnamefont {A.}~\bibnamefont
  {Kubica}}\ and\ \bibinfo {author} {\bibfnamefont {B.}~\bibnamefont
  {Yoshida}},\ }{Precise estimation of critical exponents from real-space
  renormalization group analysis},\ \href@noop {} {\ }\Eprint
  {http://arxiv.org/abs/1402.0619} {arXiv:1402.0619}  (\bibinfo {year}
  {2014})\BibitemShut {NoStop}%
\bibitem [{\citenamefont {O'Brien}\ \emph {et~al.}(2015)\citenamefont
  {O'Brien}, \citenamefont {Bartlett}, \citenamefont {Doherty},\ and\
  \citenamefont {Flammia}}]{aroon}%
  \BibitemOpen
  \bibfield  {author} {\bibinfo {author} {\bibfnamefont {A.}~\bibnamefont
  {O'Brien}}, \bibinfo {author} {\bibfnamefont {S.~D.}\ \bibnamefont
  {Bartlett}}, \bibinfo {author} {\bibfnamefont {A.~C.}\ \bibnamefont
  {Doherty}}, \ and\ \bibinfo {author} {\bibfnamefont {S.~T.}\ \bibnamefont
  {Flammia}},\ }{Symmetry-respecting real-space renormalization for the quantum
  Ashkin-Teller model},\ \href {\doibase 10.1103/PhysRevE.92.042163} {\bibfield
   {journal} {\bibinfo  {journal} {Phys. Rev. E}\ }\textbf {\bibinfo {volume}
  {92}},\ \bibinfo {pages} {042163}},\ \Eprint
  {http://arxiv.org/abs/1507.00038} {arXiv:1507.00038}  (\bibinfo {year}
  {2015})\BibitemShut {NoStop}%
\bibitem [{\citenamefont {Kennedy}\ and\ \citenamefont
  {Tasaki}(1992)}]{PhysRevB.45.304}%
  \BibitemOpen
  \bibfield  {author} {\bibinfo {author} {\bibfnamefont {T.}~\bibnamefont
  {Kennedy}}\ and\ \bibinfo {author} {\bibfnamefont {H.}~\bibnamefont
  {Tasaki}},\ }{Hidden ${\mathbb{Z}}_{2}\times{\mathbb{Z}}_{2}$ symmetry
  breaking in Haldane-gap antiferromagnets},\ \href {\doibase
  10.1103/PhysRevB.45.304} {\bibfield  {journal} {\bibinfo  {journal} {Phys.
  Rev. B}\ }\textbf {\bibinfo {volume} {45}},\ \bibinfo {pages} {304}}
  (\bibinfo {year} {1992})\BibitemShut {NoStop}%
\bibitem [{\citenamefont {Else}\ \emph {et~al.}(2013)\citenamefont {Else},
  \citenamefont {Bartlett},\ and\ \citenamefont {Doherty}}]{Else2012b}%
  \BibitemOpen
  \bibfield  {author} {\bibinfo {author} {\bibfnamefont {D.~V.}\ \bibnamefont
  {Else}}, \bibinfo {author} {\bibfnamefont {S.~D.}\ \bibnamefont {Bartlett}},
  \ and\ \bibinfo {author} {\bibfnamefont {A.~C.}\ \bibnamefont {Doherty}},\
  }{The hidden symmetry-breaking picture of symmetry-protected topological
  order},\ \href {\doibase 10.1103/PhysRevB.88.085114} {\bibfield  {journal}
  {\bibinfo  {journal} {Phys. Rev. B}\ }\textbf {\bibinfo {volume} {88}},\
  \bibinfo {pages} {085114}},\ \Eprint {http://arxiv.org/abs/1304.0783}
  {arXiv:1304.0783}  (\bibinfo {year} {2013})\BibitemShut {NoStop}%
\bibitem [{\citenamefont {Yang}(1987)}]{Yang1987}%
  \BibitemOpen
  \bibfield  {author} {\bibinfo {author} {\bibfnamefont {S.-K.}\ \bibnamefont
  {Yang}},\ }{Modular invariant partition function of the Ashkin-Teller model
  on the critical line and N= 2 superconformal invariance},\ \href {\doibase
  10.1016/0550-3213(87)90334-8} {\bibfield  {journal} {\bibinfo  {journal}
  {Nucl. Phys. B}\ }\textbf {\bibinfo {volume} {285}},\ \bibinfo {pages} {183}}
  (\bibinfo {year} {1987})\BibitemShut {NoStop}%
\bibitem [{\citenamefont {Baake}\ \emph {et~al.}(1987)\citenamefont {Baake},
  \citenamefont {von Gehlen},\ and\ \citenamefont {Rittenberg}}]{Baake1987a}%
  \BibitemOpen
  \bibfield  {author} {\bibinfo {author} {\bibfnamefont {M.}~\bibnamefont
  {Baake}}, \bibinfo {author} {\bibfnamefont {G.}~\bibnamefont {von Gehlen}}, \
  and\ \bibinfo {author} {\bibfnamefont {V.}~\bibnamefont {Rittenberg}},\
  }{Operator content of the Ashkin-Teller quantum chain-superconformal and
  Zamolodchikov-Fateev invariance: II. Boundary conditions compatible with the
  torus},\ \href {\doibase 10.1088/0305-4470/20/8/002} {\bibfield  {journal}
  {\bibinfo  {journal} {J. Phys. A: Math. Gen.}\ }\textbf {\bibinfo {volume}
  {20}},\ \bibinfo {pages} {6635}} (\bibinfo {year} {1987})\BibitemShut
  {NoStop}%
\bibitem [{\citenamefont {Alcaraz}\ \emph
  {et~al.}(1988{\natexlab{a}})\citenamefont {Alcaraz}, \citenamefont {Baake},
  \citenamefont {Grimmn},\ and\ \citenamefont {Rittenberg}}]{Alcaraz1988JPHYS}%
  \BibitemOpen
  \bibfield  {author} {\bibinfo {author} {\bibfnamefont {F.~C.}\ \bibnamefont
  {Alcaraz}}, \bibinfo {author} {\bibfnamefont {M.}~\bibnamefont {Baake}},
  \bibinfo {author} {\bibfnamefont {U.}~\bibnamefont {Grimmn}}, \ and\ \bibinfo
  {author} {\bibfnamefont {V.}~\bibnamefont {Rittenberg}},\ }{Operator content
  of the XXZ chain},\ \href {\doibase 10.1088/0305-4470/21/3/001} {\bibfield
  {journal} {\bibinfo  {journal} {J. Phys. A: Math. Gen.}\ }\textbf {\bibinfo
  {volume} {21}},\ \bibinfo {pages} {L117}} (\bibinfo {year}
  {1988}{\natexlab{a}})\BibitemShut {NoStop}%
\bibitem [{\citenamefont {Alcaraz}\ \emph
  {et~al.}(1988{\natexlab{b}})\citenamefont {Alcaraz}, \citenamefont {Barber},\
  and\ \citenamefont {Batchelor}}]{Alcaraz1988}%
  \BibitemOpen
  \bibfield  {author} {\bibinfo {author} {\bibfnamefont {F.~C.}\ \bibnamefont
  {Alcaraz}}, \bibinfo {author} {\bibfnamefont {M.~N.}\ \bibnamefont {Barber}},
  \ and\ \bibinfo {author} {\bibfnamefont {M.~T.}\ \bibnamefont {Batchelor}},\
  }{Conformal invariance, the XXZ chain and the operator content of
  two-dimensional critical systems},\ \href {\doibase
  10.1016/0003-4916(88)90015-2} {\bibfield  {journal} {\bibinfo  {journal}
  {Ann. Phys.}\ }\textbf {\bibinfo {volume} {182}},\ \bibinfo {pages} {280}}
  (\bibinfo {year} {1988}{\natexlab{b}})\BibitemShut {NoStop}%
\bibitem [{\citenamefont {Bridgeman}(2014)}]{Bridgemanmasters}%
  \BibitemOpen
  \bibfield  {author} {\bibinfo {author} {\bibfnamefont {J.~C.}\ \bibnamefont
  {Bridgeman}},\ }\emph {\bibinfo {title} {{Effective Edge States of Symmetry
  Protected Topological Systems}}},\ \href
  {http://www.physics.usyd.edu.au/~jbridge/MastersThesis.pdf} {Master's
  thesis},\ \bibinfo  {school} {Perimeter Institute} (\bibinfo {year}
  {2014})\BibitemShut {NoStop}%
\bibitem [{\citenamefont {Bridgeman}\ \emph {et~al.}(2015)\citenamefont
  {Bridgeman}, \citenamefont {O'Brien}, \citenamefont {Bartlett},\ and\
  \citenamefont {Doherty}}]{Bridgeman2015}%
  \BibitemOpen
  \bibfield  {author} {\bibinfo {author} {\bibfnamefont {J.~C.}\ \bibnamefont
  {Bridgeman}}, \bibinfo {author} {\bibfnamefont {A.}~\bibnamefont {O'Brien}},
  \bibinfo {author} {\bibfnamefont {S.~D.}\ \bibnamefont {Bartlett}}, \ and\
  \bibinfo {author} {\bibfnamefont {A.~C.}\ \bibnamefont {Doherty}},\
  }{Multiscale entanglement renormalization ansatz for spin chains with
  continuously varying criticality},\ \href {\doibase
  10.1103/PhysRevB.91.165129} {\bibfield  {journal} {\bibinfo  {journal} {Phys.
  Rev. B}\ }\textbf {\bibinfo {volume} {91}},\ \bibinfo {pages} {165129}},\
  \Eprint {http://arxiv.org/abs/1501.02817} {arXiv:1501.02817}  (\bibinfo
  {year} {2015})\BibitemShut {NoStop}%
\bibitem [{\citenamefont {de~Wild~Propitius}(1997)}]{DEWILDPROPITIUS1997297}%
  \BibitemOpen
  \bibfield  {author} {\bibinfo {author} {\bibfnamefont {M.}~\bibnamefont
  {de~Wild~Propitius}},\ }{(Spontaneously broken) Abelian Chern-Simons
  theories},\ \href {\doibase 10.1016/S0550-3213(97)00005-9} {\bibfield
  {journal} {\bibinfo  {journal} {Nucl. Phys. B}\ }\textbf {\bibinfo {volume}
  {489}},\ \bibinfo {pages} {297 }},\ \Eprint
  {http://arxiv.org/abs/hep-th/9606029} {arXiv:hep-th/9606029}  (\bibinfo
  {year} {1997})\BibitemShut {NoStop}%
\bibitem [{\citenamefont {Goff}\ \emph {et~al.}(2007)\citenamefont {Goff},
  \citenamefont {Mason},\ and\ \citenamefont {Ng}}]{Goff}%
  \BibitemOpen
  \bibfield  {author} {\bibinfo {author} {\bibfnamefont {C.}~\bibnamefont
  {Goff}}, \bibinfo {author} {\bibfnamefont {G.}~\bibnamefont {Mason}}, \ and\
  \bibinfo {author} {\bibfnamefont {S.-H.}\ \bibnamefont {Ng}},\ }{On the Gauge
  Equivalence of Twisted Quantum Doubles of Elementary Abelian and
  Extra-Special 2-Groups},\ \href {\doibase 10.1016/j.jalgebra.2006.10.022}
  {\bibfield  {journal} {\bibinfo  {journal} {J. Algebra}\ }\textbf {\bibinfo
  {volume} {312}},\ \bibinfo {pages} {849}},\ \Eprint
  {http://arxiv.org/abs/math/0603191} {arXiv:math/0603191}  (\bibinfo {year}
  {2007})\BibitemShut {NoStop}%
\bibitem [{\citenamefont {Wang}\ and\ \citenamefont {Wen}(2015)}]{Juven2014}%
  \BibitemOpen
  \bibfield  {author} {\bibinfo {author} {\bibfnamefont {J.}~\bibnamefont
  {Wang}}\ and\ \bibinfo {author} {\bibfnamefont {X.-G.}\ \bibnamefont {Wen}},\
  }{Non-Abelian String and Particle Braiding in Topological Order: Modular
  SL(3,$\mathbb{Z}$) Representation and $3+1$D Twisted Gauge Theory},\ \href
  {\doibase 10.1103/PhysRevB.91.035134} {\bibfield  {journal} {\bibinfo
  {journal} {Phys. Rev. B}\ }\textbf {\bibinfo {volume} {91}},\ \bibinfo
  {pages} {035134}},\ \Eprint {http://arxiv.org/abs/1404.7854}
  {arXiv:1404.7854}  (\bibinfo {year} {2015})\BibitemShut {NoStop}%
\bibitem [{\citenamefont {He}\ \emph {et~al.}(2017)\citenamefont {He},
  \citenamefont {Zheng},\ and\ \citenamefont {von
  Keyserlingk}}]{PhysRevB.95.035131}%
  \BibitemOpen
  \bibfield  {author} {\bibinfo {author} {\bibfnamefont {H.}~\bibnamefont
  {He}}, \bibinfo {author} {\bibfnamefont {Y.}~\bibnamefont {Zheng}}, \ and\
  \bibinfo {author} {\bibfnamefont {C.}~\bibnamefont {von Keyserlingk}},\
  }{Field theories for gauged symmetry-protected topological phases:
  Non-Abelian anyons with Abelian gauge group
  ${\mathbb{Z}}_{2}^{\ensuremath{\bigotimes}3}$},\ \href {\doibase
  10.1103/PhysRevB.95.035131} {\bibfield  {journal} {\bibinfo  {journal} {Phys.
  Rev. B}\ }\textbf {\bibinfo {volume} {95}},\ \bibinfo {pages} {035131}},\
  \Eprint {http://arxiv.org/abs/1608.05393} {arXiv:1608.05393}  (\bibinfo
  {year} {2017})\BibitemShut {NoStop}%
\bibitem [{\citenamefont {Trebst}\ \emph {et~al.}(2008)\citenamefont {Trebst},
  \citenamefont {Ardonne}, \citenamefont {Feiguin}, \citenamefont {Huse},
  \citenamefont {Ludwig},\ and\ \citenamefont
  {Troyer}}]{PhysRevLett.101.050401}%
  \BibitemOpen
  \bibfield  {author} {\bibinfo {author} {\bibfnamefont {S.}~\bibnamefont
  {Trebst}}, \bibinfo {author} {\bibfnamefont {E.}~\bibnamefont {Ardonne}},
  \bibinfo {author} {\bibfnamefont {A.}~\bibnamefont {Feiguin}}, \bibinfo
  {author} {\bibfnamefont {D.~A.}\ \bibnamefont {Huse}}, \bibinfo {author}
  {\bibfnamefont {A.~W.~W.}\ \bibnamefont {Ludwig}}, \ and\ \bibinfo {author}
  {\bibfnamefont {M.}~\bibnamefont {Troyer}},\ }{Collective States of
  Interacting Fibonacci Anyons},\ \href {\doibase
  10.1103/PhysRevLett.101.050401} {\bibfield  {journal} {\bibinfo  {journal}
  {Phys. Rev. Lett.}\ }\textbf {\bibinfo {volume} {101}},\ \bibinfo {pages}
  {050401}},\ \Eprint {http://arxiv.org/abs/0801.4602} {arXiv:0801.4602}
  (\bibinfo {year} {2008})\BibitemShut {NoStop}%
\bibitem [{\citenamefont {Gils}\ \emph {et~al.}(2009)\citenamefont {Gils},
  \citenamefont {Trebst}, \citenamefont {Kitaev}, \citenamefont {Ludwig},
  \citenamefont {Troyer},\ and\ \citenamefont {Wang}}]{gils2009topology}%
  \BibitemOpen
  \bibfield  {author} {\bibinfo {author} {\bibfnamefont {C.}~\bibnamefont
  {Gils}}, \bibinfo {author} {\bibfnamefont {S.}~\bibnamefont {Trebst}},
  \bibinfo {author} {\bibfnamefont {A.}~\bibnamefont {Kitaev}}, \bibinfo
  {author} {\bibfnamefont {A.~W.}\ \bibnamefont {Ludwig}}, \bibinfo {author}
  {\bibfnamefont {M.}~\bibnamefont {Troyer}}, \ and\ \bibinfo {author}
  {\bibfnamefont {Z.}~\bibnamefont {Wang}},\ }{Topology-driven quantum phase
  transitions in time-reversal-invariant anyonic quantum liquids},\ \href
  {\doibase 10.1038/nphys1396} {\bibfield  {journal} {\bibinfo  {journal} {Nat.
  Phys.}\ }\textbf {\bibinfo {volume} {5}},\ \bibinfo {pages} {834}},\ \Eprint
  {http://arxiv.org/abs/0906.1579} {arXiv:0906.1579}  (\bibinfo {year}
  {2009})\BibitemShut {NoStop}%
\bibitem [{\citenamefont {Bauer}\ \emph {et~al.}(2011)\citenamefont {Bauer},
  \citenamefont {Carr}, \citenamefont {Evertz}, \citenamefont {Feiguin},
  \citenamefont {Freire}, \citenamefont {Fuchs}, \citenamefont {Gamper},
  \citenamefont {Gukelberger}, \citenamefont {Gull}, \citenamefont {Guertler},
  \citenamefont {Hehn}, \citenamefont {Igarashi}, \citenamefont {Isakov},
  \citenamefont {Koop}, \citenamefont {Ma}, \citenamefont {Mates},
  \citenamefont {Matsuo}, \citenamefont {Parcollet}, \citenamefont {Pawlowski},
  \citenamefont {Picon}, \citenamefont {Pollet}, \citenamefont {Santos},
  \citenamefont {Scarola}, \citenamefont {Schollwöck}, \citenamefont {Silva},
  \citenamefont {Surer}, \citenamefont {Todo}, \citenamefont {Trebst},
  \citenamefont {Troyer}, \citenamefont {Wall}, \citenamefont {Werner},\ and\
  \citenamefont {Wessel}}]{ALPS1}%
  \BibitemOpen
  \bibfield  {author} {\bibinfo {author} {\bibfnamefont {B.}~\bibnamefont
  {Bauer}}, \bibinfo {author} {\bibfnamefont {L.~D.}\ \bibnamefont {Carr}},
  \bibinfo {author} {\bibfnamefont {H.}~\bibnamefont {Evertz}}, \bibinfo
  {author} {\bibfnamefont {A.}~\bibnamefont {Feiguin}}, \bibinfo {author}
  {\bibfnamefont {J.}~\bibnamefont {Freire}}, \bibinfo {author} {\bibfnamefont
  {S.}~\bibnamefont {Fuchs}}, \bibinfo {author} {\bibfnamefont
  {L.}~\bibnamefont {Gamper}}, \bibinfo {author} {\bibfnamefont
  {J.}~\bibnamefont {Gukelberger}}, \bibinfo {author} {\bibfnamefont
  {E.}~\bibnamefont {Gull}}, \bibinfo {author} {\bibfnamefont {S.}~\bibnamefont
  {Guertler}}, \bibinfo {author} {\bibfnamefont {A.}~\bibnamefont {Hehn}},
  \bibinfo {author} {\bibfnamefont {R.}~\bibnamefont {Igarashi}}, \bibinfo
  {author} {\bibfnamefont {S.}~\bibnamefont {Isakov}}, \bibinfo {author}
  {\bibfnamefont {D.}~\bibnamefont {Koop}}, \bibinfo {author} {\bibfnamefont
  {P.}~\bibnamefont {Ma}}, \bibinfo {author} {\bibfnamefont {P.}~\bibnamefont
  {Mates}}, \bibinfo {author} {\bibfnamefont {H.}~\bibnamefont {Matsuo}},
  \bibinfo {author} {\bibfnamefont {O.}~\bibnamefont {Parcollet}}, \bibinfo
  {author} {\bibfnamefont {G.}~\bibnamefont {Pawlowski}}, \bibinfo {author}
  {\bibfnamefont {J.}~\bibnamefont {Picon}}, \bibinfo {author} {\bibfnamefont
  {L.}~\bibnamefont {Pollet}}, \bibinfo {author} {\bibfnamefont
  {E.}~\bibnamefont {Santos}}, \bibinfo {author} {\bibfnamefont
  {V.}~\bibnamefont {Scarola}}, \bibinfo {author} {\bibfnamefont
  {U.}~\bibnamefont {Schollwöck}}, \bibinfo {author} {\bibfnamefont
  {C.}~\bibnamefont {Silva}}, \bibinfo {author} {\bibfnamefont
  {B.}~\bibnamefont {Surer}}, \bibinfo {author} {\bibfnamefont
  {S.}~\bibnamefont {Todo}}, \bibinfo {author} {\bibfnamefont {S.}~\bibnamefont
  {Trebst}}, \bibinfo {author} {\bibfnamefont {M.}~\bibnamefont {Troyer}},
  \bibinfo {author} {\bibfnamefont {M.}~\bibnamefont {Wall}}, \bibinfo {author}
  {\bibfnamefont {P.}~\bibnamefont {Werner}}, \ and\ \bibinfo {author}
  {\bibfnamefont {S.}~\bibnamefont {Wessel}},\ }{The ALPS project release 2.0:
  Open source software for strongly correlated systems},\ \href {\doibase
  10.1088/1742-5468/2011/05/P05001} {\bibfield  {journal} {\bibinfo  {journal}
  {J. Stat. Mech.}\ }\textbf {\bibinfo {volume} {2011}},\ \bibinfo {pages}
  {P05001}},\ \Eprint {http://arxiv.org/abs/1101.2646} {arXiv:1101.2646}
  (\bibinfo {year} {2011})\BibitemShut {NoStop}%
\bibitem [{\citenamefont {Dolfi}\ \emph {et~al.}(2014)\citenamefont {Dolfi},
  \citenamefont {Bauer}, \citenamefont {Keller}, \citenamefont {Kosenkov},
  \citenamefont {Ewart}, \citenamefont {Kantian}, \citenamefont {Giamarchi},\
  and\ \citenamefont {Troyer}}]{ALPS2}%
  \BibitemOpen
  \bibfield  {author} {\bibinfo {author} {\bibfnamefont {M.}~\bibnamefont
  {Dolfi}}, \bibinfo {author} {\bibfnamefont {B.}~\bibnamefont {Bauer}},
  \bibinfo {author} {\bibfnamefont {S.}~\bibnamefont {Keller}}, \bibinfo
  {author} {\bibfnamefont {A.}~\bibnamefont {Kosenkov}}, \bibinfo {author}
  {\bibfnamefont {T.}~\bibnamefont {Ewart}}, \bibinfo {author} {\bibfnamefont
  {A.}~\bibnamefont {Kantian}}, \bibinfo {author} {\bibfnamefont
  {T.}~\bibnamefont {Giamarchi}}, \ and\ \bibinfo {author} {\bibfnamefont
  {M.}~\bibnamefont {Troyer}},\ }{Matrix Product State applications for the
  ALPS project},\ \href {\doibase 10.1016/j.cpc.2014.08.019} {\bibfield
  {journal} {\bibinfo  {journal} {Comput. Phys. Commun.}\ }\textbf {\bibinfo
  {volume} {185}},\ \bibinfo {pages} {3430}},\ \Eprint
  {http://arxiv.org/abs/1407.0872} {arXiv:1407.0872}  (\bibinfo {year}
  {2014})\BibitemShut {NoStop}%
\bibitem [{\citenamefont {von Gehlen}\ and\ \citenamefont
  {Rittenberg}(1987)}]{Gehlen1987}%
  \BibitemOpen
  \bibfield  {author} {\bibinfo {author} {\bibfnamefont {G.}~\bibnamefont {von
  Gehlen}}\ and\ \bibinfo {author} {\bibfnamefont {V.}~\bibnamefont
  {Rittenberg}},\ }{The Ashkin-Teller quantum chain and conformal invariance},\
  \href {\doibase 10.1088/0305-4470/20/1/030} {\bibfield  {journal} {\bibinfo
  {journal} {J. Phys. A: Math. Gen}\ }\textbf {\bibinfo {volume} {20}},\
  \bibinfo {pages} {227}} (\bibinfo {year} {1987})\BibitemShut {NoStop}%
\bibitem [{\citenamefont {{Aasen \emph{et al.}}}()}]{daveprep}%
  \BibitemOpen
  \bibfield  {author} {\bibinfo {author} {\bibfnamefont {D.}~\bibnamefont
  {{Aasen \emph{et al.}}}},\ }\href@noop {} {\bibinfo  {journal} {in
  preparation}\ }\BibitemShut {NoStop}%
\bibitem [{\citenamefont {Buican}\ and\ \citenamefont
  {Gromov}(2017)}]{buican2017anyonic}%
  \BibitemOpen
\bibfield  {journal} {  }\bibfield  {author} {\bibinfo {author} {\bibfnamefont
  {M.}~\bibnamefont {Buican}}\ and\ \bibinfo {author} {\bibfnamefont
  {A.}~\bibnamefont {Gromov}},\ }{Anyonic Chains, Topological Defects, and
  Conformal Field Theory},\ \href@noop {} {\ }\Eprint
  {http://arxiv.org/abs/1701.02800} {arXiv:1701.02800}  (\bibinfo {year}
  {2017})\BibitemShut {NoStop}%
\bibitem [{\citenamefont {Lan}\ and\ \citenamefont {Wen}(2014)}]{Qalgebra}%
  \BibitemOpen
  \bibfield  {author} {\bibinfo {author} {\bibfnamefont {T.}~\bibnamefont
  {Lan}}\ and\ \bibinfo {author} {\bibfnamefont {X.-G.}\ \bibnamefont {Wen}},\
  }{Topological quasiparticles and the holographic bulk-edge relation in
  $(2+1)$-dimensional string-net models},\ \href {\doibase
  10.1103/PhysRevB.90.115119} {\bibfield  {journal} {\bibinfo  {journal} {Phys.
  Rev. B}\ }\textbf {\bibinfo {volume} {90}},\ \bibinfo {pages} {115119}},\
  \Eprint {http://arxiv.org/abs/1311.1784} {arXiv:1311.1784}  (\bibinfo {year}
  {2014})\BibitemShut {NoStop}%
\bibitem [{\citenamefont {{Haah}}(2016)}]{haah}%
  \BibitemOpen
  \bibfield  {author} {\bibinfo {author} {\bibfnamefont {J.}~\bibnamefont
  {{Haah}}},\ }{An invariant of topologically ordered states under local
  unitary transformations},\ \href {\doibase 10.1007/s00220-016-2594-y}
  {\bibfield  {journal} {\bibinfo  {journal} {Commun. Math. Phys.}\ }\textbf
  {\bibinfo {volume} {342}},\ \bibinfo {pages} {771}},\ \Eprint
  {http://arxiv.org/abs/quant-ph/1407.2926} {arXiv:quant-ph/1407.2926}
  (\bibinfo {year} {2016})\BibitemShut {NoStop}%
\bibitem [{\citenamefont {{Hu}}\ \emph {et~al.}(2015)\citenamefont {{Hu}},
  \citenamefont {{Geer}},\ and\ \citenamefont {{Wu}}}]{dyonicspectrum}%
  \BibitemOpen
  \bibfield  {author} {\bibinfo {author} {\bibfnamefont {Y.}~\bibnamefont
  {{Hu}}}, \bibinfo {author} {\bibfnamefont {N.}~\bibnamefont {{Geer}}}, \ and\
  \bibinfo {author} {\bibfnamefont {Y.-S.}\ \bibnamefont {{Wu}}},\ }{Full Dyon
  Excitation Spectrum in Generalized Levin-Wen Models},\ \href@noop {} {\
  }\Eprint {http://arxiv.org/abs/cond-mat.str-el/1502.03433}
  {arXiv:cond-mat.str-el/1502.03433}  (\bibinfo {year} {2015})\BibitemShut
  {NoStop}%
\bibitem [{\citenamefont {Evenbly}\ and\ \citenamefont
  {Vidal}(2015)}]{PhysRevLett.115.180405}%
  \BibitemOpen
  \bibfield  {author} {\bibinfo {author} {\bibfnamefont {G.}~\bibnamefont
  {Evenbly}}\ and\ \bibinfo {author} {\bibfnamefont {G.}~\bibnamefont
  {Vidal}},\ }{Tensor Network Renormalization},\ \href {\doibase
  10.1103/PhysRevLett.115.180405} {\bibfield  {journal} {\bibinfo  {journal}
  {Phys. Rev. Lett.}\ }\textbf {\bibinfo {volume} {115}},\ \bibinfo {pages}
  {180405}},\ \Eprint {http://arxiv.org/abs/1412.0732} {arXiv:1412.0732}
  (\bibinfo {year} {2015})\BibitemShut {NoStop}%
\bibitem [{\citenamefont {Evenbly}(2017)}]{PhysRevB.95.045117}%
  \BibitemOpen
  \bibfield  {author} {\bibinfo {author} {\bibfnamefont {G.}~\bibnamefont
  {Evenbly}},\ }{Algorithms for tensor network renormalization},\ \href
  {\doibase 10.1103/PhysRevB.95.045117} {\bibfield  {journal} {\bibinfo
  {journal} {Phys. Rev. B}\ }\textbf {\bibinfo {volume} {95}},\ \bibinfo
  {pages} {045117}},\ \Eprint {http://arxiv.org/abs/1509.07484}
  {arXiv:1509.07484}  (\bibinfo {year} {2017})\BibitemShut {NoStop}%
\bibitem [{\citenamefont {Yang}\ \emph {et~al.}(2017)\citenamefont {Yang},
  \citenamefont {Gu},\ and\ \citenamefont {Wen}}]{yang2015loop}%
  \BibitemOpen
  \bibfield  {author} {\bibinfo {author} {\bibfnamefont {S.}~\bibnamefont
  {Yang}}, \bibinfo {author} {\bibfnamefont {Z.-C.}\ \bibnamefont {Gu}}, \ and\
  \bibinfo {author} {\bibfnamefont {X.-G.}\ \bibnamefont {Wen}},\ }{Loop
  optimization for tensor network renormalization},\ \href {\doibase
  PhysRevLett.118.110504} {\bibfield  {journal} {\bibinfo  {journal} {Phys.
  Rev. Lett}\ }\textbf {\bibinfo {volume} {118}},\ \bibinfo {pages} {110504}},\
  \Eprint {http://arxiv.org/abs/1512.04938} {arXiv:1512.04938}  (\bibinfo
  {year} {2017})\BibitemShut {NoStop}%
\bibitem [{\citenamefont {Bal}\ \emph {et~al.}(2017)\citenamefont {Bal},
  \citenamefont {Mari{\"e}n}, \citenamefont {Haegeman},\ and\ \citenamefont
  {Verstraete}}]{bal2017renormalization}%
  \BibitemOpen
  \bibfield  {author} {\bibinfo {author} {\bibfnamefont {M.}~\bibnamefont
  {Bal}}, \bibinfo {author} {\bibfnamefont {M.}~\bibnamefont {Mari{\"e}n}},
  \bibinfo {author} {\bibfnamefont {J.}~\bibnamefont {Haegeman}}, \ and\
  \bibinfo {author} {\bibfnamefont {F.}~\bibnamefont {Verstraete}},\
  }{Renormalization group flows of Hamiltonians using tensor networks},\
  \href@noop {} {\ }\Eprint {http://arxiv.org/abs/1703.00365}
  {arXiv:1703.00365}  (\bibinfo {year} {2017})\BibitemShut {NoStop}%
\bibitem [{\citenamefont {Flammia}\ \emph {et~al.}(2009)\citenamefont
  {Flammia}, \citenamefont {Hamma}, \citenamefont {Hughes},\ and\ \citenamefont
  {Wen}}]{topologicalrenyi}%
  \BibitemOpen
  \bibfield  {author} {\bibinfo {author} {\bibfnamefont {S.}~\bibnamefont
  {Flammia}}, \bibinfo {author} {\bibfnamefont {A.}~\bibnamefont {Hamma}},
  \bibinfo {author} {\bibfnamefont {T.}~\bibnamefont {Hughes}}, \ and\ \bibinfo
  {author} {\bibfnamefont {X.-G.}\ \bibnamefont {Wen}},\ }{Topological
  entanglement Renyi entropy and reduced density matrix structure},\ \href
  {\doibase 10.1103/PhysRevLett.103.261601} {\bibfield  {journal} {\bibinfo
  {journal} {Phys. Rev. Lett.}\ }\textbf {\bibinfo {volume} {103}},\ \bibinfo
  {pages} {261601}},\ \Eprint {http://arxiv.org/abs/0909.3305}
  {arXiv:0909.3305}  (\bibinfo {year} {2009})\BibitemShut {NoStop}%
\bibitem [{\citenamefont {Kitaev}\ and\ \citenamefont
  {Preskill}(2006)}]{KitaevPreskill}%
  \BibitemOpen
  \bibfield  {author} {\bibinfo {author} {\bibfnamefont {A.}~\bibnamefont
  {Kitaev}}\ and\ \bibinfo {author} {\bibfnamefont {J.}~\bibnamefont
  {Preskill}},\ }{Topological Entanglement Entropy},\ \href {\doibase
  10.1103/PhysRevLett.96.110404} {\bibfield  {journal} {\bibinfo  {journal}
  {Phys. Rev. Lett.}\ }\textbf {\bibinfo {volume} {96}},\ \bibinfo {pages}
  {110404}},\ \Eprint {http://arxiv.org/abs/hep-th/0510092}
  {arXiv:hep-th/0510092}  (\bibinfo {year} {2006})\BibitemShut {NoStop}%
\bibitem [{\citenamefont {Levin}\ and\ \citenamefont
  {Wen}(2006)}]{levinwenentanglement}%
  \BibitemOpen
  \bibfield  {author} {\bibinfo {author} {\bibfnamefont {M.}~\bibnamefont
  {Levin}}\ and\ \bibinfo {author} {\bibfnamefont {X.-G.}\ \bibnamefont
  {Wen}},\ }{Detecting Topological Order in a Ground State Wave Function},\
  \href {\doibase 10.1103/PhysRevLett.96.110405} {\bibfield  {journal}
  {\bibinfo  {journal} {Phys. Rev. Lett.}\ }\textbf {\bibinfo {volume} {96}},\
  \bibinfo {pages} {110405}},\ \Eprint {http://arxiv.org/abs/cond-mat/0510613}
  {arXiv:cond-mat/0510613}  (\bibinfo {year} {2006})\BibitemShut {NoStop}%
\bibitem [{\citenamefont {Pfeifer}\ \emph {et~al.}(2014)\citenamefont
  {Pfeifer}, \citenamefont {Haegeman},\ and\ \citenamefont
  {Verstraete}}]{netcon}%
  \BibitemOpen
  \bibfield  {author} {\bibinfo {author} {\bibfnamefont {R.~N.~C.}\
  \bibnamefont {Pfeifer}}, \bibinfo {author} {\bibfnamefont {J.}~\bibnamefont
  {Haegeman}}, \ and\ \bibinfo {author} {\bibfnamefont {F.}~\bibnamefont
  {Verstraete}},\ }{Faster identification of optimal contraction sequences for
  tensor networks},\ \href {\doibase 10.1103/PhysRevE.90.033315} {\bibfield
  {journal} {\bibinfo  {journal} {Phys. Rev. E}\ }\textbf {\bibinfo {volume}
  {90}},\ \bibinfo {pages} {033315}},\ \Eprint {http://arxiv.org/abs/1304.6112}
  {arXiv:1304.6112}  (\bibinfo {year} {2014})\BibitemShut {NoStop}%
\end{thebibliography}%

\appendix
\clearpage

\onecolumngrid
\section{Conformal data in all topological sectors}\label{appendix:fullmeradata}

\begin{figure}[h!]
\includeTikz{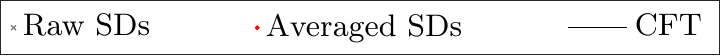}{
			\setlength\figureheight{.37\columnwidth}
			\setlength\figurewidth{.8\columnwidth}
			\input{Graphs/LocalLegend}}
\begin{tabular}{rr}
	\begin{tabular}{r}
		\includeTikz{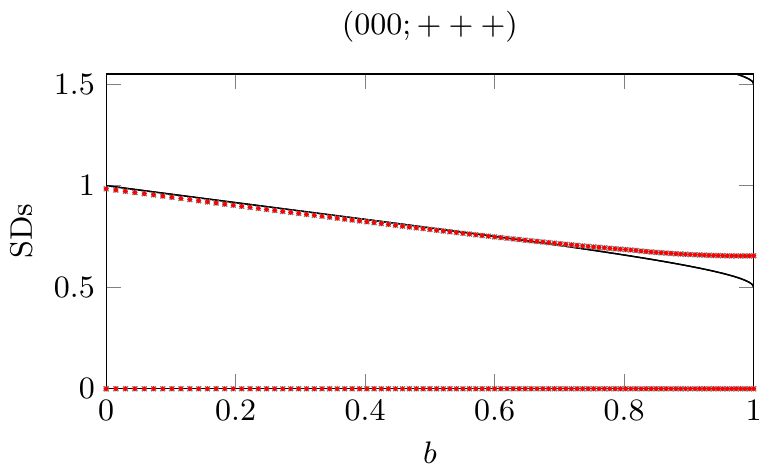}{
			\setlength\figureheight{.37\columnwidth}
			\setlength\figurewidth{.8\columnwidth}
			\input{Graphs/ScalingDims_1_1}}
		\\
		\includeTikz{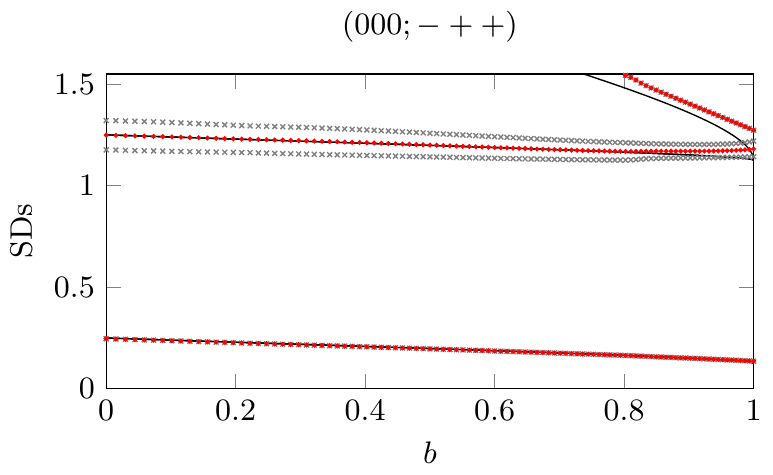}{
			\setlength\figureheight{.37\columnwidth}
			\setlength\figurewidth{.8\columnwidth}
			\input{Graphs/ScalingDims_1_2}}
		\\
		\includeTikz{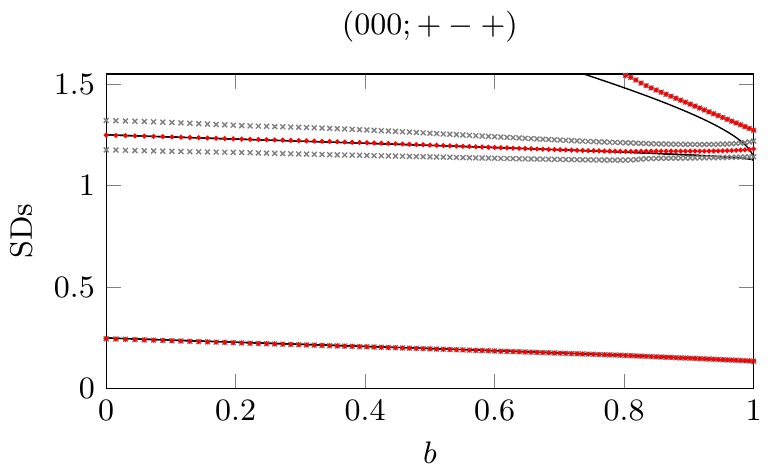}{
			\setlength\figureheight{.37\columnwidth}
			\setlength\figurewidth{.8\columnwidth}
			\input{Graphs/ScalingDims_1_3}}
		\\
		\includeTikz{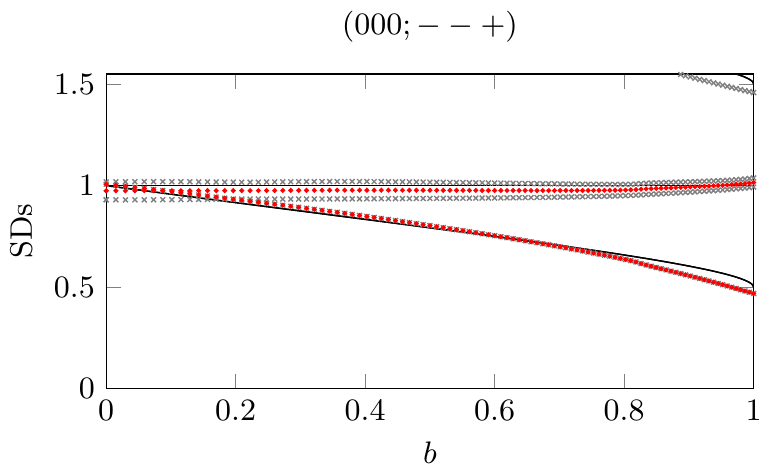}{
			\setlength\figureheight{.37\columnwidth}
			\setlength\figurewidth{.8\columnwidth}
			\input{Graphs/ScalingDims_1_4}}
	\end{tabular}
	\begin{tabular}{r}
		\includeTikz{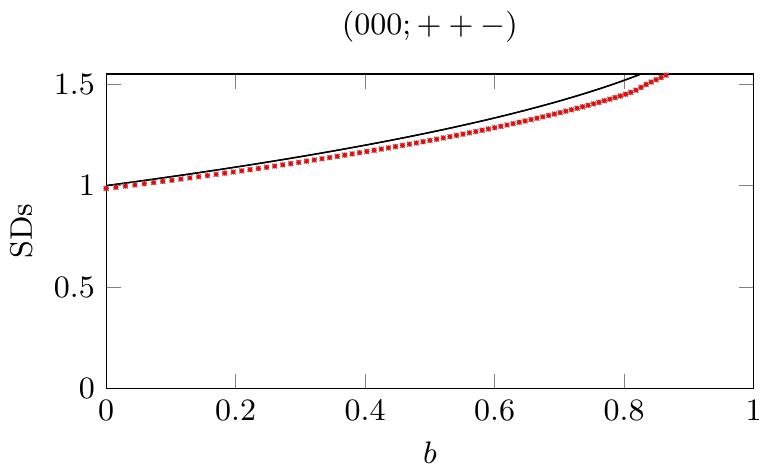}{
			\setlength\figureheight{.37\columnwidth}
			\setlength\figurewidth{.8\columnwidth}
			\input{Graphs/ScalingDims_1_5}}
		\\
		\includeTikz{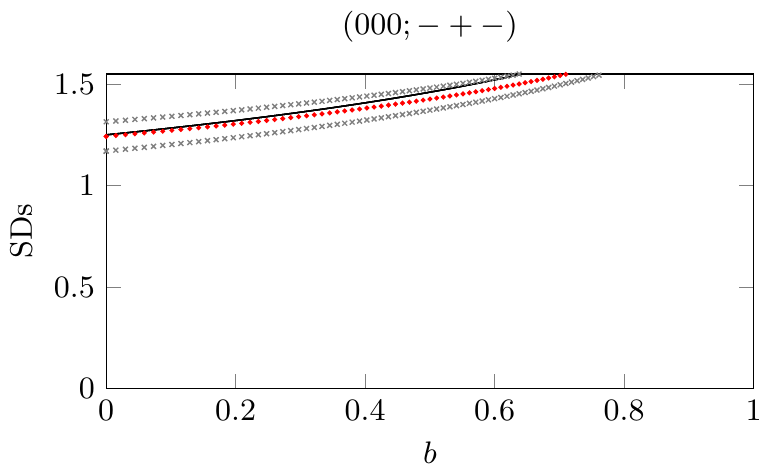}{
			\setlength\figureheight{.37\columnwidth}
			\setlength\figurewidth{.8\columnwidth}
			\input{Graphs/ScalingDims_1_6}}
		\\
		\includeTikz{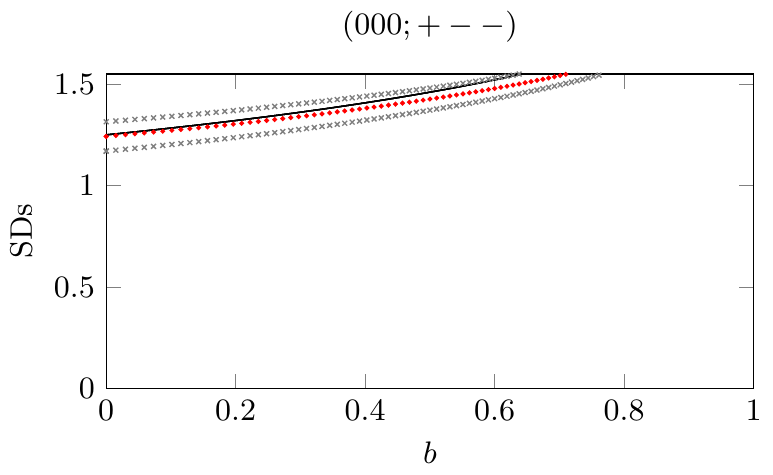}{
			\setlength\figureheight{.37\columnwidth}
			\setlength\figurewidth{.8\columnwidth}
			\input{Graphs/ScalingDims_1_7}}
		\\
		\includeTikz{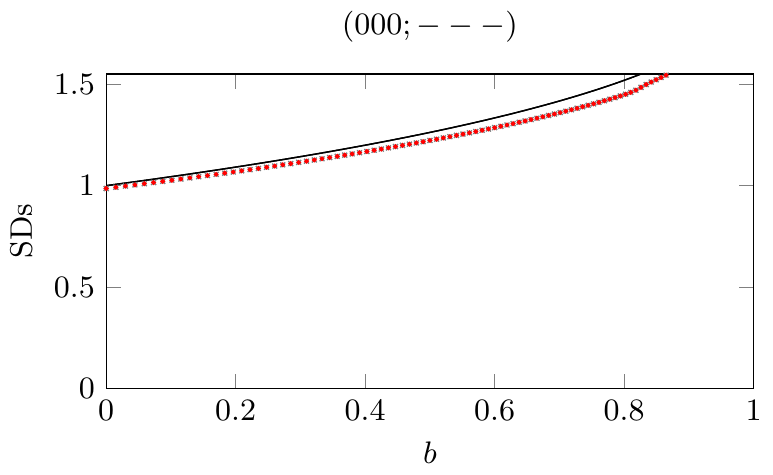}{
			\setlength\figureheight{.37\columnwidth}
			\setlength\figurewidth{.8\columnwidth}
			\input{Graphs/ScalingDims_1_8}}
	\end{tabular}
\end{tabular}
		\caption{MERA scaling dimensions for the trivial twist of the abc model along the `$b$' line. This line is symmetric under the anomalous action of $\zt^3$.\\
			Figure titles label: (twist label; irreducible representation label).\\
			Grey points are the raw data extracted from the MERA. Red points correspond to averaged data as discussed in Appendix~\ref{appendix:fullmeradata}. Black lines correspond to local fields of the compactified free boson CFT.}
			\label{fig:trivialdefect}
\end{figure}

\begin{figure}[h!]
\begin{center}
\includeTikz{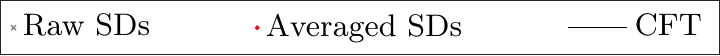}{
			\setlength\figureheight{.37\columnwidth}
			\setlength\figurewidth{.8\columnwidth}
			\input{Graphs/LocalLegend}}
\end{center}			
\begin{tabular}{rr}
	\begin{tabular}{r}
		\includeTikz{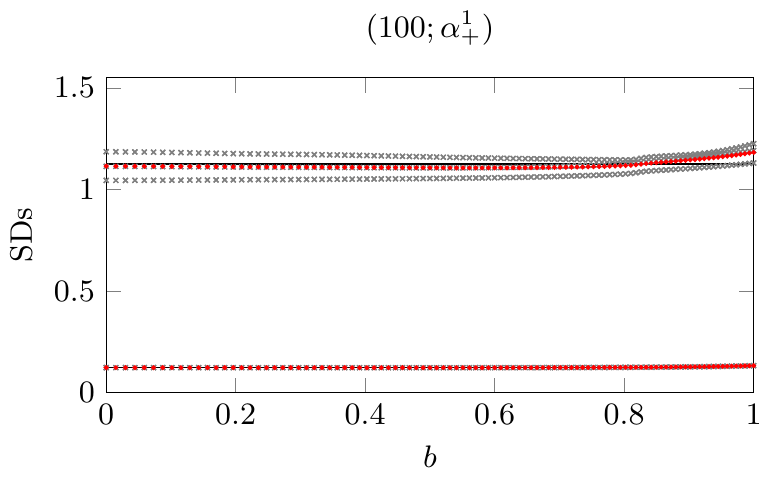}{
			\setlength\figureheight{.37\columnwidth}
			\setlength\figurewidth{.8\columnwidth}
			\input{Graphs/ScalingDims_2_1}}
			\\
		\includeTikz{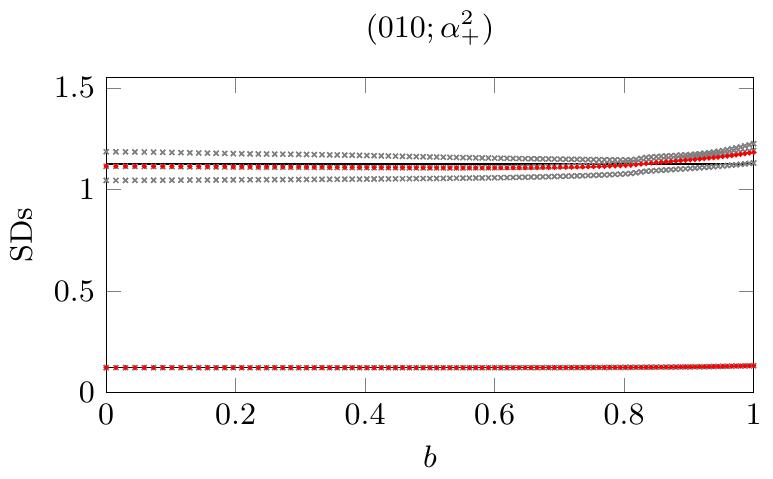}{
			\setlength\figureheight{.37\columnwidth}
			\setlength\figurewidth{.8\columnwidth}
			\input{Graphs/ScalingDims_3_1}}
			\\
		\includeTikz{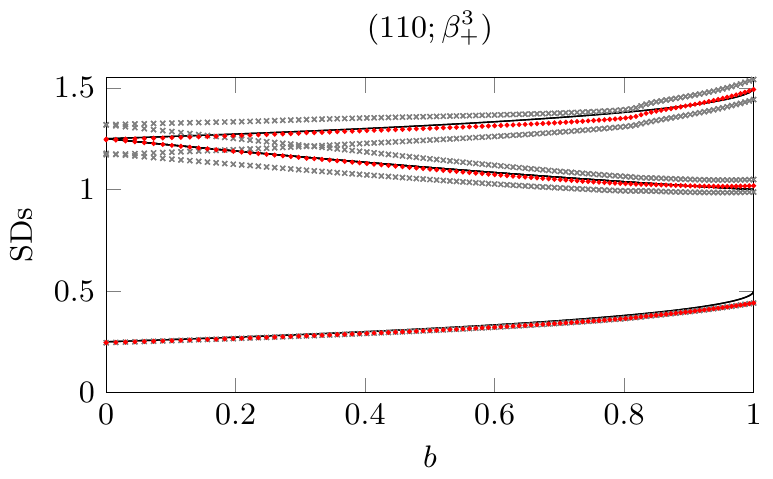}{
			\setlength\figureheight{.37\columnwidth}
			\setlength\figurewidth{.8\columnwidth}
			\input{Graphs/ScalingDims_4_1}}
	\end{tabular}
	\begin{tabular}{r}
		\includeTikz{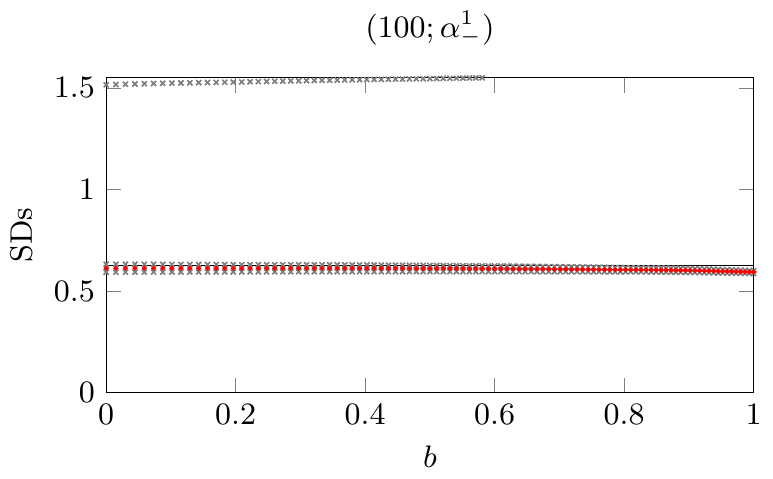}{
			\setlength\figureheight{.37\columnwidth}
			\setlength\figurewidth{.8\columnwidth}
			\input{Graphs/ScalingDims_2_2}}
			\\
		\includeTikz{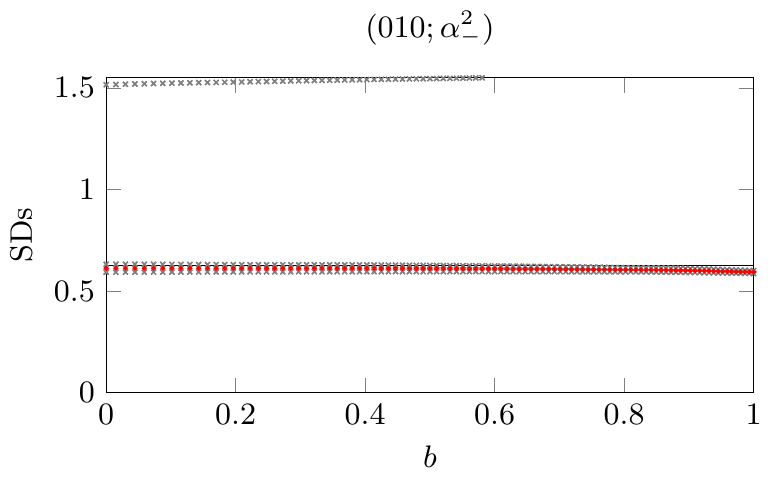}{
			\setlength\figureheight{.37\columnwidth}
			\setlength\figurewidth{.8\columnwidth}
			\input{Graphs/ScalingDims_3_2}}
			\\
		\includeTikz{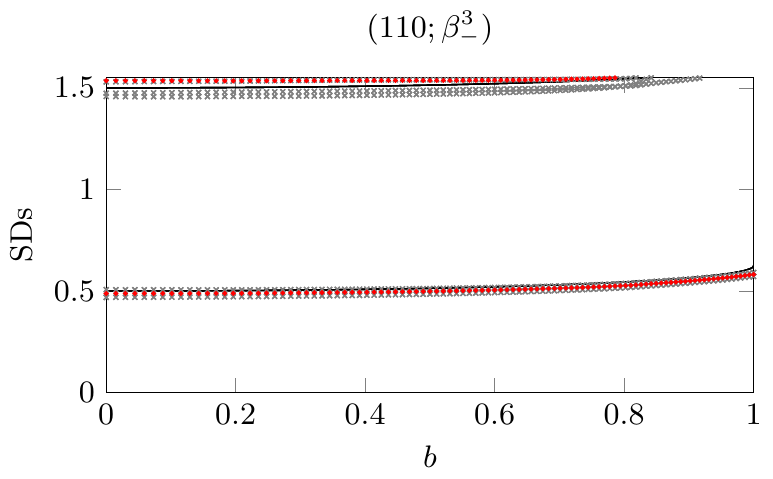}{
			\setlength\figureheight{.37\columnwidth}
			\setlength\figurewidth{.8\columnwidth}
			\input{Graphs/ScalingDims_4_2}}
	\end{tabular}
\end{tabular}
		\caption{Scaling dimensions for topological sectors with twists of the form $(x,y,0)$.\\
					Figure titles label: (twist label; irreducible projective representation label).\\
					 Grey points are the raw data extracted from the MERA. Red points correspond to averaged data as discussed in Appendix~\ref{appendix:fullmeradata}. Black lines correspond to equations in Table~\ref{table:CFTid}.}
			\label{Fig:czplusdefect}
\end{figure}

\begin{figure}[h!]
\begin{center}
\includeTikz{NonLocalLegend}{
			\setlength\figureheight{.37\columnwidth}
			\setlength\figurewidth{.8\columnwidth}
			\input{Graphs/LocalLegend}}
\end{center}			
\begin{tabular}{rr}
	\begin{tabular}{r}
		\includeTikz{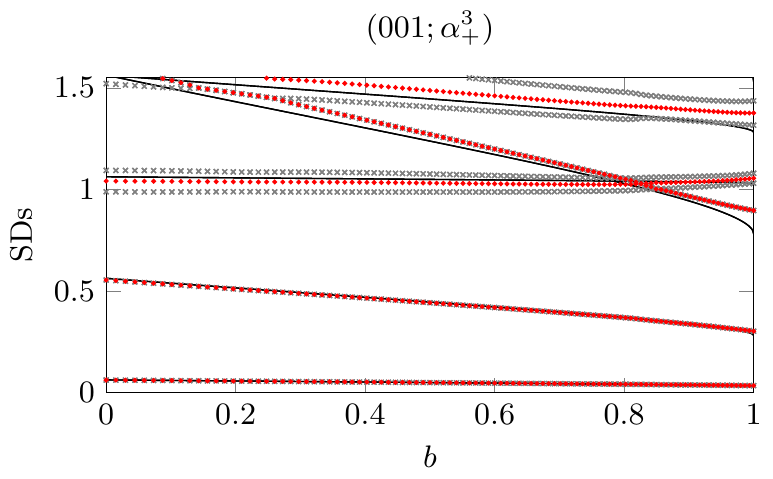}{
			\setlength\figureheight{.37\columnwidth}
			\setlength\figurewidth{.8\columnwidth}
			\input{Graphs/ScalingDims_5_1}}
			\\
		\includeTikz{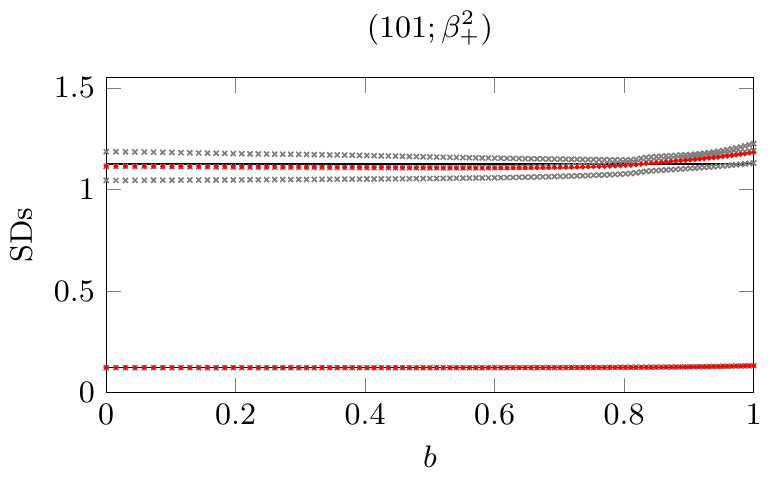}{
			\setlength\figureheight{.37\columnwidth}
			\setlength\figurewidth{.8\columnwidth}
			\input{Graphs/ScalingDims_6_1}}
			\\
		\includeTikz{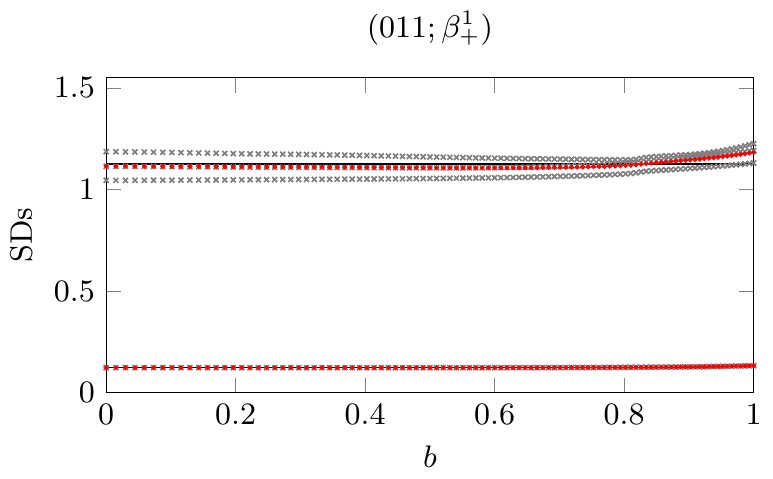}{
			\setlength\figureheight{.37\columnwidth}
			\setlength\figurewidth{.8\columnwidth}
			\input{Graphs/ScalingDims_7_1}}
			\\
		\includeTikz{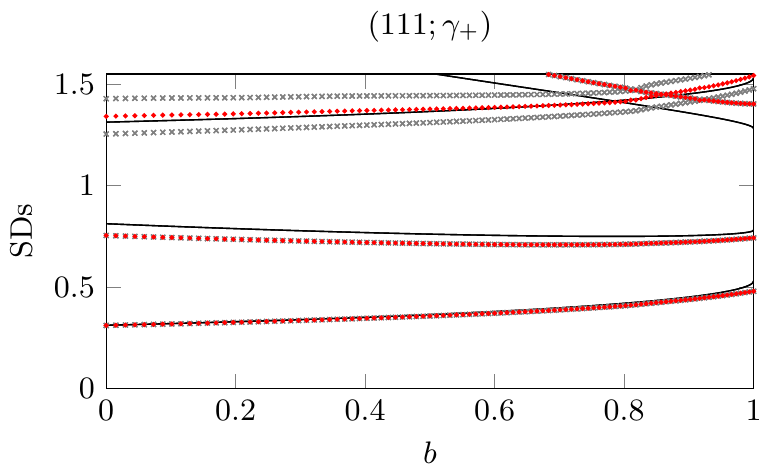}{
			\setlength\figureheight{.37\columnwidth}
			\setlength\figurewidth{.8\columnwidth}
			\input{Graphs/ScalingDims_8_1}}
	\end{tabular}
	\begin{tabular}{r}
		\includeTikz{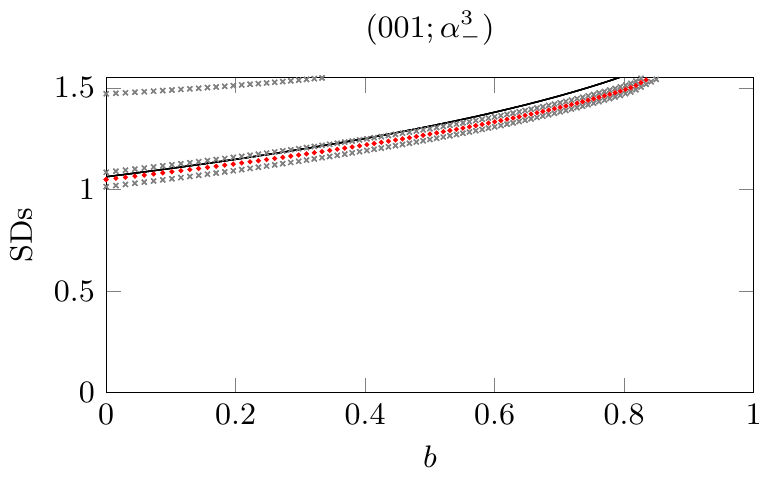}{
			\setlength\figureheight{.37\columnwidth}
			\setlength\figurewidth{.8\columnwidth}
			\input{Graphs/ScalingDims_5_2}}
			\\
		\includeTikz{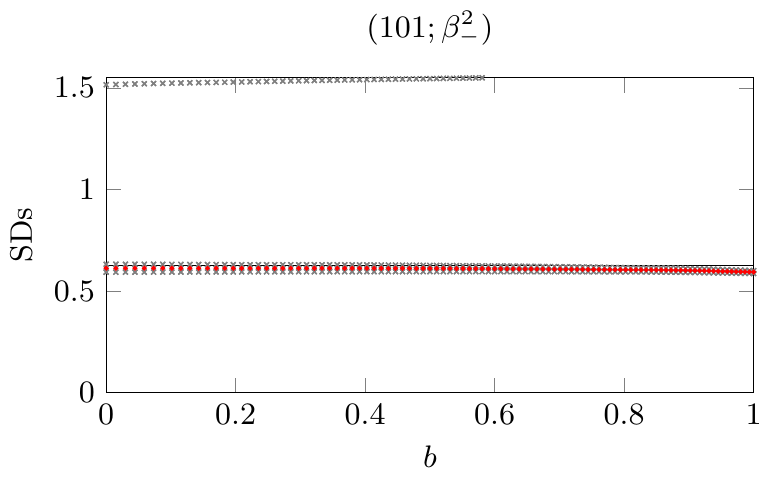}{
			\setlength\figureheight{.37\columnwidth}
			\setlength\figurewidth{.8\columnwidth}
			\input{Graphs/ScalingDims_6_2}}
			\\
		\includeTikz{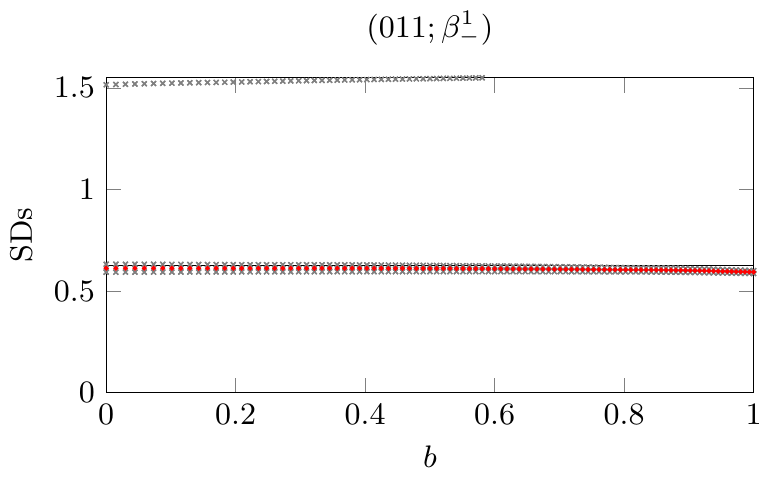}{
			\setlength\figureheight{.37\columnwidth}
			\setlength\figurewidth{.8\columnwidth}
			\input{Graphs/ScalingDims_7_2}}
			\\
		\includeTikz{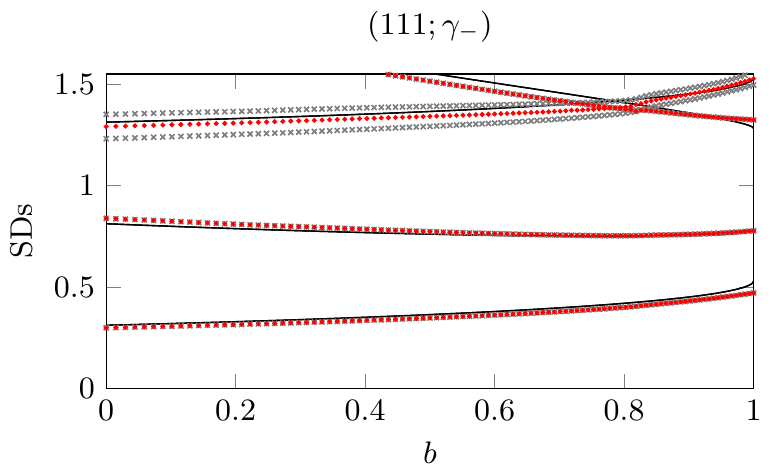}{
			\setlength\figureheight{.37\columnwidth}
			\setlength\figurewidth{.8\columnwidth}
			\input{Graphs/ScalingDims_8_2}}
	\end{tabular}
\end{tabular}
		\caption{Scaling dimensions for topological sectors with twists of the form $(x,y,1)$.\\
			Figure titles label: (twist label; irreducible projective representation label).\\
			Grey points are the raw data extracted from the MERA. Red points correspond to averaged data as discussed in Appendix~\ref{appendix:fullmeradata}. Black lines correspond to equations in Table~\ref{table:CFTid}.}
			\label{Fig:czminusdefect}
\end{figure}

In this appendix, we present the full set of scaling dimensions extracted from the bond dimension 8 MERA with full anomalous symmetry enforced. The data is shown in Fig.~\ref{fig:trivialdefect} for the trivial twist, and Fig.~\ref{Fig:czplusdefect} and Fig.~\ref{Fig:czminusdefect} for the nontrivial twists. Each subplot in these figures corresponds to a distinct topological sector.

When examining the gray points, one notices a broken degeneracy. This was previously noted in Ref.~\onlinecite{Bridgeman2015}. We conjecture that this occurs via coupling of states which, in the field theoretic limit, would be forbidden from coupling due to the full conformal symmetry. As such, we conjecture that the scaling dimensions corresponding to degenerate fields obtained from the MERA experience a splitting $\Delta_{\mathrm{MERA}}=\Delta_{\mathrm{CFT}}\pm\epsilon$, where the size of the splitting $\epsilon$ decreases with increased bond dimension as the full conformal symmetry is effectively recovered.

To combat this splitting, we average the MERA scaling dimensions in an attempt to recover the CFT values. When choosing which lines should be averaged together, we have taken all lines of similar gradient and position on the plot. The result of this procedure is indicated in red, and closely matches the CFT values.

The scaling dimensions and conformal spins in each topological sector are given in Table~\ref{table:CFTid}. Table~\ref{table:fusion} shows the fusion rules for the sectors, computed using the symmetric MERA.

\begin{table}[h!] 
	\renewcommand{\arraystretch}{1.6}
	\setlength{\tabcolsep}{0.5em}
	\begin{tabular}{|c|c||c|c|c|cl|}
		\hline
\multicolumn{2}{|c||}{Topological Sector}&\multirow{2}{*}{Topological spin} & \multirow{2}{*}{Scaling Dimension}&\multirow{2}{*}{Conformal spin}&\multicolumn{2}{|c|}{\multirow{2}{*}{Parameters}}\\
\cline{1-2}
Twist&Proj. Irrep.&&&&& \\
		\hline
		\hline
		\multirow{8}{*}{\normalsize$(000)$}&$\chi^{1}_{+}$&0&$\frac{e^2}{R^2}+\frac{m^2 R^2}{4}$&$em$&\multicolumn{2}{|c|}{$e,m\in 2\mathbb{Z}$}\\
		\cline{2-7}
		&$\chi^{4}_{+}$&0&$\{\frac{e^2}{R^2}+\frac{m^2 R^2}{4},\,\,\,1\}$&$em$&\multicolumn{2}{|c|}{$e,m\in 2\mathbb{Z},\,\,em\neq0$}\\
		\cline{2-7}
		&$\chi^{2}_{+}$&0&\multirow{6}{*}{\normalsize$\frac{e^2}{R^2}+\frac{m^2 R^2}{4}$}&\multirow{6}{*}{\normalsize$em$}&\multicolumn{2}{|c|}{$e\in 2\mathbb{Z},m\in 2\mathbb{Z}+1$}\\
		&$\chi^{3}_{+}$&0&&&\multicolumn{2}{|c|}{$e\in 2\mathbb{Z},m\in 2\mathbb{Z}+1$}\\
		&$\chi^{1}_{-}$&0&&&\multicolumn{2}{|c|}{$e\in 2\mathbb{Z}+1,m\in 2\mathbb{Z}$}\\
		&$\chi^{2}_{-}$&0&&&\multicolumn{2}{|c|}{$e\in 2\mathbb{Z}+1,m\in 2\mathbb{Z}+1$}\\
		&$\chi^{3}_{-}$&0&&&\multicolumn{2}{|c|}{$e\in 2\mathbb{Z}+1,m\in 2\mathbb{Z}+1$}\\
		&$\chi^{4}_{-}$&0&&&\multicolumn{2}{|c|}{$e\in 2\mathbb{Z}+1,m\in 2\mathbb{Z}$}\\
		\hline
		\multirow{2}{*}{\normalsize$(100)$}&$\alpha^1_+$&0&\multirow{2}{*}{\normalsize$h+\bar{h}$}&\multirow{2}{*}{\normalsize$h-\bar{h}$}&\multirow{2}{*}{\normalsize$h,\bar{h}\in\{\frac{1}{16},\frac{9}{16}\}$} &$h-\bar{h}\in\mathbb{Z}$\\
		&$\alpha^1_-$&$ \frac{1}{2} $&&&& $h-\bar{h}\in\mathbb{Z}+\frac{1}{2}$\\
		\hline
		\multirow{2}{*}{\normalsize$(010)$}&$\alpha^2_+$&0&\multirow{2}{*}{\normalsize$h+\bar{h}$}&\multirow{2}{*}{\normalsize$h-\bar{h}$}&\multirow{2}{*}{\normalsize$h,\bar{h}\in\{\frac{1}{16},\frac{9}{16}\}$}&$h-\bar{h}\in\mathbb{Z}$\\
		&$\alpha^2_-$&$ \frac{1}{2} $&&&& $h-\bar{h}\in\mathbb{Z}+\frac{1}{2}$\\
		\hline
		\multirow{2}{*}{\normalsize$(110)$}&$\beta^3_+$&0&\multirow{2}{*}{\normalsize$\frac{e^2}{R^2}+\frac{m^2 R^2}{4}$}&\multirow{2}{*}{\normalsize$em$}&\multirow{2}{*}{\normalsize$e\in\mathbb{Z}+\frac{1}{2},m\in\mathbb{Z}$}&$em\in \mathbb{Z}$\\
		&$\beta^3_-$&$ \frac{1}{2}$&&&&$em\in \mathbb{Z}+\frac{1}{2}$\\
		\hline
		\multirow{2}{*}{\normalsize$(001)$}&$\alpha^3_+$&0&\multirow{2}{*}{\normalsize$\frac{e^2}{R^2}+\frac{m^2 R^2}{4}$}&\multirow{2}{*}{\normalsize$em$}&\multirow{2}{*}{\normalsize$e\in\mathbb{Z},m\in\mathbb{Z}+\frac{1}{2}$}&$em\in \mathbb{Z}$\\
		&$\alpha^3_-$&$\frac{1}{2}$&&&&$em\in \mathbb{Z}+\frac{1}{2}$\\
		\hline
		\multirow{2}{*}{\normalsize$(101)$}&$\beta^2_+$&0&\multirow{2}{*}{\normalsize$h+\bar{h}$}&\multirow{2}{*}{\normalsize$h-\bar{h}$}&\multirow{2}{*}{$h,\bar{h}\in\{\frac{1}{16},\frac{9}{16}\}$}& $h-\bar{h}\in\mathbb{Z}$\\
		&$\beta^2_-$&$\frac{1}{2}$&&&&$h-\bar{h}\in\mathbb{Z}+\frac{1}{2}$\\
		\hline
		\multirow{2}{*}{\normalsize$(011)$}&$\beta^1_+$&0&\multirow{2}{*}{\normalsize$h+\bar{h}$}&\multirow{2}{*}{\normalsize$h-\bar{h}$}&\multirow{2}{*}{$h,\bar{h}\in\{\frac{1}{16},\frac{9}{16}\}$}& $h-\bar{h}\in\mathbb{Z}$\\
		&$\beta^1_-$&$\frac{1}{2}$&&&&$h-\bar{h}\in\mathbb{Z}+\frac{1}{2}$\\
		\hline
		\multirow{2}{*}{\normalsize$(111)$}&$\gamma_+$&$\frac{3}{4}$&\multirow{2}{*}{\normalsize$\frac{e^2}{R^2}+\frac{m^2 R^2}{4}$}&\multirow{2}{*}{\normalsize$em$}&\multirow{2}{*}{\normalsize$e,m\in\mathbb{Z}+\frac{1}{2}$}&$em\in \mathbb{Z}+\frac{3}{4}$\\
		&$\gamma_-$&$\frac{1}{4}$&&&&$em\in \mathbb{Z}+\frac{1}{4}$\\
		\hline
	\end{tabular}
	\caption{Primary fields in each topological sector labeled by a twist (an element of $\G=\mathbb{Z}_2^3$) and an irreducible (projective) representation. These sectors are the simple objects of $D^\phi(\zt^3)\cong D(\mathsf{D}_4)$.\\
		Note that the choices of $e$ and $m$ allowed for each representation under the trivial twist corresponds to $(-1)^e=\chi(001)$ and $(-1)^m=\chi(110)$, where $\chi$ is the representation being considered.\\
		Projective representations in each topological sector are indicated in Eqn.~\ref{eqn:projrep}, reproduced from Ref.~\onlinecite{propitius}.\\
		The fusion table, computed using the symmetric MERA, for these sectors is explicitly presented in Table~\ref{table:fusion}. All sectors with a nontrivial twist have quantum dimension 2, and so are nonabelian.}\label{table:CFTid}
\end{table}

The irreps are given explicitly in Eqn.~\ref{eqn:projrep}. Those below the line are nontrivial projective representations.

\begin{subequations}\label{eqn:projrep}
	\begin{alignat}{4}
	\chi^{1}_\pm(100)&=+1&&\qquad\chi^{1}_\pm(010)=+1&&\qquad\chi^{1}_\pm(001)=\pm 1\\
	\chi^{2}_\pm(100)&=-1&&\qquad\chi^{2}_\pm(010)=+1&&\qquad\chi^{2}_\pm(001)=\pm 1\\
	\chi^{3}_\pm(100)&=+1&&\qquad\chi^{3}_\pm(010)=-1&&\qquad\chi^{3}_\pm(001)=\pm 1\\
	\chi^{4}_\pm(100)&=-1&&\qquad\chi^{4}_\pm(010)=-1&&\qquad\chi^{4}_\pm(001)=\pm 1\\
	\cline{1-6}
	\alpha^1_\pm(100)&=\pm
	\openone
	&&\qquad 
	\alpha^1_\pm(010)=
	X
	&&\qquad 
	\alpha^1_\pm(001)=
	Z
	\\
	\alpha^2_\pm(100)&=
	Z
	&&\qquad 
	\alpha^2_\pm(010)=\pm
	\openone
	&&\qquad 
	\alpha^2_\pm(001)=
	X
	\\
	\alpha^3_\pm(100)&=
	X
	&&\qquad 
	\alpha^3_\pm(010)=
	Z
	&&\qquad 
	\alpha^3_\pm(001)=\pm
	\openone
	\\
	\beta^1_\pm(100)&=
	Z
	&&\qquad 
	\beta^1_\pm(010)=
	X
	&&\qquad 
	\beta^1_\pm(001)=\pm
	X
	\\
	\beta^2_\pm(100)&=\pm
	X
	&&\qquad 
	\beta^2_\pm(010)=
	Z
	&&\qquad 
	\beta^2_\pm(001)=
	X
	\\
	\beta^3_\pm(100)&=
	X
	&&\qquad 
	\beta^3_\pm(010)=\pm
	X
	&&\qquad 
	\beta^3_\pm(001)=
	Z
	\\
	\gamma_\pm(100)&=\pm
	X
	&&\qquad 
	\gamma_\pm(010)=\pm
	Y
	&&\qquad 
	\gamma_\pm(001)=\pm
	Z
	\end{alignat}
\end{subequations}

\begin{table}
\rotatebox{-90}{
\begin{tabular}{|c||c|c|c|c|c|c|c|c|c|c|c|c|c|c|c|c|c|c|c|c|c|c|}
	\hline
	\boldmath$a\times b$&\boldmath$\chi^{1}_{+}$&\boldmath$\chi^{2}_{+}$&\boldmath$\chi^{3}_{+}$&\boldmath$\chi^{4}_{+}$&\boldmath$\chi^{1}_{-}$&\boldmath$\chi^{2}_{-}$&\boldmath$\chi^{3}_{-}$&\boldmath$\chi^{4}_{-}$&\boldmath$\alpha^{1}_{+}$&\boldmath$\alpha^{1}_{-}$&\boldmath$\alpha^{2}_{+}$&\boldmath$\alpha^{2}_{-}$&\boldmath$\beta^{3}_{+}$&\boldmath$\beta^{3}_{-}$&\boldmath$\alpha^{3}_{+}$&\boldmath$\alpha^{3}_{-}$&\boldmath$\beta^{2}_{+}$&\boldmath$\beta^{2}_{-}$&\boldmath$\beta^{1}_{+}$&\boldmath$\beta^{1}_{-}$&\boldmath$\gamma_{+}$&\boldmath$\gamma_{-}$\\\hline\hline
	\boldmath$\chi^{1}_{+}$&\footnotesize$\chi^{1}_{+}$&\footnotesize$\chi^{2}_{+}$&\footnotesize$\chi^{3}_{+}$&\footnotesize$\chi^{4}_{+}$&\footnotesize$\chi^{1}_{-}$&\footnotesize$\chi^{2}_{-}$&\footnotesize$\chi^{3}_{-}$&\footnotesize$\chi^{4}_{-}$&\footnotesize$\alpha^{1}_{+}$&\footnotesize$\alpha^{1}_{-}$&\footnotesize$\alpha^{2}_{+}$&\footnotesize$\alpha^{2}_{-}$&\footnotesize$\beta^{3}_{+}$&\footnotesize$\beta^{3}_{-}$&\footnotesize$\alpha^{3}_{+}$&\footnotesize$\alpha^{3}_{-}$&\footnotesize$\beta^{2}_{+}$&\footnotesize$\beta^{2}_{-}$&\footnotesize$\beta^{1}_{+}$&\footnotesize$\beta^{1}_{-}$&\footnotesize$\gamma_{+}$&\footnotesize$\gamma_{-}$\\\hline
	\boldmath$\chi^{2}_{+}$&\footnotesize$\chi^{2}_{+}$&\footnotesize$\chi^{1}_{+}$&\footnotesize$\chi^{4}_{+}$&\footnotesize$\chi^{3}_{+}$&\footnotesize$\chi^{2}_{-}$&\footnotesize$\chi^{1}_{-}$&\footnotesize$\chi^{4}_{-}$&\footnotesize$\chi^{3}_{-}$&\footnotesize$\alpha^{1}_{-}$&\footnotesize$\alpha^{1}_{+}$&\footnotesize$\alpha^{2}_{+}$&\footnotesize$\alpha^{2}_{-}$&\footnotesize$\beta^{3}_{-}$&\footnotesize$\beta^{3}_{+}$&\footnotesize$\alpha^{3}_{+}$&\footnotesize$\alpha^{3}_{-}$&\footnotesize$\beta^{2}_{-}$&\footnotesize$\beta^{2}_{+}$&\footnotesize$\beta^{1}_{+}$&\footnotesize$\beta^{1}_{-}$&\footnotesize$\gamma_{-}$&\footnotesize$\gamma_{+}$\\\hline
	\boldmath$\chi^{3}_{+}$&\footnotesize$\chi^{3}_{+}$&\footnotesize$\chi^{4}_{+}$&\footnotesize$\chi^{1}_{+}$&\footnotesize$\chi^{2}_{+}$&\footnotesize$\chi^{3}_{-}$&\footnotesize$\chi^{4}_{-}$&\footnotesize$\chi^{1}_{-}$&\footnotesize$\chi^{2}_{-}$&\footnotesize$\alpha^{1}_{+}$&\footnotesize$\alpha^{1}_{-}$&\footnotesize$\alpha^{2}_{-}$&\footnotesize$\alpha^{2}_{+}$&\footnotesize$\beta^{3}_{-}$&\footnotesize$\beta^{3}_{+}$&\footnotesize$\alpha^{3}_{+}$&\footnotesize$\alpha^{3}_{-}$&\footnotesize$\beta^{2}_{+}$&\footnotesize$\beta^{2}_{-}$&\footnotesize$\beta^{1}_{-}$&\footnotesize$\beta^{1}_{+}$&\footnotesize$\gamma_{-}$&\footnotesize$\gamma_{+}$\\\hline
	\boldmath$\chi^{4}_{+}$&\footnotesize$\chi^{4}_{+}$&\footnotesize$\chi^{3}_{+}$&\footnotesize$\chi^{2}_{+}$&\footnotesize$\chi^{1}_{+}$&\footnotesize$\chi^{4}_{-}$&\footnotesize$\chi^{3}_{-}$&\footnotesize$\chi^{2}_{-}$&\footnotesize$\chi^{1}_{-}$&\footnotesize$\alpha^{1}_{-}$&\footnotesize$\alpha^{1}_{+}$&\footnotesize$\alpha^{2}_{-}$&\footnotesize$\alpha^{2}_{+}$&\footnotesize$\beta^{3}_{+}$&\footnotesize$\beta^{3}_{-}$&\footnotesize$\alpha^{3}_{+}$&\footnotesize$\alpha^{3}_{-}$&\footnotesize$\beta^{2}_{-}$&\footnotesize$\beta^{2}_{+}$&\footnotesize$\beta^{1}_{-}$&\footnotesize$\beta^{1}_{+}$&\footnotesize$\gamma_{+}$&\footnotesize$\gamma_{-}$\\\hline
	\boldmath$\chi^{1}_{-}$&\footnotesize$\chi^{1}_{-}$&\footnotesize$\chi^{2}_{-}$&\footnotesize$\chi^{3}_{-}$&\footnotesize$\chi^{4}_{-}$&\footnotesize$\chi^{1}_{+}$&\footnotesize$\chi^{2}_{+}$&\footnotesize$\chi^{3}_{+}$&\footnotesize$\chi^{4}_{+}$&\footnotesize$\alpha^{1}_{+}$&\footnotesize$\alpha^{1}_{-}$&\footnotesize$\alpha^{2}_{+}$&\footnotesize$\alpha^{2}_{-}$&\footnotesize$\beta^{3}_{+}$&\footnotesize$\beta^{3}_{-}$&\footnotesize$\alpha^{3}_{-}$&\footnotesize$\alpha^{3}_{+}$&\footnotesize$\beta^{2}_{-}$&\footnotesize$\beta^{2}_{+}$&\footnotesize$\beta^{1}_{-}$&\footnotesize$\beta^{1}_{+}$&\footnotesize$\gamma_{-}$&\footnotesize$\gamma_{+}$\\\hline
	\boldmath$\chi^{2}_{-}$&\footnotesize$\chi^{2}_{-}$&\footnotesize$\chi^{1}_{-}$&\footnotesize$\chi^{4}_{-}$&\footnotesize$\chi^{3}_{-}$&\footnotesize$\chi^{2}_{+}$&\footnotesize$\chi^{1}_{+}$&\footnotesize$\chi^{4}_{+}$&\footnotesize$\chi^{3}_{+}$&\footnotesize$\alpha^{1}_{-}$&\footnotesize$\alpha^{1}_{+}$&\footnotesize$\alpha^{2}_{+}$&\footnotesize$\alpha^{2}_{-}$&\footnotesize$\beta^{3}_{-}$&\footnotesize$\beta^{3}_{+}$&\footnotesize$\alpha^{3}_{-}$&\footnotesize$\alpha^{3}_{+}$&\footnotesize$\beta^{2}_{+}$&\footnotesize$\beta^{2}_{-}$&\footnotesize$\beta^{1}_{-}$&\footnotesize$\beta^{1}_{+}$&\footnotesize$\gamma_{+}$&\footnotesize$\gamma_{-}$\\\hline
	\boldmath$\chi^{3}_{-}$&\footnotesize$\chi^{3}_{-}$&\footnotesize$\chi^{4}_{-}$&\footnotesize$\chi^{1}_{-}$&\footnotesize$\chi^{2}_{-}$&\footnotesize$\chi^{3}_{+}$&\footnotesize$\chi^{4}_{+}$&\footnotesize$\chi^{1}_{+}$&\footnotesize$\chi^{2}_{+}$&\footnotesize$\alpha^{1}_{+}$&\footnotesize$\alpha^{1}_{-}$&\footnotesize$\alpha^{2}_{-}$&\footnotesize$\alpha^{2}_{+}$&\footnotesize$\beta^{3}_{-}$&\footnotesize$\beta^{3}_{+}$&\footnotesize$\alpha^{3}_{-}$&\footnotesize$\alpha^{3}_{+}$&\footnotesize$\beta^{2}_{-}$&\footnotesize$\beta^{2}_{+}$&\footnotesize$\beta^{1}_{+}$&\footnotesize$\beta^{1}_{-}$&\footnotesize$\gamma_{+}$&\footnotesize$\gamma_{-}$\\\hline
	\boldmath$\chi^{4}_{-}$&\footnotesize$\chi^{4}_{-}$&\footnotesize$\chi^{3}_{-}$&\footnotesize$\chi^{2}_{-}$&\footnotesize$\chi^{1}_{-}$&\footnotesize$\chi^{4}_{+}$&\footnotesize$\chi^{3}_{+}$&\footnotesize$\chi^{2}_{+}$&\footnotesize$\chi^{1}_{+}$&\footnotesize$\alpha^{1}_{-}$&\footnotesize$\alpha^{1}_{+}$&\footnotesize$\alpha^{2}_{-}$&\footnotesize$\alpha^{2}_{+}$&\footnotesize$\beta^{3}_{+}$&\footnotesize$\beta^{3}_{-}$&\footnotesize$\alpha^{3}_{-}$&\footnotesize$\alpha^{3}_{+}$&\footnotesize$\beta^{2}_{+}$&\footnotesize$\beta^{2}_{-}$&\footnotesize$\beta^{1}_{+}$&\footnotesize$\beta^{1}_{-}$&\footnotesize$\gamma_{-}$&\footnotesize$\gamma_{+}$\\\hline
	\boldmath$\alpha^{1}_{+}$&\footnotesize$\alpha^{1}_{+}$&\footnotesize$\alpha^{1}_{-}$&\footnotesize$\alpha^{1}_{+}$&\footnotesize$\alpha^{1}_{-}$&\footnotesize$\alpha^{1}_{+}$&\footnotesize$\alpha^{1}_{-}$&\footnotesize$\alpha^{1}_{+}$&\footnotesize$\alpha^{1}_{-}$&\footnotesize$\begin{matrix}\chi^{1}_{+}&\chi^{3}_{+}\\\chi^{1}_{-}&\chi^{3}_{-}\end{matrix}$&\footnotesize$\begin{matrix}\chi^{2}_{+}&\chi^{4}_{+}\\\chi^{2}_{-}&\chi^{4}_{-}\end{matrix}$&\footnotesize$\begin{matrix}\beta^{3}_{+}&\\&\beta^{3}_{-}\end{matrix}$&\footnotesize$\begin{matrix}\beta^{3}_{+}&\\&\beta^{3}_{-}\end{matrix}$&\footnotesize$\begin{matrix}\alpha^{2}_{+}&\\&\alpha^{2}_{-}\end{matrix}$&\footnotesize$\begin{matrix}\alpha^{2}_{+}&\\&\alpha^{2}_{-}\end{matrix}$&\footnotesize$\begin{matrix}\beta^{2}_{+}&\\&\beta^{2}_{-}\end{matrix}$&\footnotesize$\begin{matrix}\beta^{2}_{+}&\\&\beta^{2}_{-}\end{matrix}$&\footnotesize$\begin{matrix}\alpha^{3}_{+}&\\&\alpha^{3}_{-}\end{matrix}$&\footnotesize$\begin{matrix}\alpha^{3}_{+}&\\&\alpha^{3}_{-}\end{matrix}$&\footnotesize$\begin{matrix}\gamma_{+}&\\&\gamma_{-}\end{matrix}$&\footnotesize$\begin{matrix}\gamma_{+}&\\&\gamma_{-}\end{matrix}$&\footnotesize$\begin{matrix}\beta^{1}_{+}&\\&\beta^{1}_{-}\end{matrix}$&\footnotesize$\begin{matrix}\beta^{1}_{+}&\\&\beta^{1}_{-}\end{matrix}$\\\hline
	\boldmath$\alpha^{1}_{-}$&\footnotesize$\alpha^{1}_{-}$&\footnotesize$\alpha^{1}_{+}$&\footnotesize$\alpha^{1}_{-}$&\footnotesize$\alpha^{1}_{+}$&\footnotesize$\alpha^{1}_{-}$&\footnotesize$\alpha^{1}_{+}$&\footnotesize$\alpha^{1}_{-}$&\footnotesize$\alpha^{1}_{+}$&\footnotesize$\begin{matrix}\chi^{2}_{+}&\chi^{4}_{+}\\\chi^{2}_{-}&\chi^{4}_{-}\end{matrix}$&\footnotesize$\begin{matrix}\chi^{1}_{+}&\chi^{3}_{+}\\\chi^{1}_{-}&\chi^{3}_{-}\end{matrix}$&\footnotesize$\begin{matrix}\beta^{3}_{+}&\\&\beta^{3}_{-}\end{matrix}$&\footnotesize$\begin{matrix}\beta^{3}_{+}&\\&\beta^{3}_{-}\end{matrix}$&\footnotesize$\begin{matrix}\alpha^{2}_{+}&\\&\alpha^{2}_{-}\end{matrix}$&\footnotesize$\begin{matrix}\alpha^{2}_{+}&\\&\alpha^{2}_{-}\end{matrix}$&\footnotesize$\begin{matrix}\beta^{2}_{+}&\\&\beta^{2}_{-}\end{matrix}$&\footnotesize$\begin{matrix}\beta^{2}_{+}&\\&\beta^{2}_{-}\end{matrix}$&\footnotesize$\begin{matrix}\alpha^{3}_{+}&\\&\alpha^{3}_{-}\end{matrix}$&\footnotesize$\begin{matrix}\alpha^{3}_{+}&\\&\alpha^{3}_{-}\end{matrix}$&\footnotesize$\begin{matrix}\gamma_{+}&\\&\gamma_{-}\end{matrix}$&\footnotesize$\begin{matrix}\gamma_{+}&\\&\gamma_{-}\end{matrix}$&\footnotesize$\begin{matrix}\beta^{1}_{+}&\\&\beta^{1}_{-}\end{matrix}$&\footnotesize$\begin{matrix}\beta^{1}_{+}&\\&\beta^{1}_{-}\end{matrix}$\\\hline
	\boldmath$\alpha^{2}_{+}$&\footnotesize$\alpha^{2}_{+}$&\footnotesize$\alpha^{2}_{+}$&\footnotesize$\alpha^{2}_{-}$&\footnotesize$\alpha^{2}_{-}$&\footnotesize$\alpha^{2}_{+}$&\footnotesize$\alpha^{2}_{+}$&\footnotesize$\alpha^{2}_{-}$&\footnotesize$\alpha^{2}_{-}$&\footnotesize$\begin{matrix}\beta^{3}_{+}&\\&\beta^{3}_{-}\end{matrix}$&\footnotesize$\begin{matrix}\beta^{3}_{+}&\\&\beta^{3}_{-}\end{matrix}$&\footnotesize$\begin{matrix}\chi^{1}_{+}&\chi^{2}_{+}\\\chi^{1}_{-}&\chi^{2}_{-}\end{matrix}$&\footnotesize$\begin{matrix}\chi^{3}_{+}&\chi^{4}_{+}\\\chi^{3}_{-}&\chi^{4}_{-}\end{matrix}$&\footnotesize$\begin{matrix}\alpha^{1}_{+}&\\&\alpha^{1}_{-}\end{matrix}$&\footnotesize$\begin{matrix}\alpha^{1}_{+}&\\&\alpha^{1}_{-}\end{matrix}$&\footnotesize$\begin{matrix}\beta^{1}_{+}&\\&\beta^{1}_{-}\end{matrix}$&\footnotesize$\begin{matrix}\beta^{1}_{+}&\\&\beta^{1}_{-}\end{matrix}$&\footnotesize$\begin{matrix}\gamma_{+}&\\&\gamma_{-}\end{matrix}$&\footnotesize$\begin{matrix}\gamma_{+}&\\&\gamma_{-}\end{matrix}$&\footnotesize$\begin{matrix}\alpha^{3}_{+}&\\&\alpha^{3}_{-}\end{matrix}$&\footnotesize$\begin{matrix}\alpha^{3}_{+}&\\&\alpha^{3}_{-}\end{matrix}$&\footnotesize$\begin{matrix}\beta^{2}_{+}&\\&\beta^{2}_{-}\end{matrix}$&\footnotesize$\begin{matrix}\beta^{2}_{+}&\\&\beta^{2}_{-}\end{matrix}$\\\hline
	\boldmath$\alpha^{2}_{-}$&\footnotesize$\alpha^{2}_{-}$&\footnotesize$\alpha^{2}_{-}$&\footnotesize$\alpha^{2}_{+}$&\footnotesize$\alpha^{2}_{+}$&\footnotesize$\alpha^{2}_{-}$&\footnotesize$\alpha^{2}_{-}$&\footnotesize$\alpha^{2}_{+}$&\footnotesize$\alpha^{2}_{+}$&\footnotesize$\begin{matrix}\beta^{3}_{+}&\\&\beta^{3}_{-}\end{matrix}$&\footnotesize$\begin{matrix}\beta^{3}_{+}&\\&\beta^{3}_{-}\end{matrix}$&\footnotesize$\begin{matrix}\chi^{3}_{+}&\chi^{4}_{+}\\\chi^{3}_{-}&\chi^{4}_{-}\end{matrix}$&\footnotesize$\begin{matrix}\chi^{1}_{+}&\chi^{2}_{+}\\\chi^{1}_{-}&\chi^{2}_{-}\end{matrix}$&\footnotesize$\begin{matrix}\alpha^{1}_{+}&\\&\alpha^{1}_{-}\end{matrix}$&\footnotesize$\begin{matrix}\alpha^{1}_{+}&\\&\alpha^{1}_{-}\end{matrix}$&\footnotesize$\begin{matrix}\beta^{1}_{+}&\\&\beta^{1}_{-}\end{matrix}$&\footnotesize$\begin{matrix}\beta^{1}_{+}&\\&\beta^{1}_{-}\end{matrix}$&\footnotesize$\begin{matrix}\gamma_{+}&\\&\gamma_{-}\end{matrix}$&\footnotesize$\begin{matrix}\gamma_{+}&\\&\gamma_{-}\end{matrix}$&\footnotesize$\begin{matrix}\alpha^{3}_{+}&\\&\alpha^{3}_{-}\end{matrix}$&\footnotesize$\begin{matrix}\alpha^{3}_{+}&\\&\alpha^{3}_{-}\end{matrix}$&\footnotesize$\begin{matrix}\beta^{2}_{+}&\\&\beta^{2}_{-}\end{matrix}$&\footnotesize$\begin{matrix}\beta^{2}_{+}&\\&\beta^{2}_{-}\end{matrix}$\\\hline
	\boldmath$\beta^{3}_{+}$&\footnotesize$\beta^{3}_{+}$&\footnotesize$\beta^{3}_{-}$&\footnotesize$\beta^{3}_{-}$&\footnotesize$\beta^{3}_{+}$&\footnotesize$\beta^{3}_{+}$&\footnotesize$\beta^{3}_{-}$&\footnotesize$\beta^{3}_{-}$&\footnotesize$\beta^{3}_{+}$&\footnotesize$\begin{matrix}\alpha^{2}_{+}&\\&\alpha^{2}_{-}\end{matrix}$&\footnotesize$\begin{matrix}\alpha^{2}_{+}&\\&\alpha^{2}_{-}\end{matrix}$&\footnotesize$\begin{matrix}\alpha^{1}_{+}&\\&\alpha^{1}_{-}\end{matrix}$&\footnotesize$\begin{matrix}\alpha^{1}_{+}&\\&\alpha^{1}_{-}\end{matrix}$&\footnotesize$\begin{matrix}\chi^{1}_{+}&\chi^{4}_{+}\\\chi^{1}_{-}&\chi^{4}_{-}\end{matrix}$&\footnotesize$\begin{matrix}\chi^{2}_{+}&\chi^{3}_{+}\\\chi^{2}_{-}&\chi^{3}_{-}\end{matrix}$&\footnotesize$\begin{matrix}\gamma_{+}&\\&\gamma_{-}\end{matrix}$&\footnotesize$\begin{matrix}\gamma_{+}&\\&\gamma_{-}\end{matrix}$&\footnotesize$\begin{matrix}\beta^{1}_{+}&\\&\beta^{1}_{-}\end{matrix}$&\footnotesize$\begin{matrix}\beta^{1}_{+}&\\&\beta^{1}_{-}\end{matrix}$&\footnotesize$\begin{matrix}\beta^{2}_{+}&\\&\beta^{2}_{-}\end{matrix}$&\footnotesize$\begin{matrix}\beta^{2}_{+}&\\&\beta^{2}_{-}\end{matrix}$&\footnotesize$\begin{matrix}\alpha^{3}_{+}&\\&\alpha^{3}_{-}\end{matrix}$&\footnotesize$\begin{matrix}\alpha^{3}_{+}&\\&\alpha^{3}_{-}\end{matrix}$\\\hline
	\boldmath$\beta^{3}_{-}$&\footnotesize$\beta^{3}_{-}$&\footnotesize$\beta^{3}_{+}$&\footnotesize$\beta^{3}_{+}$&\footnotesize$\beta^{3}_{-}$&\footnotesize$\beta^{3}_{-}$&\footnotesize$\beta^{3}_{+}$&\footnotesize$\beta^{3}_{+}$&\footnotesize$\beta^{3}_{-}$&\footnotesize$\begin{matrix}\alpha^{2}_{+}&\\&\alpha^{2}_{-}\end{matrix}$&\footnotesize$\begin{matrix}\alpha^{2}_{+}&\\&\alpha^{2}_{-}\end{matrix}$&\footnotesize$\begin{matrix}\alpha^{1}_{+}&\\&\alpha^{1}_{-}\end{matrix}$&\footnotesize$\begin{matrix}\alpha^{1}_{+}&\\&\alpha^{1}_{-}\end{matrix}$&\footnotesize$\begin{matrix}\chi^{2}_{+}&\chi^{3}_{+}\\\chi^{2}_{-}&\chi^{3}_{-}\end{matrix}$&\footnotesize$\begin{matrix}\chi^{1}_{+}&\chi^{4}_{+}\\\chi^{1}_{-}&\chi^{4}_{-}\end{matrix}$&\footnotesize$\begin{matrix}\gamma_{+}&\\&\gamma_{-}\end{matrix}$&\footnotesize$\begin{matrix}\gamma_{+}&\\&\gamma_{-}\end{matrix}$&\footnotesize$\begin{matrix}\beta^{1}_{+}&\\&\beta^{1}_{-}\end{matrix}$&\footnotesize$\begin{matrix}\beta^{1}_{+}&\\&\beta^{1}_{-}\end{matrix}$&\footnotesize$\begin{matrix}\beta^{2}_{+}&\\&\beta^{2}_{-}\end{matrix}$&\footnotesize$\begin{matrix}\beta^{2}_{+}&\\&\beta^{2}_{-}\end{matrix}$&\footnotesize$\begin{matrix}\alpha^{3}_{+}&\\&\alpha^{3}_{-}\end{matrix}$&\footnotesize$\begin{matrix}\alpha^{3}_{+}&\\&\alpha^{3}_{-}\end{matrix}$\\\hline
	\boldmath$\alpha^{3}_{+}$&\footnotesize$\alpha^{3}_{+}$&\footnotesize$\alpha^{3}_{+}$&\footnotesize$\alpha^{3}_{+}$&\footnotesize$\alpha^{3}_{+}$&\footnotesize$\alpha^{3}_{-}$&\footnotesize$\alpha^{3}_{-}$&\footnotesize$\alpha^{3}_{-}$&\footnotesize$\alpha^{3}_{-}$&\footnotesize$\begin{matrix}\beta^{2}_{+}&\\&\beta^{2}_{-}\end{matrix}$&\footnotesize$\begin{matrix}\beta^{2}_{+}&\\&\beta^{2}_{-}\end{matrix}$&\footnotesize$\begin{matrix}\beta^{1}_{+}&\\&\beta^{1}_{-}\end{matrix}$&\footnotesize$\begin{matrix}\beta^{1}_{+}&\\&\beta^{1}_{-}\end{matrix}$&\footnotesize$\begin{matrix}\gamma_{+}&\\&\gamma_{-}\end{matrix}$&\footnotesize$\begin{matrix}\gamma_{+}&\\&\gamma_{-}\end{matrix}$&\footnotesize$\begin{matrix}\chi^{1}_{+}&\chi^{2}_{+}\\\chi^{3}_{+}&\chi^{4}_{+}\end{matrix}$&\footnotesize$\begin{matrix}\chi^{1}_{-}&\chi^{2}_{-}\\\chi^{3}_{-}&\chi^{4}_{-}\end{matrix}$&\footnotesize$\begin{matrix}\alpha^{1}_{+}&\\&\alpha^{1}_{-}\end{matrix}$&\footnotesize$\begin{matrix}\alpha^{1}_{+}&\\&\alpha^{1}_{-}\end{matrix}$&\footnotesize$\begin{matrix}\alpha^{2}_{+}&\\&\alpha^{2}_{-}\end{matrix}$&\footnotesize$\begin{matrix}\alpha^{2}_{+}&\\&\alpha^{2}_{-}\end{matrix}$&\footnotesize$\begin{matrix}\beta^{3}_{+}&\\&\beta^{3}_{-}\end{matrix}$&\footnotesize$\begin{matrix}\beta^{3}_{+}&\\&\beta^{3}_{-}\end{matrix}$\\\hline
	\boldmath$\alpha^{3}_{-}$&\footnotesize$\alpha^{3}_{-}$&\footnotesize$\alpha^{3}_{-}$&\footnotesize$\alpha^{3}_{-}$&\footnotesize$\alpha^{3}_{-}$&\footnotesize$\alpha^{3}_{+}$&\footnotesize$\alpha^{3}_{+}$&\footnotesize$\alpha^{3}_{+}$&\footnotesize$\alpha^{3}_{+}$&\footnotesize$\begin{matrix}\beta^{2}_{+}&\\&\beta^{2}_{-}\end{matrix}$&\footnotesize$\begin{matrix}\beta^{2}_{+}&\\&\beta^{2}_{-}\end{matrix}$&\footnotesize$\begin{matrix}\beta^{1}_{+}&\\&\beta^{1}_{-}\end{matrix}$&\footnotesize$\begin{matrix}\beta^{1}_{+}&\\&\beta^{1}_{-}\end{matrix}$&\footnotesize$\begin{matrix}\gamma_{+}&\\&\gamma_{-}\end{matrix}$&\footnotesize$\begin{matrix}\gamma_{+}&\\&\gamma_{-}\end{matrix}$&\footnotesize$\begin{matrix}\chi^{1}_{-}&\chi^{2}_{-}\\\chi^{3}_{-}&\chi^{4}_{-}\end{matrix}$&\footnotesize$\begin{matrix}\chi^{1}_{+}&\chi^{2}_{+}\\\chi^{3}_{+}&\chi^{4}_{+}\end{matrix}$&\footnotesize$\begin{matrix}\alpha^{1}_{+}&\\&\alpha^{1}_{-}\end{matrix}$&\footnotesize$\begin{matrix}\alpha^{1}_{+}&\\&\alpha^{1}_{-}\end{matrix}$&\footnotesize$\begin{matrix}\alpha^{2}_{+}&\\&\alpha^{2}_{-}\end{matrix}$&\footnotesize$\begin{matrix}\alpha^{2}_{+}&\\&\alpha^{2}_{-}\end{matrix}$&\footnotesize$\begin{matrix}\beta^{3}_{+}&\\&\beta^{3}_{-}\end{matrix}$&\footnotesize$\begin{matrix}\beta^{3}_{+}&\\&\beta^{3}_{-}\end{matrix}$\\\hline
	\boldmath$\beta^{2}_{+}$&\footnotesize$\beta^{2}_{+}$&\footnotesize$\beta^{2}_{-}$&\footnotesize$\beta^{2}_{+}$&\footnotesize$\beta^{2}_{-}$&\footnotesize$\beta^{2}_{-}$&\footnotesize$\beta^{2}_{+}$&\footnotesize$\beta^{2}_{-}$&\footnotesize$\beta^{2}_{+}$&\footnotesize$\begin{matrix}\alpha^{3}_{+}&\\&\alpha^{3}_{-}\end{matrix}$&\footnotesize$\begin{matrix}\alpha^{3}_{+}&\\&\alpha^{3}_{-}\end{matrix}$&\footnotesize$\begin{matrix}\gamma_{+}&\\&\gamma_{-}\end{matrix}$&\footnotesize$\begin{matrix}\gamma_{+}&\\&\gamma_{-}\end{matrix}$&\footnotesize$\begin{matrix}\beta^{1}_{+}&\\&\beta^{1}_{-}\end{matrix}$&\footnotesize$\begin{matrix}\beta^{1}_{+}&\\&\beta^{1}_{-}\end{matrix}$&\footnotesize$\begin{matrix}\alpha^{1}_{+}&\\&\alpha^{1}_{-}\end{matrix}$&\footnotesize$\begin{matrix}\alpha^{1}_{+}&\\&\alpha^{1}_{-}\end{matrix}$&\footnotesize$\begin{matrix}\chi^{2}_{+}&\chi^{4}_{+}\\\chi^{1}_{-}&\chi^{3}_{-}\end{matrix}$&\footnotesize$\begin{matrix}\chi^{1}_{+}&\chi^{3}_{+}\\\chi^{2}_{-}&\chi^{4}_{-}\end{matrix}$&\footnotesize$\begin{matrix}\beta^{3}_{+}&\\&\beta^{3}_{-}\end{matrix}$&\footnotesize$\begin{matrix}\beta^{3}_{+}&\\&\beta^{3}_{-}\end{matrix}$&\footnotesize$\begin{matrix}\alpha^{2}_{+}&\\&\alpha^{2}_{-}\end{matrix}$&\footnotesize$\begin{matrix}\alpha^{2}_{+}&\\&\alpha^{2}_{-}\end{matrix}$\\\hline
	\boldmath$\beta^{2}_{-}$&\footnotesize$\beta^{2}_{-}$&\footnotesize$\beta^{2}_{+}$&\footnotesize$\beta^{2}_{-}$&\footnotesize$\beta^{2}_{+}$&\footnotesize$\beta^{2}_{+}$&\footnotesize$\beta^{2}_{-}$&\footnotesize$\beta^{2}_{+}$&\footnotesize$\beta^{2}_{-}$&\footnotesize$\begin{matrix}\alpha^{3}_{+}&\\&\alpha^{3}_{-}\end{matrix}$&\footnotesize$\begin{matrix}\alpha^{3}_{+}&\\&\alpha^{3}_{-}\end{matrix}$&\footnotesize$\begin{matrix}\gamma_{+}&\\&\gamma_{-}\end{matrix}$&\footnotesize$\begin{matrix}\gamma_{+}&\\&\gamma_{-}\end{matrix}$&\footnotesize$\begin{matrix}\beta^{1}_{+}&\\&\beta^{1}_{-}\end{matrix}$&\footnotesize$\begin{matrix}\beta^{1}_{+}&\\&\beta^{1}_{-}\end{matrix}$&\footnotesize$\begin{matrix}\alpha^{1}_{+}&\\&\alpha^{1}_{-}\end{matrix}$&\footnotesize$\begin{matrix}\alpha^{1}_{+}&\\&\alpha^{1}_{-}\end{matrix}$&\footnotesize$\begin{matrix}\chi^{1}_{+}&\chi^{3}_{+}\\\chi^{2}_{-}&\chi^{4}_{-}\end{matrix}$&\footnotesize$\begin{matrix}\chi^{2}_{+}&\chi^{4}_{+}\\\chi^{1}_{-}&\chi^{3}_{-}\end{matrix}$&\footnotesize$\begin{matrix}\beta^{3}_{+}&\\&\beta^{3}_{-}\end{matrix}$&\footnotesize$\begin{matrix}\beta^{3}_{+}&\\&\beta^{3}_{-}\end{matrix}$&\footnotesize$\begin{matrix}\alpha^{2}_{+}&\\&\alpha^{2}_{-}\end{matrix}$&\footnotesize$\begin{matrix}\alpha^{2}_{+}&\\&\alpha^{2}_{-}\end{matrix}$\\\hline
	\boldmath$\beta^{1}_{+}$&\footnotesize$\beta^{1}_{+}$&\footnotesize$\beta^{1}_{+}$&\footnotesize$\beta^{1}_{-}$&\footnotesize$\beta^{1}_{-}$&\footnotesize$\beta^{1}_{-}$&\footnotesize$\beta^{1}_{-}$&\footnotesize$\beta^{1}_{+}$&\footnotesize$\beta^{1}_{+}$&\footnotesize$\begin{matrix}\gamma_{+}&\\&\gamma_{-}\end{matrix}$&\footnotesize$\begin{matrix}\gamma_{+}&\\&\gamma_{-}\end{matrix}$&\footnotesize$\begin{matrix}\alpha^{3}_{+}&\\&\alpha^{3}_{-}\end{matrix}$&\footnotesize$\begin{matrix}\alpha^{3}_{+}&\\&\alpha^{3}_{-}\end{matrix}$&\footnotesize$\begin{matrix}\beta^{2}_{+}&\\&\beta^{2}_{-}\end{matrix}$&\footnotesize$\begin{matrix}\beta^{2}_{+}&\\&\beta^{2}_{-}\end{matrix}$&\footnotesize$\begin{matrix}\alpha^{2}_{+}&\\&\alpha^{2}_{-}\end{matrix}$&\footnotesize$\begin{matrix}\alpha^{2}_{+}&\\&\alpha^{2}_{-}\end{matrix}$&\footnotesize$\begin{matrix}\beta^{3}_{+}&\\&\beta^{3}_{-}\end{matrix}$&\footnotesize$\begin{matrix}\beta^{3}_{+}&\\&\beta^{3}_{-}\end{matrix}$&\footnotesize$\begin{matrix}\chi^{3}_{+}&\chi^{4}_{+}\\\chi^{1}_{-}&\chi^{2}_{-}\end{matrix}$&\footnotesize$\begin{matrix}\chi^{1}_{+}&\chi^{2}_{+}\\\chi^{3}_{-}&\chi^{4}_{-}\end{matrix}$&\footnotesize$\begin{matrix}\alpha^{1}_{+}&\\&\alpha^{1}_{-}\end{matrix}$&\footnotesize$\begin{matrix}\alpha^{1}_{+}&\\&\alpha^{1}_{-}\end{matrix}$\\\hline
	\boldmath$\beta^{1}_{-}$&\footnotesize$\beta^{1}_{-}$&\footnotesize$\beta^{1}_{-}$&\footnotesize$\beta^{1}_{+}$&\footnotesize$\beta^{1}_{+}$&\footnotesize$\beta^{1}_{+}$&\footnotesize$\beta^{1}_{+}$&\footnotesize$\beta^{1}_{-}$&\footnotesize$\beta^{1}_{-}$&\footnotesize$\begin{matrix}\gamma_{+}&\\&\gamma_{-}\end{matrix}$&\footnotesize$\begin{matrix}\gamma_{+}&\\&\gamma_{-}\end{matrix}$&\footnotesize$\begin{matrix}\alpha^{3}_{+}&\\&\alpha^{3}_{-}\end{matrix}$&\footnotesize$\begin{matrix}\alpha^{3}_{+}&\\&\alpha^{3}_{-}\end{matrix}$&\footnotesize$\begin{matrix}\beta^{2}_{+}&\\&\beta^{2}_{-}\end{matrix}$&\footnotesize$\begin{matrix}\beta^{2}_{+}&\\&\beta^{2}_{-}\end{matrix}$&\footnotesize$\begin{matrix}\alpha^{2}_{+}&\\&\alpha^{2}_{-}\end{matrix}$&\footnotesize$\begin{matrix}\alpha^{2}_{+}&\\&\alpha^{2}_{-}\end{matrix}$&\footnotesize$\begin{matrix}\beta^{3}_{+}&\\&\beta^{3}_{-}\end{matrix}$&\footnotesize$\begin{matrix}\beta^{3}_{+}&\\&\beta^{3}_{-}\end{matrix}$&\footnotesize$\begin{matrix}\chi^{1}_{+}&\chi^{2}_{+}\\\chi^{3}_{-}&\chi^{4}_{-}\end{matrix}$&\footnotesize$\begin{matrix}\chi^{3}_{+}&\chi^{4}_{+}\\\chi^{1}_{-}&\chi^{2}_{-}\end{matrix}$&\footnotesize$\begin{matrix}\alpha^{1}_{+}&\\&\alpha^{1}_{-}\end{matrix}$&\footnotesize$\begin{matrix}\alpha^{1}_{+}&\\&\alpha^{1}_{-}\end{matrix}$\\\hline
	\boldmath$\gamma_{+}$&\footnotesize$\gamma_{+}$&\footnotesize$\gamma_{-}$&\footnotesize$\gamma_{-}$&\footnotesize$\gamma_{+}$&\footnotesize$\gamma_{-}$&\footnotesize$\gamma_{+}$&\footnotesize$\gamma_{+}$&\footnotesize$\gamma_{-}$&\footnotesize$\begin{matrix}\beta^{1}_{+}&\\&\beta^{1}_{-}\end{matrix}$&\footnotesize$\begin{matrix}\beta^{1}_{+}&\\&\beta^{1}_{-}\end{matrix}$&\footnotesize$\begin{matrix}\beta^{2}_{+}&\\&\beta^{2}_{-}\end{matrix}$&\footnotesize$\begin{matrix}\beta^{2}_{+}&\\&\beta^{2}_{-}\end{matrix}$&\footnotesize$\begin{matrix}\alpha^{3}_{+}&\\&\alpha^{3}_{-}\end{matrix}$&\footnotesize$\begin{matrix}\alpha^{3}_{+}&\\&\alpha^{3}_{-}\end{matrix}$&\footnotesize$\begin{matrix}\beta^{3}_{+}&\\&\beta^{3}_{-}\end{matrix}$&\footnotesize$\begin{matrix}\beta^{3}_{+}&\\&\beta^{3}_{-}\end{matrix}$&\footnotesize$\begin{matrix}\alpha^{2}_{+}&\\&\alpha^{2}_{-}\end{matrix}$&\footnotesize$\begin{matrix}\alpha^{2}_{+}&\\&\alpha^{2}_{-}\end{matrix}$&\footnotesize$\begin{matrix}\alpha^{1}_{+}&\\&\alpha^{1}_{-}\end{matrix}$&\footnotesize$\begin{matrix}\alpha^{1}_{+}&\\&\alpha^{1}_{-}\end{matrix}$&\footnotesize$\begin{matrix}\chi^{1}_{+}&\chi^{4}_{+}\\\chi^{2}_{-}&\chi^{3}_{-}\end{matrix}$&\footnotesize$\begin{matrix}\chi^{2}_{+}&\chi^{3}_{+}\\\chi^{1}_{-}&\chi^{4}_{-}\end{matrix}$\\\hline
	\boldmath$\gamma_{-}$&\footnotesize$\gamma_{-}$&\footnotesize$\gamma_{+}$&\footnotesize$\gamma_{+}$&\footnotesize$\gamma_{-}$&\footnotesize$\gamma_{+}$&\footnotesize$\gamma_{-}$&\footnotesize$\gamma_{-}$&\footnotesize$\gamma_{+}$&\footnotesize$\begin{matrix}\beta^{1}_{+}&\\&\beta^{1}_{-}\end{matrix}$&\footnotesize$\begin{matrix}\beta^{1}_{+}&\\&\beta^{1}_{-}\end{matrix}$&\footnotesize$\begin{matrix}\beta^{2}_{+}&\\&\beta^{2}_{-}\end{matrix}$&\footnotesize$\begin{matrix}\beta^{2}_{+}&\\&\beta^{2}_{-}\end{matrix}$&\footnotesize$\begin{matrix}\alpha^{3}_{+}&\\&\alpha^{3}_{-}\end{matrix}$&\footnotesize$\begin{matrix}\alpha^{3}_{+}&\\&\alpha^{3}_{-}\end{matrix}$&\footnotesize$\begin{matrix}\beta^{3}_{+}&\\&\beta^{3}_{-}\end{matrix}$&\footnotesize$\begin{matrix}\beta^{3}_{+}&\\&\beta^{3}_{-}\end{matrix}$&\footnotesize$\begin{matrix}\alpha^{2}_{+}&\\&\alpha^{2}_{-}\end{matrix}$&\footnotesize$\begin{matrix}\alpha^{2}_{+}&\\&\alpha^{2}_{-}\end{matrix}$&\footnotesize$\begin{matrix}\alpha^{1}_{+}&\\&\alpha^{1}_{-}\end{matrix}$&\footnotesize$\begin{matrix}\alpha^{1}_{+}&\\&\alpha^{1}_{-}\end{matrix}$&\footnotesize$\begin{matrix}\chi^{2}_{+}&\chi^{3}_{+}\\\chi^{1}_{-}&\chi^{4}_{-}\end{matrix}$&\footnotesize$\begin{matrix}\chi^{1}_{+}&\chi^{4}_{+}\\\chi^{2}_{-}&\chi^{3}_{-}\end{matrix}$\\\hline
\end{tabular}
}
\caption{Fusion rules for $D^\phi(\zt^3)$ sectors computed from symmetric MERA. Cell entries denote the allowed fusion outcome sectors for $a\times b$. The OPE coefficients defined in Eqn.~\ref{eqn:OPE} are zero if the resultant field $c$ does not lie in an allowed sector.}\label{table:fusion}
\end{table}

\clearpage
\section{MPO group representations and third cohomology}
\label{appendix:thirdcoho}
In this appendix we recount the definition of the third cohomology class of an injective MPO representation of a finite group $\G$, as first introduced in Ref.~\onlinecite{czxmodel}. MPO representations appear in the study of $(2+1)$D SPT tensor network states and it was shown in Ref.~\onlinecite{williamson2014matrix} that they are always injective. 
The presence of such an MPO symmetry has an important physical consequence; all short range entangled states must break the symmetry, either explicitly or spontaneously.
For details about group cohomology theory in the context of SPT order we refer the reader to Ref.~\onlinecite{chen2013symmetry}. 
 
In an MPO representation of $\G$, multiplying a pair of MPOs labeled by the group elements $g_0$ and $g_1$ is equal to the MPO labeled by $g_0g_1$ for every length. For injective MPOs there exists a gauge transformation on the virtual indices that brings both representations into the same canonical form~\cite{Fannes92,MPSrepresentations,Cirac2017100}. This implies that there exists an operator (the reduction tensor) $X(g_0,g_1):(\mathbb{C}^{\chi})^{\otimes 2}\rightarrow \mathbb{C}^{ \chi}$ such that
\begin{align}
\label{mporeduction}
	\begin{array}{c}
		\includeTikz{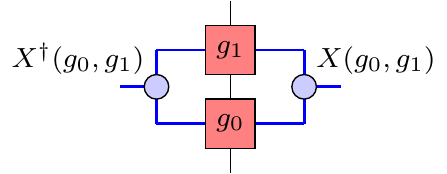}{
		\begin{tikzpicture}[scale=.25]
			\def\x{1}
			\draw[thick,blue,shift={(0,1.5*\x)}](-3*\x,0)--(3*\x,0);
			\draw[thick,blue,shift={(0,-1.5*\x)}](-3*\x,0)--(3*\x,0);
			\draw[thick,blue] (3*\x,1.5*\x)--(3*\x,-1.5*\x);
			\draw[thick,blue] (-3*\x,1.5*\x)--(-3*\x,-1.5*\x);
			\draw[thick,blue] (3*\x,0)--(4.5*\x,0);
			\draw[thick,blue] (-3*\x,0)--(-4.5*\x,0);
			\draw (0,3.5*\x)--(0,-3.5*\x);
			\filldraw[tenred,shift={(0,1.5*\x)}] (-\x,-\x)--(\x,-\x)--(\x,\x)--(-\x,\x)--cycle;\node at(0,1.5*\x) {\textcolor{black}{\footnotesize$g_1$}};
			\filldraw[tenred,shift={(0,-1.5*\x)}] (-\x,-\x)--(\x,-\x)--(\x,\x)--(-\x,\x)--cycle;\node at(0,-1.5*\x) {\textcolor{black}{\footnotesize$g_0$}};
			\filldraw[ten](3*\x,0) circle (\x/2);\node[above right] at (3*\x,0) {\textcolor{black}{\footnotesize$X(g_0,g_1)$}};
			\filldraw[ten](-3*\x,0) circle (\x/2);\node[above left] at(-3*\x,0) {\textcolor{black}{\footnotesize$X^\dagger(g_0,g_1)$}};
		\end{tikzpicture}}
	\end{array}
	&=
	\begin{array}{c}
		\includeTikz{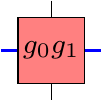}{
		\begin{tikzpicture}[scale=.25]
			\def\x{1.35}
			\draw[thick,blue,shift={(0,0)}](-1.5*\x,0)--(1.5*\x,0);
			\draw (0,1.5*\x)--(0,-1.5*\x);
			\filldraw[tenred,shift={(0,0)}] (-\x,-\x)--(\x,-\x)--(\x,\x)--(-\x,\x)--cycle;\node at(0,0) {\textcolor{black}{\footnotesize$g_0g_1$}};
		\end{tikzpicture}}
	\end{array},
\end{align}
where $X(g_0,g_1)$ is only defined up to multiplication by a complex phase $\beta(g_0,g_1)$.

If we now multiply three MPOs labeled by $g_0$, $g_1$ and $g_2$ there are two ways to reduce the multiplied MPOs to the MPO labeled by $g_0g_1g_2$. When only acting on the right virtual indices these two reductions are equivalent up to a complex phase
\begin{align}
\label{associator}
	\begin{array}{c}
		\includeTikz{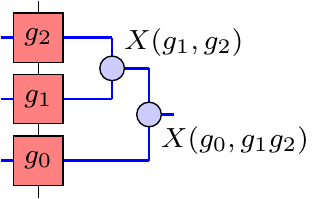}{
		\begin{tikzpicture}[scale=.25]
			\begin{scope}[yscale=-1]
				\def\x{1}
				\draw[thick,blue,shift={(0,2.5*\x)}](-1.5*\x,0)--(4.5*\x,0);
				\draw[thick,blue,shift={(0,0)}](-1.5*\x,0)--(3*\x,0);
				\draw[thick,blue,shift={(0,-2.5*\x)}](-1.5*\x,0)--(3*\x,0);
				\draw[thick,blue] (3*\x,0)--(3*\x,-2.5*\x);
				\draw[thick,blue] (3*\x,-1.25*\x)--(4.5*\x,-1.25*\x);
				\draw[thick,blue] (4.5*\x,-1.25*\x)--(4.5*\x,2.5*\x);
				\draw[thick,blue] (4.5*\x,0.625*\x)--(5.5*\x,0.625*\x);
				\draw (0,4*\x)--(0,-4*\x);
				\filldraw[tenred,shift={(0,2.5*\x)}] (-\x,-\x)--(\x,-\x)--(\x,\x)--(-\x,\x)--cycle;\node at(0,2.5*\x) {\textcolor{black}{\footnotesize$g_0$}};
				\filldraw[tenred,shift={(0,0)}] (-\x,-\x)--(\x,-\x)--(\x,\x)--(-\x,\x)--cycle;\node at(0,0) {\textcolor{black}{\footnotesize$g_1$}};
				\filldraw[tenred,shift={(0,-2.5*\x)}] (-\x,-\x)--(\x,-\x)--(\x,\x)--(-\x,\x)--cycle;\node at(0,-2.5*\x) {\textcolor{black}{\footnotesize$g_2$}};
				\filldraw[ten](3*\x,-1.25*\x) circle (\x/2);
				\node[above right] at (3*\x,-1.25*\x) {\textcolor{black}{\footnotesize$X(g_1,g_2)$}};
				\filldraw[ten](4.5*\x,0.625*\x) circle (\x/2);\node[below right] at(4.5*\x,0.625*\x) {\textcolor{black}{\footnotesize$X(g_0,g_1g_2)$}};
			\end{scope}
		\end{tikzpicture}}
	\end{array}
	&=\phi(g_0,g_1,g_2)
	\begin{array}{c}
		\includeTikz{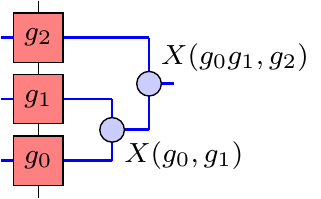}{
		\begin{tikzpicture}[scale=.25]
			\def\x{1}
			\draw[thick,blue,shift={(0,2.5*\x)}](-1.5*\x,0)--(4.5*\x,0);
			\draw[thick,blue,shift={(0,0)}](-1.5*\x,0)--(3*\x,0);
			\draw[thick,blue,shift={(0,-2.5*\x)}](-1.5*\x,0)--(3*\x,0);
			\draw[thick,blue] (3*\x,0)--(3*\x,-2.5*\x);
			\draw[thick,blue] (3*\x,-1.25*\x)--(4.5*\x,-1.25*\x);
			\draw[thick,blue] (4.5*\x,-1.25*\x)--(4.5*\x,2.5*\x);
			\draw[thick,blue] (4.5*\x,0.625*\x)--(5.5*\x,0.625*\x);
			\draw (0,4*\x)--(0,-4*\x);
			\filldraw[tenred,shift={(0,2.5*\x)}] (-\x,-\x)--(\x,-\x)--(\x,\x)--(-\x,\x)--cycle;\node at(0,2.5*\x) {\textcolor{black}{\footnotesize$g_2$}};
			\filldraw[tenred,shift={(0,0)}] (-\x,-\x)--(\x,-\x)--(\x,\x)--(-\x,\x)--cycle;\node at(0,0) {\textcolor{black}{\footnotesize$g_1$}};
			\filldraw[tenred,shift={(0,-2.5*\x)}] (-\x,-\x)--(\x,-\x)--(\x,\x)--(-\x,\x)--cycle;\node at(0,-2.5*\x) {\textcolor{black}{\footnotesize$g_0$}};
			\filldraw[ten](3*\x,-1.25*\x) circle (\x/2);
			\node[below right] at (3*\x,-1.25*\x) {\textcolor{black}{\footnotesize$X(g_0,g_1)$}};
			\filldraw[ten](4.5*\x,0.625*\x) circle (\x/2);\node[above right] at(4.5*\x,0.625*\x) {\textcolor{black}{\footnotesize$X(g_0g_1,g_2)$}};
		\end{tikzpicture}}
	\end{array}.
\end{align}

When multiplying four MPOs, one observes that $\phi$ has to obey certain consistency conditions. By performing a series of moves (changing order of reduction), one can achieve the same reduction 

\begin{align}
	\begin{array}{c}
		\includeTikz{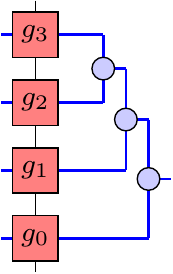}{
			\begin{tikzpicture}[scale=.23]
			\def\x{1}
			\begin{scope}[yscale=-1]
			\draw[thick,blue](-1.5,4.5*\x)--(5*\x,4.5*\x);
			\draw[thick,blue](-1.5,1.5*\x)--(4*\x,1.5*\x);
			\draw[thick,blue](-1.5,-1.5*\x)--(3*\x,-1.5*\x);
			\draw[thick,blue](-1.5,-4.5*\x)--(3*\x,-4.5*\x);
			\draw[thick,blue](3*\x,-4.5*\x)--(3*\x,-1.5*\x);\draw[thick,blue](3*\x,-3*\x)--(4*\x,-3*\x);
			\draw[thick,blue](4*\x,-3*\x)--(4*\x,1.5*\x);\draw[thick,blue](4*\x,-.75*\x)--(5*\x,-.75*\x);
			\draw[thick,blue](5*\x,-.75*\x)--(5*\x,4.5*\x);\draw[thick,blue](5*\x,1.875*\x)--(6*\x,1.875*\x);
			\filldraw[ten](3*\x,-3*\x) circle (\x/2);
			\filldraw[ten](4*\x,-.75*\x) circle (\x/2);
			\filldraw[ten](5*\x,1.875*\x) circle (\x/2);
			\end{scope}
			\draw (0,6*\x)--(0,-6*\x);
			\filldraw[tenred,shift={(0,4.5*\x)}] (-\x,-\x)--(\x,-\x)--(\x,\x)--(-\x,\x)--cycle;\node at(0,4.5*\x) {\textcolor{black}{\footnotesize$g_3$}};
			\filldraw[tenred,shift={(0,1.5*\x)}] (-\x,-\x)--(\x,-\x)--(\x,\x)--(-\x,\x)--cycle;\node at(0,1.5*\x) {\textcolor{black}{\footnotesize$g_2$}};
			\filldraw[tenred,shift={(0,-1.5*\x)}] (-\x,-\x)--(\x,-\x)--(\x,\x)--(-\x,\x)--cycle;\node at(0,-1.5*\x) {\textcolor{black}{\footnotesize$g_1$}};
			\filldraw[tenred,shift={(0,-4.5*\x)}] (-\x,-\x)--(\x,-\x)--(\x,\x)--(-\x,\x)--cycle;\node at(0,-4.5*\x) {\textcolor{black}{\footnotesize$g_0$}};
			\end{tikzpicture}}
	\end{array}
	&=\phi(g_1,g_2,g_3)
	\begin{array}{c}
		\includeTikz{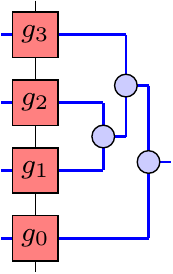}{
			\begin{tikzpicture}[scale=.23]
			\def\x{1}
			\begin{scope}
			\draw[thick,blue](-1.5,4.5*\x)--(4*\x,4.5*\x);
			\draw[thick,blue](-1.5,1.5*\x)--(3*\x,1.5*\x);
			\draw[thick,blue](-1.5,-1.5*\x)--(3*\x,-1.5*\x);
			\draw[thick,blue](-1.5,-4.5*\x)--(5*\x,-4.5*\x);
			\draw[thick,blue](3*\x,-1.5*\x)--(3*\x,1.5*\x);\draw[thick,blue](3*\x,0*\x)--(4*\x,0*\x);
			\draw[thick,blue](4*\x,4.5*\x)--(4*\x,0*\x);\draw[thick,blue](4*\x,2.25*\x)--(5*\x,2.25*\x);
			\draw[thick,blue](5*\x,2.25*\x)--(5*\x,-4.5*\x);\draw[thick,blue](5*\x,-1.125*\x)--(6*\x,-1.125*\x);
			\filldraw[ten](3*\x,0*\x) circle (\x/2);
			\filldraw[ten](4*\x,2.25*\x) circle (\x/2);
			\filldraw[ten](5*\x,-1.125*\x) circle (\x/2);
			\end{scope}
			\draw (0,6*\x)--(0,-6*\x);
			\filldraw[tenred,shift={(0,4.5*\x)}] (-\x,-\x)--(\x,-\x)--(\x,\x)--(-\x,\x)--cycle;\node at(0,4.5*\x) {\textcolor{black}{\footnotesize$g_3$}};
			\filldraw[tenred,shift={(0,1.5*\x)}] (-\x,-\x)--(\x,-\x)--(\x,\x)--(-\x,\x)--cycle;\node at(0,1.5*\x) {\textcolor{black}{\footnotesize$g_2$}};
			\filldraw[tenred,shift={(0,-1.5*\x)}] (-\x,-\x)--(\x,-\x)--(\x,\x)--(-\x,\x)--cycle;\node at(0,-1.5*\x) {\textcolor{black}{\footnotesize$g_1$}};
			\filldraw[tenred,shift={(0,-4.5*\x)}] (-\x,-\x)--(\x,-\x)--(\x,\x)--(-\x,\x)--cycle;\node at(0,-4.5*\x) {\textcolor{black}{\footnotesize$g_0$}};
			\end{tikzpicture}}
	\end{array}	
	=\phi(g_1,g_2,g_3)\phi(g_0,g_1g_2,g_3)
	\begin{array}{c}
		\includeTikz{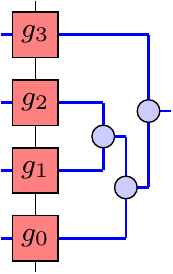}{
			\begin{tikzpicture}[scale=.23]
			\def\x{1}
			\begin{scope}
			\draw[thick,blue](-1.5,4.5*\x)--(5*\x,4.5*\x);
			\draw[thick,blue](-1.5,1.5*\x)--(3*\x,1.5*\x);
			\draw[thick,blue](-1.5,-1.5*\x)--(3*\x,-1.5*\x);
			\draw[thick,blue](-1.5,-4.5*\x)--(4*\x,-4.5*\x);
			\draw[thick,blue](3*\x,-1.5*\x)--(3*\x,1.5*\x);\draw[thick,blue](3*\x,0*\x)--(4*\x,0*\x);
			\draw[thick,blue](4*\x,-4.5*\x)--(4*\x,0*\x);\draw[thick,blue](4*\x,-2.25*\x)--(5*\x,-2.25*\x);
			\draw[thick,blue](5*\x,-2.25*\x)--(5*\x,4.5*\x);\draw[thick,blue](5*\x,1.125*\x)--(6*\x,1.125*\x);
			\filldraw[ten](3*\x,0*\x) circle (\x/2);
			\filldraw[ten](4*\x,-2.25*\x) circle (\x/2);
			\filldraw[ten](5*\x,1.125*\x) circle (\x/2);
			\end{scope}
			\draw (0,6*\x)--(0,-6*\x);
			\filldraw[tenred,shift={(0,4.5*\x)}] (-\x,-\x)--(\x,-\x)--(\x,\x)--(-\x,\x)--cycle;\node at(0,4.5*\x) {\textcolor{black}{\footnotesize$g_3$}};
			\filldraw[tenred,shift={(0,1.5*\x)}] (-\x,-\x)--(\x,-\x)--(\x,\x)--(-\x,\x)--cycle;\node at(0,1.5*\x) {\textcolor{black}{\footnotesize$g_2$}};
			\filldraw[tenred,shift={(0,-1.5*\x)}] (-\x,-\x)--(\x,-\x)--(\x,\x)--(-\x,\x)--cycle;\node at(0,-1.5*\x) {\textcolor{black}{\footnotesize$g_1$}};
			\filldraw[tenred,shift={(0,-4.5*\x)}] (-\x,-\x)--(\x,-\x)--(\x,\x)--(-\x,\x)--cycle;\node at(0,-4.5*\x) {\textcolor{black}{\footnotesize$g_0$}};
			\end{tikzpicture}}
	\end{array}	\nonumber\\
	&=\phi(g_1,g_2,g_3)\phi(g_0,g_1g_2,g_3)\phi(g_0,g_1,g_2)
	\begin{array}{c}
		\includeTikz{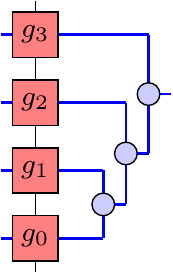}{
			\begin{tikzpicture}[scale=.23]
			\def\x{1}
			\begin{scope}
			\draw[thick,blue](-1.5,4.5*\x)--(5*\x,4.5*\x);
			\draw[thick,blue](-1.5,1.5*\x)--(4*\x,1.5*\x);
			\draw[thick,blue](-1.5,-1.5*\x)--(3*\x,-1.5*\x);
			\draw[thick,blue](-1.5,-4.5*\x)--(3*\x,-4.5*\x);
			\draw[thick,blue](3*\x,-4.5*\x)--(3*\x,-1.5*\x);\draw[thick,blue](3*\x,-3*\x)--(4*\x,-3*\x);
			\draw[thick,blue](4*\x,-3*\x)--(4*\x,1.5*\x);\draw[thick,blue](4*\x,-.75*\x)--(5*\x,-.75*\x);
			\draw[thick,blue](5*\x,-.75*\x)--(5*\x,4.5*\x);\draw[thick,blue](5*\x,1.875*\x)--(6*\x,1.875*\x);
			\filldraw[ten](3*\x,-3*\x) circle (\x/2);
			\filldraw[ten](4*\x,-.75*\x) circle (\x/2);
			\filldraw[ten](5*\x,1.875*\x) circle (\x/2);
			\end{scope}
			\draw (0,6*\x)--(0,-6*\x);
			\filldraw[tenred,shift={(0,4.5*\x)}] (-\x,-\x)--(\x,-\x)--(\x,\x)--(-\x,\x)--cycle;\node at(0,4.5*\x) {\textcolor{black}{\footnotesize$g_3$}};
			\filldraw[tenred,shift={(0,1.5*\x)}] (-\x,-\x)--(\x,-\x)--(\x,\x)--(-\x,\x)--cycle;\node at(0,1.5*\x) {\textcolor{black}{\footnotesize$g_2$}};
			\filldraw[tenred,shift={(0,-1.5*\x)}] (-\x,-\x)--(\x,-\x)--(\x,\x)--(-\x,\x)--cycle;\node at(0,-1.5*\x) {\textcolor{black}{\footnotesize$g_1$}};
			\filldraw[tenred,shift={(0,-4.5*\x)}] (-\x,-\x)--(\x,-\x)--(\x,\x)--(-\x,\x)--cycle;\node at(0,-4.5*\x) {\textcolor{black}{\footnotesize$g_0$}};
			\end{tikzpicture}}
	\end{array}
	=\frac{\phi(g_1,g_2,g_3)\phi(g_0,g_1g_2,g_3)\phi(g_0,g_1,g_2)}{\phi(g_0g_1,g_2,g_3)}
	\begin{array}{c}
		\includeTikz{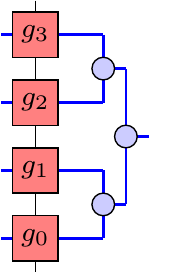}{
			\begin{tikzpicture}[scale=.23]
			\def\x{1}
			\begin{scope}
			\draw[thick,blue](-1.5,4.5*\x)--(3*\x,4.5*\x);
			\draw[thick,blue](-1.5,1.5*\x)--(3*\x,1.5*\x);
			\draw[thick,blue](-1.5,-1.5*\x)--(3*\x,-1.5*\x);
			\draw[thick,blue](-1.5,-4.5*\x)--(3*\x,-4.5*\x);
			\draw[thick,blue](3*\x,-4.5*\x)--(3*\x,-1.5*\x);\draw[thick,blue](3*\x,3*\x)--(4*\x,3*\x);
			\draw[thick,blue](3*\x,4.5*\x)--(3*\x,1.5*\x);\draw[thick,blue](3*\x,-3*\x)--(4*\x,-3*\x);
			\draw[thick,blue](4*\x,-3*\x)--(4*\x,3*\x);\draw[thick,blue](4*\x,0*\x)--(5*\x,0*\x);
			\filldraw[ten](3*\x,3*\x) circle (\x/2);
			\filldraw[ten](3*\x,-3*\x) circle (\x/2);
			\filldraw[ten](4*\x,0*\x) circle (\x/2);
			\draw[draw=none](5*\x,0)--(6*\x,0);
			\end{scope}
			\draw (0,6*\x)--(0,-6*\x);
			\filldraw[tenred,shift={(0,4.5*\x)}] (-\x,-\x)--(\x,-\x)--(\x,\x)--(-\x,\x)--cycle;\node at(0,4.5*\x) {\textcolor{black}{\footnotesize$g_3$}};
			\filldraw[tenred,shift={(0,1.5*\x)}] (-\x,-\x)--(\x,-\x)--(\x,\x)--(-\x,\x)--cycle;\node at(0,1.5*\x) {\textcolor{black}{\footnotesize$g_2$}};
			\filldraw[tenred,shift={(0,-1.5*\x)}] (-\x,-\x)--(\x,-\x)--(\x,\x)--(-\x,\x)--cycle;\node at(0,-1.5*\x) {\textcolor{black}{\footnotesize$g_1$}};
			\filldraw[tenred,shift={(0,-4.5*\x)}] (-\x,-\x)--(\x,-\x)--(\x,\x)--(-\x,\x)--cycle;\node at(0,-4.5*\x) {\textcolor{black}{\footnotesize$g_0$}};
			\end{tikzpicture}}
	\end{array}\nonumber\\
	&=\frac{\phi(g_1,g_2,g_3)\phi(g_0,g_1g_2,g_3)\phi(g_0,g_1,g_2)}{\phi(g_0g_1,g_2,g_3)\phi(g_0,g_1,g_2g_3)}
	\begin{array}{c}
		\includeTikz{fourmpoE}{
			\begin{tikzpicture}[scale=.23]
			\def\x{1}
			\begin{scope}[yscale=-1]
			\draw[thick,blue](-1.5,4.5*\x)--(5*\x,4.5*\x);
			\draw[thick,blue](-1.5,1.5*\x)--(4*\x,1.5*\x);
			\draw[thick,blue](-1.5,-1.5*\x)--(3*\x,-1.5*\x);
			\draw[thick,blue](-1.5,-4.5*\x)--(3*\x,-4.5*\x);
			\draw[thick,blue](3*\x,-4.5*\x)--(3*\x,-1.5*\x);\draw[thick,blue](3*\x,-3*\x)--(4*\x,-3*\x);
			\draw[thick,blue](4*\x,-3*\x)--(4*\x,1.5*\x);\draw[thick,blue](4*\x,-.75*\x)--(5*\x,-.75*\x);
			\draw[thick,blue](5*\x,-.75*\x)--(5*\x,4.5*\x);\draw[thick,blue](5*\x,1.875*\x)--(6*\x,1.875*\x);
			\filldraw[ten](3*\x,-3*\x) circle (\x/2);
			\filldraw[ten](4*\x,-.75*\x) circle (\x/2);
			\filldraw[ten](5*\x,1.875*\x) circle (\x/2);
			\end{scope}
			\draw (0,6*\x)--(0,-6*\x);
			\filldraw[tenred,shift={(0,4.5*\x)}] (-\x,-\x)--(\x,-\x)--(\x,\x)--(-\x,\x)--cycle;\node at(0,4.5*\x) {\textcolor{black}{\footnotesize$g_3$}};
			\filldraw[tenred,shift={(0,1.5*\x)}] (-\x,-\x)--(\x,-\x)--(\x,\x)--(-\x,\x)--cycle;\node at(0,1.5*\x) {\textcolor{black}{\footnotesize$g_2$}};
			\filldraw[tenred,shift={(0,-1.5*\x)}] (-\x,-\x)--(\x,-\x)--(\x,\x)--(-\x,\x)--cycle;\node at(0,-1.5*\x) {\textcolor{black}{\footnotesize$g_1$}};
			\filldraw[tenred,shift={(0,-4.5*\x)}] (-\x,-\x)--(\x,-\x)--(\x,\x)--(-\x,\x)--cycle;\node at(0,-4.5*\x) {\textcolor{black}{\footnotesize$g_0$}};
			\end{tikzpicture}}
	\end{array},
\end{align}
	implying that
		\begin{equation}
			\frac{\phi(g_0,g_1,g_2)\phi(g_0,g_1g_2,g_3)\phi(g_1,g_2,g_3)}{\phi(g_0g_1,g_2,g_3)\phi(g_0,g_1,g_2g_3)} = 1. \label{cocycle}
		\end{equation}
		This condition is known as the 3-cocycle conditions and identifies $\phi$ as a 3-cocycle. As mentioned above $X(g_0,g_1)$ is only defined up to a complex phase $\beta(g_0,g_1)$. This freedom can change the $\phi$, giving the equivalence relation
		\begin{equation}
			\phi'(g_0,g_1,g_2) = \phi(g_0,g_1,g_2)\frac{\beta(g_1,g_2)\beta(g_0,g_1g_2)}{\beta(g_0,g_1)\beta(g_0g_1,g_2)},
		\end{equation}
		so $\phi$ is only defined up to a 3-coboundary. For this reason the single block MPO group representation is endowed with the label $[\phi]$ from the third cohomology group $\coho{3}$. One can check that multiplying any larger number of MPOs does not give additional conditions/equivalences on $\phi$.

One can use a similar argument to demonstrate that no injective MPS can possess an anomalous symmetry. Assume an injective MPS with tensor $A$ is symmetric under an MPO symmetry for all lengths, similar reasoning that lead to Eqn.~\ref{mporeduction} implies the existence of another reduction tensor $Y(g)$ satisfying
\begin{align}
\begin{array}{c}
\includeTikz{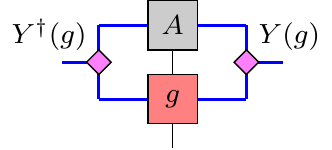}{
	\begin{tikzpicture}[scale=.25]
	\def\x{1}
	\draw[thick,blue,shift={(0,1.5*\x)}](-3*\x,0)--(3*\x,0);
	\draw[thick,blue,shift={(0,-1.5*\x)}](-3*\x,0)--(3*\x,0);
	\draw[thick,blue] (3*\x,1.5*\x)--(3*\x,-1.5*\x);
	\draw[thick,blue] (-3*\x,1.5*\x)--(-3*\x,-1.5*\x);
	\draw[thick,blue] (3*\x,0)--(4.5*\x,0);
	\draw[thick,blue] (-3*\x,0)--(-4.5*\x,0);
	\draw (0,2.5*\x)--(0,-3.5*\x);
	\filldraw[tengrey,shift={(0,1.5*\x)}] (-\x,-\x)--(\x,-\x)--(\x,\x)--(-\x,\x)--cycle;\node at(0,1.5*\x) {\textcolor{black}{\footnotesize$A$}};
	\filldraw[tenred,shift={(0,-1.5*\x)}] (-\x,-\x)--(\x,-\x)--(\x,\x)--(-\x,\x)--cycle;\node at(0,-1.5*\x) {\textcolor{black}{\footnotesize$g$}};
\node[above right] at (3*\x,0) {\textcolor{black}{\footnotesize$Y(g)$}};
\filldraw[tenpurp,shift={(3*\x,0)}](0,-\x/2)--(-\x/2,0)--(0,\x/2)--(\x/2,0)--cycle;
\filldraw[tenpurp,shift={(-3*\x,0)}](0,-\x/2)--(-\x/2,0)--(0,\x/2)--(\x/2,0)--cycle;
	\node[above left] at(-3*\x,0) {\textcolor{black}{\footnotesize$Y^\dagger(g)$}};
	\end{tikzpicture}}
\end{array}
&=
\begin{array}{c}
\includeTikz{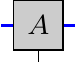}{
	\begin{tikzpicture}[scale=.25]
	\def\x{1}
	\draw[thick,blue,shift={(0,0)}](-1.5*\x,0)--(1.5*\x,0);
	\draw (0,.5*\x)--(0,-1.5*\x);
	\filldraw[tengrey,shift={(0,0)}] (-\x,-\x)--(\x,-\x)--(\x,\x)--(-\x,\x)--cycle;\node at(0,0) {\textcolor{black}{\footnotesize$A$}};
	\end{tikzpicture}}
\end{array}.
\end{align}

Similar to Eqn.~\ref{associator} we find that acting with multiple group elements leads to a complex phase $\beta(g_0,g_1)$
\begin{align}
\begin{array}{c}
\includeTikz{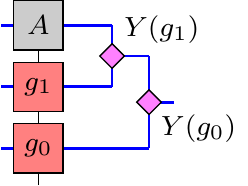}{
	\begin{tikzpicture}[scale=.25]
	\begin{scope}[yscale=-1]
	\def\x{1}
	\draw[thick,blue,shift={(0,2.5*\x)}](-1.5*\x,0)--(4.5*\x,0);
	\draw[thick,blue,shift={(0,0)}](-1.5*\x,0)--(3*\x,0);
	\draw[thick,blue,shift={(0,-2.5*\x)}](-1.5*\x,0)--(3*\x,0);
	\draw[thick,blue] (3*\x,0)--(3*\x,-2.5*\x);
	\draw[thick,blue] (3*\x,-1.25*\x)--(4.5*\x,-1.25*\x);
	\draw[thick,blue] (4.5*\x,-1.25*\x)--(4.5*\x,2.5*\x);
	\draw[thick,blue] (4.5*\x,0.625*\x)--(5.5*\x,0.625*\x);
	\draw (0,4*\x)--(0,-3*\x);
	\filldraw[tenred,shift={(0,2.5*\x)}] (-\x,-\x)--(\x,-\x)--(\x,\x)--(-\x,\x)--cycle;\node at(0,2.5*\x) {\textcolor{black}{\footnotesize$g_0$}};
	\filldraw[tenred,shift={(0,0)}] (-\x,-\x)--(\x,-\x)--(\x,\x)--(-\x,\x)--cycle;\node at(0,0) {\textcolor{black}{\footnotesize$g_1$}};
	\filldraw[tengrey,shift={(0,-2.5*\x)}] (-\x,-\x)--(\x,-\x)--(\x,\x)--(-\x,\x)--cycle;\node at(0,-2.5*\x) {\textcolor{black}{\footnotesize$A$}};
	\filldraw[tenpurp,shift={(3*\x,-1.25*\x)}](0,-\x/2)--(-\x/2,0)--(0,\x/2)--(\x/2,0)--cycle;
	\node[above right] at (3*\x,-1.25*\x) {\textcolor{black}{\footnotesize$Y(g_1)$}};
	\filldraw[tenpurp,shift={(4.5*\x,0.625*\x)}](0,-\x/2)--(-\x/2,0)--(0,\x/2)--(\x/2,0)--cycle;
	\node[below right] at(4.5*\x,0.625*\x) {\textcolor{black}{\footnotesize$Y(g_0)$}};
	\end{scope}
	\end{tikzpicture}}
\end{array}
&=\beta(g_0,g_1)
\begin{array}{c}
\includeTikz{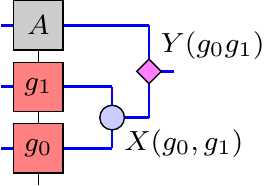}{
	\begin{tikzpicture}[scale=.25]
	\def\x{1}
	\draw[thick,blue,shift={(0,2.5*\x)}](-1.5*\x,0)--(4.5*\x,0);
	\draw[thick,blue,shift={(0,0)}](-1.5*\x,0)--(3*\x,0);
	\draw[thick,blue,shift={(0,-2.5*\x)}](-1.5*\x,0)--(3*\x,0);
	\draw[thick,blue] (3*\x,0)--(3*\x,-2.5*\x);
	\draw[thick,blue] (3*\x,-1.25*\x)--(4.5*\x,-1.25*\x);
	\draw[thick,blue] (4.5*\x,-1.25*\x)--(4.5*\x,2.5*\x);
	\draw[thick,blue] (4.5*\x,0.625*\x)--(5.5*\x,0.625*\x);
	\draw (0,3*\x)--(0,-4*\x);
	\filldraw[tengrey,shift={(0,2.5*\x)}] (-\x,-\x)--(\x,-\x)--(\x,\x)--(-\x,\x)--cycle;\node at(0,2.5*\x) {\textcolor{black}{\footnotesize$A$}};
	\filldraw[tenred,shift={(0,0)}] (-\x,-\x)--(\x,-\x)--(\x,\x)--(-\x,\x)--cycle;\node at(0,0) {\textcolor{black}{\footnotesize$g_1$}};
	\filldraw[tenred,shift={(0,-2.5*\x)}] (-\x,-\x)--(\x,-\x)--(\x,\x)--(-\x,\x)--cycle;\node at(0,-2.5*\x) {\textcolor{black}{\footnotesize$g_0$}};
	\filldraw[ten](3*\x,-1.25*\x) circle (\x/2);
	\node[below right] at (3*\x,-1.25*\x) {\textcolor{black}{\footnotesize$X(g_0,g_1)$}};
\node[above right] at(4.5*\x,0.625*\x) {\textcolor{black}{\footnotesize$Y(g_0g_1)$}};
\filldraw[tenpurp,shift={(4.5*\x,0.625*\x)}](0,-\x/2)--(-\x/2,0)--(0,\x/2)--(\x/2,0)--cycle;
	\end{tikzpicture}}
\end{array}.
\end{align}	
	
We now consider the application of three group elements
		
\begin{align}
	\begin{array}{c}
	\includeTikz{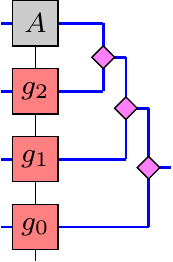}{
		\begin{tikzpicture}[scale=.23]
		\def\x{1}
		\begin{scope}[yscale=-1]
		\draw[thick,blue](-1.5,4.5*\x)--(5*\x,4.5*\x);
		\draw[thick,blue](-1.5,1.5*\x)--(4*\x,1.5*\x);
		\draw[thick,blue](-1.5,-1.5*\x)--(3*\x,-1.5*\x);
		\draw[thick,blue](-1.5,-4.5*\x)--(3*\x,-4.5*\x);
		\draw[thick,blue](3*\x,-4.5*\x)--(3*\x,-1.5*\x);\draw[thick,blue](3*\x,-3*\x)--(4*\x,-3*\x);
		\draw[thick,blue](4*\x,-3*\x)--(4*\x,1.5*\x);\draw[thick,blue](4*\x,-.75*\x)--(5*\x,-.75*\x);
		\draw[thick,blue](5*\x,-.75*\x)--(5*\x,4.5*\x);\draw[thick,blue](5*\x,1.875*\x)--(6*\x,1.875*\x);
\filldraw[tenpurp,shift={(3*\x,-3*\x)}](0,-\x/2)--(-\x/2,0)--(0,\x/2)--(\x/2,0)--cycle;
\filldraw[tenpurp,shift={(4*\x,-.75*\x)}](0,-\x/2)--(-\x/2,0)--(0,\x/2)--(\x/2,0)--cycle;
\filldraw[tenpurp,shift={(5*\x,1.875*\x)}](0,-\x/2)--(-\x/2,0)--(0,\x/2)--(\x/2,0)--cycle;
		\end{scope}
		\draw (0,5*\x)--(0,-6*\x);
		\filldraw[tengrey,shift={(0,4.5*\x)}] (-\x,-\x)--(\x,-\x)--(\x,\x)--(-\x,\x)--cycle;\node at(0,4.5*\x) {\textcolor{black}{\footnotesize$A$}};
		\filldraw[tenred,shift={(0,1.5*\x)}] (-\x,-\x)--(\x,-\x)--(\x,\x)--(-\x,\x)--cycle;\node at(0,1.5*\x) {\textcolor{black}{\footnotesize$g_2$}};
		\filldraw[tenred,shift={(0,-1.5*\x)}] (-\x,-\x)--(\x,-\x)--(\x,\x)--(-\x,\x)--cycle;\node at(0,-1.5*\x) {\textcolor{black}{\footnotesize$g_1$}};
		\filldraw[tenred,shift={(0,-4.5*\x)}] (-\x,-\x)--(\x,-\x)--(\x,\x)--(-\x,\x)--cycle;\node at(0,-4.5*\x) {\textcolor{black}{\footnotesize$g_0$}};
		\end{tikzpicture}}
	\end{array}
	&=\beta(g_1,g_2)
	\begin{array}{c}
	\includeTikz{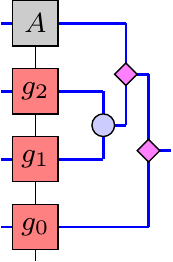}{
		\begin{tikzpicture}[scale=.23]
		\def\x{1}
		\begin{scope}
		\draw[thick,blue](-1.5,4.5*\x)--(4*\x,4.5*\x);
		\draw[thick,blue](-1.5,1.5*\x)--(3*\x,1.5*\x);
		\draw[thick,blue](-1.5,-1.5*\x)--(3*\x,-1.5*\x);
		\draw[thick,blue](-1.5,-4.5*\x)--(5*\x,-4.5*\x);
		\draw[thick,blue](3*\x,-1.5*\x)--(3*\x,1.5*\x);\draw[thick,blue](3*\x,0*\x)--(4*\x,0*\x);
		\draw[thick,blue](4*\x,4.5*\x)--(4*\x,0*\x);\draw[thick,blue](4*\x,2.25*\x)--(5*\x,2.25*\x);
		\draw[thick,blue](5*\x,2.25*\x)--(5*\x,-4.5*\x);\draw[thick,blue](5*\x,-1.125*\x)--(6*\x,-1.125*\x);
		\filldraw[ten](3*\x,0*\x) circle (\x/2);
\filldraw[tenpurp,shift={(4*\x,2.25*\x)}](0,-\x/2)--(-\x/2,0)--(0,\x/2)--(\x/2,0)--cycle;
\filldraw[tenpurp,shift={(5*\x,-1.125*\x)}](0,-\x/2)--(-\x/2,0)--(0,\x/2)--(\x/2,0)--cycle;
		\end{scope}
		\draw (0,5*\x)--(0,-6*\x);
		\filldraw[tengrey,shift={(0,4.5*\x)}] (-\x,-\x)--(\x,-\x)--(\x,\x)--(-\x,\x)--cycle;\node at(0,4.5*\x) {\textcolor{black}{\footnotesize$A$}};
		\filldraw[tenred,shift={(0,1.5*\x)}] (-\x,-\x)--(\x,-\x)--(\x,\x)--(-\x,\x)--cycle;\node at(0,1.5*\x) {\textcolor{black}{\footnotesize$g_2$}};
		\filldraw[tenred,shift={(0,-1.5*\x)}] (-\x,-\x)--(\x,-\x)--(\x,\x)--(-\x,\x)--cycle;\node at(0,-1.5*\x) {\textcolor{black}{\footnotesize$g_1$}};
		\filldraw[tenred,shift={(0,-4.5*\x)}] (-\x,-\x)--(\x,-\x)--(\x,\x)--(-\x,\x)--cycle;\node at(0,-4.5*\x) {\textcolor{black}{\footnotesize$g_0$}};
		\end{tikzpicture}}
	\end{array}	
	=\beta(g_1,g_2)\beta(g_0,g_1g_2)
	\begin{array}{c}
	\includeTikz{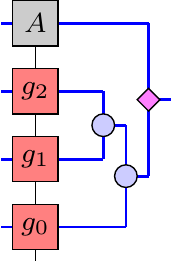}{
		\begin{tikzpicture}[scale=.23]
		\def\x{1}
		\begin{scope}
		\draw[thick,blue](-1.5,4.5*\x)--(5*\x,4.5*\x);
		\draw[thick,blue](-1.5,1.5*\x)--(3*\x,1.5*\x);
		\draw[thick,blue](-1.5,-1.5*\x)--(3*\x,-1.5*\x);
		\draw[thick,blue](-1.5,-4.5*\x)--(4*\x,-4.5*\x);
		\draw[thick,blue](3*\x,-1.5*\x)--(3*\x,1.5*\x);\draw[thick,blue](3*\x,0*\x)--(4*\x,0*\x);
		\draw[thick,blue](4*\x,-4.5*\x)--(4*\x,0*\x);\draw[thick,blue](4*\x,-2.25*\x)--(5*\x,-2.25*\x);
		\draw[thick,blue](5*\x,-2.25*\x)--(5*\x,4.5*\x);\draw[thick,blue](5*\x,1.125*\x)--(6*\x,1.125*\x);
		\filldraw[ten](3*\x,0*\x) circle (\x/2);
		\filldraw[ten](4*\x,-2.25*\x) circle (\x/2);
\filldraw[tenpurp,shift={(5*\x,1.125*\x)}](0,-\x/2)--(-\x/2,0)--(0,\x/2)--(\x/2,0)--cycle;
		\end{scope}
		\draw (0,5*\x)--(0,-6*\x);
		\filldraw[tengrey,shift={(0,4.5*\x)}] (-\x,-\x)--(\x,-\x)--(\x,\x)--(-\x,\x)--cycle;\node at(0,4.5*\x) {\textcolor{black}{\footnotesize$A$}};
		\filldraw[tenred,shift={(0,1.5*\x)}] (-\x,-\x)--(\x,-\x)--(\x,\x)--(-\x,\x)--cycle;\node at(0,1.5*\x) {\textcolor{black}{\footnotesize$g_2$}};
		\filldraw[tenred,shift={(0,-1.5*\x)}] (-\x,-\x)--(\x,-\x)--(\x,\x)--(-\x,\x)--cycle;\node at(0,-1.5*\x) {\textcolor{black}{\footnotesize$g_1$}};
		\filldraw[tenred,shift={(0,-4.5*\x)}] (-\x,-\x)--(\x,-\x)--(\x,\x)--(-\x,\x)--cycle;\node at(0,-4.5*\x) {\textcolor{black}{\footnotesize$g_0$}};
		\end{tikzpicture}}
	\end{array}	\nonumber\\
	&=\beta(g_1,g_2)\beta(g_0,g_1g_2)\phi(g_0,g_1,g_2)
	\begin{array}{c}
	\includeTikz{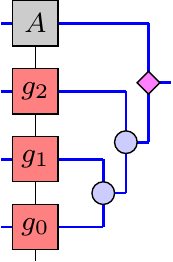}{
		\begin{tikzpicture}[scale=.23]
		\def\x{1}
		\begin{scope}
		\draw[thick,blue](-1.5,4.5*\x)--(5*\x,4.5*\x);
		\draw[thick,blue](-1.5,1.5*\x)--(4*\x,1.5*\x);
		\draw[thick,blue](-1.5,-1.5*\x)--(3*\x,-1.5*\x);
		\draw[thick,blue](-1.5,-4.5*\x)--(3*\x,-4.5*\x);
		\draw[thick,blue](3*\x,-4.5*\x)--(3*\x,-1.5*\x);\draw[thick,blue](3*\x,-3*\x)--(4*\x,-3*\x);
		\draw[thick,blue](4*\x,-3*\x)--(4*\x,1.5*\x);\draw[thick,blue](4*\x,-.75*\x)--(5*\x,-.75*\x);
		\draw[thick,blue](5*\x,-.75*\x)--(5*\x,4.5*\x);\draw[thick,blue](5*\x,1.875*\x)--(6*\x,1.875*\x);
		\filldraw[ten](3*\x,-3*\x) circle (\x/2);
		\filldraw[ten](4*\x,-.75*\x) circle (\x/2);
		\filldraw[tenpurp,shift={(5*\x,1.875*\x)}](0,-\x/2)--(-\x/2,0)--(0,\x/2)--(\x/2,0)--cycle;
		\end{scope}
		\draw (0,5*\x)--(0,-6*\x);
		\filldraw[tengrey,shift={(0,4.5*\x)}] (-\x,-\x)--(\x,-\x)--(\x,\x)--(-\x,\x)--cycle;\node at(0,4.5*\x) {\textcolor{black}{\footnotesize$A$}};
		\filldraw[tenred,shift={(0,1.5*\x)}] (-\x,-\x)--(\x,-\x)--(\x,\x)--(-\x,\x)--cycle;\node at(0,1.5*\x) {\textcolor{black}{\footnotesize$g_2$}};
		\filldraw[tenred,shift={(0,-1.5*\x)}] (-\x,-\x)--(\x,-\x)--(\x,\x)--(-\x,\x)--cycle;\node at(0,-1.5*\x) {\textcolor{black}{\footnotesize$g_1$}};
		\filldraw[tenred,shift={(0,-4.5*\x)}] (-\x,-\x)--(\x,-\x)--(\x,\x)--(-\x,\x)--cycle;\node at(0,-4.5*\x) {\textcolor{black}{\footnotesize$g_0$}};
		\end{tikzpicture}}
	\end{array}
	=\frac{\beta(g_1,g_2)\beta(g_0,g_1g_2)\phi(g_0,g_1,g_2)}{\beta(g_0g_1,g_2)}
	\begin{array}{c}
	\includeTikz{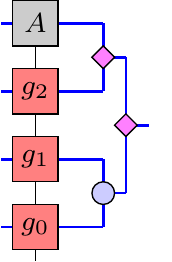}{
		\begin{tikzpicture}[scale=.23]
		\def\x{1}
		\begin{scope}
		\draw[thick,blue](-1.5,4.5*\x)--(3*\x,4.5*\x);
		\draw[thick,blue](-1.5,1.5*\x)--(3*\x,1.5*\x);
		\draw[thick,blue](-1.5,-1.5*\x)--(3*\x,-1.5*\x);
		\draw[thick,blue](-1.5,-4.5*\x)--(3*\x,-4.5*\x);
		\draw[thick,blue](3*\x,-4.5*\x)--(3*\x,-1.5*\x);\draw[thick,blue](3*\x,3*\x)--(4*\x,3*\x);
		\draw[thick,blue](3*\x,4.5*\x)--(3*\x,1.5*\x);\draw[thick,blue](3*\x,-3*\x)--(4*\x,-3*\x);
		\draw[thick,blue](4*\x,-3*\x)--(4*\x,3*\x);\draw[thick,blue](4*\x,0*\x)--(5*\x,0*\x);
		\filldraw[ten](3*\x,-3*\x) circle (\x/2);
	\filldraw[tenpurp,shift={(3*\x,3*\x)}](0,-\x/2)--(-\x/2,0)--(0,\x/2)--(\x/2,0)--cycle;
		\filldraw[tenpurp,shift={(4*\x,0*\x)}](0,-\x/2)--(-\x/2,0)--(0,\x/2)--(\x/2,0)--cycle;
		\draw[draw=none](5*\x,0)--(6*\x,0);
		\end{scope}
		\draw (0,5*\x)--(0,-6*\x);
		\filldraw[tengrey,shift={(0,4.5*\x)}] (-\x,-\x)--(\x,-\x)--(\x,\x)--(-\x,\x)--cycle;\node at(0,4.5*\x) {\textcolor{black}{\footnotesize$A$}};
		\filldraw[tenred,shift={(0,1.5*\x)}] (-\x,-\x)--(\x,-\x)--(\x,\x)--(-\x,\x)--cycle;\node at(0,1.5*\x) {\textcolor{black}{\footnotesize$g_2$}};
		\filldraw[tenred,shift={(0,-1.5*\x)}] (-\x,-\x)--(\x,-\x)--(\x,\x)--(-\x,\x)--cycle;\node at(0,-1.5*\x) {\textcolor{black}{\footnotesize$g_1$}};
		\filldraw[tenred,shift={(0,-4.5*\x)}] (-\x,-\x)--(\x,-\x)--(\x,\x)--(-\x,\x)--cycle;\node at(0,-4.5*\x) {\textcolor{black}{\footnotesize$g_0$}};
		\end{tikzpicture}}
	\end{array}\nonumber\\
	&=\frac{\beta(g_1,g_2)\beta(g_0,g_1g_2)\phi(g_0,g_1,g_2)}{\beta(g_0g_1,g_2)\beta(g_0,g_1)}
	\begin{array}{c}
	\includeTikz{reduceMPSE}{}
	\end{array},	
\end{align}
which leads to a consistency equation
\begin{align}
	\phi(g_0,g_1,g_2)=\frac{\beta(g_0g_1,g_2)\beta(g_0,g_1)}{\beta(g_1,g_2)\beta(g_0,g_1g_2)},
\end{align}
implying $\phi$ is a coboundary. Therefore $\phi\sim 1$, is in the trivial cohomology class. 
Hence no injective MPS can be symmetric under an anomalous MPO symmetry. This leaves open the possibility of a non-injective MPS, describing a state which spontaneously breaks the symmetry. Alternatively a symmetric state may be gapless and hence have no MPS description (with a fixed bond dimension). 

\clearpage
\section{Ansatz for MERA tensors with type-III $\zN^3$ symmetry}
\label{appendix:generalansatz}

In this appendix, we describe an ansatz for the tensors in a MERA with type-III $\zN^3$ symmetry. Let $\G=\zN^3$, with action as defined in Eqn.~\ref{eqn:symmetrygroup}{. Let $T$ be an isometric tensor with $2A$ upper indices and $2B$ ($B\geq A$) lower indices
\begin{align}
T&:(\mathbb{C}^N)^{\otimes2A}\to(\mathbb{C}^N)^{\otimes2B},\\
T^\dagger T&=\openone_N^{\otimes 2A}.
\end{align}
Define the decoupling circuit on $2K$ indices as
\begin{align}
\mathcal{D}_{2K}&=\prod\limits_{j=1}^{K-1}CX_{1,2j+1}CX_{2K,2j}.
\end{align}
Allowed MERA tensors are those given by
\begin{align}
T&=\mathcal{D}_{2B}^\dagger \left(\openone_N\otimes t\otimes\openone_N\right) \mathcal{D}_{2A},
\end{align}
where
\begin{align}
t&:(\mathbb{C}^N)^{\otimes2(A-1)}\to(\mathbb{C}^N)^{\otimes2(B-1)},\\
t^\dagger t&=\openone_N^{\otimes 2(A-1)}.
\end{align}

The $X$ portion of the symmetry is automatically enforced by this circuit. To enforce the $CZ$ part, one must ensure that
\begin{align}
\left(\prod_{j=1}^{B-1}CZ^\dagger_{2j-1,2j}\right)\left(\prod_{j=1}^{B-2}CZ_{2j,2j+1}\right) t &=t\left(\prod_{j=1}^{A-1}CZ^\dagger_{2j-1,2j}\right)\left(\prod_{j=1}^{A-2}CZ_{2j,2j+1}\right).
\end{align}

\subsection{4:2 MERA}
For clarity, we now include the form of the constraint on the 4:2 MERA (introduced in Fig.~\ref{Fig:fourtotwoMERA}) with bond dimension $N$, $N^2$ and $N^3$:

\begin{subequations}
	\begin{alignat}{2}
\begin{array}{c}
    \includeTikz{FourToTwoMERATensor}{}
  \end{array}
&=
 \begin{array}{c}
    \includeTikz{FourtoTwoMERADecomp}{}
  \end{array},\\
\begin{array}{c}
    \includeTikz{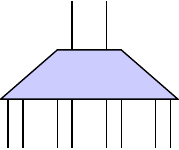}{
    \begin{tikzpicture}[scale=1]
      \def\x{.5}
      \def\dx{.15}
      \def\xi{.25}
      \def\xj{.75}
      \def\ymin{.5}
      \def\ymax{1}
      \filldraw[ten](\xi-\dx/2,0)--(3*\x+\xi+1.5*\dx,0)--(2*\x+\xi+1*\dx,.5)--(1*\x+\xi,.5)--cycle;
      \foreach \i in {0,...,3}{
      \draw(\i*\x+0*\dx+\xi,0)--(\i*\x+0*\dx+\xi,-\ymin);
      \draw(\i*\x+1*\dx+\xi,0)--(\i*\x+1*\dx+\xi,-\ymin);
      }
     \draw (\x+\xi+\dx,.5)--(\x+\xi+\dx,\ymax);
     \draw (2*\x+\xi,.5)--(2*\x+\xi,\ymax);
    \end{tikzpicture}
    }
  \end{array}
&=
 \begin{array}{c}
    \includeTikz{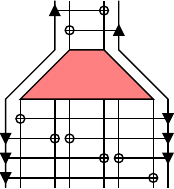}{
    \begin{tikzpicture}[scale=1]
      \def\x{.5}
      \def\dx{.15}
      \def\xi{.25}
      \def\xj{.75}
      \def\ymin{.9}
      \def\ymax{1}
      \def\dn{.2}
      \filldraw[tenred](\xi+\dx,0)--(3*\x+\xi,0)--(2*\x+\xi,.5)--(1*\x+\xi+1*\dx,.5)--cycle;
      \foreach \i in {0,...,3}{
      \draw(\i*\x+0*\dx+\xi,0)--(\i*\x+0*\dx+\xi,-\ymin);
      \draw(\i*\x+1*\dx+\xi,0)--(\i*\x+1*\dx+\xi,-\ymin);
      }
     \draw (\xi,-\ymin)--(\xi,0)--(\x+\xi,.5)--(\x+\xi,\ymax);
     \draw (\x+\xi+\dx,.5)--(\x+\xi+\dx,\ymax);
     \draw (2*\x+\xi,.5)--(2*\x+\xi,\ymax);
     \draw (3*\x+\xi+\dx,-\ymin)--(3*\x+\xi+\dx,0)--(2*\x+\xi+\dx,.5)--(2*\x+\xi+\dx,\ymax);
     \begin{scope}[shift={(0,.7)}]
     \def\n{0};
     \filldraw[black,shift={(2*\x+\xi+1*\dx,\dn*\n)}](-.05,-.05)--(.05,-.05)--(0,.05)--cycle;
     \draw (1*\x+\xi+.7*\dx,\dn*\n)--(2*\x+1*\dx+\xi,\dn*\n);
     \draw (1*\x+\dx+\xi,\dn*\n) circle (.3*\dx);
      \def\n{1};
      \filldraw[black,shift={(\x+\xi+0*\dx,\dn*\n)}](-.05,-.05)--(.05,-.05)--(0,.05)--cycle;
      \draw (1*\x+\xi,\dn*\n)--(2*\x+0.3*\dx+\xi,\dn*\n);
      \draw (2*\x+0*\dx+\xi,\dn*\n) circle (.3*\dx);
     \end{scope}
      \begin{scope}[shift={(0,-.2)}]
	      \def\n{0};
	      \filldraw[black,shift={(3*\x+\xi+1*\dx,-\dn*\n)},yscale=-1](-.05,-.05)--(.05,-.05)--(0,.05)--cycle;
	      \draw (0*\x+\xi+.7*\dx,-\dn*\n)--(3*\x+1*\dx+\xi,-\dn*\n);
	      \draw (0*\x+\dx+\xi,-\dn*\n) circle (.3*\dx);
	      \begin{scope}
		      \def\n{1};
		      \filldraw[black,shift={(\xi+0*\dx,-\dn*\n)},yscale=-1](-.05,-.05)--(.05,-.05)--(0,.05)--cycle;
		      \draw (\xi,-\dn*\n)--(\x+0*\dx+\xi+.3*\dx,-\dn*\n);
		      \draw (\x+0*\dx+\xi,-\dn*\n) circle (.3*\dx);
		      \filldraw[black,shift={(3*\x+\xi+1*\dx,-\dn*\n)},yscale=-1](-.05,-.05)--(.05,-.05)--(0,.05)--cycle;
		      \draw (\x+\xi+.7*\dx,-\dn*\n)--(3*\x+1*\dx+\xi,-\dn*\n);
		      \draw (\x+\xi+\dx,-\dn*\n) circle (.3*\dx);
	      \end{scope}
	       \begin{scope}
	       	      \def\n{2};
	       	      \filldraw[black,shift={(\xi+0*\dx,-\dn*\n)},yscale=-1](-.05,-.05)--(.05,-.05)--(0,.05)--cycle;
	       	      \draw (\xi,-\dn*\n)--(2*\x+0*\dx+\xi+.3*\dx,-\dn*\n);
	       	      \draw (2*\x+0*\dx+\xi,-\dn*\n) circle (.3*\dx);
	       	      \filldraw[black,shift={(3*\x+\xi+1*\dx,-\dn*\n)},yscale=-1](-.05,-.05)--(.05,-.05)--(0,.05)--cycle;
	       	      \draw (2*\x+\xi+.7*\dx,-\dn*\n)--(3*\x+1*\dx+\xi,-\dn*\n);
	       	      \draw (2*\x+\dx+\xi,-\dn*\n) circle (.3*\dx);
	       \end{scope}
	       \def\n{3};
	       \filldraw[black,shift={(\xi+0*\dx,-\dn*\n)},yscale=-1](-.05,-.05)--(.05,-.05)--(0,.05)--cycle;
	       \draw (\xi,-\dn*\n)--(3*\x+\xi+.3*\dx,-\dn*\n);
	       \draw (3*\x+0*\dx+\xi,-\dn*\n) circle (.3*\dx);
       \end{scope}
    \end{tikzpicture}
    }
  \end{array},\\
  \begin{array}{c}
\includeTikz{TwelvetoSixMERATensor}{}
  \end{array}
  &=
    \begin{array}{c}
  \includeTikz{TwelvetoSixMERADecomp}{}
    \end{array}.
\end{alignat}
\end{subequations}

\subsection{Ternary MERA}

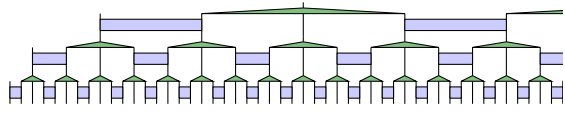
\begin{figure}[h!]
	\includeTikz{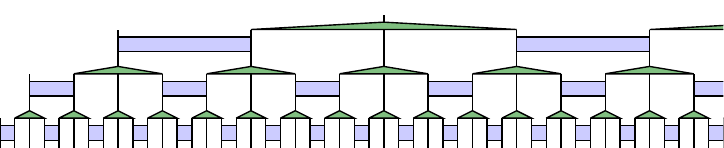}{\begin{tikzpicture}[scale=.15]
		\begin{scope}
		\clip (0,0) rectangle (49,10);
		\def\y{0};\def\s{1};
		\foreach \x in {0,1,...,80}{
			\draw[shift={(\s*\x,\y)}] (0,0)--(0,2);
		};
		\foreach \x in {0,3,...,79}{
			\filldraw[u,shift={(\s*\x,\y)}](0,.5)--(\s,.5)--(\s,1.5)--(0,1.5)--(0,.5);
		};
		\foreach \x in {0,3,...,79}{
			\filldraw[w,shift={(\s*\x,\y)}](1*\s,2)--(3*\s,2)--(2*\s,2.5)--(1*\s,2);
			\draw[shift={(\s*\x,\y)}](2*\s,2.5)--(2*\s,3);
		};
		\def\y{3};\def\s{3};
		\foreach \x in {0,1,...,20}{
			\draw[shift={(\s*\x+\s-1,\y)}] (0,0)--(0,2);
		};
		\foreach \x in {0,3,...,19}{
			\filldraw[u,shift={(\s*\x+\s-1,\y)}](0,.5)--(\s,.5)--(\s,1.5)--(0,1.5)--(0,.5);
		};
		\foreach \x in {0,3,...,18}{
			\filldraw[w,shift={(\s*\x+\s-1,\y)}](1*\s,2)--(3*\s,2)--(2*\s,2.5)--(1*\s,2);
			\draw[shift={(\s*\x+\s-1,\y)}](2*\s,2.5)--(2*\s,3);
		};
		\def\y{6};\def\s{9};
		\foreach \x in {0,1,...,10}{
			\draw[shift={(\s*\x+\s-1,\y)}] (0,0)--(0,2);
		};
		\foreach \x in {0,3,...,9}{
			\filldraw[u,shift={(\s*\x+\s-1,\y)}](0,.5)--(\s,.5)--(\s,1.5)--(0,1.5)--(0,.5);
		};
		\foreach \x in {0,3,...,9}{
			\filldraw[w,shift={(\s*\x+\s-1,\y)}](1*\s,2)--(3*\s,2)--(2*\s,2.5)--(1*\s,2);
			\draw[shift={(\s*\x+\s-1,\y)}](2*\s,2.5)--(2*\s,3);
		};
		\end{scope}
		\end{tikzpicture}}
	\caption{The ternary MERA represents a quantum state using two types of tensors; unitary `disentanglers' (rectangles) and isometric tensors (triangles).}\label{Fig:TernaryMERA_appendix}
\end{figure}

For completeness, we show how our ansatz is applied to the ternary MERA shown in Fig.~\ref{Fig:TernaryMERA_appendix}. The ternary ansatz is commonly seen in the literature due to its relatively low optimization cost. A ternary MERA is built from two kinds of tensors; unitary `disentanglers' $v$ (rectangles in \fref{TernaryMERA_appendix}) and isometric tensors $w$ (triangles in \fref{TernaryMERA_appendix}). In the general case, these tensors may all contain distinct coefficients, although symmetries such as scale invariance can be imposed by, for example, forcing the tensors on each layer to be identical.

For bond dimension $N^2$ and $N^4$, the constraint on the tensors is

\begin{subequations}
	\begin{alignat}{2}
	\begin{array}{c}
	\includeTikz{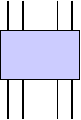}{
		\begin{tikzpicture}[scale=1]
		\def\x{.5}
		\def\dx{.15}
		\def\xi{.25}
		\def\xj{.75}
		\def\ymin{.4}
		\def\ymax{.8}
		\foreach \i in {0,...,1}{
			\draw(\i*\x+0*\dx+\xi,\ymax)--(\i*\x+0*\dx+\xi,-\ymin);
			\draw(\i*\x+1*\dx+\xi,\ymax)--(\i*\x+1*\dx+\xi,-\ymin);
		}
		\filldraw[ten](\xi-\dx/2,0)--(1*\x+\xi+1.5*\dx,0)--(1*\x+\xi+1.5*\dx,.5)--(\xi-\dx/2,.5)--cycle;
		\end{tikzpicture}
	}
	\end{array}
	&=
	\begin{array}{c}
	\includeTikz{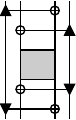}{
		\begin{tikzpicture}[scale=1]
		\def\x{.5}
		\def\dx{.15}
		\def\xi{.25}
		\def\xj{.75}
		\def\ymin{.4}
		\def\ymax{.8}
		\def\dn{.2}
		\filldraw[tengrey,yscale=1](0*\x+\xi+\dx,0)--(1*\x+\xi,0)--(1*\x+\xi,.3)--(0*\x+\xi+\dx,.3)--cycle;
		\foreach \i in {0,...,1}{
			\draw(\i*\x+0*\dx+\xi,0)--(\i*\x+0*\dx+\xi,-\ymin);
			\draw(\i*\x+1*\dx+\xi,0)--(\i*\x+1*\dx+\xi,-\ymin);
			\draw(\i*\x+0*\dx+\xi,0.3)--(\i*\x+0*\dx+\xi,\ymax);
			\draw(\i*\x+1*\dx+\xi,0.3)--(\i*\x+1*\dx+\xi,\ymax);
		}
		\draw (\xi,-\ymin)--(\xi,\ymax);
		\draw (\x+\xi+\dx,-\ymin)--(\x+\xi+\dx,\ymax);
		\begin{scope}[shift={(0,-.1)}]
		\def\n{0};
		\filldraw[black,shift={(1*\x+\xi+\dx,-\dn*\n)},yscale=-1](-.05,-.05)--(.05,-.05)--(0,.05)--cycle;
		\draw (0*\x+\xi+.7*\dx,-\dn*\n)--(1*\x+\xi+\dx,-\dn*\n);
		\draw (0*\x+\xi+\dx,-\dn*\n) circle (.3*\dx);
		\def\n{1};
		\filldraw[black,shift={(\xi+0*\dx,-\dn*\n)},yscale=-1](-.05,-.05)--(.05,-.05)--(0,.05)--cycle;
		\draw (\xi,-\dn*\n)--(1*\x+\xi+.3*\dx,-\dn*\n);
		\draw (1*\x+0*\dx+\xi,-\dn*\n) circle (.3*\dx);
		\end{scope}
		\begin{scope}[shift={(0,.5)},yscale=-1]
		\def\n{0};
		\filldraw[black,shift={(1*\x+\xi+\dx,-\dn*\n)},yscale=-1](-.05,-.05)--(.05,-.05)--(0,.05)--cycle;
		\draw (0*\x+\xi+.7*\dx,-\dn*\n)--(1*\x+\xi+\dx,-\dn*\n);
		\draw (0*\x+\xi+\dx,-\dn*\n) circle (.3*\dx);
		\def\n{1};
		\filldraw[black,shift={(\xi+0*\dx,-\dn*\n)},yscale=-1](-.05,-.05)--(.05,-.05)--(0,.05)--cycle;
		\draw (\xi,-\dn*\n)--(1*\x+\xi+.3*\dx,-\dn*\n);
		\draw (1*\x+0*\dx+\xi,-\dn*\n) circle (.3*\dx);
		\end{scope}
		\end{tikzpicture}
	}
	\end{array},&
	\begin{array}{c}
	\includeTikz{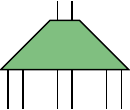}{
		\begin{tikzpicture}[scale=1]
		\def\x{.5}
		\def\dx{.15}
		\def\xi{.25}
		\def\xj{.75}
		\def\ymin{.4}
		\def\ymax{.7}
		\filldraw[tengreen](\xi-\dx/2,0)--(2*\x+\xi+1.5*\dx,0)--(1*\x+\xi+1.5*\dx,.5)--(1*\x+\xi-0.5*\dx,.5)--cycle;
		\foreach \i in {0,...,2}{
			\draw(\i*\x+0*\dx+\xi,0)--(\i*\x+0*\dx+\xi,-\ymin);
			\draw(\i*\x+1*\dx+\xi,0)--(\i*\x+1*\dx+\xi,-\ymin);
		}
		\draw (\x+\xi,.5)--(\x+\xi,\ymax);
		\draw (\x+\xi+\dx,.5)--(\x+\xi+\dx,\ymax);
		\end{tikzpicture}
	}
	\end{array}
	&=
	\begin{array}{c}
	\includeTikz{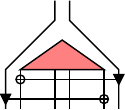}{
		\begin{tikzpicture}[scale=1]
		\def\x{.5}
		\def\dx{.15}
		\def\xi{.25}
		\def\xj{.75}
		\def\ymin{.4}
		\def\ymax{.7}
		\def\dn{.2}
		\filldraw[tenred,yscale=1](0*\x+\xi+\dx,0)--(2*\x+\xi,0)--(1*\x+\xi+\dx/2,.3)--cycle;
		\foreach \i in {0,...,2}{
			\draw(\i*\x+0*\dx+\xi,0)--(\i*\x+0*\dx+\xi,-\ymin);
			\draw(\i*\x+1*\dx+\xi,0)--(\i*\x+1*\dx+\xi,-\ymin);
		}
		\draw (\xi,-\ymin)--(\xi,0)--(\x+\xi,.5)--(\x+\xi,\ymax);
		\draw (2*\x+\xi+\dx,-\ymin)--(2*\x+\xi+\dx,0)--(1*\x+\xi+\dx,.5)--(1*\x+\xi+\dx,\ymax);
		\begin{scope}[shift={(0,-.1)}]
		\def\n{0};
		\filldraw[black,shift={(2*\x+\xi+\dx,-\dn*\n)},yscale=-1](-.05,-.05)--(.05,-.05)--(0,.05)--cycle;
		\draw (0*\x+\xi+.7*\dx,-\dn*\n)--(2*\x+\xi+\dx,-\dn*\n);
		\draw (0*\x+\xi+\dx,-\dn*\n) circle (.3*\dx);
		\def\n{1};
		\filldraw[black,shift={(\xi+0*\dx,-\dn*\n)},yscale=-1](-.05,-.05)--(.05,-.05)--(0,.05)--cycle;
		\draw (\xi,-\dn*\n)--(2*\x+\xi+.3*\dx,-\dn*\n);
		\draw (2*\x+0*\dx+\xi,-\dn*\n) circle (.3*\dx);
		\end{scope}
		\end{tikzpicture}
	}
	\end{array},\\
	\label{fourspindecomp}
	\begin{array}{c}
	\includeTikz{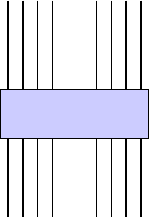}{
		\begin{tikzpicture}[scale=1]
		\def\x{.9}
		\def\dx{.15}
		\def\xi{.25}
		\def\xj{.75}
		\def\ymin{.8}
		\def\ymax{1.4}
		\foreach \i in {0,...,1}{
			\draw(\i*\x+0*\dx+\xi,\ymax)--(\i*\x+0*\dx+\xi,-\ymin);
			\draw(\i*\x+1*\dx+\xi,\ymax)--(\i*\x+1*\dx+\xi,-\ymin);
			\draw(\i*\x+2*\dx+\xi,\ymax)--(\i*\x+2*\dx+\xi,-\ymin);
			\draw(\i*\x+3*\dx+\xi,\ymax)--(\i*\x+3*\dx+\xi,-\ymin);
		}
		\filldraw[ten](\xi-\dx/2,0)--(1*\x+\xi+3.5*\dx,0)--(1*\x+\xi+3.5*\dx,.5)--(\xi-\dx/2,.5)--cycle;
		\end{tikzpicture}
	}
	\end{array}
	&=
	\begin{array}{c}
	\includeTikz{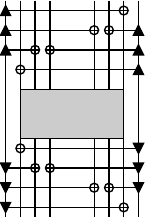}{
		\begin{tikzpicture}[scale=1]
		\def\x{.9}
		\def\dx{.15}
		\def\xi{.25}
		\def\xj{.75}
		\def\ymin{.8}
		\def\ymax{1.4}
		\def\dn{.2}
		\foreach \i in {0,...,1}{
			\draw(\i*\x+0*\dx+\xi,\ymax)--(\i*\x+0*\dx+\xi,-\ymin);
			\draw(\i*\x+1*\dx+\xi,\ymax)--(\i*\x+1*\dx+\xi,-\ymin);
			\draw(\i*\x+2*\dx+\xi,\ymax)--(\i*\x+2*\dx+\xi,-\ymin);
			\draw(\i*\x+3*\dx+\xi,\ymax)--(\i*\x+3*\dx+\xi,-\ymin);
		}
		\filldraw[tengrey](\xi+\dx,0)--(1*\x+\xi+2*\dx,0)--(1*\x+\xi+2*\dx,.5)--(\xi+\dx,.5)--cycle;
		\begin{scope}[shift={(0,.7)},yscale=-1]
		\def\n{0};
		\filldraw[black,shift={(1*\x+\xi+3*\dx,-\dn*\n)},yscale=-1](-.05,-.05)--(.05,-.05)--(0,.05)--cycle;
		\draw (0*\x+\xi+.7*\dx,-\dn*\n)--(1*\x+3*\dx+\xi,-\dn*\n);
		\draw (0*\x+\dx+\xi,-\dn*\n) circle (.3*\dx);
		\begin{scope}
		\def\n{1};
		\filldraw[black,shift={(\xi+0*\dx,-\dn*\n)},yscale=-1](-.05,-.05)--(.05,-.05)--(0,.05)--cycle;
		\draw (\xi,-\dn*\n)--(0*\x+2*\dx+\xi+.3*\dx,-\dn*\n);
		\draw (0*\x+2*\dx+\xi,-\dn*\n) circle (.3*\dx);
		\filldraw[black,shift={(1*\x+\xi+3*\dx,-\dn*\n)},yscale=-1](-.05,-.05)--(.05,-.05)--(0,.05)--cycle;
		\draw (0*\x+\xi+2.7*\dx,-\dn*\n)--(1*\x+3*\dx+\xi,-\dn*\n);
		\draw (0*\x+\xi+3*\dx,-\dn*\n) circle (.3*\dx);
		\end{scope}
		\begin{scope}
		\def\n{2};
		\filldraw[black,shift={(\xi+0*\dx,-\dn*\n)},yscale=-1](-.05,-.05)--(.05,-.05)--(0,.05)--cycle;
		\draw (\xi,-\dn*\n)--(1*\x+0*\dx+\xi+.3*\dx,-\dn*\n);
		\draw (1*\x+0*\dx+\xi,-\dn*\n) circle (.3*\dx);
		\filldraw[black,shift={(1*\x+\xi+3*\dx,-\dn*\n)},yscale=-1](-.05,-.05)--(.05,-.05)--(0,.05)--cycle;
		\draw (1*\x+\xi+0.7*\dx,-\dn*\n)--(1*\x+3*\dx+\xi,-\dn*\n);
		\draw (1*\x+1*\dx+\xi,-\dn*\n) circle (.3*\dx);
		\end{scope}
		\def\n{3};
		\filldraw[black,shift={(\xi+0*\dx,-\dn*\n)},yscale=-1](-.05,-.05)--(.05,-.05)--(0,.05)--cycle;
		\draw (\xi,-\dn*\n)--(1*\x+\xi+2.3*\dx,-\dn*\n);
		\draw (1*\x+2*\dx+\xi,-\dn*\n) circle (.3*\dx);
		\end{scope}
		\begin{scope}[shift={(0,-.1)}]
		\def\n{0};
		\filldraw[black,shift={(1*\x+\xi+3*\dx,-\dn*\n)},yscale=-1](-.05,-.05)--(.05,-.05)--(0,.05)--cycle;
		\draw (0*\x+\xi+.7*\dx,-\dn*\n)--(1*\x+3*\dx+\xi,-\dn*\n);
		\draw (0*\x+\dx+\xi,-\dn*\n) circle (.3*\dx);
		\begin{scope}
		\def\n{1};
		\filldraw[black,shift={(\xi+0*\dx,-\dn*\n)},yscale=-1](-.05,-.05)--(.05,-.05)--(0,.05)--cycle;
		\draw (\xi,-\dn*\n)--(0*\x+2*\dx+\xi+.3*\dx,-\dn*\n);
		\draw (0*\x+2*\dx+\xi,-\dn*\n) circle (.3*\dx);
		\filldraw[black,shift={(1*\x+\xi+3*\dx,-\dn*\n)},yscale=-1](-.05,-.05)--(.05,-.05)--(0,.05)--cycle;
		\draw (0*\x+\xi+2.7*\dx,-\dn*\n)--(1*\x+3*\dx+\xi,-\dn*\n);
		\draw (0*\x+\xi+3*\dx,-\dn*\n) circle (.3*\dx);
		\end{scope}
		\begin{scope}
		\def\n{2};
		\filldraw[black,shift={(\xi+0*\dx,-\dn*\n)},yscale=-1](-.05,-.05)--(.05,-.05)--(0,.05)--cycle;
		\draw (\xi,-\dn*\n)--(1*\x+0*\dx+\xi+.3*\dx,-\dn*\n);
		\draw (1*\x+0*\dx+\xi,-\dn*\n) circle (.3*\dx);
		\filldraw[black,shift={(1*\x+\xi+3*\dx,-\dn*\n)},yscale=-1](-.05,-.05)--(.05,-.05)--(0,.05)--cycle;
		\draw (1*\x+\xi+0.7*\dx,-\dn*\n)--(1*\x+3*\dx+\xi,-\dn*\n);
		\draw (1*\x+1*\dx+\xi,-\dn*\n) circle (.3*\dx);
		\end{scope}
		\def\n{3};
		\filldraw[black,shift={(\xi+0*\dx,-\dn*\n)},yscale=-1](-.05,-.05)--(.05,-.05)--(0,.05)--cycle;
		\draw (\xi,-\dn*\n)--(1*\x+\xi+2.3*\dx,-\dn*\n);
		\draw (1*\x+2*\dx+\xi,-\dn*\n) circle (.3*\dx);
		\end{scope}
		\end{tikzpicture}
	}
	\end{array},&\qquad\qquad
	\begin{array}{c}
	\includeTikz{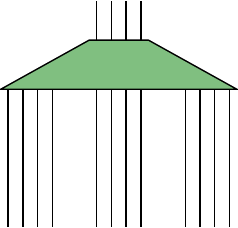}{
		\begin{tikzpicture}[scale=1]
		\def\x{.9}
		\def\dx{.15}
		\def\xi{.25}
		\def\xj{.75}
		\def\ymin{1.4}
		\def\ymax{.9}
		\filldraw[tengreen](\xi-\dx/2,0)--(2*\x+\xi+3.5*\dx,0)--(1*\x+\xi+3.5*\dx,.5)--(1*\x+\xi-.5*\dx,.5)--cycle;
		\foreach \i in {0,...,2}{
			\draw(\i*\x+0*\dx+\xi,0)--(\i*\x+0*\dx+\xi,-\ymin);
			\draw(\i*\x+1*\dx+\xi,0)--(\i*\x+1*\dx+\xi,-\ymin);
			\draw(\i*\x+2*\dx+\xi,0)--(\i*\x+2*\dx+\xi,-\ymin);
			\draw(\i*\x+3*\dx+\xi,0)--(\i*\x+3*\dx+\xi,-\ymin);
		}
		\draw(\x+0*\dx+\xi,0.5)--(\x+0*\dx+\xi,\ymax);
		\draw(\x+1*\dx+\xi,0.5)--(\x+1*\dx+\xi,\ymax);
		\draw(\x+2*\dx+\xi,0.5)--(\x+2*\dx+\xi,\ymax);
		\draw(\x+3*\dx+\xi,0.5)--(\x+3*\dx+\xi,\ymax);
		\end{tikzpicture}
	}
	\end{array}
	&=
	\begin{array}{c}
	\includeTikz{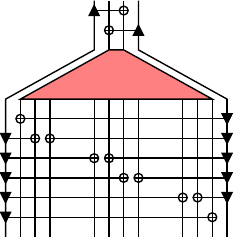}{
		\begin{tikzpicture}[scale=1]
		\def\x{.9}
		\def\dx{.15}
		\def\xi{.25}
		\def\xj{.75}
		\def\ymin{1.4}
		\def\ymax{1}
		\def\dn{.2}
		\filldraw[tenred](\xi+\dx,0)--(2*\x+\xi+2*\dx,0)--(1*\x+\xi+2*\dx,.5)--(1*\x+\xi+1*\dx,.5)--cycle;
		\foreach \i in {0,...,2}{
			\draw(\i*\x+0*\dx+\xi,0)--(\i*\x+0*\dx+\xi,-\ymin);
			\draw(\i*\x+1*\dx+\xi,0)--(\i*\x+1*\dx+\xi,-\ymin);
			\draw(\i*\x+2*\dx+\xi,0)--(\i*\x+2*\dx+\xi,-\ymin);
			\draw(\i*\x+3*\dx+\xi,0)--(\i*\x+3*\dx+\xi,-\ymin);
		}
		\draw (\xi,-\ymin)--(\xi,0)--(\x+\xi,.5)--(\x+\xi,\ymax);
		\draw (\x+\xi+\dx,.5)--(\x+\xi+\dx,\ymax);
		\draw (1*\x+\xi+2*\dx,.5)--(1*\x+\xi+2*\dx,\ymax);
		\draw (2*\x+\xi+3*\dx,-\ymin)--(2*\x+\xi+3*\dx,0)--(1*\x+\xi+3*\dx,.5)--(1*\x+\xi+3*\dx,\ymax);
		\begin{scope}[shift={(0,.7)}]
		\def\n{0};
		\filldraw[black,shift={(1*\x+\xi+3*\dx,\dn*\n)}](-.05,-.05)--(.05,-.05)--(0,.05)--cycle;
		\draw (1*\x+\xi+.7*\dx,\dn*\n)--(1*\x+3*\dx+\xi,\dn*\n);
		\draw (1*\x+\dx+\xi,\dn*\n) circle (.3*\dx);
		\def\n{1};
		\filldraw[black,shift={(\x+\xi+0*\dx,\dn*\n)}](-.05,-.05)--(.05,-.05)--(0,.05)--cycle;
		\draw (1*\x+\xi,\dn*\n)--(1*\x+2.3*\dx+\xi,\dn*\n);
		\draw (1*\x+2*\dx+\xi,\dn*\n) circle (.3*\dx);
		\end{scope}
		\begin{scope}[shift={(0,-.2)}]
		\def\n{0};
		\filldraw[black,shift={(2*\x+\xi+3*\dx,-\dn*\n)},yscale=-1](-.05,-.05)--(.05,-.05)--(0,.05)--cycle;
		\draw (0*\x+\xi+.7*\dx,-\dn*\n)--(2*\x+3*\dx+\xi,-\dn*\n);
		\draw (0*\x+\dx+\xi,-\dn*\n) circle (.3*\dx);
		\begin{scope}
		\def\n{1};
		\filldraw[black,shift={(\xi+0*\dx,-\dn*\n)},yscale=-1](-.05,-.05)--(.05,-.05)--(0,.05)--cycle;
		\draw (\xi,-\dn*\n)--(0*\x+2*\dx+\xi+.3*\dx,-\dn*\n);
		\draw (0*\x+2*\dx+\xi,-\dn*\n) circle (.3*\dx);
		\filldraw[black,shift={(2*\x+\xi+3*\dx,-\dn*\n)},yscale=-1](-.05,-.05)--(.05,-.05)--(0,.05)--cycle;
		\draw (0*\x+\xi+2.7*\dx,-\dn*\n)--(2*\x+3*\dx+\xi,-\dn*\n);
		\draw (0*\x+\xi+3*\dx,-\dn*\n) circle (.3*\dx);
		\end{scope}
		\begin{scope}
		\def\n{2};
		\filldraw[black,shift={(\xi+0*\dx,-\dn*\n)},yscale=-1](-.05,-.05)--(.05,-.05)--(0,.05)--cycle;
		\draw (\xi,-\dn*\n)--(1*\x+0*\dx+\xi+.3*\dx,-\dn*\n);
		\draw (1*\x+0*\dx+\xi,-\dn*\n) circle (.3*\dx);
		\filldraw[black,shift={(2*\x+\xi+3*\dx,-\dn*\n)},yscale=-1](-.05,-.05)--(.05,-.05)--(0,.05)--cycle;
		\draw (1*\x+\xi+0.7*\dx,-\dn*\n)--(2*\x+3*\dx+\xi,-\dn*\n);
		\draw (1*\x+1*\dx+\xi,-\dn*\n) circle (.3*\dx);
		\end{scope}
		\begin{scope}
		\def\n{3};
		\filldraw[black,shift={(\xi+0*\dx,-\dn*\n)},yscale=-1](-.05,-.05)--(.05,-.05)--(0,.05)--cycle;
		\draw (\xi,-\dn*\n)--(1*\x+2*\dx+\xi+.3*\dx,-\dn*\n);
		\draw (1*\x+2*\dx+\xi,-\dn*\n) circle (.3*\dx);
		\filldraw[black,shift={(2*\x+\xi+3*\dx,-\dn*\n)},yscale=-1](-.05,-.05)--(.05,-.05)--(0,.05)--cycle;
		\draw (1*\x+\xi+2.7*\dx,-\dn*\n)--(2*\x+3*\dx+\xi,-\dn*\n);
		\draw (1*\x+3*\dx+\xi,-\dn*\n) circle (.3*\dx);
		\end{scope}
		\begin{scope}
		\def\n{4};
		\filldraw[black,shift={(\xi+0*\dx,-\dn*\n)},yscale=-1](-.05,-.05)--(.05,-.05)--(0,.05)--cycle;
		\draw (\xi,-\dn*\n)--(2*\x+0*\dx+\xi+.3*\dx,-\dn*\n);
		\draw (2*\x+0*\dx+\xi,-\dn*\n) circle (.3*\dx);
		\filldraw[black,shift={(2*\x+\xi+3*\dx,-\dn*\n)},yscale=-1](-.05,-.05)--(.05,-.05)--(0,.05)--cycle;
		\draw (2*\x+\xi+0.7*\dx,-\dn*\n)--(2*\x+3*\dx+\xi,-\dn*\n);
		\draw (2*\x+1*\dx+\xi,-\dn*\n) circle (.3*\dx);
		\end{scope}
		\def\n{5};
		\filldraw[black,shift={(\xi+0*\dx,-\dn*\n)},yscale=-1](-.05,-.05)--(.05,-.05)--(0,.05)--cycle;
		\draw (\xi,-\dn*\n)--(2*\x+\xi+2.3*\dx,-\dn*\n);
		\draw (2*\x+2*\dx+\xi,-\dn*\n) circle (.3*\dx);
		\end{scope}
		\end{tikzpicture}
	}
	\end{array},
	\end{alignat}
\end{subequations}
with the obvious generalization to other bond dimensions.

We remark that although our examples drawn here map $\chi$ dimensional sites to $\chi$ dimensional sites, this can be relaxed. This allows the effective dimension of the sites to be increased as desired.
\clearpage

\section{Generalized $\zN$ CZX model and its gapless boundary theory}
\label{appendix:gczx}

The CZX model was introduced in Ref.~\onlinecite{czxmodel} as a simple exactly solvable representative of the nontrivial $\zt$ SPT phase in two spatial dimensions. In this paper we have considered the larger symmetry group $\zt^3$ of the model for which it is a representative of the $\zt^3$ type-III SPT phase. 
In this appendix we describe a simple generalization of the CZX model to a Hamiltonian with $\zN^3$ symmetry that is a representative of the root type-III $\zN^3$ SPT. We then outline how this fits into the more general setting of $(1+1)$D $\G$-SPT dualities at the edge of a particular $\G\times\coho{2}$-SPT bulk in $(2+1)$D.

\subsection{Definitions}

The model is defined on a two dimensional square lattice with four $\zN$ spins per site. For concreteness we label them counterclockwise as follows
\begin{align}
\begin{array}{c}
\includeTikz{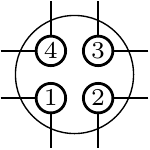}{
	\tikzset{blacksite/.style={fill=black!20!white,draw=black}}
	\tikzset{whitesite/.style={fill=white,draw=black}}
	\tikzset{internalsite/.style={fill=white,draw=black,thick}}
	\begin{tikzpicture}[scale=1.5]
	\def\r{.4};
	\def\ri{\r/4};
	\def\off{.4};
	\def\maxx{0};
	\def\maxy{0};
	\def\len{.5};
	\begin{scope}
	\filldraw[whitesite](0,0)circle (\r);
	\foreach \pma in {-1,1}{
		\foreach \pmb in {-1,1}{
			\draw(\pma*\len,\pmb*\off*\r)--(\pma*\off*\r,\pmb*\off*\r)--(\pma*\off*\r,\pmb*\len);
			\filldraw[internalsite](\pma*\off*\r,\pmb*\off*\r) circle (\ri);
		}
	}
	\node at (-1*\off*\r,+1*\off*\r){\textcolor{black}{\scriptsize$4$}};
	\node at (-1*\off*\r,-1*\off*\r){\textcolor{black}{\scriptsize$1$}};
	\node at (+1*\off*\r,-1*\off*\r){\textcolor{black}{\scriptsize$2$}};
	\node at (+1*\off*\r,+1*\off*\r){\textcolor{black}{\scriptsize$3$}};
	\end{scope}	
	\end{tikzpicture}} 
\end{array}.
\end{align}
Before stating the Hamiltonian, ground-state, and symmetries of the model we establish some definitions:

\begin{align}
P_2 &=\sum_{i=0}^{N-1} \ket{i}^{\otimes 2} \bra{i}^{\otimes 2} \\
X_4 &=\sum_{i=0}^{N-1} \ket{i+1}^{\otimes 4}  \bra{i}^{\otimes 4} \\
\ghz{4} &= \frac{1}{\sqrt{N}} \sum_{i=0}^{N-1} \ket{i}^{\otimes 4} \\
u_X^- &= X_1\otimes X_3 \\
u_X^+ &= X_2 \otimes X_4 \\
u_{CZ} &= CZ_{12}CZ_{23}^\dagger CZ_{34}CZ_{41}^\dagger ,
\end{align}
where $X,CZ$ are defined in Section~\ref{section:class}.

\subsection{Hamiltonian and ground state}

The Hamiltonian is a sum of local terms acting on each plaquette of a square lattice $H=\sum_p h_p$. The terms are given by 
\begin{align}
\label{czxham}
h_p=-\sum_{i=0}^{N-1} X_4^i \otimes P_2 \otimes P_2 \otimes P_2 \otimes P_2,
\end{align}
which act on the lattice as
\begin{align}\label{czxhamlattice}
\begin{array}{c}
\includeTikz{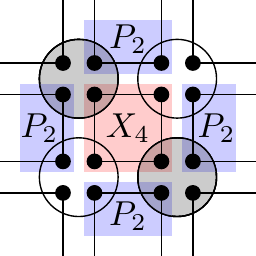}{
	\tikzset{blacksite/.style={fill=black!20!white,draw=black}}
	\tikzset{whitesite/.style={fill=white,draw=black}}
	\tikzset{internalsite/.style={fill=black,draw=none}}
	\begin{tikzpicture}[scale=1]
	\def\r{.4};
	\def\ri{\r/5};
	\def\off{.4};
	\def\maxx{1};
	\def\maxy{1};
	\def\len{.8};
	\fill[red!20!white](0.05,.05) rectangle (.95,.95);
	\begin{scope}
	\clip (-2*\r,-2*\r) rectangle (\maxx+2*\r,\maxy+2*\r);
	\fill[blue!20!white](.05,1.05) rectangle (.95,1.6);\fill[blue!20!white](.05,-.05) rectangle (.95,-.6);
	\fill[blue!20!white](1.05,0.05) rectangle (1.6,.95);\fill[blue!20!white](-.05,0.05) rectangle (-.6,.95);
	\foreach \x in {0,1,...,\maxx}{
		\foreach \y in {0,1,...,\maxy}{
			\pgfmathparse{Mod(\x+\y,2)==0?1:0};
			\ifnum\pgfmathresult>0
			\else
			\filldraw[blacksite,shift={(\x,\y)}](0,0)circle (\r);
			\fi
			\begin{scope}
			\clip (0,1) circle (\r);
			\fill[red!20!white!80!black](.05,.05) rectangle (.95,.95);
			\fill[blue!20!white!80!black](.05,1.05) rectangle (.95,1.6);
			\fill[blue!20!white!80!black](-.05,0.05) rectangle (-.6,.95);
			\end{scope}
			\begin{scope}
			\clip (1,0) circle (\r);
			\fill[red!20!white!80!black](.05,.05) rectangle (.95,.95);
			\fill[blue!20!white!80!black](.05,-.05) rectangle (.95,-.6);
			\fill[blue!20!white!80!black](1.05,0.05) rectangle (1.6,.95);
			\end{scope}
			\pgfmathparse{Mod(\x+\y,2)==0?1:0};
			\ifnum\pgfmathresult>0
			\else
			\draw[draw=black,shift={(\x,\y)}](0,0)circle (\r);
			\fi
		}
	}
	\foreach \x in {0,1,...,\maxx}{
		\foreach \y in {0,1,...,\maxy}{
			\draw[draw=black,shift={(\x,\y)}](0,0)circle (\r);
			\foreach \pma in {-1,1}{
				\foreach \pmb in {-1,1}{
					\filldraw[internalsite,shift={(\x,\y)}](\pma*\off*\r,\pmb*\off*\r) circle (\ri);
					\draw[shift={(\x,\y)}](\pma*\len,\pmb*\off*\r)--(\pma*\off*\r,\pmb*\off*\r)--(\pma*\off*\r,\pmb*\len);
				}
			}
		}
	}
	\end{scope}
	\node[anchor=center] at (.5,.5) {$X_4$};
	\node[anchor=center] at (.5,-.4) {$P_2$};\node[anchor=center] at (.5,1.4) {$P_2$};
	\node[anchor=center] at (-.4,.5) {$P_2$};\node[anchor=center] at (1.4,.5) {$P_2$};
	\end{tikzpicture}}
\end{array} .
\end{align}
The ground state is unique for closed boundary conditions and is given by a tensor product of the state $\ghz{4}$ on the four spins around each plaquette
\begin{align}
\ket{\Psi_{\text{GS}}}=\bigotimes_p \ket{GHZ_4}.
\end{align}
Note that this ground state is not a product state with respect the locality structure we have chosen by our grouping of spins into sites (if sites were instead defined to group the spins around each plaquette it would be a product state).

\subsection{Symmetry}

To describe the $\zN^3$ symmetry of the Hamiltonian in Eqn.~\ref{czxham} we first bipartition the lattice into black ($b$) and white ($w$) sites, as indicated in Fig.~\ref{fig:czxlattice}. The generators are then given by
\begin{align}
U_{\rx{}} &= \bigotimes_{b} u_X^- \bigotimes_{w} u_X^+
\\
U_{\bx{}} &= \bigotimes_{b} u_X^+ \bigotimes_{w} u_X^-
\\
U_{CZ} &= \bigotimes_{b} u_{CZ}^\dagger \bigotimes_{w} u_{CZ} .
\end{align}
One can verify that each of these operators is of order $N$ and that they mutually commute. Furthermore each local Hamiltonian term commutes with all symmetries and they leave the ground state invariant. Note the $U_{CZ}$ symmetry is an on-site symmetry for our definition of site but would not be if sites were instead defined by grouping the spins around each plaquette.

\subsection{Boundary theory}

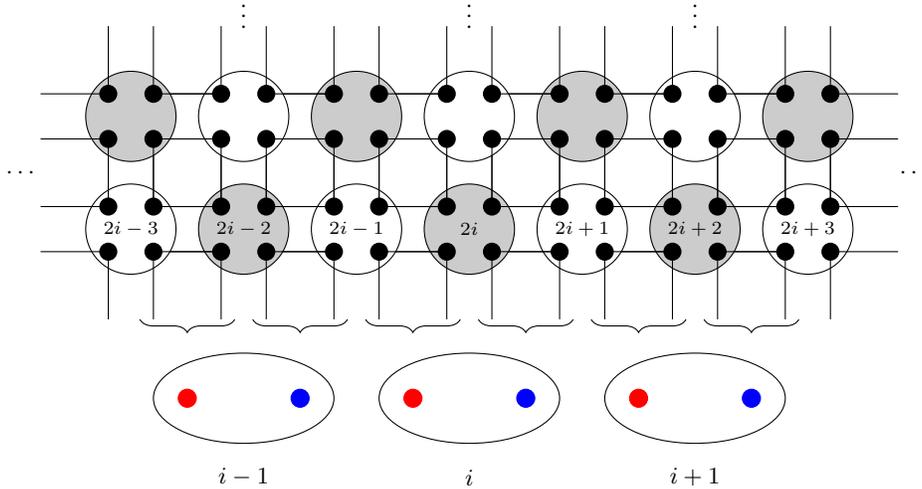
\begin{figure}[h!]
	\includeTikz{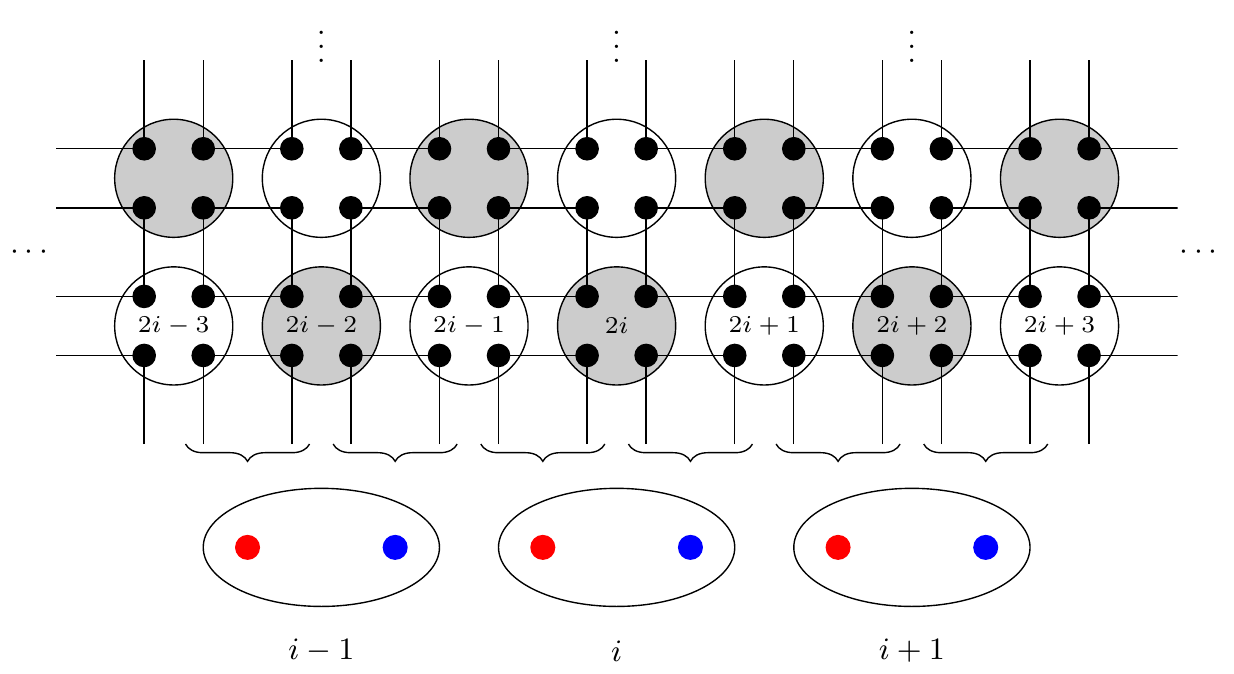}{
		\tikzset{blacksite/.style={fill=black!20!white,draw=black}}
		\tikzset{whitesite/.style={fill=white,draw=black}}
		\tikzset{internalsite/.style={fill=black,draw=none}}
		\begin{tikzpicture}[rotate=90,scale=1.5]
		\def\r{.4};
		\def\ri{\r/5};
		\def\off{.5};
		\def\maxx{1};
		\def\maxy{6};
		\def\len{.8};
		\begin{scope}
		\clip (-2*\r,-2*\r) rectangle (\maxx+2*\r,\maxy+2*\r);
		\foreach \x in {0,1,...,\maxx}{
			\foreach \y in {0,1,...,\maxy}{
				\pgfmathparse{Mod(\x+\y,2)==0?1:0};
				\ifnum\pgfmathresult>0
				\filldraw[whitesite,shift={(\x,\y)}](0,0)circle (\r);
				\else
				\filldraw[blacksite,shift={(\x,\y)}](0,0)circle (\r);
				\fi
				\foreach \pma in {-1,1}{
					\foreach \pmb in {-1,1}{
						\filldraw[internalsite,shift={(\x,\y)}](\pma*\off*\r,\pmb*\off*\r) circle (\ri);
						\draw[shift={(\x,\y)}](\pma*\len,\pmb*\off*\r)--(\pma*\off*\r,\pmb*\off*\r)--(\pma*\off*\r,\pmb*\len);
					}
				}
			}
		}
		\end{scope}
		\node at (\maxx+1.2*\len,\maxy/6) {$\vdots$};
		\node at (\maxx+1.2*\len,\maxy/2) {$\vdots$};
		\node at (\maxx+1.2*\len,5*\maxy/6) {$\vdots$};
		\node at (\maxx/2,-1.2*\len) {$\cdots$};
		\node at (\maxx/2,\maxy+1.2*\len) {$\cdots$};
		\foreach \y in {0,1,...,\maxy}{
			\ifnum\y>0
			\draw [decorate,decoration={brace,amplitude=5pt},shift={(-\len,\y-.5)}](0,-0.42) -- (0,.42);
			\pgfmathparse{Mod(\y,2)==0?1:0};
			\ifnum\pgfmathresult>0
			\filldraw[red,shift={(0,\y)}](-1.5,-.5) circle (\ri);
			\draw[shift={(0,\y-1)}] (-1.5,0) ellipse (.4 and {\len});
			\else
			\filldraw[blue,shift={(0,\y)}](-1.5,-.5) circle (\ri);
			\fi
			\fi}
		\node at (-2.2,\maxy-1) {$i-1$};	
		\node at (-2.2,\maxy/2) {$i$};	
		\node at (-2.2,1) {$i+1$};	
		\node at (0,6) {\scriptsize$2i-3$};
		\node at (0,5) {\scriptsize$2i-2$};
		\node at (0,4) {\scriptsize$2i-1$};
		\node at (0,3) {\scriptsize$2i$};
		\node at (0,2) {\scriptsize$2i+1$};
		\node at (0,1) {\scriptsize$2i+2$};
		\node at (0,0) {\scriptsize$2i+3$};
		\end{tikzpicture}}
	\caption{Identification of the edge degrees of freedom.}
	\label{fig:czxlattice}
\end{figure}

In the presence of an open boundary 
the bulk Hamiltonian is extensively degenerate as it only projects pairs of spins along the edge into the support subspace of $P_2$. 
We identify effective $\zN$ edge spins with the $N$ states in this subspace via the projector $\sum\limits_i \ket{i}\bra{ii}$. This identification is indicated by \rotatebox{90}{$\{$} in Fig.~\ref{fig:czxlattice}. An edge site is formed by a pair of these spins, as shown in Fig.~\ref{fig:czxlattice}. 
 This identification provides an exact mapping from bulk operators to the boundary. The symmetry acts on the edge as follows
\begin{align}
U_{\rx{}} &\mapsto \bigotimes_j \rx{j}
\\
U_{\bx{}} &\mapsto \bigotimes_j \bx{j}
\\
U_{CZ} &\mapsto \mathcal{C} = \begin{array}{c}\includeTikz{CZcircuit}{}\end{array} .
\end{align}
Due to the grouping of edge spins into sites only the subgroup generated by $U_{\rx{}}$ and $U_{\bx{}}$ acts on-site.

The bulk to boundary mapping can be used to find the edge action of certain operators that leave no residual effect on the bulk of the ground state. In particular
\begin{align}
(Z_1)^b_{2i} &\mapsto \rz{i}
\\
(Z_2)^w_{2i-1} &\mapsto \rz{i}
\\
(Z_1)^w_{2i+1} &\mapsto \bz{i}
\\
(Z_2)^b_{2i} &\mapsto \bz{i}
\\
( X_2)^w_{2i-1}(X_1)^b_{2i} &\mapsto \rx{i}
\\
(X_2)^b_{2i} (X_1 )^w_{2i+1} &\mapsto \bx{i},
\end{align}
where the numbering is indicated in Fig.~\ref{fig:czxlattice}. 
We find an effective edge Hamiltonian by considering symmetric perturbations in the bulk with minimal support. 
\begin{align}
(Z_1^\dagger Z_3)^b_{2i} (Z_2 Z_4^\dagger)^w_{2i+1} &\mapsto \rzd{i} \rz{i+1}
\\
(Z_1 Z_3^\dagger)^w_{2i+1} (Z_2^\dagger Z_4)^b_{2i+2} &\mapsto \bz{i}\bzd{i+1}
\\
( X_2)^w_{2i-1} (X_1)^b_{2i} + ( Z_1  X_2 Z_3^\dagger)^w_{2i-1}  (X_1 Z_2^\dagger Z_4)^b_{2i} &\mapsto \rx{i} +  \bz{i-1}\rx{i}\bzd{i}
\\
(X_2)^b_{2i} (X_1)^w _{2i+1} + (Z_1^\dagger X_2 Z_3)^b_{2i} (X_1 Z_2 Z_4^\dagger)^w_{2i+1} &\mapsto \bx{i} + \rzd{i}\bx{i}\rz{i+1}.
\end{align}
The edge Hamiltonian is given by 
\begin{align}
    H_{\text{Edge}}=-&\sum_i \sum_{k=0}^{N-1} c_k \sum_{j=0}^{N-1} (\rzd{i}^{jk}\bx{i}^{j}\rz{i+1}^{jk}+\bz{i-1}^{jk}\rx{i}^{j}\bzd{i}^{jk}) 
    -  \sum_i \sum_{k=0}^{N-1}  b_k (\rzd{i}^k\rz{i+1}^k+\bz{i}^k\bzd{i+1}^k).
\end{align}
where $b_k=b_{N-k}$. The Hamiltonian is fully symmetric under $U_{\rx{}}$ and $U_{\bx{}}$ while the parameters transform as follows under $\mathcal{C}$
\begin{align}
 c_k &\mapsto c_{k-1},
\\
 b_k &\mapsto b_k.
\end{align} 
When $c_k$ is the only nonzero parameter the Hamiltonian is in the $[k]\in\coho{2}$ SPT phase, while for $b_k=b_{N-k}$ the only nonzero parameters it describes a symmetry broken phase. Hence the $\mathcal{C}$ operator cycles the SPT phases $[k]\mapsto [k+1]$ and the Hamiltonian is fully symmetric when all $c_k=c_0$. This may correspond to an SPT critical point or a symmetry breaking point depending upon the relative strength of the $b_k$ parameters.

\subsection{General $(1+1)$D $\G$ SPT duality at the edge of a $(2+1)$D $\G\times\coho{2}$ SPT}

The above construction for $\zN^3$ is a specific instance of a general connection between duality of $(1+1)$D edge $\G$ SPT phases and a $(2+1)$D bulk $\G\times\coho{2}$ SPT phase. This connection may be of independent interest. The action of the bulk $\coho{2}$ symmetry can be though of as pumping $\G$ SPTs onto the edge. 

Similarly to the case above, the Hilbert space of each spin is given by $\mathbb{C}[\G]$ and 4 spins are grouped per site of a square lattice. $R_g$ denotes the right regular representation, we fix a choice of representative for a set of generators of $\coho{2} \cong \prod_k \mathbb{Z}_{N_k}$ (their products fix all other representatives) and
\begin{align}
R_g^{\otimes 4} P_4 &:=\sum_{g\in\G} \ket{hg^{-1}}^{\otimes 4}  \bra{h}^{\otimes 4} \\
C\omega_{12} &:= \sum_{g_0,g_1} \omega({g_0g_1^{-1},g_1}) \ket{g_0,g_1}\bra{g_0,g_1}
\\
u_{\omega} &:= C\omega_{12}C\omega_{23} C\omega_{34}C\omega_{41}
\end{align}
for $[\omega]\in\coho{2}$. 

The local Hamiltonian terms are given by 
\begin{align}
h_p=-\sum_{g\in\G}  {R_g^{\otimes 4}} P_4 \otimes P_2 \otimes P_2 \otimes P_2 \otimes P_2
\end{align}
acting on the square lattice similarly to the term in Eqn.~\ref{czxhamlattice}.
The ground state is again given by
\begin{align}
\ket{\Psi_{\text{GS}}}=\bigotimes_p \ket{GHZ_4}.
\end{align}

The global on-site symmetry is generated by 
\begin{align}
U_{g} &= \bigotimes R_g^{\otimes 4}
\\
U_{\omega} &= \bigotimes_{b} u_{\omega}^\dagger \bigotimes_{w} u_{\omega} 
\end{align}
which can be seen to mutually commute and also commute with $h_p$. These symmetries also leave the ground state invariant. 

As above, the effective edge spins are identified with the ground state subspace of plaquettes crossing the boundary, via the projector $\sum\limits_g \ket{g}\bra{gg}$. 
The action of the symmetry on the edge is given by 

\begin{align}
U_{g} &\mapsto  \bigotimes_i R_g
\\
U_{\omega} &\mapsto \prod_i C\omega_{2i,2i+1} C\omega_{2i-1,2i}^\dagger.
\end{align}
This forms a matrix product operator representation of $\G\times\coho{2}$ with 3-cocycle 
\begin{align}\alpha((g_0,\omega_0),(g_1,\omega_1),(g_2,\omega_2))=\omega_2(g_0,g_1).\end{align}
The edge action of $U_\omega$ maps a $\G$ SPT phase $[\beta]$ to $[\beta+\omega]$. This can be seen by examining the effect of $U_\omega$ on a fixed point local Hamiltonian such as the $\G$-paramagnet
\begin{align}
H=-\sum_{v} \sum_g (R_g)_v.
\end{align}
 Alternatively, note the edge action of $U_\omega$ restricted to an open chain is an MPO with two dangling virtual indices associated to its boundaries. Denote this MPO $M_\omega$.  $M_\omega$ obeys the following commutation rules $R_g^{\otimes L}M_\omega R_g^{\dagger \otimes L} = V_g M_\omega V_g^\dagger$. Here $V_g$ is a projective representation of $\G$, with cocycle $\omega$, given by 
\begin{align}
V_g= \sum_h  \omega(h,g) \ket{hg}\bra{h},
\end{align}
which acts on one dangling virtual bond of the MPO.
Hence applying $M_\omega$ to a unique symmetric ground state, such as $\ket{+}^{\otimes N}$, maps it to a state in the SPT phase $[\omega]$.

\end{document}